\documentclass[acmsmall]{acmart}
\usepackage{amsmath}
\usepackage{amssymb}
\usepackage{amsthm}
\usepackage{booktabs}
\usepackage{graphicx}
\usepackage{url}
\usepackage[english]{babel}
\usepackage{xargs}
\usepackage[colorinlistoftodos,prependcaption,textsize=normalsize]{todonotes}
\usepackage{hyperref}
\usepackage{makecell}
\usepackage{tabularx}
\usepackage{csquotes}
\colorlet{linecol}{black!75}
\usepackage[export]{adjustbox}
\usepackage[hyperfirst=false]{glossaries}
\usepackage{supertabular}
\usepackage{adjustbox}
\usepackage{makecell}
\usepackage{tabularx}
\usepackage{ltablex}
\usepackage[skins]{tcolorbox}
\usepackage{subcaption}

\def\BibTeX{{\rm B\kern-.05em{\sc i\kern-.025em b}\kern-.08emT\kern-.1667em\lower.7ex\hbox{E}\kern-.125emX}}

\newacronym{qos}{QoS}{Quality of Service}
\newacronym{sla}{SLA}{Service Level Agreement}
\newacronym{slo}{SL}{Service Level Objective}
\newacronym{iaas}{IaaS}{Infrastructure as a Service}
\newacronym{paas}{PaaS}{Platform as a Service}
\newacronym{saas}{SaaS}{Software as a Service}
\newacronym{faas}{FaaS}{Function as a Service}
\newacronym{baas}{BaaS}{Backend as a Service}
\newacronym{dag}{DAG}{Directed Acyclic Graph}
\newacronym{bpmn}{BPMN}{Business Process Model and Notation}
\newacronym{yawl}{YAWL}{Yet Another Workflow Language}
\newacronym{cwl}{CWL}{Common Workflow Language}
\newacronym{pn}{PN}{Petri net}
\newacronym{dpia}{DPIA}{Data Protection Impact Assessment}
\newacronym{nfr}{NFR}{non-functional requirement}
\newacronym{tfidf}{TF-IDF}{Term Frequency -- Inverse Data Frequency}
\newacronym{vm}{VM}{Virtual Machine}

\newcommand{\vcutS}{\vspace*{-0cm}}
\newcommand{\vcutM}{\vspace*{-0cm}}
\newcommand{\vcutL}{\vspace*{-0cm}}

\definecolor{KamPurple}{HTML}{907C97}
\newcommandx{\add}[1]{\todo[inline,linecolor=black,backgroundcolor=orange!25]{Add: #1}}
\newcommandx{\rewrite}[1]{\todo[inline,linecolor=black,backgroundcolor=blue!25]{Rewrite:
		#1}}
\newcommandx{\proofread}{\todo[inline,linecolor=black,backgroundcolor=green!25]{}}
\newcommandx{\findplace}[1]{\todo[inline,linecolor=black,backgroundcolor=purple!25]{Need a place for #1}}
\newcommandx{\shouldhave}[1]{\todo[inline,linecolor=black,backgroundcolor=yellow!25]{Should have: #1}}
\newcommandx{\idea}[1]{\todo[inline,linecolor=black,backgroundcolor=lime!25]{Idea: #1}}

\newcounter{queryid}
\newcommand\executedqueries[2]{
\refstepcounter{queryid}\label{#1}
\begin{tcolorbox}[colback=KamPurple!15, colframe=KamPurple!90!black,enhanced,sharp corners,boxsep=-1mm]%
{\small \textbf{Query~\arabic{queryid}}: \texttt{#2}}
\end{tcolorbox}
}
\newcommand{\refquery}[1]{\textbf{Query~\ref{#1}}}

\newcounter{observationid}
\newcommand{\observation}[2]{\refstepcounter{observationid}\label{#1}
\ifnum\value{observationid}=1 %
    \item[Observation-\arabic{observationid} (O-\arabic{observationid}):] #2
  \else 
    \item[O-\arabic{observationid}:] #2 \noindent
\fi

}

\title{A Survey and Annotated Bibliography of Workflow Scheduling in Computing Infrastructures: Community, Keyword, and Article Reviews -- Extended Technical Report}

\author{Laurens Versluis}
\email{l.f.d.versluis@vu.nl}
\affiliation{%
	\institution{Vrije Universiteit Amsterdam}
}

\author{Alexandru Iosup}
\email{a.iosup@vu.nl}
\affiliation{%
	\institution{Vrije Universiteit Amsterdam}
}

\copyrightyear{2019}
\acmYear{2019}
\setcopyright{acmlicensed}

\begin{document}

\begin{abstract}
	
	\textit{Workflows} are prevalent in today's computing infrastructures.
The workflow model support various different domains, from machine learning to finance and from astronomy to chemistry.
Different \gls{qos} requirements and other desires of both users and providers makes \textit{workflow scheduling} a tough problem, especially since resource providers need to be as efficient as possible with their resources to be competitive.

To a newcomer or even an experienced researcher, sifting through the vast amount of articles can be a daunting task.
Questions regarding the difference techniques, policies, emerging areas, and opportunities arise.

Surveys are an excellent way to cover these questions, yet surveys rarely publish their tools and data on which it is based.
Moreover, the communities that are behind these articles are rarely studied.
We attempt to address these shortcomings in this work.

We focus on four areas within workflow scheduling: 1) the workflow formalism, 2) workflow allocation, 3) resource provisioning, and 4) applications and services.
Each part features one or more taxonomies, a view of the community, important and emerging keywords, and directions for future work.
We introduce and make open-source an instrument we used to combine and store article meta-data.
Using this meta-data, we 1) obtain important keywords overall and per year, per community, 2) identify keywords growing in importance, 3) get insight into the structure and relations within each community, and 4) perform a systematic literature survey per part to validate and complement our taxonomies.

\end{abstract}

\maketitle

\section{Introduction}

\begin{figure}[thb]
	\includegraphics[max width=\linewidth, keepaspectratio]{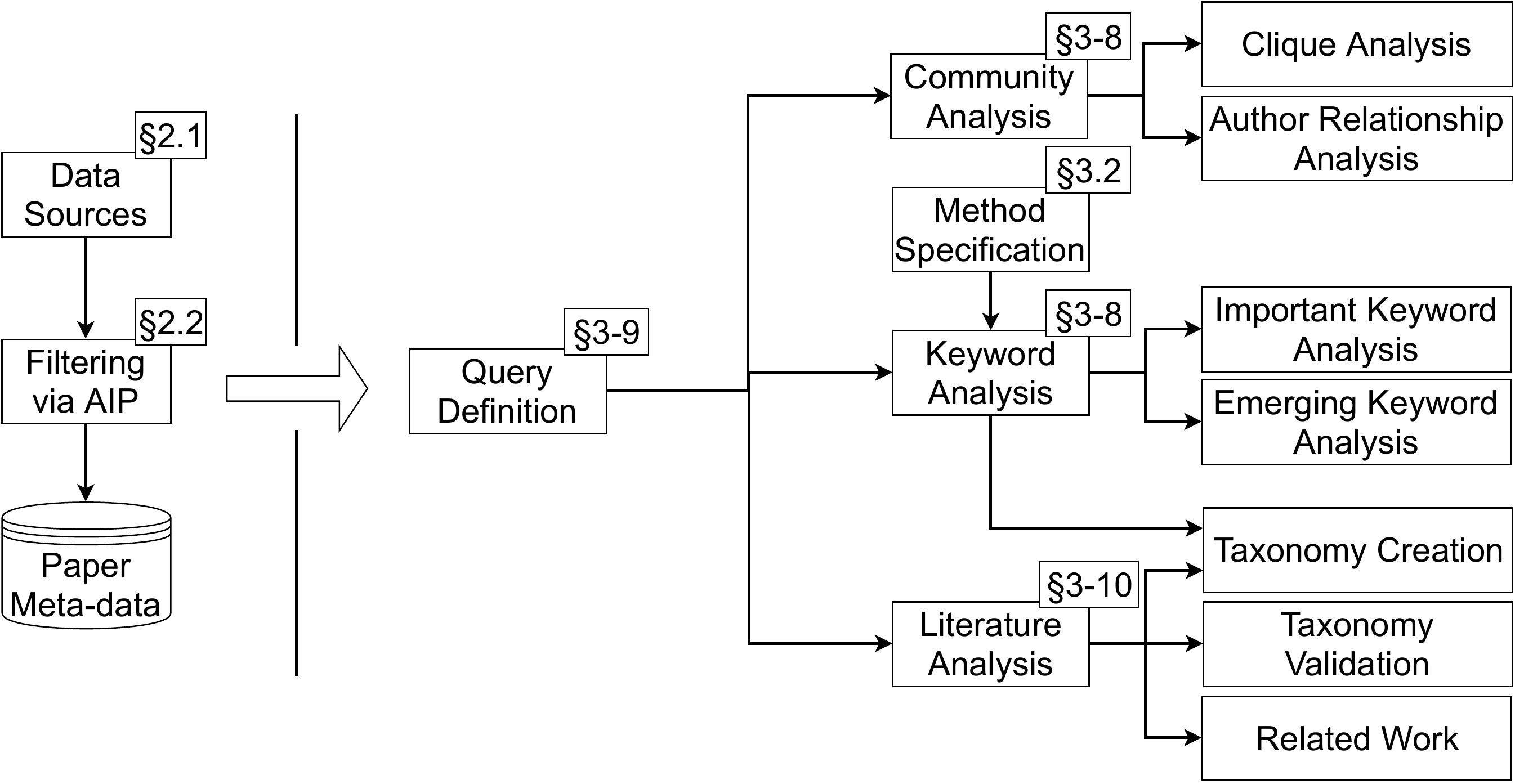}
	\vcutL
	\caption{The structure of and process applied in this survey. Left: obtaining a database of paper meta-data. Right: the usage of this meta-data for four type of analyses.}
	\label{fig:overview-process-survey}
	\vcutL
\end{figure}

Datacenters and cloud providers are increasingly becoming the go-to point for leasing additional computing power.
Both industry and academia are embracing this new paradigm of computation on demand, ranging from financial institutes \cite{ron2011cloud} to bioinformatics research communities \cite{DBLP:conf/ccgrid/BlytheJDGVMK05, stein2010case}.

Nowadays, many applications submitted to these resources are \textit{workflows}.
These workflows can be modeled differently.
The simplest model is the that of Coffman and Graham~\cite{DBLP:journals/acta/CoffmanG72}.
In this model, a workflow is represented as a Directed Acyclic Graph (DAG) where each vertex represents a task (i.e., a unit of work) and an edge a computation/data constraint.
Other formalisms such as BPMN allow for cycles and human-in-the-loop elements.
These workflows origin from various domains, ranging from bioinformatics to finance and from geology to astronomy.
Datacenters, clusters, and clouds receive tens to millions of such workflows per hour~\cite{2019arXiv190607471V}.
A key component in executing these workflows efficiently is the \textit{scheduler}~\cite{DBLP:conf/sc/AndreadisVMI18}.
Scheduling these workflows to make efficient use of the available resources is a challenging task, demonstrated by the sheer amount of proposed scheduling systems and policies. \glsreset{qos}
Moreover, nowadays, the resource providers must adhere to different \gls{qos} requirements that can differ per workflow.

Performing well in workflow scheduling requires keeping up with the most recent advances in workflow scheduling.
Especially with the recent developments in edge and serverless computing~\cite{shi2016edge, van2018serverless}, it is important to remain an up-to-date overview.
The accelerating growth of the number of workflow scheduling articles published makes it a daunting task to get insights into the attempts made by research to solve these complex challenges.
Semantic Scholar also underlines this challenge: "The rate of scientific publication is increasing every year, with more than 3 million papers published across 42,500 journals in 2018 alone. This unprecedented flow of information makes staying up-to-date with the scientific literature an increasingly pressing challenge for scholars."\footnote{\url{https://pages.semanticscholar.org/about-us}}.

Questions arise such as Which different techniques are being used nowadays to schedule workflow or resources?
What structures do these schedulers have?
What is currently important in the community? Which areas and topics are emerging? and which opportunities for research are there?

Surveys are an excellent way to get answers to such questions.
They provide an overview of the current field by using taxonomies and other means such as tables to present and compare approaches, enumerate emerging topics and list challenges and possible directions for future work.
Yet, survey articles rarely publish the tools and data on which they are based.
This data is important to reproduce the survey's findings, verify the completeness, and use as a base for extensions.

In this work, we address these issues by following the process visualized in Figure~\ref{fig:overview-process-survey}.
We start by introducing three datasets of article meta-data which we parse to filter and unify using an instrument we developed.

We demonstrate parsing and unifying these datasets is non-trivial due to issues encountered in both the data and their formats.
Next, using the filtered article meta-data, we perform several different analyses to obtain insights into different areas of workflow scheduling.

The instrument, database, and other tools used in this work are offered as open-source artifacts for the community to use.
The database is suitable for similar studies on different topics.
The instrument and tools are extensible and can be added to to include more information sources and capture more properties.

Overall, we make the following five main contributions in this work:
\begin{enumerate}
	\item We assemble a unique dataset of relevant articles and create a specialized instrument to process it~(Section~\ref{sct:survey-community-and-tool}).
	Our dataset combines data from three major curators 
	into a comprehensive dataset for the computer systems community. 
	We develop AIP, an open-source software
	for combining, filtering, and analyzing bibliographic datasets.
	
    \item We perform a novel survey of the workflow scheduling community~(Section~\ref{sct:analysis-paper-meta-data}).
    We propose a method for analyzing article meta-data, resulting in important insights into the structure of the workflow scheduling community and its emerging keywords. 
    
	\item Using results from Section~\ref{sct:analysis-paper-meta-data}, we propose a taxonomy that focuses on four areas in workflow scheduling~(Sections~\ref{sct:taxonomy}-\ref{sct:applications-and-services}).
	Our work proposes novel taxonomic aspects and significantly extends state-of-the-art taxonomies~\cite{DBLP:journals/sigmod/YuB05, DBLP:journals/csur/Kardani-Moghaddam19, DBLP:journals/fgcs/SmanchatV15} on workflow scheduling in areas such as 
	formalisms for workflow specification;
	workflow allocation policies, strategies, and structures; 
	elasticity and serverless in resource provisioning;
	and various types of resources used in current applications and services.
	For each area, we also make observations about community, trending keywords, and emerging trends in the timespan 2009-2018.
	
	\item We validate our taxonomies by mapping well-cited and recently introduced workflow allocation and resource provisioning policies using a systematic survey (Section~\ref{sct:workflow-allocation-policies}). We map several elements of allocation and resource provisioning policies to our taxonomies to validate our taxonomies contain these elements.
	\item We provide an annotated bibliography in this work, summarizing each article.
\end{enumerate}

\section{Data Sources and Parsing Instrument}
\label{sct:survey-community-and-tool}

The number of peer-reviewed articles published is growing rapidly.
Searching and in particular systematically searching through this number of articles is therefore also growing in importance.
Current information sources do not cover the spectrum of the systems community entirely.
For example, DBLP~\cite{dblpSnapshotNovember2019} -- which specifically focuses on computer science articles -- lacks certain venues and does not record article abstracts.
Other datasets such as Semantic Scholar and AMiner have similar and other limitations.
Moreover, these datasets also overlap, yet contain important information the others do not offer; they are disjoint.
Our approach is to parse each dataset and filter and unify the information provided.
Using this information, we then can perform keyword analysis, investigate author-paper relationships, perform systematic surveys, etc.

Identifying keywords is regarded as a useful way to obtain important topics within text~\cite{onan2016ensemble}, and by extension, communities.
By investigating author-paper relationships, we can map the structure of a community. 
Healthy communities should demonstrate various collaborations, rather than isolated islands of authors.
Systematic surveys allow for reproducibility of the survey, which is a topic growing in importance~\cite{6171147}.
To perform such comprehensive analyses and surveys, we require a vast database containing paper meta-data to query.

In this section we describe which article information sources we use and introduce an instrument named Article Information Parser (AIP).
AIP is developed to parse these sources, extract paper meta-data of articles of interest published in top-tier venues and store this information in a database.

\subsection{Data Sources}

\begin{table}[]
	\caption{An overview of article data sources used.}
	\vcutM
	\label{tbl:article-datasources}
	\begin{adjustbox}{max width=\columnwidth}
		\begin{tabular}{@{}lllrrlccc@{}}
			\toprule
			Dataset          & Version & Topics &  \#Articles  & \#Systems Articles &  Title & Abstract & \#Citations \\ \midrule
			DBLP             & 2019-11-01 & Comp. Sci. & 4,673,333   & 110,926 & \checkmark      &                    &                     \\
			Semantic Scholar & 2019-11-01 & General & 38,219,709  & 93,314 & \checkmark      & \checkmark        & \checkmark           \\
			AMiner           & OAG v2 & General & 172,209,563  & 9,463 & \checkmark      & \checkmark        & \checkmark           \\ \midrule
			Our dataset           & 2019-11-20 & Systems & 117,058 & 117,058 & \checkmark      & \checkmark        & \checkmark           \\ \bottomrule
		\end{tabular}
	\end{adjustbox}
	\vcutL
\end{table}

To analyze articles, we use three data sources: DBLP~\cite{dblpSnapshotNovember2019}, Semantic Scholar\cite{ammar:18}, and AMiner\cite{DBLP:conf/kdd/TangZYLZS08}, which we filter and store in a SQLite database.
DBLP is a well-known European archive that focuses on computer science and features all the top-level venues (journals and conferences).
Semantic Scholar is an American project created by the Allen Institute for AI.
The project aims to analyze and extract important data from scientific publications.
AMiner is an Asian project that aims to provide a knowledge graph for mining academic social networks.
Both AMiner and Semantic Scholar have incorporated Microsoft's Academic Graph (MAG) in their datasets nowadays.
As can be observed from Table~\ref{tbl:article-datasources}, the three data sources feature different elements and sizes. 
The corpus of Semantic Scholar that we use didn't have MAG included yet when we created the database, yet AMiner did, so we assume we have MAG included as well.
Using only one or two of these datasets would lead to smaller database, possibly missing relevant work.

As we are interesting in the computer systems domain, we only extract paper information from these data sources when we recognize the venue.
In total, by unifying all datasets, we extract paper information of 117,058 articles accepted at top-tier venues in the systems communities.
For each article, we store the title, abstract, authors, venue, year, volume, DOI, and number of citations.
Next to article meta-data we store co-author relationship data.
We only use author data from DBLP as they make a significant effort to differentiate between homonym persons, which is a hard problem~\cite{ley2009dblp}.
We do not use author information from other data sources as these only include the names of authors, which we are unable to map accurately to DBLP or another data source.
Semantic Scholar and AMiner both contain citation data; if they report different numbers, we use the highest citation count.
Due to discrepancies between the data sources, our database has citation information on 94,98\% of the articles and author information on 83,62\% of them.

Parsing to filter and unifying these data sources is non-trivial.
These data sources all have their own format and represent data differently.
For example, one difference is how the venue is provided. DBLP may provide an abbreviation whereas Semantic Scholar might give the venue name in full.
We also find discrepancies in titles; DBLP ends all titles with a dot whereas Semantic Scholar and AMiner do not.
Finally, we find several cases where text is encoded.

To address these issues, we create an instrument that parses and filters all these data sources.

\subsection{AIP: An Instrument for Parsing Article-Information}

To unify all data sources and filter articles of interest, we developed an instrument called Article Information Parser (AIP).
AIP tackles several non-trivial challenges in unifying these datasets:
\begin{enumerate}
	\item Data discrepancies between sources. For example, titles in DBLP end with a dot, whereas they do not in the Semantic Scholar and AMiner corpuses, causing exact matching to fail.
	\item Titles and abstracts may contain encoded characters leading to mismatching articles that are in fact the same.
	\item Despite all data sources having a format specified, we encountered several instances where the format specified is not adhered to, or the data is malformed.
	\item Venue strings being different among these sources. Some sources use an abbreviation, some use a BibTeX string, etc. AIP maps all these occurrences to the same abbreviation.
	\item Complementing existing entries. For example, DBLP does not offer abstracts whilst Semantic Scholar and AMiner do.
\end{enumerate}

Cases of malformed data and format issues have been reported to the respective providers.
We will cover three main parts of AIP: the data source parsers, the database manager, and the venue mapper.
The venue mapper is developed as a separate module as it can be used in other projects.
Both AIP and the venue mapper are offered as open-source artifacts.

\subsubsection{Data Source Parsers}
AIP features for each data source a specific parser.
This is required as each data source has its own data and file format.
Moreover, we found several issues with different formats.
These issues include:
\begin{enumerate}
	\item Keys specified in the format not being present in the actual data,
	\item Keys in the data that are not specified in the format,
	\item Unspecified elements such as nested JSON objects that contain important data,
	\item Malformed JSON entries,
	\item Inconsistent data representation. For example, year being expressed as \enquote{1990} or \enquote{'90} and the venue either being the abbreviation or written in full.
\end{enumerate}

AIP uses a tailored strategy per data source.
For example, DBLP uses XML whereas Semantic Scholar and AMiner offer data in JSON format.
Furthermore, by keeping the parser separate from the internals of AIP, new data sources can be easily added in the future.

\subsubsection{Data Manager}
The database manager (DM) is responsible for creating the database, managing updates to its structure, and handling entries from data sources.

When an entry is provided, the DM first attempts to match the entry to an already existing entry in the database.
First, unique identifiers such as the DOI are tried. 
If this doesn't provide any matches, inexact matches based on the title are attempted.
Finally, if no match can be found, the entry is inserted as a new article.

Once the article has been matched, the author-article information is inserted into the database.
As discussed previously, only author data from DBLP is used.

\subsubsection{Venue Mapper}
A crucial element in determining if an entry is of interest is by determining the venue of the article.
As the venue is described differently per data source, we developed a module that matches a raw venue string to a venue abbreviation through string matching and regex.
The reason for developing it as a separate module is that we perceive this module also being useful in other projects.

If the venue can be determined, we insert the entry into the database.
By unifying all venue strings to the same abbreviation, searching for articles from a specific venue also becomes easier.
\section{Analysis of Article Meta-data}
\label{sct:analysis-paper-meta-data}

In this article, using the database introduced in Section~\ref{sct:survey-community-and-tool}, we make use of four different analyses:
\begin{enumerate}
    \item Community analysis, where we inspect the structure of the community and its characteristics. This can be useful for community managers and organizers.
	\item Overall trend analysis, where we obtain the top-\textit{n} most important keywords.
	\item Analysis of trends over time, where we obtain the top-\textit{n} most important keywords per year.
	\item Emerging keyword analysis, where we attempt to identify \textit{new} and \textit{rising} keywords.
\end{enumerate}

In this section, we demonstrate each analysis separately using articles on workflow scheduling, the main theme of this article, published in timespan 2009-2018. We focus on this decade as not all articles published in 2019 have been recorded yet by the data sources we use.

\subsection{Analysis of the Workflow Scheduling Community}

\begin{figure}
	\centering
	\begin{subfigure}[t]{0.49\textwidth}
		\includegraphics[width=\textwidth]{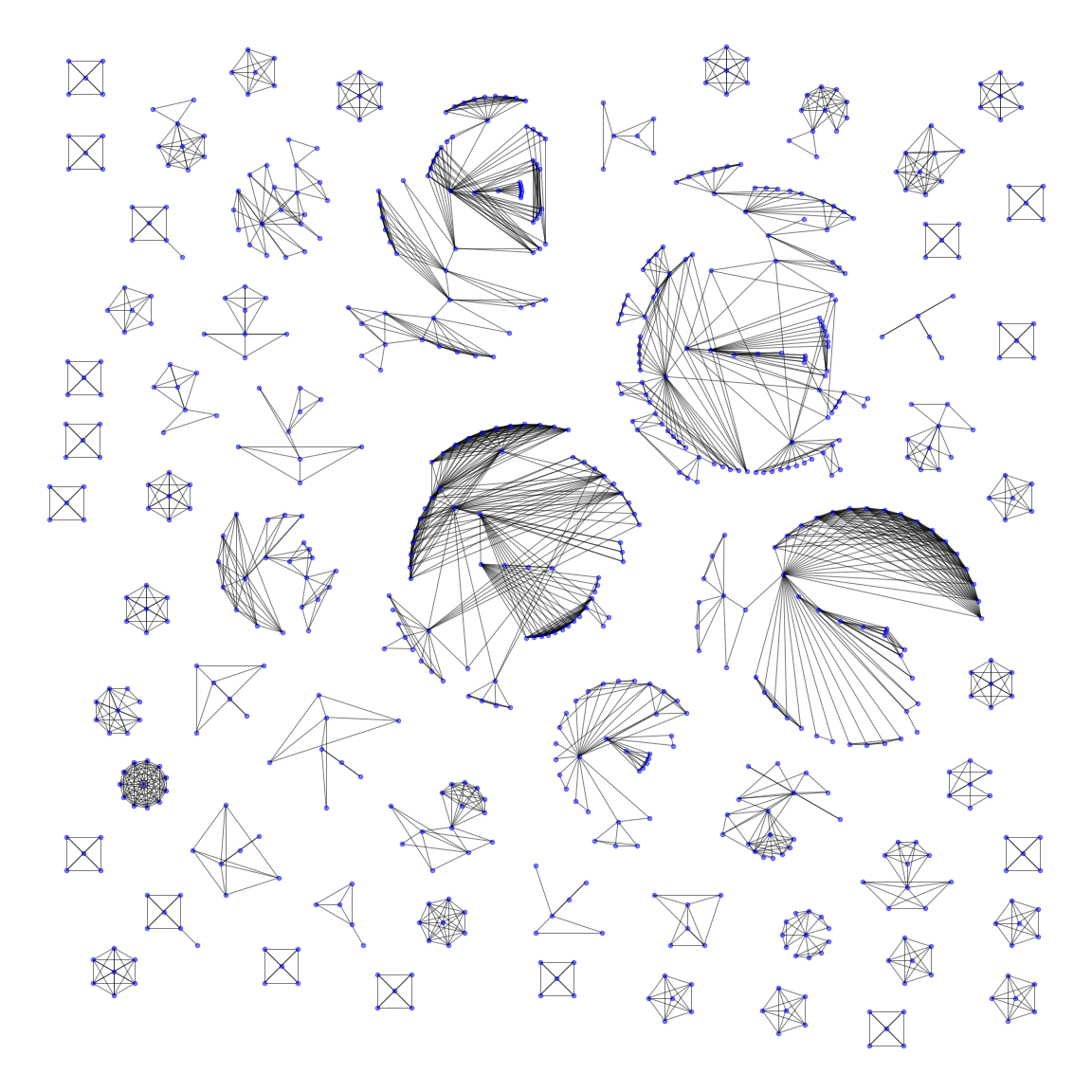}
		\caption{A visual representation of the workflow scheduling community. Only components with cardinality 5 or higher are shown.}
		\label{fig:overview-workflow-scheduling-community}
		\vcutM
	\end{subfigure}~
	\begin{subfigure}[t]{0.5\textwidth}
		\includegraphics[width=\textwidth]{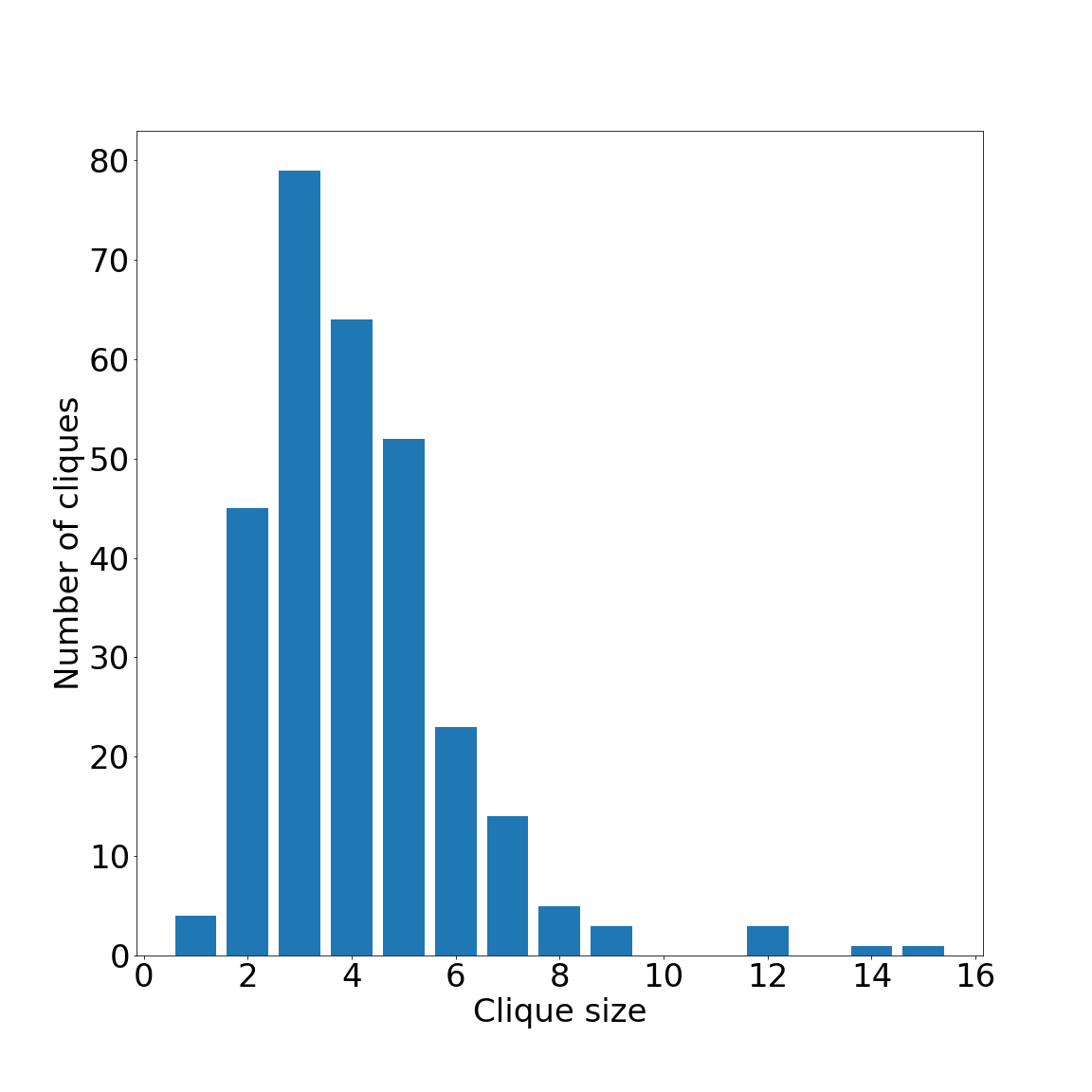}
		\caption{A bar plot depicting the size and number of cliques within the workflow scheduling community.}
		\label{fig:cliques-workflow-schedulung-community}
		\vcutM
	\end{subfigure}
	\caption{An overview of the workflow scheduling community.}\label{fig:workflow-scheduling-community}
	\vcutL
\end{figure}

\begin{figure}
	\centering
	\begin{subfigure}[t]{0.66\textwidth}
		\includegraphics[width=\textwidth,keepaspectratio]{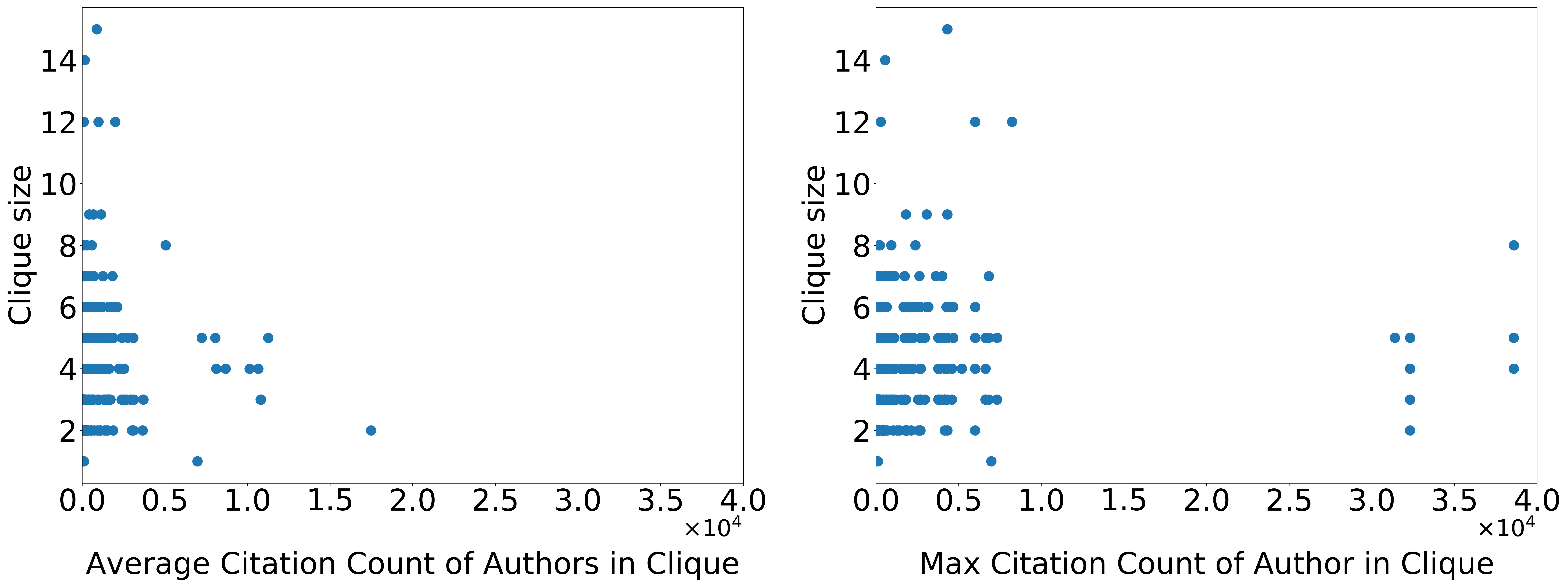}
		\caption{A scatterplot depicting the average (left) and maximum (right) citations of the authors for each clique in the workflow scheduling community.}
		\label{fig:citations-vs-clique-size-workflow-scheduling-community}
		\vcutM
	\end{subfigure}~
	\begin{subfigure}[t]{0.33\textwidth}
		\includegraphics[width=\textwidth,keepaspectratio]{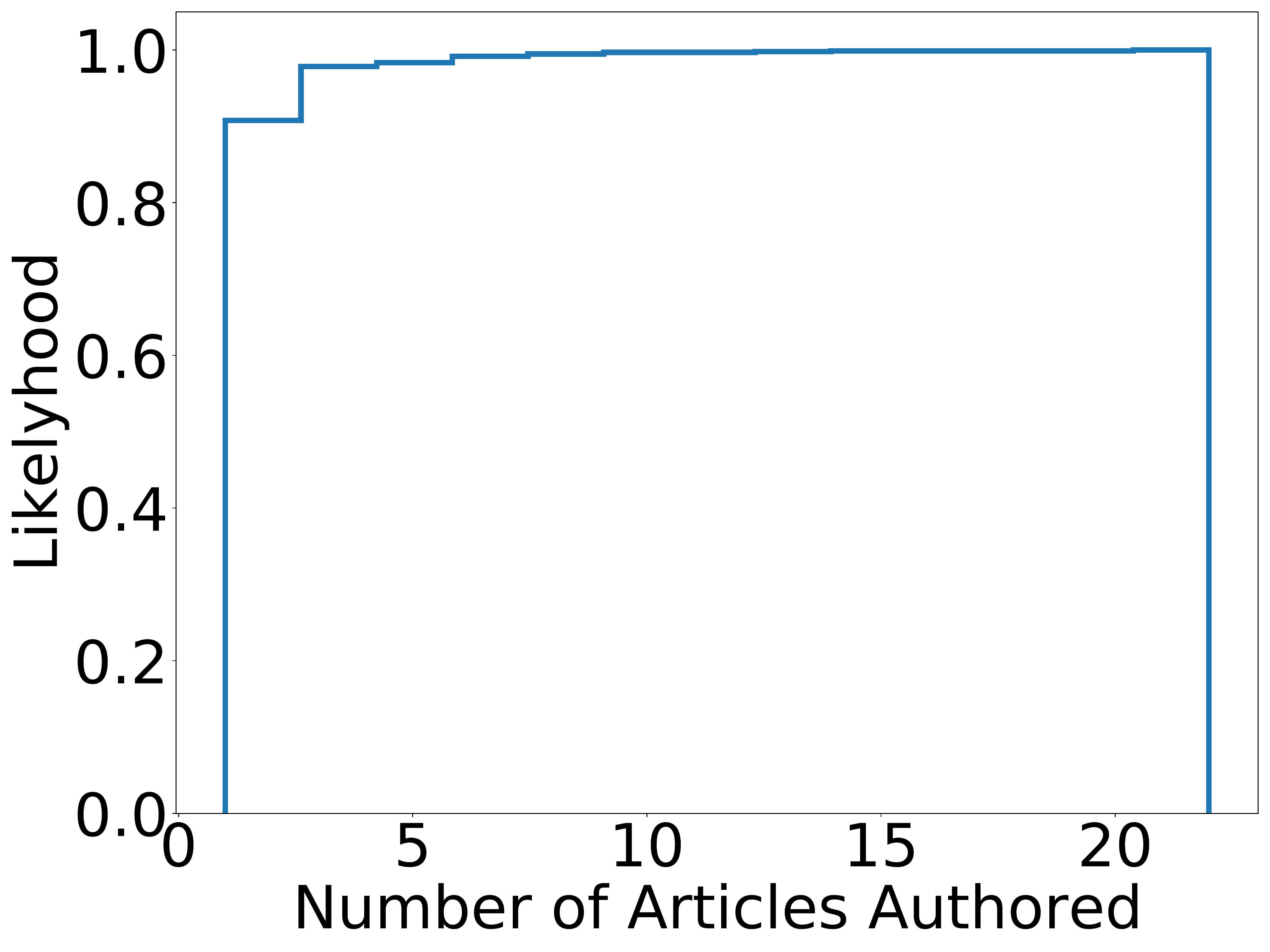}
		\caption{A CDF of number of articles authored per author in the workflow scheduling community.}
		\label{fig:number-articles-co-authored-workflow-scheduling-community}
		\vcutM
	\end{subfigure}
	\caption{Clique size vs. author citation count (average and max) and a CDF of number of articles authored.}
	\vcutL
\end{figure}

We take a look at the workflow scheduling community by visualizing it and look at how many cliques are within that community and their sizes, the amount of citations per author on average and at maximum per clique, and how often authors co-author together.
When ranking authors or communities, self-isolated cliques are often seen as less desirable~\cite{liu2005co}.
Hence, a community with a lot of interaction between different (groups of) authors are signs of a \textit{healthy} community.

\executedqueries{workflow-scheduling-articles-past-decade}{SELECT * FROM publications WHERE year BETWEEN 2009 AND 2018 AND (lower(title) LIKE '\%workflow\%' OR lower(abstract) LIKE '\%workflow\%') AND (lower(title) LIKE '\%schedul\%' OR lower(abstract) LIKE '\%schedul\%')}

To get insights into the workflow scheduling community, we analyzed the author and citation information of articles returned by \refquery{workflow-scheduling-articles-past-decade}.
Figure~\ref{fig:overview-workflow-scheduling-community} present the structure of the community.
From this figure we observe many large collaborating communities, which indicates a dynamic and collaborating community.
Additionally, we observe plenty of authors forming \enquote{bridges} between two or more groups.
We believe such bridges are positive, as they may facilitate individuals in these respective groups gaining knowledge from the other groups through such individuals.

Figure~\ref{fig:cliques-workflow-schedulung-community} shows the number of cliques per clique size.
As we can observe, most cliques are of size 2-5, which is a normal set of authors in a single article, forming a clique per definition.
Overall, there are only a few large cliques. 
Together with the visual of the community, it seems that the community is of a more collaborative nature than forming tightly connected, yet closed groups.

If we look at clique size versus average and maximum citation count per clique, visualized in Figure~\ref{fig:citations-vs-clique-size-workflow-scheduling-community}, we observe that well-cited authors, both in maxima and on average, are not forming or participating in large closed groups.
This further adds to the intuition of a collaborative community.

Finally, Figure~\ref{fig:number-articles-co-authored-workflow-scheduling-community} shows a CDF of the number of articles published per author in the workflow scheduling community.
Roughly 90\% of the authors publish a single article in the workflow scheduling domain in the span 2009-2018, with a long tail having authors publish up to twenty articles.

\subsection{Method for Keyword Analysis}
\label{ssct:survey-methodology}

Identifying keywords is an effective way to obtain important topics within text~\cite{onan2016ensemble}. What keywords are important given a set of articles? How often do we see the same keyword appear? Does the importance of keywords change over time?
To obtain important keywords from articles matching a certain scope, defined by a database query, we apply the following process to sanitize and refine the data and then use \gls{tfidf}.
\gls{tfidf} is a commonly applied technique in the information retrieval domain to obtain important keywords from text~\cite{zhang2008tfidf}.
The process we apply is as follows.

First, two queries are defined.
The first query is to construct a corpus that we will use to compare articles against using \gls{tfidf}.
The second query is to fetch articles of interest, e.g., articles having certain keywords in their title or abstract. 

Text of articles that match these queries are preprocessed using a process similarly to~\cite{gupta2009survey}:
\begin{enumerate}
	\item All text is converted to lower case.
	\item Each character that is not an alphanumeric character, underscore, or white space is removed.
	\item Words are lemmatized to obtain their \enquote{root}, avoiding variations of the same word.
	\item Stop words defined defined by the NLTK library are removed.
	\item We create a count vector that tracks word frequencies and automatically discards words that occur in 90\% or more articles.
\end{enumerate} 

Next, we compute keywords unique per article using \gls{tfidf}.
This method ranks words unique to the article vs. the entire corpus.

We extract the top-50 most important keywords per article based on their \gls{tfidf} values.
We count and store in a list the occurrences (term frequency) of the top-50 keywords found to obtain a ranking for all articles that match our second query.
We limit the number of words we extract per article, otherwise we would end up counting all words in all articles matching the second query, defeating the purpose of \gls{tfidf}.

We can then use this list to, e.g., create a top-$n$ of keywords.
If this top-$n$ contains words that have no significant meaning, we manually filter them out and take the next meaningful word.

Scikit-learn 0.21.3 was used to compute the \gls{tfidf} vector, TextBlob 0.15.3 was used for Lemmatization, Pandas 0.25.1 was used for computing and cleaning data, and NLTK 3.4.5 was used for stop word filtering along with a list of custom stop words.

To support reproducibility and FAIR data, all instruments and tools, scripts, and the database containing article meta-data used in this article are available as open-source artifacts.

\subsection{Analysis of Keywords in Workflow Scheduling Articles}

\begin{table}[t]
	\caption{Top-10 keywords in articles on scheduling workflow published between 2009 and 2018.}
	\label{tbl:top-10-scheduling-workflow-overall}
	\vspace{-0.2cm}
	\begin{adjustbox}{max width=\columnwidth}
		\begin{tabular}{lllllllllll}
			\toprule
			Rank & 1 & 2 & 3 & 4 & 5 & 6 & 7 & 8 & 9 & 10 \\ \midrule
			Word & workflow & scheduling & cloud & task & algorithm & application & data & time & cost & deadline \\ \bottomrule
		\end{tabular}
	\end{adjustbox}
	\vcutL
\end{table}

\begin{description}
	\observation{workflow-scheduling-observation}{The keywords \enquote{task}, \enquote{time}, \enquote{cost}, and \enquote{deadline} are often mentioned in articles on workflow scheduling, highlighting the focus of the community on these topics.}
\end{description}

To inspect what the focus is of articles working on scheduling and workflows, we first look at the top-10 most important keywords using the method described in Section~\ref{ssct:survey-methodology}.
We inspect articles returned by \refquery{workflow-scheduling-articles-past-decade}.

The results are in Table~\ref{tbl:top-10-scheduling-workflow-overall}.
From this table we observe, besides the \enquote{workflow} and \enquote{scheduling}, that the notion of clouds is the important keyword. 
This makes sense as most articles on workflow scheduling target either public or private cloud settings.
Consequently, \enquote{tasks}, \enquote{application}, \enquote{computing}, and \enquote{cost} keywords are popular as these are closely aligned with workflow scheduling in clouds.
In particular, plenty of scheduling policies focuses on the duo \enquote{cost} and \enquote{deadline}.
This can also be observed in our keyword analysis for workflow allocation (Section~\ref{ssct:workflow-allocation-community-keyword-analysis}) and in the mapping in Section~\ref{sct:workflow-allocation-policies}.
As mentioned prior, \enquote{data} is a keyword that we did expect.
Plenty of research on workflow scheduling produces or relies on data, from characterization to simulation studies.
Scientific workflow applications are often used as a use-case for experimentation, but rarely is the data provided as auxiliary data for reproducibility purposes~\cite{2019arXiv190607471V}.

\subsection{Analysis of Trends over Time}
\label{ssct:trends-over-time-workflow-scheduling}

\begin{description}
	\observation{cloud-more-important}{The keyword \enquote{cloud} grew in importance over time, and remained in the top 5 since 2013, highlighting the importance of this topic.}
	\observation{cost-deadline}{The ranks of \enquote{task}, \enquote{deadline}, \enquote{data}, \enquote{cost}, and \enquote{algorithm} fluctuate, yet are often in the top-10 during the years 2009-2018.}
\end{description}

\begin{figure}[t]
	\includegraphics[max width=\linewidth]{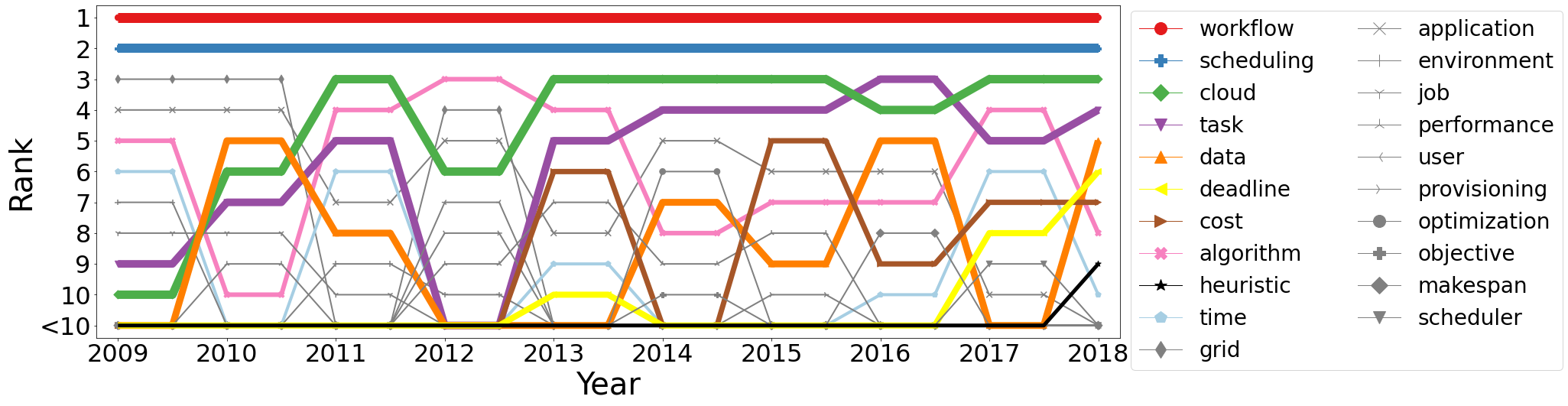}
	\caption{Top-10 keywords in scheduling workflow articles in the past decade per year. Coloured lines are part of 2018's top-10 and have a decreasing line thickness for better visual tracking. The legend shows the coloured curves in order, and grey curves out of order.}
	\label{fig:top-10-scheduling-workflow-per-year}
	\vcutL
\end{figure}

To see how the focus of the community working on scheduling workflows shifted, we visualize in Figure~\ref{fig:top-10-scheduling-workflow-per-year} the top-10 keywords per year.
We take the output of \refquery{workflow-scheduling-articles-past-decade} per year, and apply the same method as described in Section~\ref{ssct:survey-methodology}.

From this figure, we observe that \enquote{cloud} already became a popular term in 2009. 
Since 2013, the term has been in the top-5 consistently.
We also see a clear shift in focus, where grids were popular before the cloud, \enquote{grids} disappeared completely from the top-10 after 2010.
The keywords \enquote{cost}, \enquote{deadline}, and \enquote{heuristic} are keywords that have been rising in importance in recent years.
Other keywords such as \enquote{algorithm}, \enquote{task}, and \enquote{data} appear to be consistently important, which makes sense provided they are general concepts and building blocks in workflow scheduling. 

\subsection{Discussion on Emerging trends}
\label{ssct:emerging-trends}

\begin{description}
	\observation{deadline-growing-in-importance}{Performance and in particular deadline-aware scheduling are growing in importance.}
\end{description}

Emerging trends are often a good topic for research as they are deemed interesting by the community.
Due to the many open issues -- and possibly low-hanging fruits -- plenty of opportunities for new research are present.
We attempt to detect emerging trends in two ways:
\begin{description}
	\item[New keywords:] Keywords that were found by our method to be among the most common/important keywords in recent years, but did not come up in previous years. This type of analysis highlights keywords that previously were not common/important or are new and gaining traction fast. In this survey, we compare half of the selected time span, i.e. 2014 - 2018, with the remainder (2009-2013).
	\item[Rising keywords]: Keywords that throughout the years kept monotonically increasing in rank since their appearance. This type of analysis finds keywords of two categories: \begin{enumerate}
		\item Keywords that are receiving more (or the same amount of) attention each year and thus indicates an interest. 
		\item Keywords that became emerging in the last year of the timespan checked.
	\end{enumerate}
\end{description}

For each year that we investigate, we apply the method outlined in Section~\ref{ssct:survey-methodology}, and take the top-10 most important/frequent keywords.

\subsubsection{Emerging trends in workflow scheduling}

We attempt to discover new and emerging keywords using articles that match \refquery{workflow-scheduling-articles-past-decade}.

The keywords found during \textit{new keywords} analysis are as follows.

\begin{tcolorbox}[colback=blue!15, colframe=blue!90!black,enhanced,sharp corners,boxsep=-1mm]
	\enquote{deadline}, \enquote{optimization}, \enquote{performance}
\end{tcolorbox}

Besides some false positives, we observe the keywords \enquote{deadline} and \enquote{performance} research the top 10 in the last 5 years.
One possibility for this could be that simply scheduling workflows is no longer good enough, managing deadlines or be more performant is critical when serving customers and becoming more efficient overall.

If we look at \textit{rising keywords}, we obtain the following.
\begin{tcolorbox}[colback=blue!15, colframe=blue!90!black,enhanced,sharp corners,boxsep=-1mm]
\enquote{deadline}, \enquote{workflow}
\end{tcolorbox}

Again, we see the keyword \enquote{deadline}, yet this time alongside \enquote{workflow}.
\enquote{workflow} is expected as it is the focus of our query.
We observed from Figure \ref{fig:top-10-scheduling-workflow-per-year} that it is consistently at number one, thus being monotonically increasing.
As for the keyword \enquote{deadline}, this indicates that besides its appearance in the top-5 as of recent, it also is rising in importance within that time frame.

\subsection{Future Research Directions Inspired by Meta-Data Analysis}

Both the important keyword section and the emerging trend section suggest that deadline-aware scheduling remains and is growing as a hot topic.
Our conjecture is that being deadline-aware whilst optimizing for other metrics such as costs, and energy consumption will continue to grow in importance and is an excellent topic for future work.
Further refining these analyses and introducing new angles of investigation is another interesting item for future work.
\section{A Taxonomy of Workflow Scheduling}
\label{sct:taxonomy}

\begin{figure}[htb]
	\adjustbox{max width=.9\linewidth}{
	\includegraphics{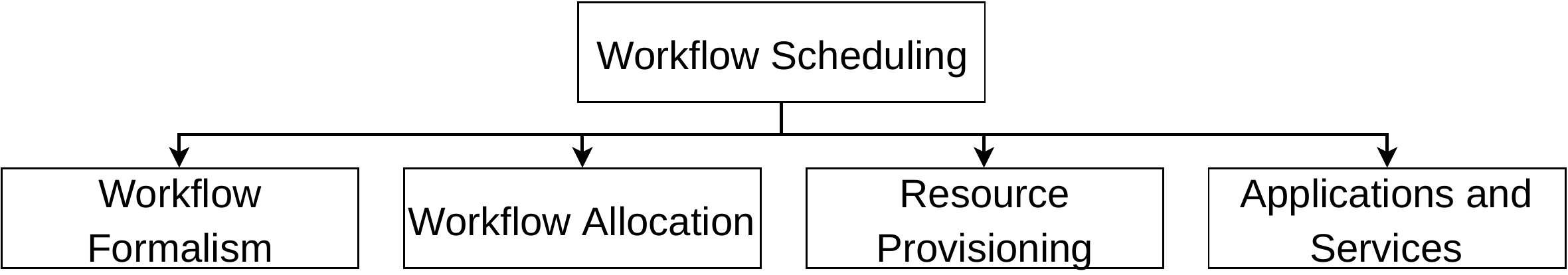}
	}
	\caption{Taxonomy of workflow scheduling.} \label{tax:workflow-scheduling}
	\vcutM
\end{figure}

\begin{table}[t]
	\caption{Possible relationships between all keywords in Figure~\ref{fig:top-10-scheduling-workflow-per-year} and the areas related to workflow scheduling.}
	\label{tbl:relationship-keywords-sub-comunities-workflow-scheduling}
	\vcutL
	\begin{tabularx}{\linewidth}{@{}lX@{}}
		\toprule
		Area              & Keywords (found as important in Section~\ref{ssct:trends-over-time-workflow-scheduling})                                                                                                                                              \\ \midrule
		Workflow Formalism        & task, data, graph, application, environment, job, user                                                                                                 \\
		Workflow Allocation       & task, deadline,  cost, algorithm, heuristic, graph, grid, application, time, environment, job, performance, user, optimization, objective, makespan    \\
		Resource Provisioning     & task, cost, algorithm, heuristic, graph, grid, application, time, environment, job, performance, user, provisioning, optimization, objective, makespan \\
		Applications and Services & data, cost,  graph, grid, application, time, environment, job, performance, user, provisioning, optimization, makespan, scheduler                      \\ \bottomrule
	\end{tabularx}
	\vcutL
\end{table}

Taxonomies provide a structured and detailed decomposition of a certain topic and/or field.
These decompositions allow for a good overview of possible and attempted avenues to tackle challenges and attempt to find a feasible or optimal solution to challenges.
These taxonomies can also provide new ideas for methods and/or combinations not attempted yet.

To limit the scope of this survey, we focus on four areas within workflow scheduling, depicted in Figure~\ref{tax:workflow-scheduling}: Workflow Formalism, Workflow Allocation, Resource Provisioning, and Applications and Services.
We select these four as they relate closely to the keywords found in Section~\ref{ssct:trends-over-time-workflow-scheduling}.
Table~\ref{tbl:relationship-keywords-sub-comunities-workflow-scheduling} shows our selection of keywords, per area.

\begin{description}
	\item[Formalisms] describe the way workflows are represented, and the possible features they can have, i.e., different formalisms support different notions of computation.
	Not many surveys focus on this aspect, yet we believe it is important as the formalism defines what properties can and cannot be captured.
	\item[Workflow allocation] is the problem of assigning units of work to the available resources to adhere to the various \gls{qos} constraints set, while potentially attempting to improve other aspects such as resource utilization or power consumption. 
	\item[Resource provisioning] covers the research of when to allocate resources and how many given current and predicted demand. Adding the right amount of resources is crucial in lowering costs and improving the overall resource utilization, while avoiding slowdowns and other issues in the system. 
	\item[Application and services] cover the different type of resources, the execution model, and services that are considered in literature and available today.
\end{description}

These four main elements will be covered in the next four sections, each with their respective (sub-)taxonomies.
\section{Workflow Formalism}
\label{ssct:workflow_model}

\begin{figure}[t]
	\centering
	\begin{subfigure}[t]{0.66\textwidth}
		\centering
		\adjustbox{max width=\linewidth, max height=6cm}{
			\includegraphics{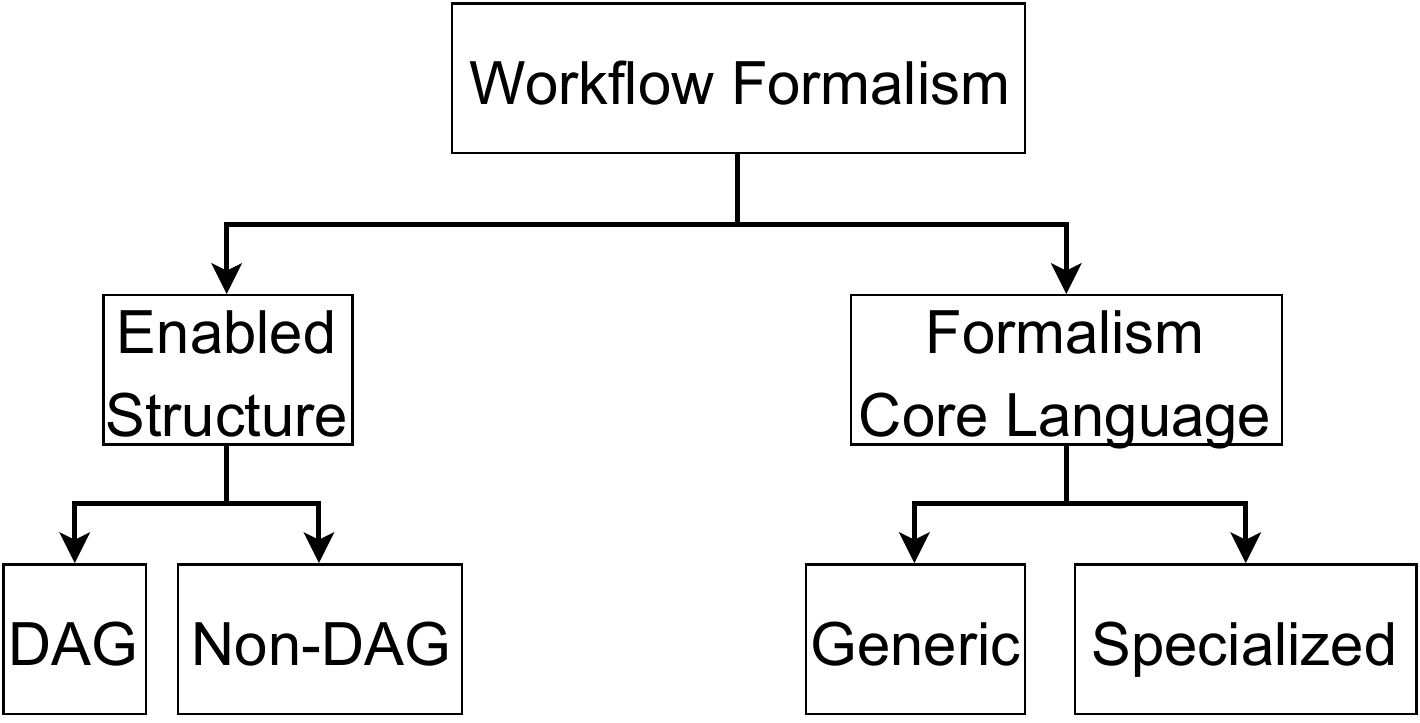}
		}
		\caption{}
		\label{tax:workflow-formalism}
	\end{subfigure}~
	\begin{subfigure}[t]{0.34\textwidth}
		\centering
		\includegraphics[width=\linewidth]{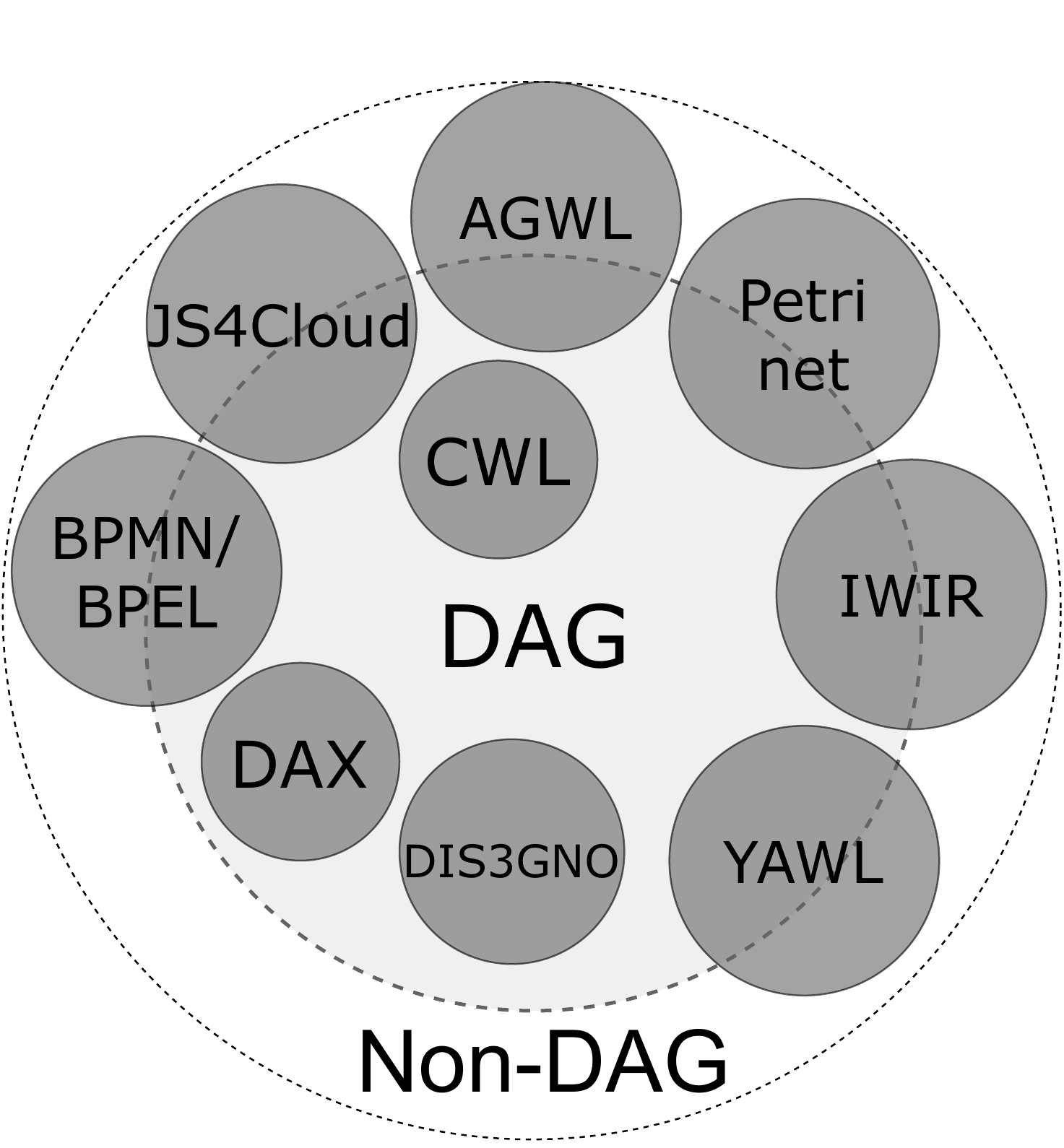}
		\caption{}
		\label{tax:workflow_formalisms}
	\end{subfigure}
	\caption{Left: the taxonomy of workflow formalisms. Right: a Venn diagram showing how workflow formalisms relate to the abstract formalism of (non-)\acrshort{dag}.}
	\label{fig:workflow-formalism-overview}
	\vcutL
\end{figure}

\begin{figure}
	\centering
	\begin{subfigure}[t]{0.49\textwidth}
		\includegraphics[width=\textwidth]{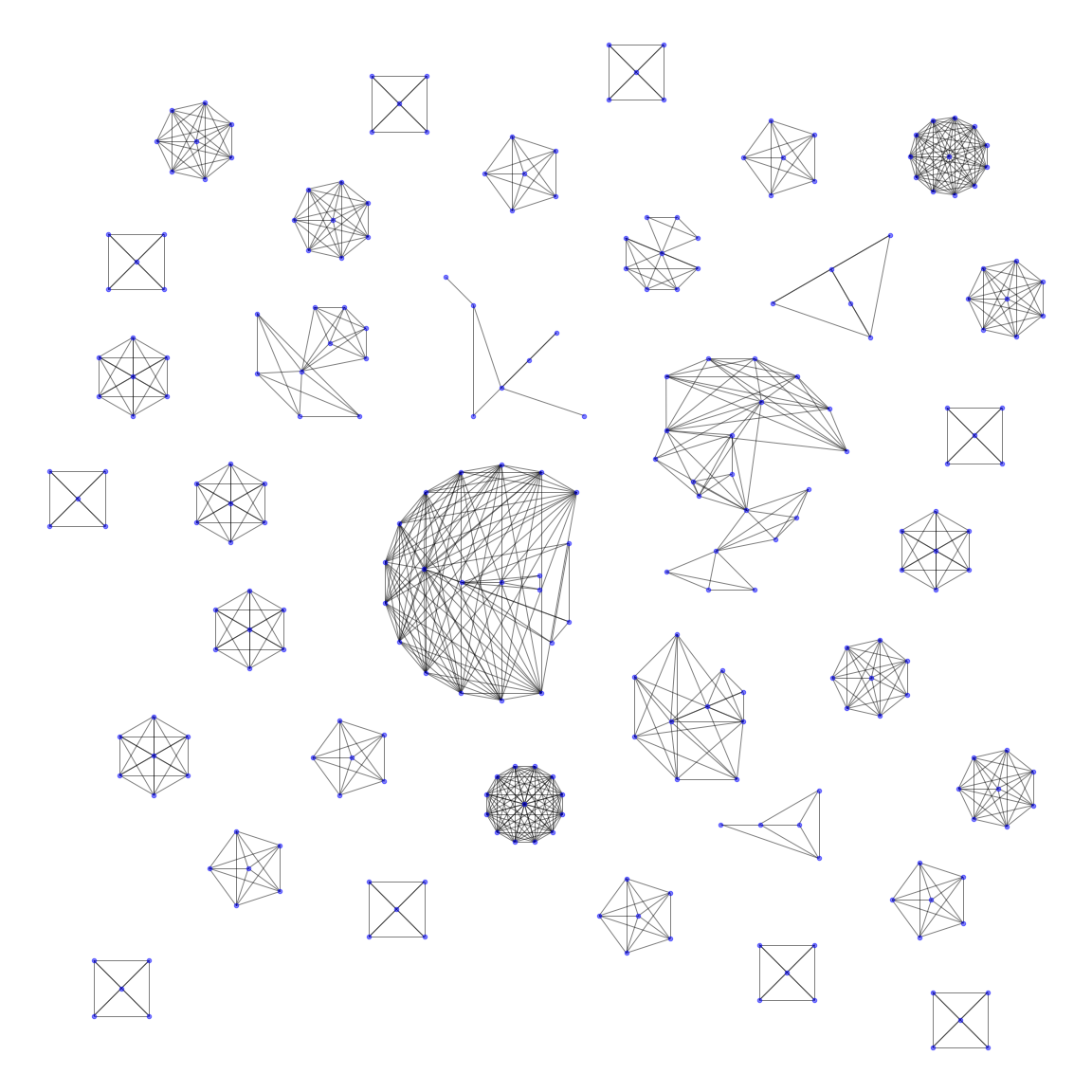}
		\caption{A visual representation of the workflow formalism community. Only components with cardinality 5 or higher are shown.}
		\label{fig:overview-workflow-formalisms-community}
	\end{subfigure}~
	\begin{subfigure}[t]{0.49\textwidth}
		\includegraphics[width=\textwidth]{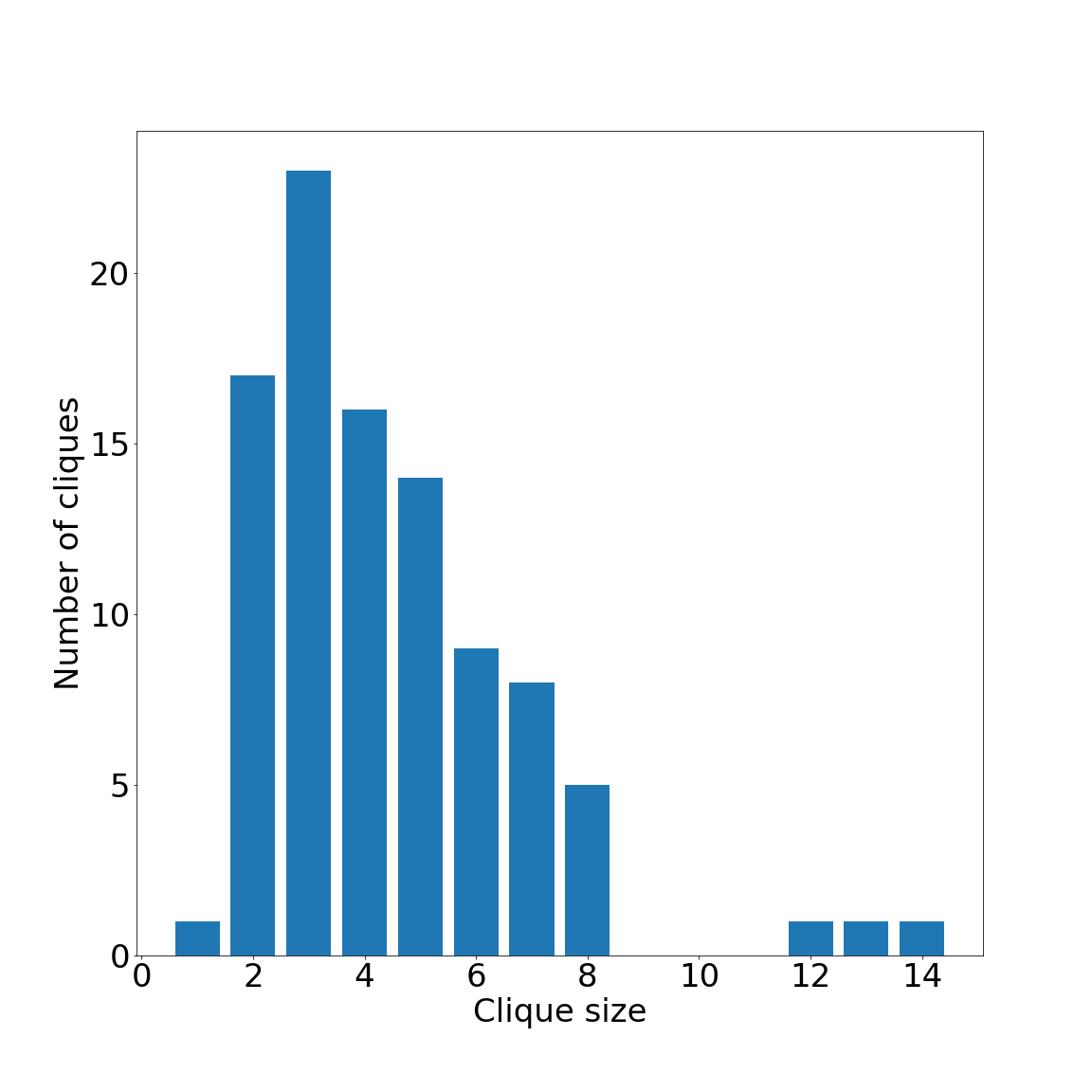}
		\caption{A bar plot depicting the size and number of cliques within the workflow formalism community.}
		\label{fig:cliques-workflow-formalisms-community}
	\end{subfigure}
	\caption{An overview of the workflow formalism community.}\label{fig:workflow-formalisms-community}
\end{figure}

\begin{figure}
	\centering
	\begin{subfigure}[t]{0.66\textwidth}
		\includegraphics[width=\textwidth,keepaspectratio]{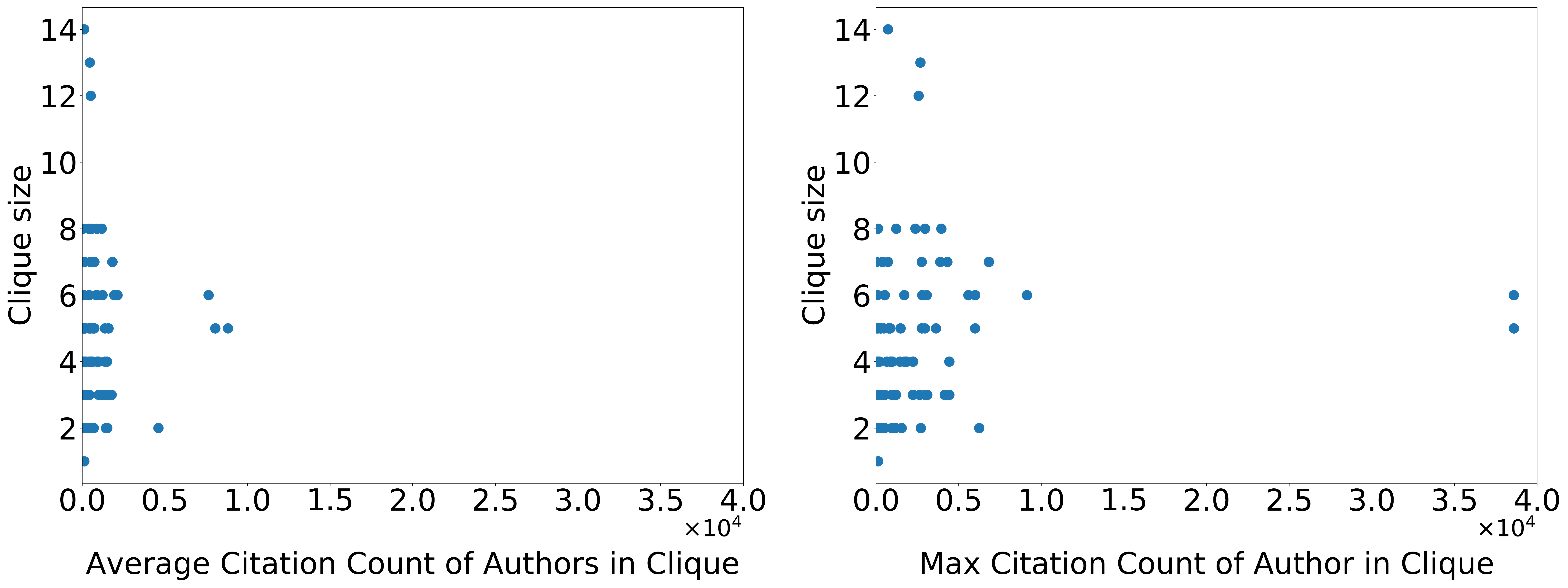}
		\caption{A scatterplot depicting the average (left) and maximum (right) citations of the authors for each clique in the workflow formalism community.}
		\label{fig:citations-vs-clique-size-workflow-formalisms-community}
	\end{subfigure}~
	\begin{subfigure}[t]{0.33\textwidth}
		\includegraphics[width=\textwidth,keepaspectratio]{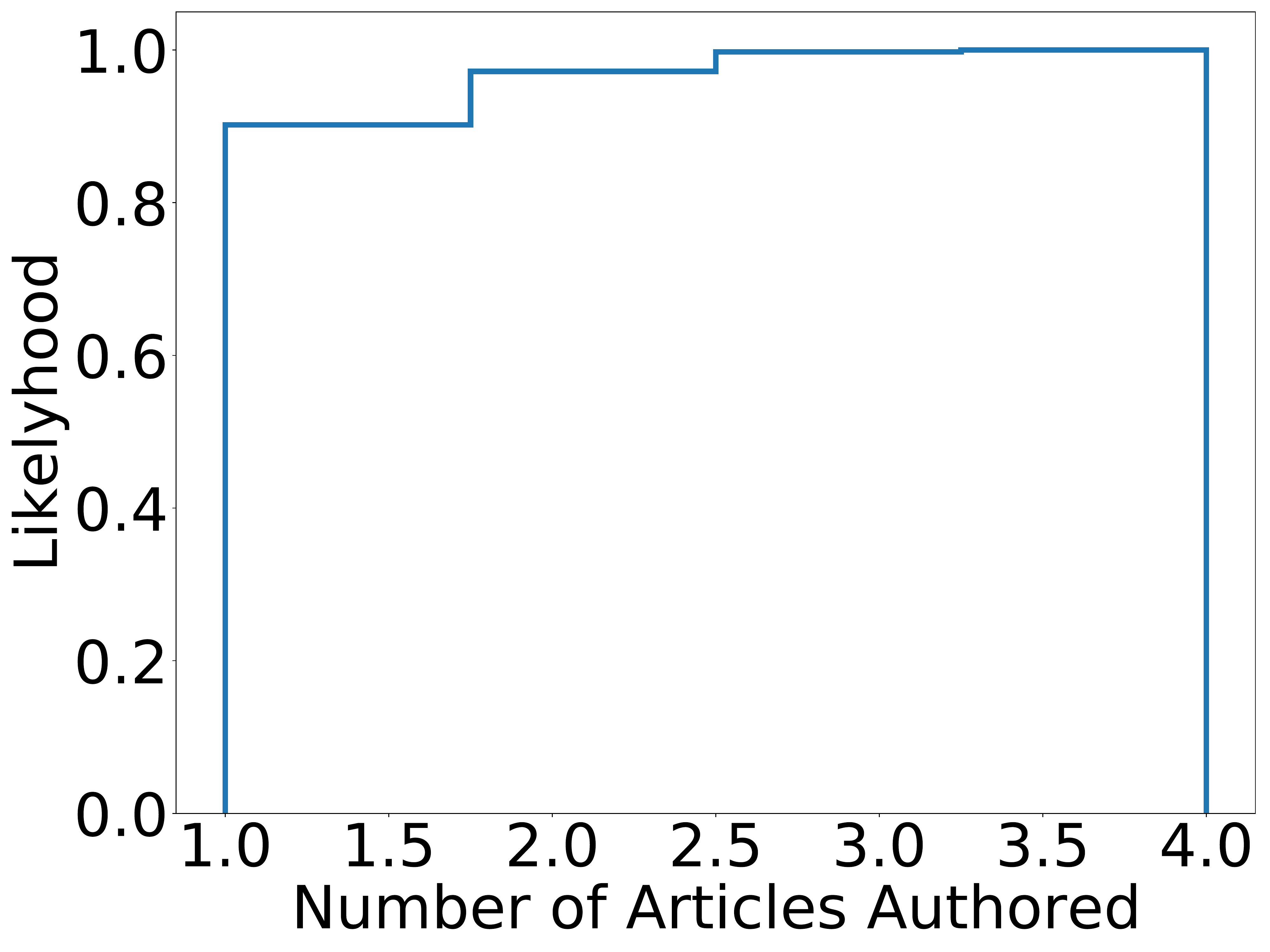}
		\caption{A CDF of number of articles authored per author in the workflow formalism community.}
		\label{fig:number-articles-co-authored-workflow-formalisms-community}
	\end{subfigure}
	\caption{Clique size vs. author citation count (average and max) and a CDF of number of articles authored.}
\end{figure}

\begin{figure}[thb]
	\includegraphics[max width=\linewidth]{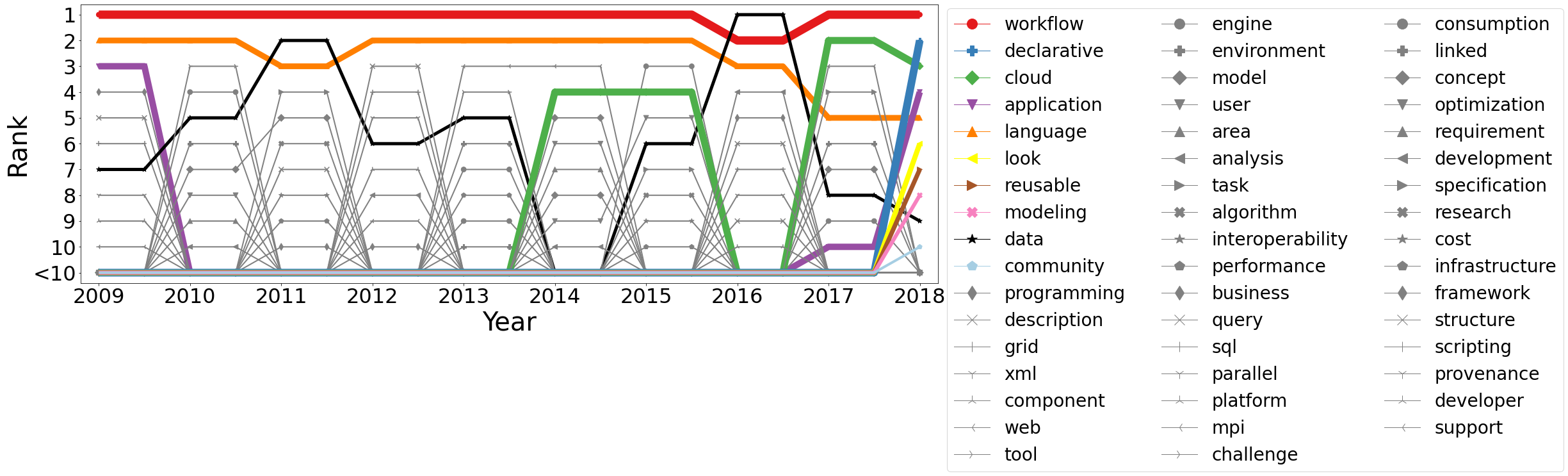}
	\caption{Top-10 keywords in workflow formalism articles in the past decade per year.}
	\label{fig:top-10-workflow-formalisms-per-year}
	\vcutS
\end{figure}

\executedqueries{workflow-formalism-query}{SELECT * FROM publications WHERE year BETWEEN 2009 AND 2018 AND (lower(title) LIKE '\%workflow\%' OR lower(abstract) LIKE '\%workflow\%') AND ((lower(title) LIKE '\%formalism\%' OR lower(abstract) LIKE '\%formalism\%') OR (lower(title) LIKE '\%language\%' OR lower(abstract) LIKE '\%language\%'))}

A workflow formalism provides a language to construct workflows with.
To create an overview and taxonomy of formalisms commonly used with workflow scheduling, we perform a systematic search to find articles on this topic and complement it with our experience.
The query used to obtain articles on workflow formalisms for our systematic search is visible in \refquery{workflow-formalism-query}.

Modifying the workflow structure taxonomy of Yu et al.~\cite{DBLP:journals/sigmod/YuB05}, our taxonomy of workflow formalism is presented in Figure~\ref{tax:workflow-formalism} and consists of two main branches: the enabled structure and what we call the core language, covered in Sections~\ref{ssct:workflow-formalism-enabled-structure} and~\ref{ssct:workflow-formalism-core-language}, respectively.

\subsection{Community Analysis}

\begin{description}
	\observation{formalism-community-small-healthy}{The formalism community is a small yet healthy community. A few large components exist and most author-relations are one-time.}
	\observation{formalisms-focus-users}{Many of the emerging trend keywords indicate that users and convenience of use are growing in importance.}
	\observation{formalisms-larger-cliques-smaller-avg}{Larger cliques have a lower citation author citation count both on average and in maxima.}
	\observation{formalisms-mostly-singly-publication}{Over 80\% author a single article.}
	
\end{description}

\begin{table}[]
\caption{An overview of the workflow formalism community in numbers.}
\label{tbl:workflow-formalism-community-overview}
\begin{tabular}{@{}rrrrrr@{}}
\toprule
Articles & Authors & Co-authorship Relations & Unique Relations & Cliques & Largest Clique \\ \midrule
455 & 401     & 1032      & 978              & 96      & 14             \\ \bottomrule
\end{tabular}
\end{table}

We summarize the main characteristics of the community in Table~\ref{tbl:workflow-formalism-community-overview}.
The first observation is that the number of articles is the least of all selected sub-communities.
This matches our experience as the formalism is rarely discussed in articles. 
Interestingly, compared to the second smallest community in terms of articles -- resource provisioning -- the number of authors is substantially higher.
This means that creating a formalism tends to involve more authors and possibly parties.
We did not detect many duplicate co-author relationships, meaning that most relationships are one-time.

The visualization of the community depicted in Figure~\ref{fig:overview-workflow-formalisms-community} and the distribution of cliques in Figure~\ref{fig:cliques-workflow-formalisms-community} show this community looks healthy, yet somewhat segregated.
many larger groups exist, yet few \enquote{bridges} exist, i.e., a person that operates within two groups.
Authors that only publish by themselves are rare.
Most authors are in a clique of size 2-5, which could be inter-department teams or smaller yet tight collaboration groups.

In Figure~\ref{fig:citations-vs-clique-size-workflow-formalisms-community} we plot the average and maximum number of citations per clique in the community.
Here, we observe some cliques where the average citations of the authors is quite high, around one thousand citations. These cliques are moderately sized; having a size of five or six.
If we look at the maxima per clique, we observe some outliers where moderately sized cliques work together with a well-cited author.
Interestingly, as the clique size grows, both the average and maximum drop significantly compared to moderately sized and smaller cliques.
Our data is too little to draw any correlating conclusions, but based on this data we conjecture that 1) larger cliques may involve more junior researchers leading to a lower average, and 2) well-cited authors don't engage with sub-communities in such a way that they co-author with all members of that sub-community.

Figure~\ref{fig:number-articles-co-authored-workflow-formalisms-community} presents a CDF of the number of articles that get published within this community per author.
From this figure we observe over 80\% of the authors in the workflow formalism community author only once a paper in this community.
The highest number being four indicates this area does not receive much (follow-up) work authored by the same author(s).

\subsubsection{Analysis of Trends over Time}
We use the method detailed in Section~\ref{ssct:trends-over-time-workflow-scheduling} combined with the output of \refquery{query-workflow-allocation-articles}.
The output of the trend analysis is visible in Figure~\ref{fig:top-10-workflow-formalisms-per-year}.
From this figure, we observe a lot of variation.
\enquote{Cloud} emerged as important in 2014 and has been in the top-10 fairly consistently since then.
\enquote{Language} remained a very important keyword until 2018, where it dropped out of the top-10.
This might be due to an increased focus on the ease of use, reproducibility, and expressiveness of the formalisms itself in 2018, yet a more in-depth study should be done to draw any definitive conclusions.
We do observe quite a few keywords describing the usefulness of a formalism as of late.
In particular, \enquote{declarative}, \enquote{modeling}, \enquote{convenient}, and \enquote{reusable} could be related.
Furthermore, we see that the top-10 of the workflow formalism community varies greatly per year.
This indicates that the formalism community focuses on several directions over time rather than working towards one single goal.

\subsubsection{Emerging Trends}
We use the method detailed in Section~\ref{ssct:emerging-trends} combined with the output of \refquery{query-workflow-allocation-articles}.
We find the following \textit{new keywords}.
\begin{tcolorbox}[colback=blue!15, colframe=blue!90!black,enhanced,sharp corners,boxsep=-1mm]%
	\enquote{cloud}, \enquote{concept}, \enquote{convenient}, \enquote{declarative}, \enquote{description}, \enquote{framework}, \enquote{model}, \enquote{modeled}, \enquote{modeling}, \enquote{structure}, \enquote{support}, \enquote{task}, \enquote{user}
\end{tcolorbox}

This list of keywords is quite diverse.
First of all, we observe that modeling workflows is important as of late.
This might be due to attempts to better model the applications to be executed.
The second observation is that there are hints of \enquote{ease of use} in the list.
The keywords \enquote{convenient}, \enquote{framework}, \enquote{support}, and \enquote{user} could indicate that recent focus is on how to better support the requirements of the users.
Finally, the notion of \enquote{cloud} and \enquote{task} in the context of workflows is expected, given the popularity of clouds as environment and tasks being the atomic units of workflows. 

As for \textit{rising keywords}, the output is as follows.
\begin{tcolorbox}[colback=blue!15, colframe=blue!90!black,enhanced,sharp corners,boxsep=-1mm]%
\enquote{convenient}, \enquote{modeling}, \enquote{declarative}, \enquote{predefined}
\end{tcolorbox}

Again, we see a hint of \enquote{ease of use} here, and a hint of the importance of the (workflow) model in formalisms.

\subsection{Taxonomy of Enabled Structures}
\label{ssct:workflow-formalism-enabled-structure}
The enabled structure refers to the constructs possible within the workflow.
We differentiate between \gls{dag} and non-\gls{dag}.
Due to the constraints between tasks and to prevent increasing complexity when having to deal with (complex) loops, most papers use the \gls{dag} formalism to represent workflows with \cite{DBLP:conf/wosp/VersluisEI18}.
The \gls{dag} formalism is a simple and general concept often used in other fields.
However, since this formalism is abstract, many implementations exist that allow developers to express their programs as a \gls{dag}.

Non-\glspl{dag} have the same entities as \gls{dag}, yet offer one additional instruction: iteration (or looping)~\cite{li2012trust}.
Some workflow management systems support the non-DAG formalism, yet most well-known systems use \glspl{dag}. 
Many formalisms implementing a (non-)\gls{dag} formalism exist that are used by various other systems.
Bastos et al.~\cite{bastos2015scientific} look at the different structures of workflow formalisms for interchanging specification between workflow management systems.

Figure~\ref{fig:workflow-formalism-overview} (right) presents a non-exhaustive overview of how workflow formalisms relate to the common abstracts of DAG and Non-DAG formalisms.

\subsection{Taxonomy of Core Languages}
\label{ssct:workflow-formalism-core-language}
Next to the structure enabled by the formalism, we use the term core language of the formalism to depict the language used to construct these workflows.
We introduce this term to avoid ambiguity between the terms \enquote{formalism}, and \enquote{language} which are used interchangeably in literature.
Core languages can be generic purpose languages such as CSV, XML, YAML, and JSON.
Example of formalisms based on generic core languages include AGWL, JS4Cloud, DAX, DIS3GNO, and CWL.

The AGWL is a formalism from the grid era based on XML~\cite{fahringer2005specification}.
The language explicitly models parallelism, loops, and forks such as if-else statements.

JS4Cloud is a JavaScript based workflow formalism for defining and executing data analysis workflows~\cite{DBLP:journals/concurrency/MarozzoTT15}.
It has been implemented in the data mining cloud framework.

The Pegasus project uses an abstract workflow formalism called DAX.
A DAX file describes a workflow as a \gls{dag} in XML format.

Cesario et al.~\cite{DBLP:journals/concurrency/CesarioLTT13} introduce a DAG-based workflow formalism for designing and executing distributed knowledge discovery workflows in their workflow executing framework named DIS3GNO.

The \gls{cwl} is a formalism to describe command line tools and create a workflow out of them~\cite{amstutz2016common}.
The formalism focuses on portability.

Formalisms that use a specialized core language include BPMN, Petri net, YAWL, and UML.

\gls{bpmn} is a formalism commonly used in businesses to outline workflows or processes within a company~\cite{borger2012approaches}.
The formalism is comprehensive as it features over 100 symbols, including support for cycles and human interaction in workflows.

\gls{pn} is a formalism commonly used in chemistry to model chemical processes and reactions.
It is similar to the \gls{dag} formalism, yet uses tokens and weights on links to describe dependencies~\cite{DBLP:journals/jcsc/Aalst98}.
Many variations of the original Petri net formalism have been introduced to enhance its capabilities, for example the use of colored Petri nets~\cite{van1994modelling}.
Hoheisel et al. show how Petri nets can be used to model \glspl{dag}~\cite{hoheisel2003dynamic}.

\gls{yawl} is a formalism inspired by \glspl{pn}~\cite{van2005yawl}. 
It features similar constructs to \gls{bpmn} yet is more simple in its constructs.
YAWL can be used to construct \glspl{dag}~\cite{wust2015generation}.
The formalism supports dynamicity and has extensive support for (unexpected) error handling.

Other formals such as UML and control and data flow approaches are also used~\cite{DBLP:journals/fgcs/DeelmanGST09}. 

\subsection{Future Directions}
There are several future directions that we believe are worth pursuing in the context of workflow formalisms.
We believe \glspl{nfr} can be better incorporated in the formalisms.
Our preliminary investigation~\cite{DBLP:conf/wosp/VersluisEI18} found the \enquote{\gls{dag}}-based solutions are the most common and the most simple to extend, but further investigation is required.
Another interesting direction is to allow the environment to give hints or suggestions to the workflow management system through the formalism.
The CWL-project is investigating incorporating this aspect into their formalism through the form of splitters\footnote{\url{https://github.com/common-workflow-language/common-workflow-language/issues/446}}, where an \enquote{executor} can provide tools to split and combine chunks.

Capturing provenance related elements has been a focus for a while by the community.
In general, reproducibility has received attention as of late.
We believe incorporating provenance details in the formalism, such as input parameters, file names, hardware details, and other (potentially) important elements deserves more attention.
\section{Workflow Allocation}
\label{sct:workflow-allocation}

\begin{figure}[htb]
	\centering
	\adjustbox{max width=\linewidth}{
	\includegraphics{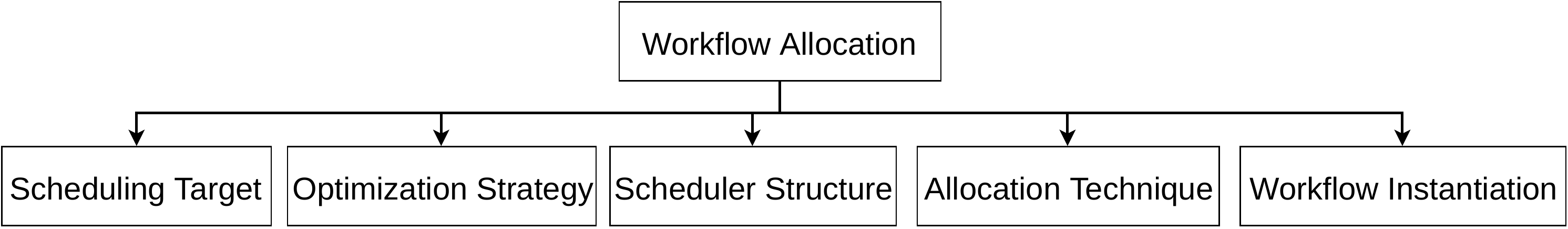}
	}
	\caption{Workflow allocation taxonomy.} \label{tax:workflow-allocation}
\end{figure}

\begin{figure}
	\centering
	\begin{subfigure}[t]{0.49\textwidth}
		\includegraphics[width=\textwidth]{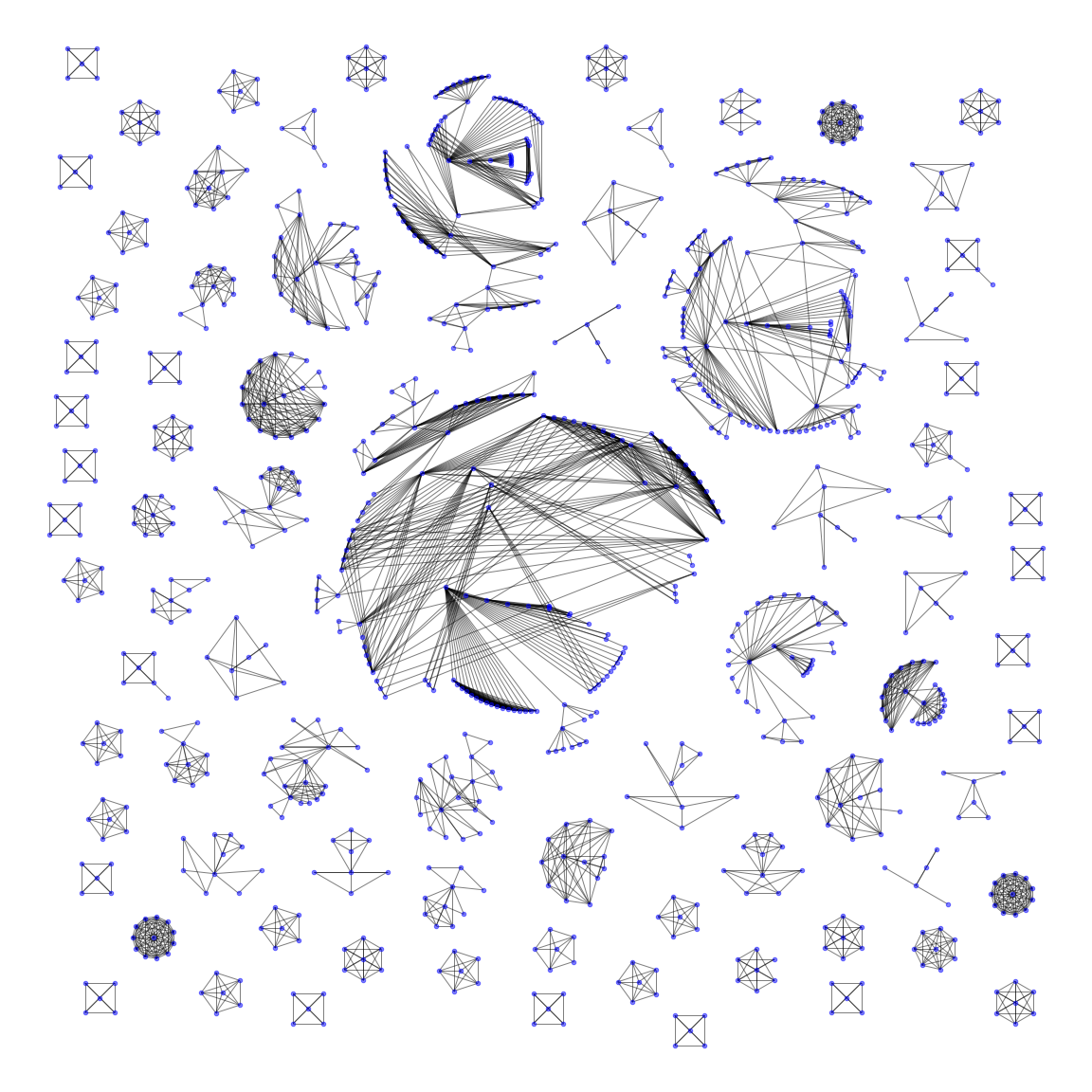}
		\caption{A visual representation of the workflow allocation community. Only components with cardinality 5 or higher are shown.}
		\label{fig:overview-workflow-allocation-community}
	\end{subfigure}~
	\begin{subfigure}[t]{0.49\textwidth}
		\includegraphics[width=\textwidth]{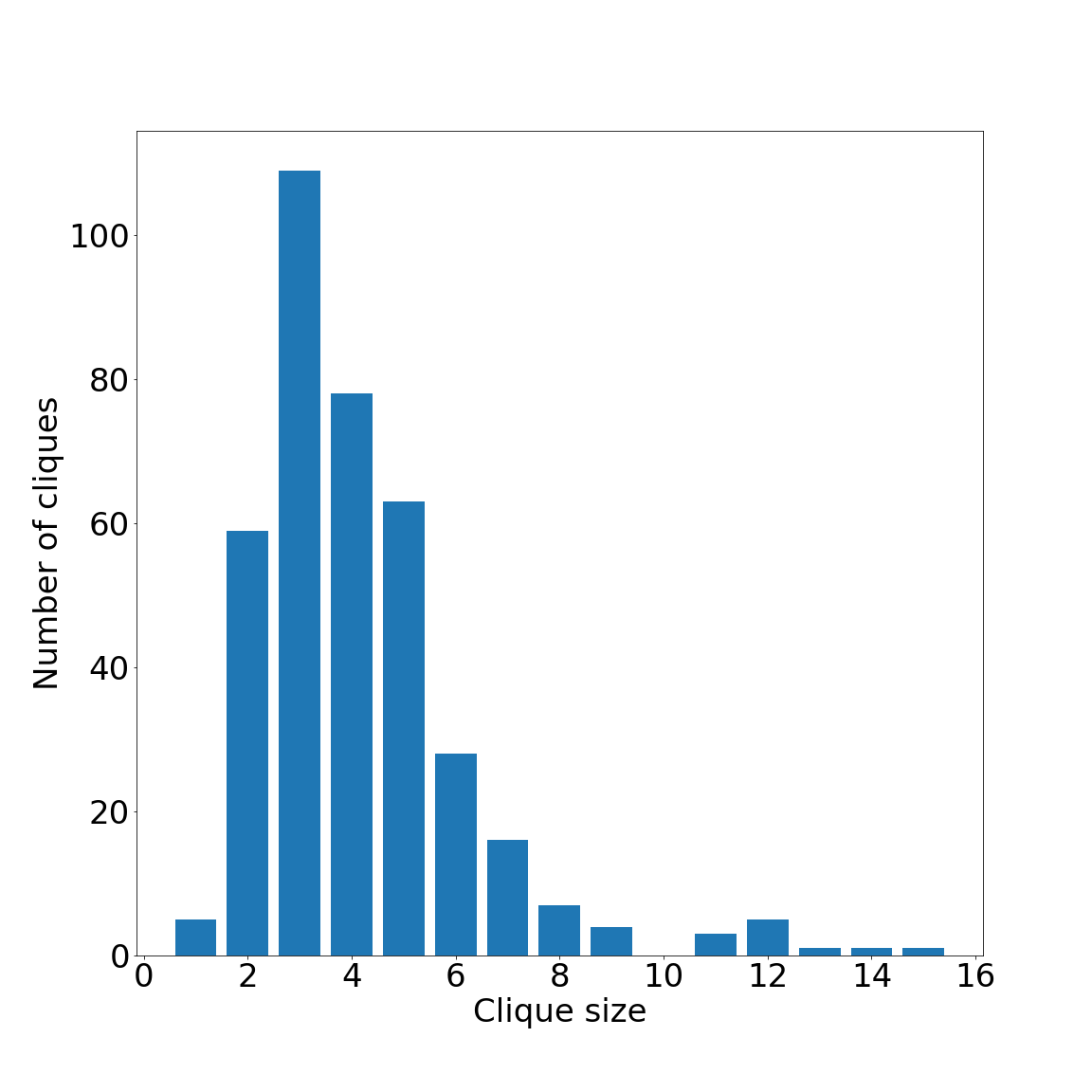}
		\caption{A bar plot depicting the size and number of cliques within the workflow allocation community.}
		\label{fig:cliques-workflow-allocation-community}
	\end{subfigure}
	\caption{An overview of the workflow allocation community.}\label{fig:workflow-allocation-community}
\end{figure}

\begin{figure}
	\centering
	\begin{subfigure}[t]{0.66\textwidth}
		\includegraphics[width=\textwidth,keepaspectratio]{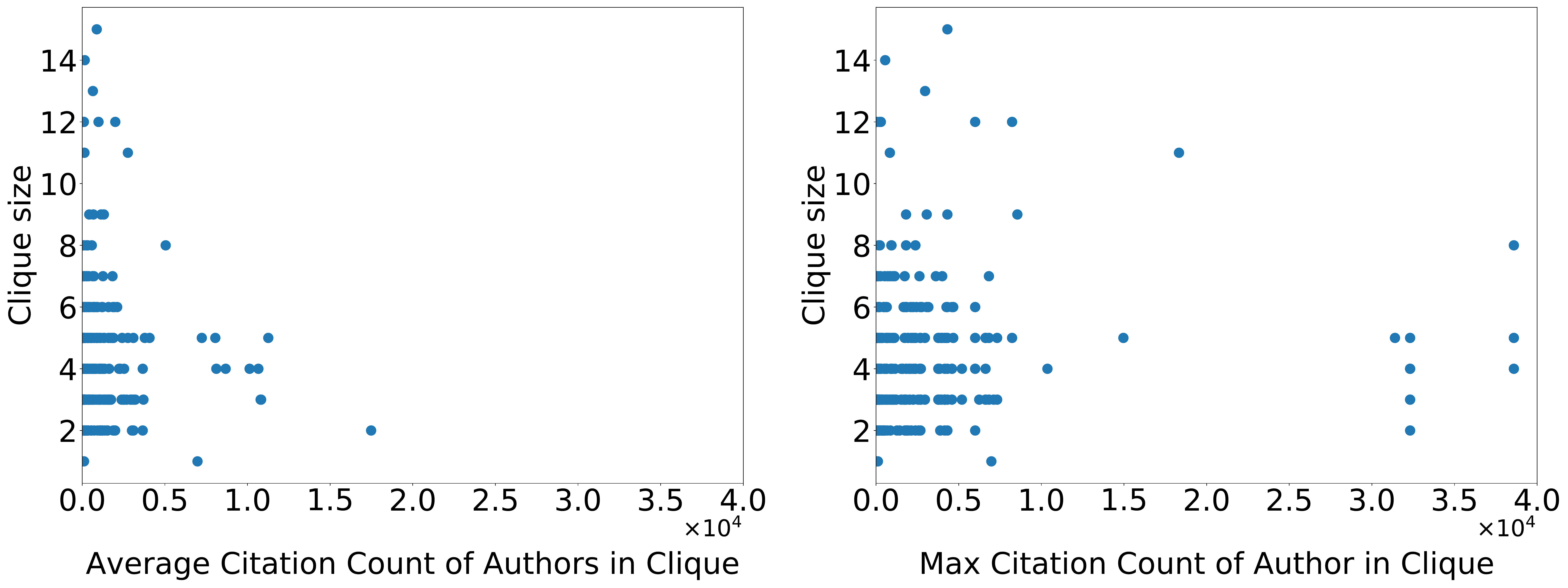}
		\caption{A scatterplot depicting the average (left) and maximum (right) citations of the authors for each clique in the workflow allocation community.}
		\label{fig:citations-vs-clique-size-workflow-allocation-community}
	\end{subfigure}~
	\begin{subfigure}[t]{0.33\textwidth}
		\includegraphics[width=\textwidth,keepaspectratio]{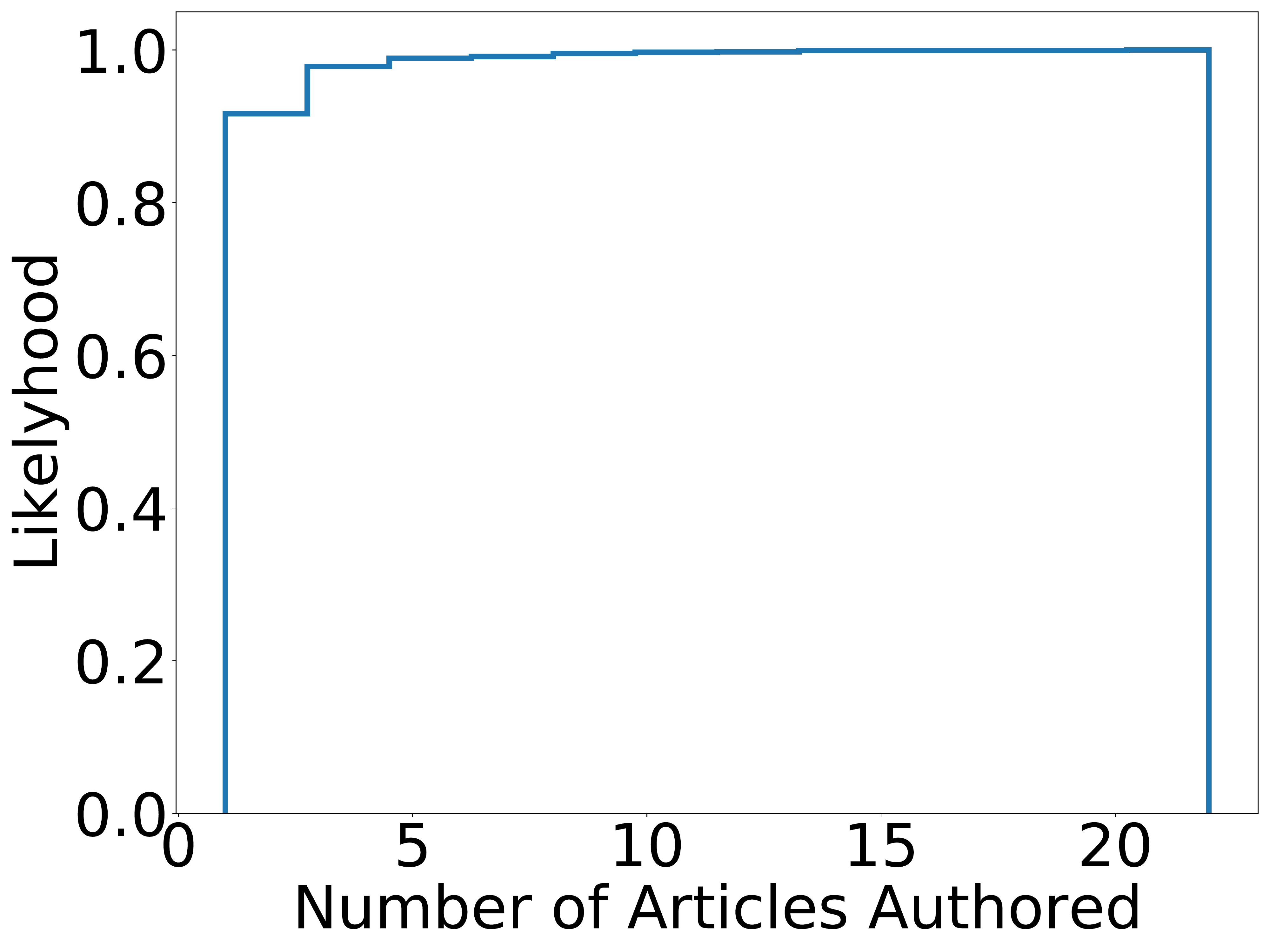}
		\caption{A CDF of number of articles authored per author in the workflow allocation community.}
		\label{fig:number-articles-co-authored-workflow-allocation-community}
	\end{subfigure}
	\caption{Clique size vs. author citation count (average and max) and a CDF of number of articles authored.}
\end{figure}

\begin{figure}[thb]
	\includegraphics[max width=\linewidth]{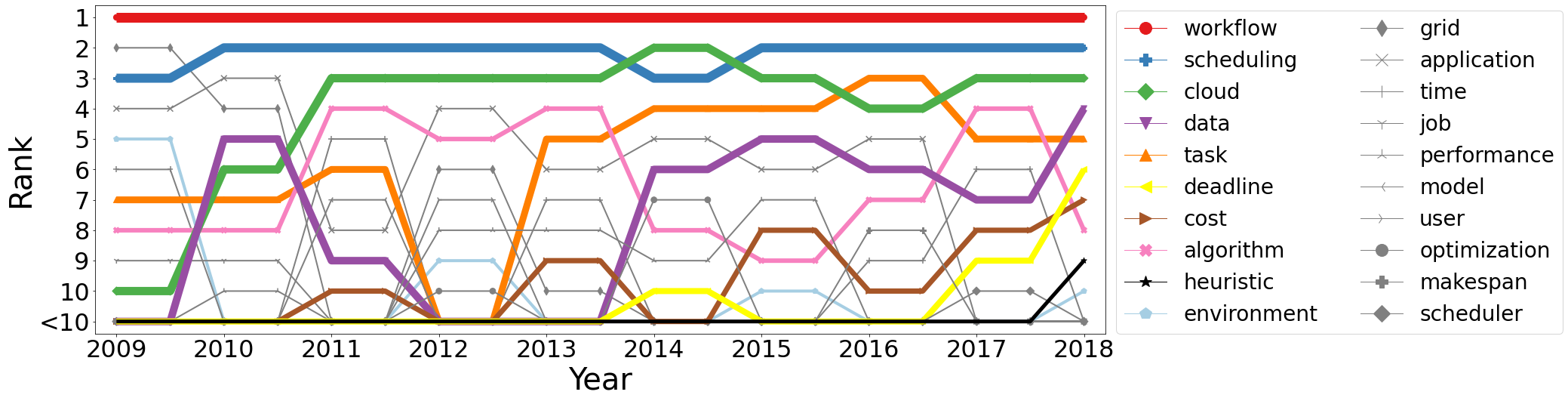}
	\caption{Top-10 keywords in workflow allocation articles in the past decade per year.}
	\label{fig:top-10-workflow-allocation-per-year}
	\vcutS
\end{figure}

Workflow allocation is the process of placing the workflows onto available resources by analyzing its structure, i.e., its tasks and dependencies, and putting the tasks on the available resources in such a way that the scheduling target(s) is(are) being achieved (see Section~\ref{ssct:optimization_target}) while not violating any constraint.
The placement of workflows is done (preferably) as close to optimal as possible, with respect to the target(s).
To achieve this, a workflow scheduler needs to take into consideration both global and/or local constraints and focus on a single or multiple criteria, see Section~\ref{ssct:optimization-strategy}.

In this section, we focus on the diverse sub-parts of workflow allocation, see the taxonomy in Figure~\ref{tax:workflow-allocation}.
Each of the sub-parts will be discussed with their respective sub-taxonomies.

\executedqueries{query-workflow-allocation-articles}{SELECT * FROM publications WHERE year BETWEEN 2009 AND 2018 AND (lower(title) LIKE '\%workflow\%' OR lower(abstract) LIKE '\%workflow\%') AND (lower(title) LIKE '\%schedul\%' OR lower(abstract) LIKE '\%schedul\%' OR lower(title) LIKE '\%plan\%' OR lower(abstract) LIKE '\%plan\%' OR lower(title) LIKE '\%allocat\%' OR lower(abstract) LIKE '\%allocat\%')}

\refquery{query-workflow-allocation-articles} is used to find articles related to workflow allocating, which in turn are used to verify the completeness of our taxonomies regarding workflow allocation.
Performing such systematic literature search to verify completeness adds a novel attribute to our work.

\subsection{Community Analysis}
\label{ssct:workflow-allocation-community-keyword-analysis}
\begin{description}
	\observation{workflow-allocation-community-many-relations-one-time}{The workflow allocation community is reasonably big. Many relationships are one-time. Plenty of authors exists that \enquote{bridge} two groups, i.e., being the single link between two connected components.}
	\observation{workflow-allocation-deadline-important}{\enquote{Deadline} and \enquote{cost} are emerging and important topics within the workflow allocation community. We observed the same for workflow scheduling articles, and as allocation is more common focus than resource provisioning (based on number of articles and community sizes), this result makes sense.}
	\observation{allocation-larger-cliques-smaller-avg}{Similar to the workflow formalism community, larger cliques tend to have a lower average citation count among the authors. Different from the workflow formalism community, the maxima can be found in both larger and smaller sized cliques.}
	\observation{allocation-mostly-singly-publication}{Over 90\% of authors author a single article, yet a small number authored up to twenty papers in the timespan 2009-2018.}
\end{description}

\begin{table}[]
\caption{An overview of the workflow allocation community in numbers.}
\label{tbl:workflow-allocation-community-overview}
\begin{tabular}{@{}rrrrrr@{}}
\toprule
Articles & Authors & Co-authorship Relations & Unique Relations & Cliques & Largest Clique \\ \midrule
1800 & 1273     & 3603      & 3177              & 380      & 15             \\ \bottomrule
\end{tabular}
\end{table}

We summarize the main characteristics of the community in Table~\ref{tbl:workflow-allocation-community-overview}.
We observe the community has a reasonable amount of articles, along with plenty of authors.
Similar to the workflow formalism community, a lot of the co-author relationships are one-time.
Out of the 3603 co-author relationships, 3177 are seen once (88.2\%).

The community structure outlined in Figure~\ref{fig:overview-workflow-allocation-community} depicts several large connected components, depicting several individuals that collaborate with multiple groups, forming \enquote{bridges} between them.
Observing Figure~\ref{fig:cliques-workflow-allocation-community}, most authors are in a clique of size two to five.
This generally points to small groups working on articles, perhaps from the same institute. 
The figure also shows authors that only publish by themselves are rare.

In Figure~\ref{fig:citations-vs-clique-size-workflow-allocation-community} we plot the average and maximum number of citations per clique in the workflow allocation community.
Similar to the workflow formalism community, we observe small and moderately sized cliques have the highest citation count on average.
If we look at the maxima per clique, we observe more outliers where moderately sized cliques work together with a well-cited author.
Similarly, as the clique size grows, both the average and maximum drop significantly compared to moderately sized and smaller cliques, yet smaller than observed in the workflow formalism community.

Figure~\ref{fig:number-articles-co-authored-workflow-allocation-community} presents a CDF of the number of articles that get published within this community per author.
From this figure we observe roughly 90\% of the authors in the workflow formalism community author only once.
Contrary to the workflow formalism community, some authors are quite active, having authored up to 20 articles in the 10 years we investigate.

\subsubsection{Analysis of Trends over Time}
We use the method detailed in Section~\ref{ssct:trends-over-time-workflow-scheduling} combined with the output of \refquery{query-workflow-allocation-articles}.
The output of the trend analysis is visible in Figure~\ref{fig:top-10-workflow-allocation-per-year}.
From this figure, we observe \enquote{workflow}, \enquote{scheduling}, and \enquote{cloud} are frequently in the top-3, highlighting the importance of these keywords to the workflow allocation community.
We also see that \enquote{task}, \enquote{data}, and \enquote{algorithm} are important keywords.
These keywords makes sense, as workflow allocation articles often introduce algorithms that schedule per task.
Data is one of the most common targets as it is inherently tied to the payment model of clouds and data transfers can impact the runtime of a workflow significantly.
Other keywords such as \enquote{deadline}, \enquote{cost}, and \enquote{heuristic} are also expected, as a lot of allocation policies focus on cost and/or deadlines.
Heuristics are commonly applied by this community due to the complexity of workflow allocation.
The keywords outside the top-10 in 2018 are also unsurprising; grids used to be popular before the cloud-era and the other keywords are commonly found in workflow allocation articles.

\subsubsection{Emerging Trends}
We use the method detailed in Section~\ref{ssct:emerging-trends} combined with the output of \refquery{query-workflow-allocation-articles}.
We find the following \textit{new keywords}.
\begin{tcolorbox}[colback=blue!15, colframe=blue!90!black,enhanced,sharp corners,boxsep=-1mm]%
	\enquote{cost}, \enquote{deadline}, \enquote{optimization}
\end{tcolorbox}

It is somewhat surprising the keywords \enquote{cost} and \enquote{deadline} do not appear in the top-10 in the first five years of the decade we inspect.
Both keywords were in the top-10 for workflow scheduling articles, as observed in Section~\ref{ssct:trends-over-time-workflow-scheduling}.
However, it appears that for workflow allocation articles these keywords became more popular later.

If we look at the \textit{rising keywords}, we again obtain these two keywords:
\begin{tcolorbox}[colback=blue!15, colframe=blue!90!black,enhanced,sharp corners,boxsep=-1mm]%
\enquote{cost}, \enquote{deadline}, \enquote{workflow}
\end{tcolorbox}

This indicates that these \glspl{nfr} are deemed particularly important as of late.
The fact that \enquote{workflow} also comes up, highlights the growing importance of this type of workload.

\subsection{Taxonomy of Scheduling Targets}
\label{ssct:optimization_target}

\begin{figure}[htb]
	\centering
	\adjustbox{max width=.85\linewidth}{
		\includegraphics{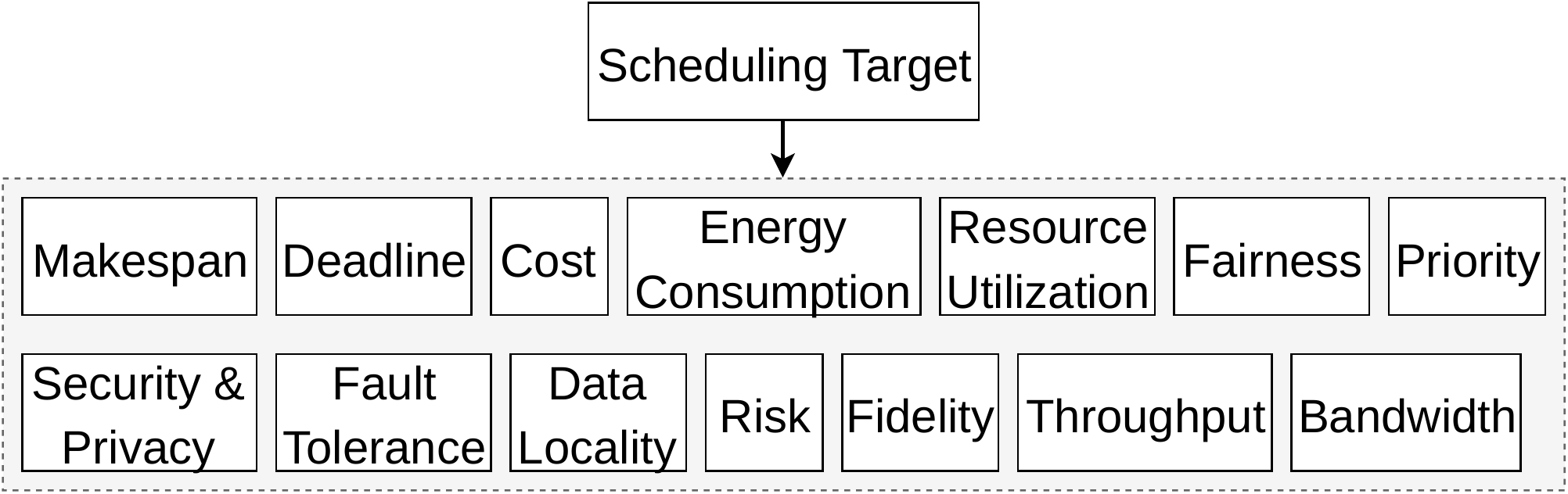}
	}
	\caption{The optimization target taxonomy. The grey box present a collection and does not impose any difference between the elements.} \label{tax:optimization_target}
	\vcutM
\end{figure}

In this section we cover the optimization metrics described in Taxonomy~\ref{tax:optimization_target}, which significantly extends the taxonomy of Yu et al.~\cite{DBLP:journals/sigmod/YuB05}.
There are likely more optimization targets that policies target, yet the taxonomy discussed here covers a significant portion of them.

\subsubsection{Makespan}
\textit{Makespan} (or \textit{runtime}) is a common targeted metric when scheduling jobs.
Makespan is the total time elapsed between the start and finish of the entire job. 
Several techniques have been used to minimize the makespan of jobs, including Particle Swarm optimization~\cite{zhan2012improved}, simulated annealing~\cite{yu2008workflow}, and min-cut/max-flow~\cite{DBLP:conf/osdi/GogSGWH16}.
Dealing with latency sensitive applications also may require low makespans.
An example of such a system is provided by Bonvin et al.~\cite{DBLP:conf/ccgrid/BonvinPA11}.

\subsubsection{Deadline}
Related to makespan as scheduling target, deadlines are a more strict and may require different decisions of a scheduling system.
Deadlines cover the total turnaround time, which is composed of wait time(s), makespan, and latency of submitting and obtaining a response~\cite{abrishami2013deadline}.

\subsubsection{Costs}
Another common, yet important target for optimizing is costs.
Cloud providers offer a pay-as-you-go model for leasing resources.
Traditionally billing would be on an hourly basis, however, several cloud providers have moved towards a second-based billing granularity~\cite{barr_2018, peterson_2017}.

Cost is closely related to resource utilization (see Section~\ref{sssct:resource_utilization}).
For example, autoscalers already are concerned with costs since resources pricing schemes differ per cloud provider.

Alkhanak et al.~\cite{alkhanak2015cost} provide an extensive overview and taxonomy of cost-aware approaches of workflow scheduling in cloud environments.

\subsubsection{Energy consumption}
\label{sssct:energy_consumption}

With the growing importance of green computing, energy-aware scheduling is emerging, with new approaches and techniques being introduced.
In 2014, datacenters already accounted for 2\% of energy consumption in the US~\cite{datacenterknowledgeenergy}.
Datacenter operators are focusing on becoming energy-neutral, including Amazon~\cite{physamazon}, Google~\cite{google2019renewable}, and Microsoft~\cite{microsoft2019with}.
Especially in high performance computing, the number of flops per watt has become increasingly important\footnote{Keynote CCGrid 2018.}.

Articles in this domain focus on least-loaded machines~\cite{li2011energy}, trade-offs between makespan and energy efficiency~\cite{durillo2013multi}, Pareto-based scheduling~\cite{durillo2014multi}, dynamic voltage and frequency scaling~\cite{pietri2014energy}, power minimization in networks and protocols, and self-adaptive systems~\cite{berl2010energy}.
These techniques are sometimes combined as demonstrated in~\cite{DBLP:conf/ipps/BenoitRR10}.

\subsubsection{Resource utilization}
\label{sssct:resource_utilization}

Resource utilization denotes the efficient use of allocated resources.
With the growth of cloud popularity, this metric is becoming increasingly important for cloud operators.
Resource utilization levels of 70\% are possible in domains such as super computing~\cite{DBLP:conf/ipps/JonesN99}, yet the utilization of clouds is as low as 6-12\% are reported~\cite{DBLP:conf/hpca/VasanSSSS10, DBLP:conf/debs/HeinzePJF14, DBLP:journals/fgcs/DoughertyWS12}.

Cloud providers employ autoscalers (i.e. provisioning policies) to automatically scale resources based on the resource demand of the client. 
Autoscalers minimize under- and overprovisioning to improve resource utilization while not violating any \gls{qos} of the client.
Especially when facing challenging, e.g., bursty or unpredictable workloads, autoscalers tend to perform differently~\cite{DBLP:journals/tompecs/IlyushkinAHBPEI18, DBLP:conf/ccgrid/VersluisNI18}.
The interplay between allocation and provisioning then becomes increasingly important to make sure resources are utilized properly.

\subsubsection{Fairness}

The notion of fairness can have multiple definitions.%
Fairness can relate to an equal share of resources~\cite{DBLP:journals/grid/BittencourtM10}, sharing resources~\cite{ghodsi2011dominant}, equal slowdown~\cite{DBLP:conf/ipps/ZhaoS06}, fair use in multi-resources with placement constraints~\cite{DBLP:conf/sc/WangLLL16}, and slowdown~\cite{genez2016flexible}.

Quang et al.~\cite{DBLP:conf/ccgrid/QuangKRKKHMB15} present a comparative analysis of two scheduling mechanisms for virtual screening workflows sharing the same infrastructure.
They focus on fairness, overall system throughput, and response time.

\subsubsection{(User-defined) Priority}
Some workflow execution systems support (user-defined) priorities.
Deng et al. mention the use of priority-driven execution in their scheme~\cite{deng1997scheme}.
An example of a policy taking priorities into account is PISA~\cite{wu2012priority}.
PISA differentiates between user priority levels e.g. free-tier and pay-tier users when scheduling.
The level of priority determines the speed at which the user will be assigned the required resources when resource contention occurs.
In the cluster traces released by Google, priority is also exposed as one of the field used to schedule~\cite{clusterdata:Wilkes2011}.

\subsubsection{Security \& Privacy}

As public clouds are freely accessible by design, for some applications security is a desirable goal.
Trust-based scheduling and result verification are necessary in these situations.
Proposed solutions range include quiz systems~\cite{Zhao2004ResultVA}, risk rate constraints~\cite{li2016security}, secure key sharing and fine-grained access control~\cite{DBLP:journals/tpds/ZhuJ16}.

Shishido et al.~\cite{shishido2018cloudsim} propose an extension to measure security overhead in CloudSim.

Following the recently enforced GDPR legislation in the European Union (EU), processing and storing data of EU citizens must happen on systems located in Europe. 
\glspl{dpia} methods are employed to identify and risks and rights of entity regarding data~\cite{alnemr2015data}. 
Countries such as Russia have similar legislation~\cite{lawamendments}.

\subsubsection{Fault tolerance}

Fault tolerance is of vital importance when running business critical applications and required at several levels when running workflows.
Replication, preemption, and checkpointing are common techniques to fault tolerance when executing tasks in datacenters.

Fault-tolerance has been investigated when using spot and on-demand instances~\cite{poola2014fault}, using task replication and other resubmission techniques~\cite{ramakrishnan2009vgrads, jayadivya2012fault, zhu2016fault}, challenges and tools~\cite{bala2012fault}.

A survey on the topic of fault-tolerance and taxonomies is provided by Poola et al.~\cite{poola2017taxonomy}.

\subsubsection{Data locality}
With IO-intensive workflows, data locality can reduce the runtime and cost of workflows.
It is especially important when sending data to and from cloud environments. Typically, sending data within the same cluster is free, yet communication to and from the cluster is not.
Being data locality aware may also help in reducing costs by not having to send data.

Especially in the Map-Reduce domain, IO-intensive workflows are common.
A well-know article on this topic is Xie et al.~\cite{xie2010improving} who introduce a data placement scheme for MapReduce applications running on heterogeneous nodes.
Articles such as~\cite{guo2012investigation} and~\cite{wang2016maptask} also focus on data-locality in MapReduce applications.
The study by Wang et al.~\cite{wang2016maptask} has similarities to e.g. Duro et al.~\cite{duro2014exploiting} who focus on the trade-off between data-locality and load balancing when executing generic workflow applications.
More articles on workflow scheduling with data locality exist, e.g.~\cite{choi2017data} and~\cite{tanaka2012workflow}.

\subsubsection{Risk}

Risk relates to allowing resource contention within acceptable bounds to reduce costs whilst still meeting the \gls{qos} requirements set by the customers.
Van Beek et al.~\cite{vancpu} describe risk based on CPU contention while running business-critical workloads.
While focusing primarily on security, Li et al.~\cite{li2016security} compute the risk rate proportional to the security levels and the distribution of risk to judge if it is within bounds.

\subsubsection{Fidelity}

Fidelity relates to the quality of output of a workflow~\cite{DBLP:journals/sigmod/YuB05}.
Cardoso et al. refer to fidelity as a function of effective design and in intrinsic property or characteristic of a good produced or service rendered~\cite{cardoso2002workflow}.
Video streaming is a good example where fidelity versus computation power can be a trade-off.
Another example is using dynamic voltage and frequency scaling to trade-off quality with power consumption~\cite{stavrinides2019energy}.

\subsubsection{Throughput}
Throughput focuses on completing as many tasks in an as short as possible timespan.
Different from focusing on makespan, throughput related approaches may not attempt to speed up the duration of tasks themselves by running them on, e.g., special hardware.
Simply running more tasks in parallel could already be an feasible approach to improve throughput.

\subsubsection{Bandwidth}

Related to data locality, yet different, bandwidth can be a target as well.
Some scheduling strategies involve messaging between components, where reducing bandwidth becomes important.
An example of such work is that of Momenzadeh et al.~\cite{momenzadeh2019workflow}.
Their work focuses on workflow segmentation to execute workflows on multiple \glspl{vm}. 
Bandwidth, due to the message communication, becomes an important metric in such systems.

\subsection{Taxonomy of Optimization Strategies}
\label{ssct:optimization-strategy}

\begin{figure}[htb]
	\adjustbox{max width=.9\linewidth}{
		\includegraphics{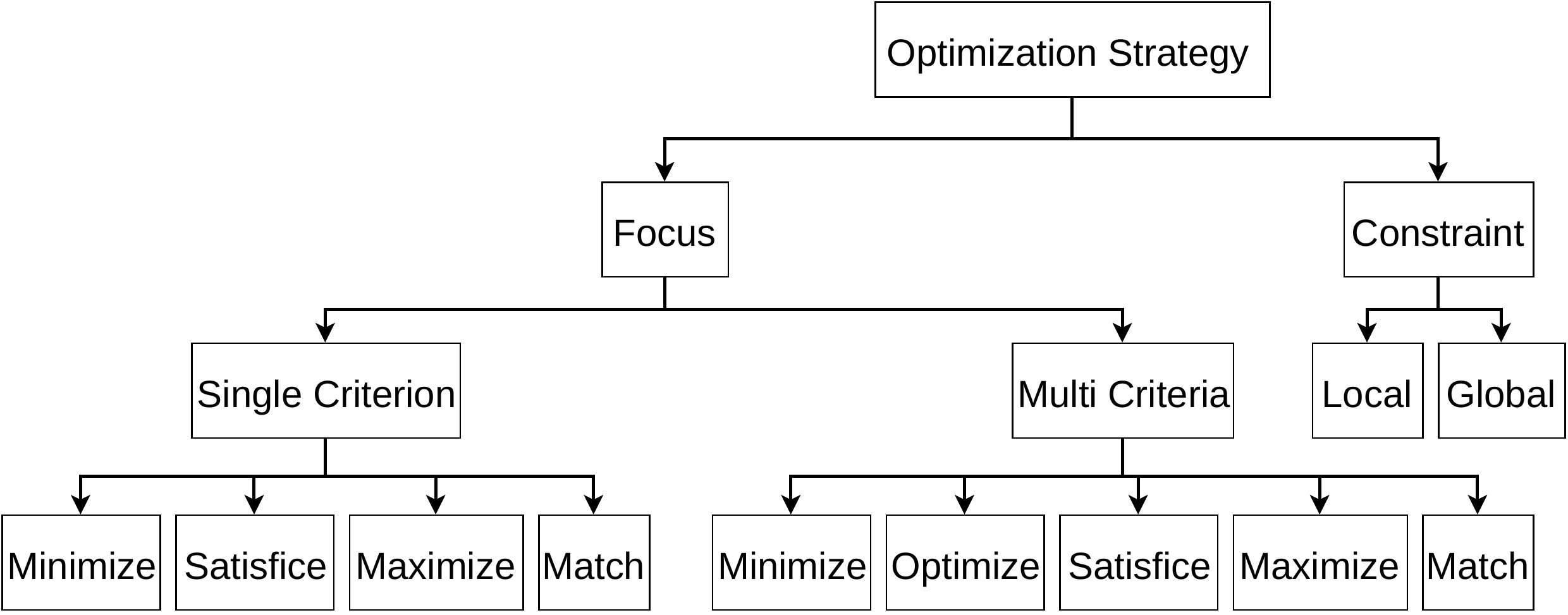}
	}
	\caption{Optimization strategy taxonomy.} \label{tax:optimization_strategy}
	\vcutS
\end{figure}

The optimization strategy of policies varies in both \textit{focus} and \textit{constraint}.
Figure~\ref{tax:optimization_strategy} presents the taxonomy for optimization strategies.
The focus can be on a single criterion or multiple criteria.
Popular single criterion for minimization are cost~\cite{bittencourt2011hcoc} and makespan~\cite{genez2016flexible}.
Examples where policies must match a specific value can be found for example in Galaxy's workflow scheduler having to match tool versions~\cite{goecks2010galaxy}, and \gls{faas} instances relying on specific library versions~\cite{abad2018package}.
Maximization policies focus on e.g. throughput \cite{liu2008throughput} or fairness~\cite{genez2016flexible}.
Satisficing is about generating \enquote{good enough} solutions.
The term was introduced by Herbert A. Simon~\cite{simon1956rational}.
Jaeger et al.~\cite{jaeger1994framework} use satisficing to modify business information system models using AI techniques.
Zhang et al.~\cite{DBLP:conf/cloudcom/ZhangCHW11} use iterative ordinal optimization to schedule scientific workflows in elastic cloud computing environments that satisfice the problem. 

Multi-criteria policies consider multiple metrics at once.
Similar to single criterion policies, multi criteria policies can both maximize, minimize and match certain criteria.
New to multi-criteria is optimize.
With optimize, two or more metrics are trade-off to create an overall better outcome.
A popular combination of criteria to optimize is cost while meeting deadlines such as~\cite{yu2006scheduling}.
Recently, energy-aware workflow scheduling is also becoming more prevalent, see e.g.~\cite{goiri2011greenslot, yassa2013multi}.

Constraints can be both local and global.
Local constraints focus on the single resource environment, whereas global constraints can span many.
Policies can use a mixture of both.

\subsection{Taxonomy of Scheduler Structures}

\begin{figure}[htb]
	\adjustbox{max width=\linewidth}{
		\includegraphics{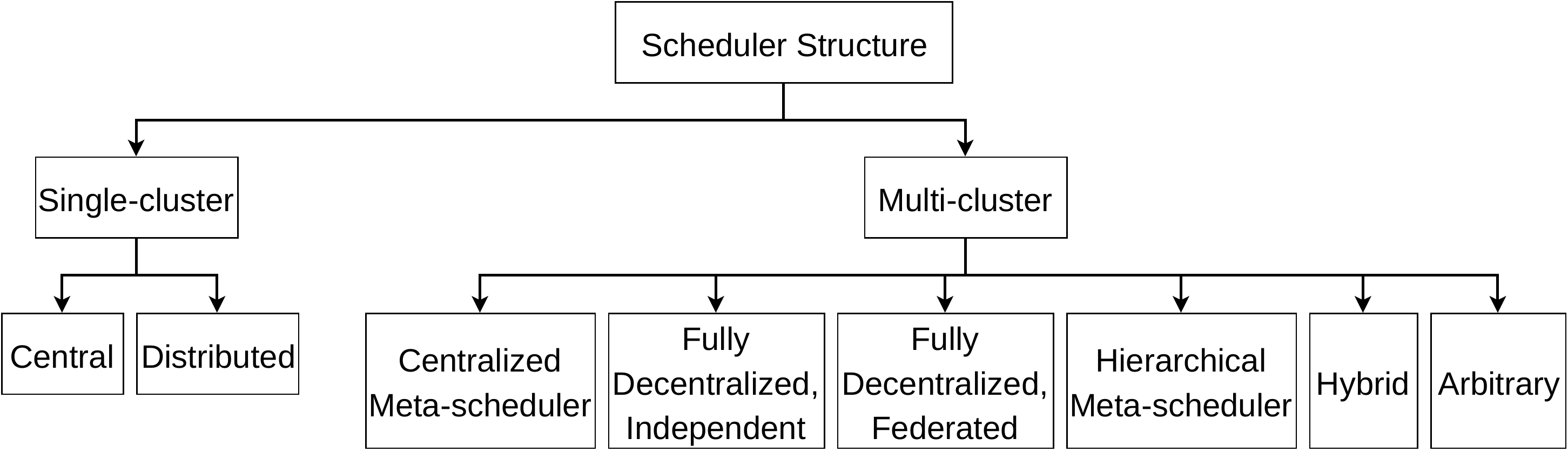}
	}
	\caption{Taxonomy of scheduler structures.} \label{tax:scheduler_structure}
	\vcutM
\end{figure}

In the past decades, several scheduling \textit{structures} have been proposed.
Figure~\ref{tax:scheduler_structure} presents the taxonomy of different structures.
Often, scheduler structures are divided into centralized, decentralized, and hierarchical structures, which is quite coarse grained.

The first differentiation is to be made between single-cluster and multi-cluster architectures.
Next, for each of these levels, we focus on multiple different structures within these levels.
This extends significantly the characterization done by of Moghaddam et al.~\cite{DBLP:journals/csur/Kardani-Moghaddam19}.

\subsubsection{Single-cluster Architectures}
For a single-cluster scheduling architecture, one can use either a centralized or a decentralized architecture in which the bootstrapping problem is solved centrally.
Both of these architectures have their trade-offs.

A centralized scheduler, sometimes called a headnode, acts as a coordinator and keeps track of the global state.
Using such architecture makes it easier to manage resource yet may become a bottleneck or single point of failure.
An example of a centralized single cluster scheduling architecture is that of Kubernetes.
Centralized systems are generally monoliths and can evolve into sophisticated systems that are hard to change, e.g., the situation at Google prior to Omega~\cite{schwarzkopf2013omega}.

In a decentralized architecture, schedulers are responsible for a part of the resources.
A decentralized scheduler is more resilient to failures, yet keeping a global state is more difficult and comes at the cost of having to communicate between the schedulers.
HTCondor is an example of a distributed scheduler in which a central matchmaker delegates the work to nodes.
Due to this centralized component, the bootstrapping problem is avoided.

\subsubsection{Multi-cluster Architectures}
With a centralized meta-scheduler, a centralized headnode sends jobs to different clusters.
Using such architecture makes it easier to manage resource yet may become a bottleneck or single point of failure.
Firmament~\cite{DBLP:conf/osdi/GogSGWH16} is an example of a centralized workflow scheduler that is fast, even at large scale.

In a fully decentralized setting, costumers can check compare each cluster and independently submit to these clusters (observational scheduling).
When load-sharing happens, this can turn into a fully decentralized, federated architecture, for example, OurGrid~\cite{DBLP:conf/jsspp/AndradeCBR03}.

Hierarchical architectures attempt to combine some of the benefits centralized and decentralized architectures offer.
In a hierarchical architecture, tasks often pass multiple schedulers in a layer fashion~\cite{DBLP:conf/rtss/DengL97}.
Examples of hierarchical schedulers are PUNCH, CCS, and Moab/Torque.
PUNCH and CSS were one of the first tools to use a hierarchical architecture for scheduling in large-scale distributed environments with CSS being able to operate both in cluster and supercomputer environments~\cite{iosup2009framework}.
Moab/Torque is a commercial scheduler that to date is still used in distributed environments.
For example, the Los Alamos National Laboratory uses Moab, as can be observed from recent workflow traces that originate from this lab~\cite{DBLP:conf/usenix/AmvrosiadisPGGB18}.

With delegated matchmaking, instead of performing load-balancing by sending jobs to other clusters, clusters resources usage rights are delegated, rather than jobs~\cite{DBLP:journals/sp/IosupTFEL08}.
Questions regarding who controls the delegation and notions of fairness arise in such architectures.

Shared-state scheduling is a special kind of fully decentralized, federated approach.
It involves a situation in which multiple schedulers have an overview of and can manage and lay claim to \textit{all} resources in a given environment.
To schedule using this structure, you need to provide a unified notion of when resource allocations are permitted and a notion of precedence (who wins when competing). 
The Omega scheduler~\cite{schwarzkopf2013omega} is a well-known example of using shared-state scheduling.

Other arbitrary architectures includes architectures such as by a TAGS-based policy in which clusters serve different job depending on their runtime.
This is implemented in KOALA-C in where jobs that run too long are preempted and moved to another queue for execution.

\subsection{Taxonomy of Allocation Techniques}
\label{ssct:technique}

\begin{figure}[htb]
	\adjustbox{max width=\linewidth}{
		\includegraphics{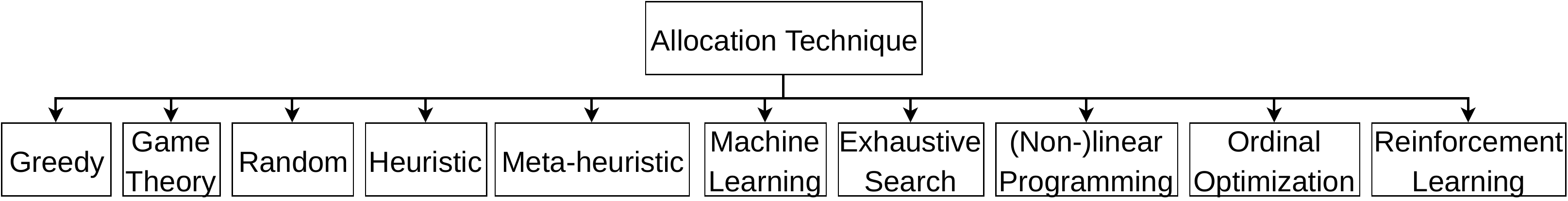}
	}
	\caption{Allocation policy technique taxonomy.} \label{tax:task-placement-technique}
	\vcutM
\end{figure}

The \emph{technique} used by a scheduler determines how the schedule is created.
It has been shown that computing the optimal schedule is an NP-hard problem.
Computing the optimal schedule is therefore infeasible in terms of time, especially with the ever-increasing dynamic workload.

To this end, many different techniques have been proposed to generate an optimal or near optimal schedule in feasible time.
In this section we will discuss various techniques employed for task placement.
Figure~\ref{tax:task-placement-technique} presents the taxonomy of task placement techniques.

\subsubsection{Greedy}
\label{sssct:greedy}

Selecting jobs and/or resources greedily sometimes helps in reducing time required to compute a schedule, or push a solution to a local optimum.
Greedy algorithms are often used to generate \enquote{good enough} solutions within a timely manner.

Xiang et al.~\cite{xiang2017greedy} introduce a greedy ant colony optimization algorithm that performs greedy machine allocation with low overhead.
Yu et al.~\cite{yu2009deadline} mention several greedy algorithms for scheduling workflows in grid environments.

\subsubsection{Game Theory}
\label{sssct:game-theory}

Game Theory is another technique employed by schedulers to meet scheduling targets.
The ICENI scheduler uses game theory for scheduling workflows, for example~\cite{mcgough2004workflow}. 
Yaghoobi et al.~\cite{yaghoobi2013non} use a game theory approach for scheduling workflows in grid environments to minimize turnaround time and cost.
Duan et al.~\cite{duan2014multi} introduce a workflow scheduling policy based on game theory that attempts to optimize for both makespan and cost while taking network bandwidth and storage into account. 

\subsubsection{Random}
\label{sssct:random}

Scheduling eligible tasks randomly is often done to obtain a baseline when experimenting.
First-come-first-serve (FIFO) is usually used as an allocation policy in this case.
For example, Wu et al.~\cite{wu2012priority} use the FIFO sequence for a random baseline.

Another random method is lottery ticket scheduling~\cite{waldspurger1994lottery}.
Resources get assigned a certain amount of tickets and for each task, a ticket is drawn at random.
The task is then assigned to the corresponding resource if it fits, else a new ticket is drawn.

\subsubsection{Heuristic}
\label{sssct:heuristic}

Heuristic approaches apply best-effort methods that work well for a given setting, e.g. workload of workflows, or use specific elements from a domain.
Since scheduling tasks is a NP-hard problem, many policies rely on heuristic for task placement decisions.
Examples of specific domain properties related to workflow scheduling can be task runtime, task size, etc.

Examples of policies that use heuristics are SJF~\cite{yu2008workflow} and HEFT~\cite{zhao2003experimental}.

\subsubsection{Meta-heuristic}
\label{sssct:meta-heuristic}

Meta-heuristics are the class of heuristic that are problem independent.
Examples of such meta-heuristics applied for workflow scheduling are ant colony optimization~\cite{DBLP:journals/tsmc/ChenZ09}, cat swarm optimization~\cite{bilgaiyan2014workflow}, Shuffled Frog Leaping~\cite{kaur2017resource}, evolutionary algorithms such as genetic algorithms~\cite{DBLP:conf/ipps/SongKH05}, and simulated annealing~\cite{Scholar:conf/ahm/YoungL2003}.
Wu et al.~\cite{wu2010revised} present a revised particle swarm optimization approach.

\subsubsection{Machine learning}
\label{sssct:neural-network}

Machine learning describes the notion where a system can make decisions based on prior seen (similar) situations.
These systems require prior training before being able to classify new cases.
Due to the agnostic principle of machine learning, it can be used for many different purposes, including scheduling.
With the recent surge in interest for machine learning, approaches for scheduling and autoscaling using this techniqure has increased too.

Vukmirovic et al.~\cite{vukmirovic2012optimal} use artificial neural networks for dynamically executing scheduling algorithms.
Bauer et al. use machine learning for autoscaling multi-tier micro services~\cite{bauer2018chameleon}.

\subsubsection{Exhaustive Search}
\label{sssct:exhaustive-search}

Exhaustive search algorithms compute the optimal planning given a workload and resource settings.
However, as scheduling workflows is NP-hard, in practice such solutions are infeasible.
One of the most famous examples of scheduling is bin packing.
Exhaustive search algorithms are infeasible in practice where real-time decisions need to be taken.

Examples of exhaustive search algorithms are ILAO and COLAO~\cite{malik2017co}.

\subsubsection{(Non-)linear Programming}
(Non-)linear programming can be used to construct mathematical models in which requirements can be specified by (non-)linear relationships.
Such models can then be used to compute the optimal outcomes given the constraints.
(Non-)linear programming has been used to construct workflow schedules, often with \gls{slo} defined as constraints.
Such schedules can be either used for verification or benchmarking purposes (comparison), or to use as scheduler (functionality).

The applicability of (non-)linear programming ranges from executing workflows in grid environments~\cite{afzal2006qos}, multi cloud environments~\cite{genez2013using}, to scheduling pipelined workflows~\cite{benoit2013survey}.

\subsubsection{Ordinal optimization}
\label{ssct:ordinal-optimization}
Ordinal optimization was introduced by Ho et al.~\cite{DBLP:conf/cloudcom/ZhangCHW11} to effectively generate local-optimal solutions (or \enquote{good enough}) to NP-Hard problems.
Zhang et al.~\cite{DBLP:conf/cloudcom/ZhangCHW11} extended ordinal optimization and included an iterative approach to reduce the search space and overhead.
Ordinal optimization has also been used in combination with other techniques.
El-Zarif et al.~\cite{el2012ordinal} use ordinal optimization to improve parameter selection for their genetic algorithm approach. While this work is not in the context of (workflow) scheduling, this approach could be tested since genetic algorithm approaches have been introduced by related work (see Section~\ref{sssct:meta-heuristic}).

\subsubsection{Reinforcement learning}
\label{ssct:reinforcement-learning}

With reinforcement learning, the system contains a feedback loop that tunes parameters according to the feedback obtained.
If implemented correctly, a system containing such a feedback loop will correct itself to changes in environment and workload.
Examples of workflow scheduling using reinforcement learning are Ma et al.~\cite{DBLP:conf/icac/MaISI17} who combine a Q-learning approach to portfolio scheduling, Wang et al.~\cite{wang2019multi} apply a Deep-Q-network in a multi-agent reinforcement learning setting to improve both workflow makespan and cost, 

\subsection{Workflow Instantiation}
\label{ssct:scheduling_planning}
Yu et al. define the notion of abstract and concrete workflows~\cite{DBLP:conf/icppw/YuS08}.
An abstract workflow defines the tasks and their dependencies, yet lacks the detail of where each task will be ran and where the data should be read from and written to, which the concrete workflow entails.
The concrete workflows is therefore an instantiation of the abstract workflow.

The process of instantiation can be done statically or dynamically.
In the static case, concrete workflow plans are generated before execution in accordance to the latest state of the system.
Any dynamic changes in this state are not taken into account.
Dynamic schemes do make use of the dynamics in state as well as static information beforehand to make scheduling decisions at runtime.

User-directed and simulation-based scheduling are common when creating static schemes.
In user-directed scheduling, the consumers themselves emulate the scheduling process and assign resources to tasks, or modify workflows themselves~\cite{DBLP:journals/sigmod/YuB05}.
This process is often done by human experts who rely on their knowledge and can also incorporate preferences and/or other \gls{qos} criteria such as performance or availability.
In simulation based scheduling, the \enquote{best} schedule is picked after simulating the workload on a set of resources according to defined metrics.

\subsubsection{Partitioning Technique}

Other aligned methods arose to instantiate workflows. For example in graph processing where partitioning techniques are applied to the graphs first.
The graph is partitioned based on the graph itself and the algorithm (often workflows) to be applied~\cite{guo2017modeling}.
Similarly to workflow allocation policies managing data locality, the partitioning of the graph determines how the workflow is instantiated and which tasks of the workflows run where and on what data.

\subsubsection{Workflow and Task Optimization}

Several workflow management systems perform optimization steps when having created an concrete workflow.
An example of such a system is Pegasus~\cite{deelman2006pegasus}.
First, the Pegasus Mapper holds the abstract workflow.
The Mapper can e.g. reorder, group or prioritize tasks to improve performance.
This workflow is then passed on to DAGMan, which turns the abstract workflow into a concrete workflow, by e.g. determining where each task will run and where the data will reside.
DAGMan then monitors the execution of the workflow and tracks if task dependencies have been met.
Finally, the HTCondor scheduler executed the workflow on the targeted resources.

\subsection{Future Research Directions Inspired by Meta-Data Analysis}

Cost and deadline-aware scheduling remains an important and growing topic within the workflow allocation community as seen in Section~\ref{ssct:workflow-allocation-community-keyword-analysis}.
Given the recent emerging topics of edge/fog, IoT, and serverless computing, we believe there are plenty of opportunities within this domain.

Metrics such as Risk and Fidelity are less studied in the context of workflow allocation.
We believe especially for business critical workflows, these metrics are important.
More work on these topics, especially in emerging areas as edge/fog etc.   

Another topic we believe will grow in importance is green computing, i.e., a focus on reducing power consumption while adhering to all functional and non-functional requirements.
Sustainable energy sources are already invested in heavily.
With datacenters being major power consumers, and the power consumption is likely to rise, it is worthwhile investing into this topic~\cite{whitney2014data}.

Finally, we believe policies with multiple criteria or objectives will become the norm.
Already, we see many policies focusing on more than one metric (see Section~\ref{sct:workflow-allocation-policies}).

\section{Resource Provisioning}
\label{sct:provisioning}

\begin{figure}
	\centering
	\begin{subfigure}[t]{0.49\textwidth}
		\includegraphics[width=\textwidth]{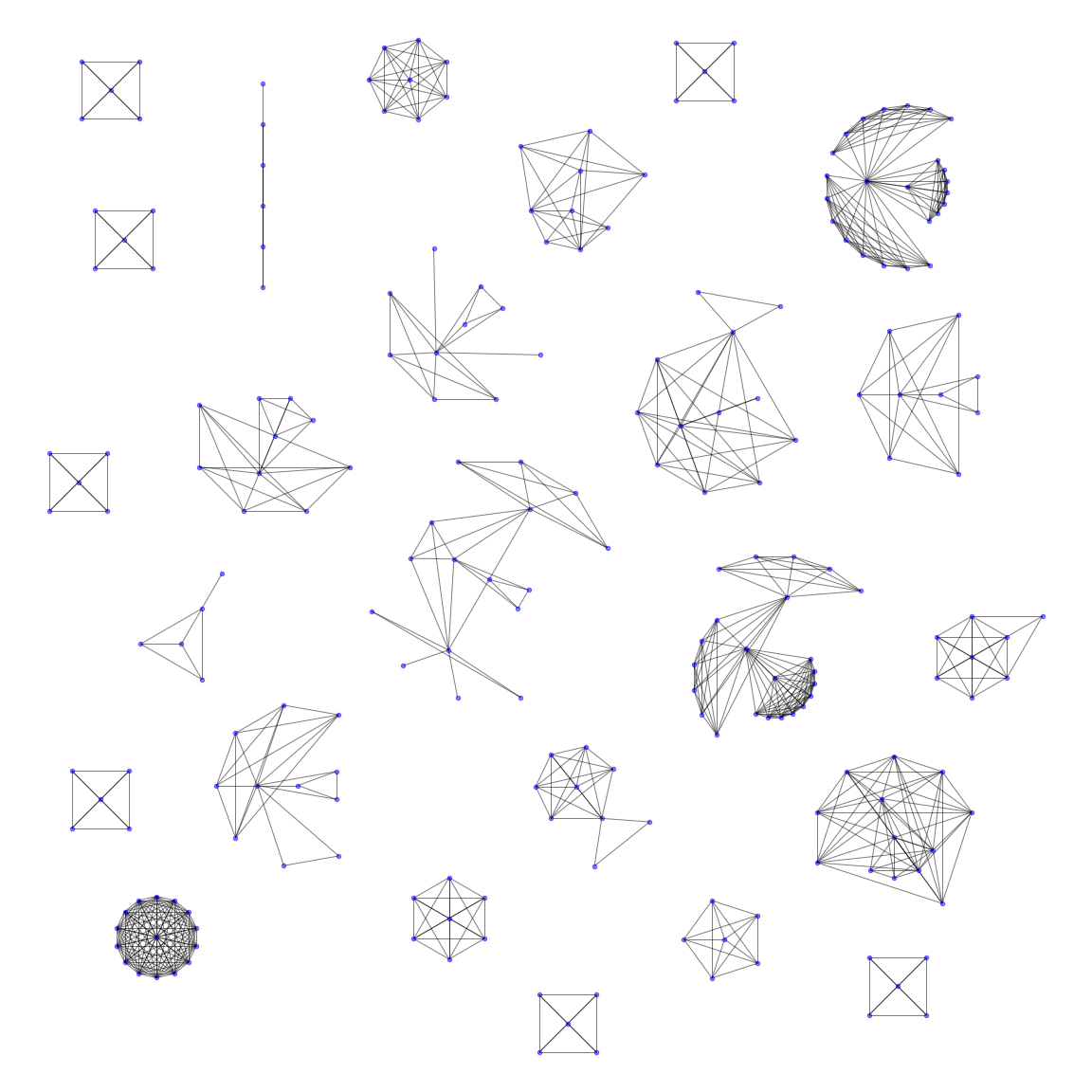}
		\caption{A visual representation of the resource provisioning community. Only components with cardinality 5 or higher are shown.}
		\label{fig:overview-resource-provisioning-community}
	\end{subfigure}~
	\begin{subfigure}[t]{0.49\textwidth}
		\includegraphics[width=\textwidth]{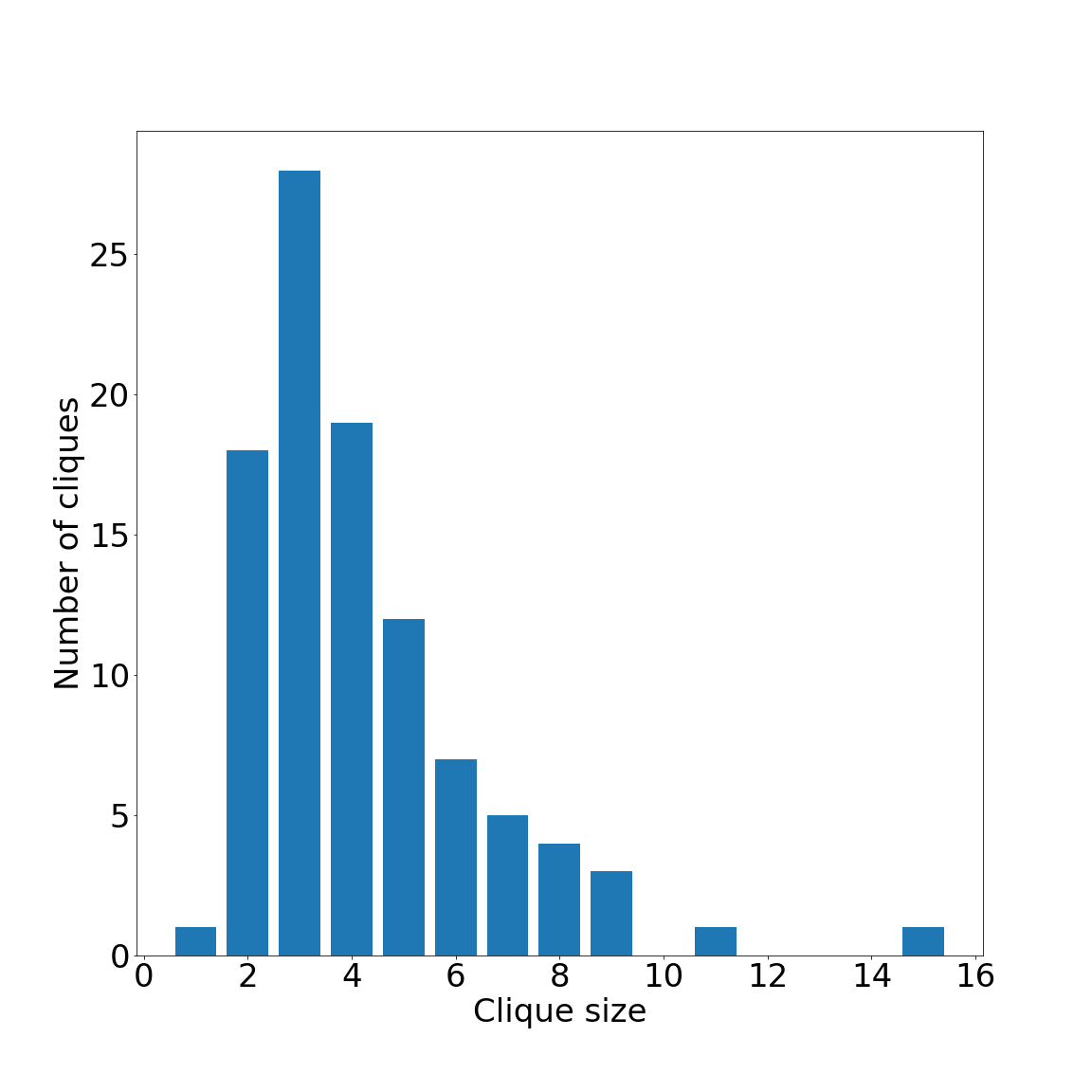}
		\caption{A bar plot depicting the size and number of cliques within the resource provisioning community.}
		\label{fig:cliques-resource-provisioning-community}
	\end{subfigure}
	\caption{An overview of the resource provisioning community.}\label{fig:resource-provisioning-community}
\end{figure}

\begin{figure}
	\centering
	\begin{subfigure}[t]{0.66\textwidth}
		\includegraphics[width=\textwidth,keepaspectratio]{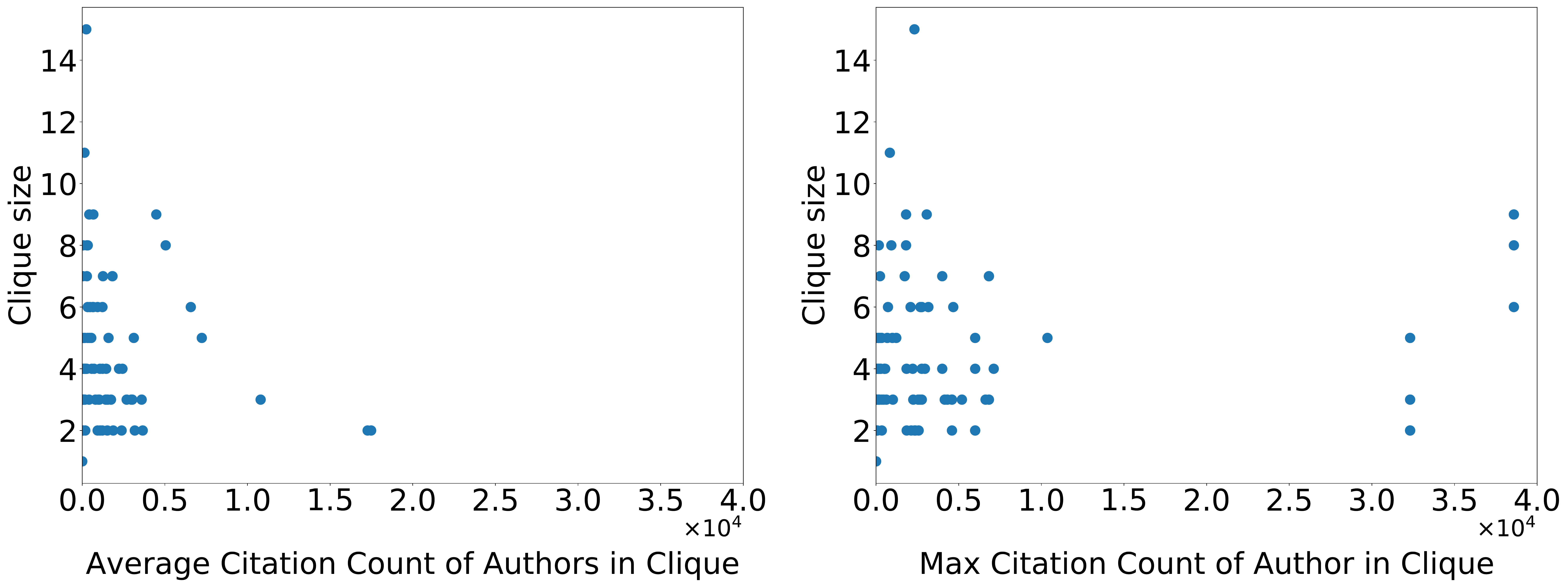}
		\caption{A scatterplot depicting the average (left) and maximum (right) citations of the authors for each clique in the resource provisioning community.}
		\label{fig:citations-vs-clique-size-resource-provisioning-community}
	\end{subfigure}~
	\begin{subfigure}[t]{0.33\textwidth}
		\includegraphics[width=\textwidth,keepaspectratio]{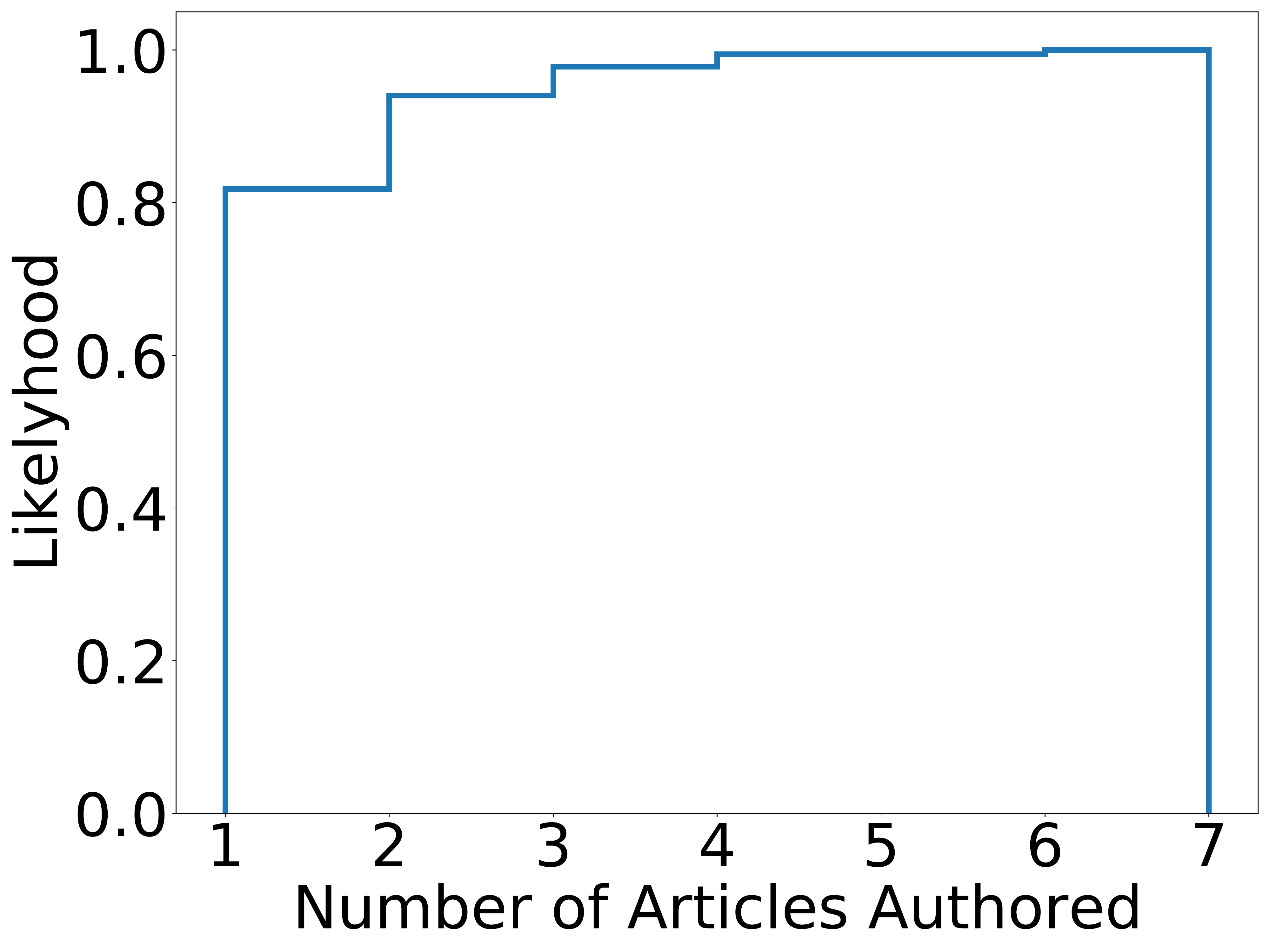}
		\caption{A CDF of number of articles authored per author in the resource provisioning community.}
		\label{fig:number-articles-co-authored-resource-provisioning-community}
	\end{subfigure}
	\caption{Clique size vs. author citation count (average and max) and a CDF of number of articles authored.}
\end{figure}

\begin{figure}[thb]
	\includegraphics[max width=\linewidth]{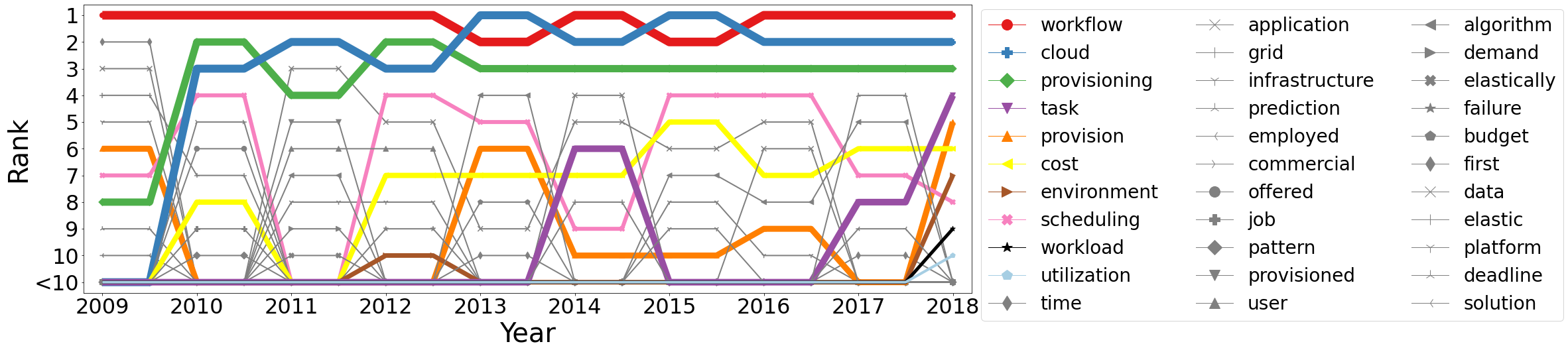}
	\caption{Top-10 keywords in resource provisioning articles in the past decade per year.}
	\label{fig:top-10-resource-provisioning-per-year}
	\vcutS
\end{figure}

In this section, we discuss resource provisioning. 
The scope of this section is limited as this topic deserves a survey itself.
As workflows do play an important role in several autoscalers, we will cover both provisioning in general and autoscalers.

\executedqueries{query-workflow-provisioning-articles}{SELECT * FROM publications WHERE year BETWEEN 2009 AND 2018 AND (lower(title) LIKE '\%workflow\%' OR lower(abstract) LIKE '\%workflow\%') AND (lower(title) LIKE '\%provision\%' OR lower(abstract) LIKE '\%provision\%' OR lower(title) LIKE '\%autoscal\%' OR lower(abstract) LIKE '\%autoscal\%')}

\refquery{query-workflow-provisioning-articles} is used to find articles related to workflow resource provisioning, which in turn are used to verify the completeness of our taxonomies.

\subsection{Community Analysis}
\begin{description}
	\observation{resource-provisioning-community-many-relations-one-time}{The resource provisioning community is relatively small. Many authorship relationships are one-time. The connected components of size greater or equal to five are quite diverse and dynamic.}
	\observation{resource-provisioning-elastic}{Elasticity and the environment seem to be emerging and trending keywords within the resource provisioning community.}
	\observation{provisoning-larger-cliques-smaller-avg}{The average and maximum citation count per clique situation is very similar to that of the workflow allocation community. The only difference is that there are less outliers.}
	\observation{provisioning-mostly-singly-publication}{Similar to the workflow formalism community, around 80\% of authors author a single article. The highest number of co-authored papers by a single author was seven in the year span 2009-2018.}
\end{description}

\begin{table}[]
\caption{An overview of the resource provisioning community in numbers.}
\label{tbl:resource-provisioning-community-overview}
\begin{tabular}{@{}rrrrrr@{}}
\toprule
Articles & Authors & Co-authorship Relations & Unique Relations & Cliques & Largest Clique \\ \midrule
464 & 368     & 965      & 876              & 99      & 15             \\ \bottomrule
\end{tabular}
\end{table}

We summarize the main characteristics of the community in Table~\ref{tbl:resource-provisioning-community-overview}.
From this table, we observe the amount of articles on this topic is similar to that of workflow formalisms; a smaller community.
The number of authors is significantly less than other communities, indicating this community is small.
From the table and Figure~\ref{fig:cliques-resource-provisioning-community}, we observe most authors work in groups of size 2-4, and generally only once as the number of unique relationships is close to the number of total relationships.
Interestingly, the groups that are visualized in Figure~\ref{fig:overview-resource-provisioning-community} do show quite diverse structures, rather than fully connected components, highlighting a dynamic community in which several authors collaborate with different groups.

In Figure~\ref{fig:citations-vs-clique-size-resource-provisioning-community} we plot the average and maximum number of citations per clique in the resource provisioning community.
The pattern is very similar to that of the workflow allocation community. 
The small and moderately sized cliques have the highest citation count on average and at maximum.
This community has the same trend as the previous two investigated where larger cliques tend to have less cited authors both on average and at maximum.

Figure~\ref{fig:number-articles-co-authored-resource-provisioning-community} presents a CDF of the number of articles that get published within the resource provisioning community per author.
From this figure we observe roughly 80\% of the authors in the workflow formalism community author only once, very similar to that of the workflow formalism community.
Another roughly 10\% co-author two papers, and another 5\% three.
The maximum number of articles co-authored by a single author is seven.

\subsubsection{Analysis of Trends over Time}
We use the method detailed in Section~\ref{ssct:trends-over-time-workflow-scheduling} combined with the output of \refquery{query-workflow-provisioning-articles}.
The output of the trend analysis is visible in Figure~\ref{fig:top-10-resource-provisioning-per-year}.
In this figure, we see that \enquote{workflow}, \enquote{cloud}, \enquote{provisioning} are frequently or consistently in the top-3, which makes sense given the selection of articles.
We observe more fluctuating patterns for other keywords.
\enquote{Task}, \enquote{cost}, \enquote{environment}, and \enquote{provision} were expected keywords.
Most autoscalers are task-based and focus on the cost, naturally through the environment that they manage.
The \enquote{workload} and \enquote{utilization} of the system are often important sources of information on which scaling decisions are based.
Most other keywords out of the top-10 in 2018 are also expected, such as \enquote{job}, \enquote{prediction}, and \enquote{elastic}.
We discuss these keywords more in-depth in Section~\ref{ssct:resource-provisioning-taxonomy}.

\subsubsection{Emerging Trends}
We use the method detailed in Section~\ref{ssct:emerging-trends} combined with the output of \refquery{query-workflow-provisioning-articles}.
We find the following \textit{new keywords}.
\begin{tcolorbox}[colback=blue!15, colframe=blue!90!black,enhanced,sharp corners,boxsep=-1mm]%
\enquote{cost}, \enquote{data}, \enquote{deadline}, \enquote{elastic}, \enquote{environment}, \enquote{system}, \enquote{task}
\end{tcolorbox}

The output highlights that in the last five years of the decade we investigate, \enquote{deadline}, \enquote{cost}, and \enquote{elasticity} are emerging keywords.
This matches the work of Ilyushkin et al.~\cite{DBLP:journals/tompecs/IlyushkinAHBPEI18} where they show the need for elasticity and accompanying metrics is required.
The other keywords might indicate the environment and tasks along with their constraints are increasingly focused on, which makes sense in the context of workflows resource provisioning.

The output for \textit{rising keywords} is as follows.
\begin{tcolorbox}[colback=blue!15, colframe=blue!90!black,enhanced,sharp corners,boxsep=-1mm]%
\enquote{environment}
\end{tcolorbox}

From this output, the only conjecture we can make is that the environment the autoscaler is operating in is considered increasingly important.
With the different resource types and metrics that can be focused upon (see Section~\ref{sct:workflow-allocation}), the environment indeed is playing a big role.

\subsection{Taxonomy of Provisioning}
\label{ssct:resource-provisioning-taxonomy}

\begin{figure}[!htb]
	\adjustbox{max width=\linewidth}{
		\includegraphics{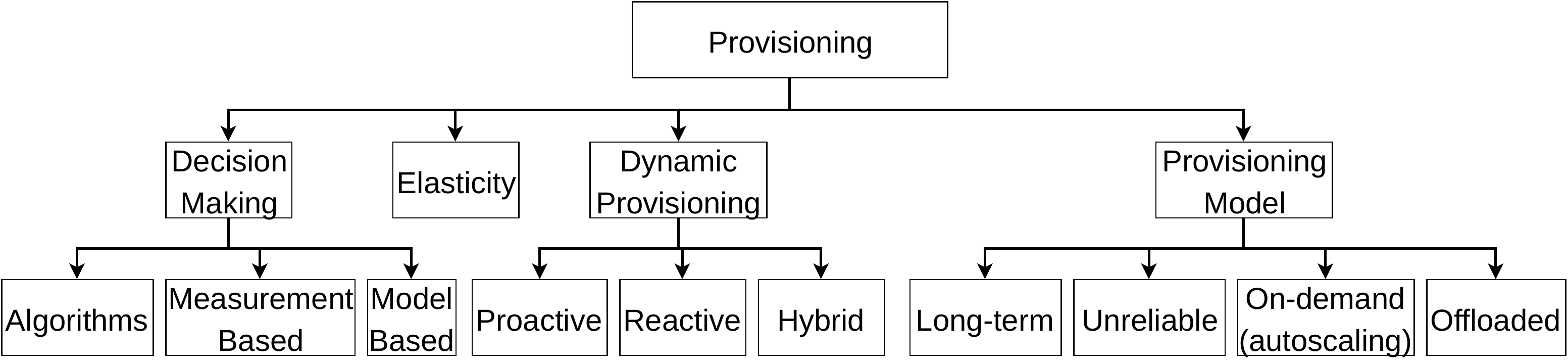}
	}
	\caption{Autoscaler taxonomy.} \label{tax:provisioning-model}
	\vcutM
\end{figure}

We select from and extend the provisioning taxonomies of Smanchat et al.~\cite{DBLP:journals/fgcs/SmanchatV15} and Shoaib et al.~\cite{DBLP:journals/corr/ShoaibD14a}.
We only focus on the provisioning model, decision making, elasticity, and dynamic provisioning strategies as they relate closely to our scope of workflow scheduling.

Provisioning decisions are divided into three categories: algorithms that make decisions, (static) scaling decisions based on measurements, and scaling decisions based on models~\cite{DBLP:journals/corr/ShoaibD14a}.

Algorithms typically consider multiple parameters, including deadlines, costs of resources, the workload, etc.
These algorithms may incorporate models to make decisions.

Scaling decisions based on measurements are more simplistic.
For example, when scheduling a bag-of-tasks, assuming that each task requires one CPU core, the amount of allocated machines might be straight forward.

Resource provisioning systems can also rely on performance models to make decisions.
Shoaib et al.~\cite{DBLP:journals/corr/ShoaibD14a} refer to several in their survey.

Elasticity defines how well a provisioning approach scales with the need for resources.
Ilyushkin et al.~\cite{DBLP:journals/tompecs/IlyushkinAHBPEI18} define several novel metrics for system elasticity.
We cover this topic in Section~\ref{ssct:elasticity}.

Dynamic provisioning refers how autoscalers respond to changes in resource requirements.
Proactive approaches predict changes and act accordingly, to attempt to avoid over- and underprovisioning scenarios.
The danger of such an approach is miscalculating the required resources.
Reactive approaches respond to changes that have already taken place or are taking place.
Reactive approaches might lead to short periods of over- and underprovisioning, yet follow changes in demand more closely albeit delayed.
Finally, hybrid approaches combine both techniques.
A typical combination is changing resources on a reactive basis approach when facing, e.g., bursts, yet be proactive with common patterns such as diurnal use of resources.

Finally, the provisioning model can vary.
Long-term, unreliable, and on-demand provisioning are covered by Smanchat et al.~\cite{DBLP:journals/fgcs/SmanchatV15}.
Long-term resources are rented for extended periods of time, up to years.
Unreliable provisioning relates to resources that may not always be available at a certain price, or simply available at all.
Amazon's Spot Instances is an example of such unreliable resources.
On-demand provisioning is the model of getting resources from a (usually) a fixed list from cloud providers.
When scaling resources up and down to deal with sudden flash crowds, typically on-demand provisioning models using autoscalers are used.
We cover this topic more in-depth in Section~\ref{ssct:autoscaler}.
Finally, we add to this category the offloaded model.
In this case, the provisioning of resources is managed by the resource provider.
An example of this is using the autoscalers of Amazon's autoscaling service.

\subsection{Taxonomy of Autoscalers}
\label{ssct:autoscaler}

\begin{figure}[!htb]
	\centering
	\adjustbox{max width=\linewidth, max height=3.5cm}{
		\includegraphics{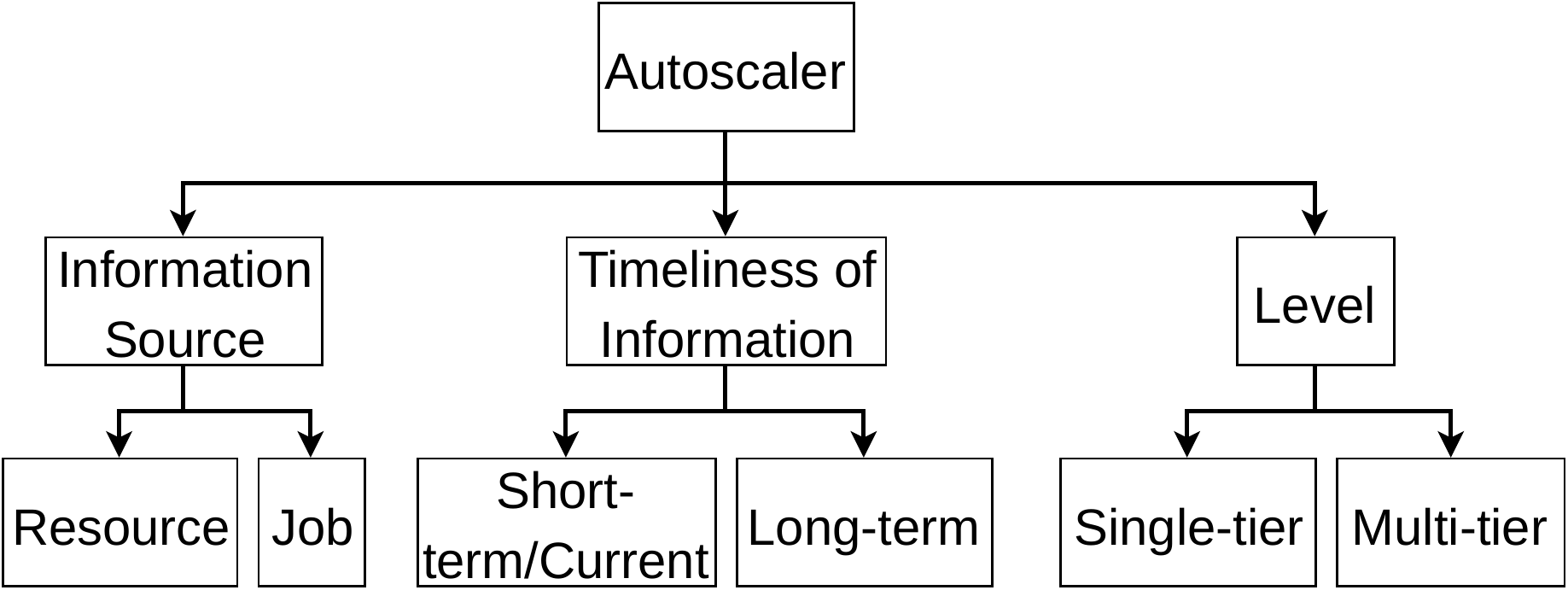}
	}
	\caption{Autoscaler taxonomy.} \label{tax:autoscalers}
	\vcutL
\end{figure}

Autoscalers are employed to scale resources during changing resource requirements of workflows following a \textit{provisioning policy}.
The provisioning policy dictates how many and when resources are (de)allocated.
Figure~\ref{tax:autoscalers} presents the taxonomy of autoscalers based on Ilyushkin et al.~\cite{DBLP:journals/tompecs/IlyushkinAHBPEI18}.

In workflow execution, autoscalers that solely monitor server-level information as their information source for making scaling decisions are agnostic to the workload.
Examples of informative metrics used are current throughput, length of the task queue, and amount of available resources.

Workflow-specific autoscalers exploit the structure of workflows to improve provisioning decisions.
Examples include estimating the level of parallelism in workflows~\cite{DBLP:conf/ccgrid/IlyushkinGE15} and construction partial execution plans for eligible tasks~\cite{DBLP:journals/tompecs/IlyushkinAHBPEI18}.
In our recent work~\cite{DBLP:conf/ccgrid/VersluisNI18}, we demonstrated that autoscaler performance varies as the workload, environment, and other system components change.
This indicates that careful benchmarking and proper identification of strengths and weaknesses of autoscalers is required.
Such new insights can be exploited into new approaches and possibly new scheduler designs where allocation and provisioning are co-designed.

Timeliness of information refers to how recent the data is that is being used to estimate incoming workloads.
There are two classes in this branch: short-term or current information and long-term.
Autoscalers that only operate on the current incoming amount of workload or very recent information.
There is no clear definition on how \enquote{recent} information should be, generally this differs per autoscaler.
Long-term information spans days to even months or years where diurnal or even seasonal patterns can be observed.

The level at which autoscalers can operate is either single-tier or multi-tier. 
Single-tier autoscalers typically manage the resources for a single application.
Multi-tier autoscalers such as Chamulteon~\cite{bauer2019chamulteon} and FAHP~\cite{khorsand2018fahp} scale resources for multiple, different applications. 

\subsection{Elasticity}
\label{ssct:elasticity}
When additional resources are required, the autoscaler must obtain enough resources as fast as possible, potentially in accordance with other \glspl{nfr} such as adhering to a budget.
Similarly, when resources are no longer required, they should be deallocated to avoid unnecessary costs.
\textit{Elasticity} defines how well a system responds to changes in resource requirements without looking at the secondary requirements.
The work of Ilyushkin et al.\cite{DBLP:journals/tompecs/IlyushkinAHBPEI18} use and introduce several metrics for elasticity.
Among other elements, these metrics capture \textit{overprovisioning}, i.e. the time and amount of resources that were idle and \textit{underprovisioning}, i.e. the time and amount of resources that were required but not provisioned. 

\subsection{Allocation and Provisioning Policy Interplay}
The allocation policy can have a direct impact on the performance of an autoscaler (and the provisioning policy that goes with it).
Versluis et al.~\cite{DBLP:conf/ccgrid/VersluisNI18} demonstrate that without task preemption resources may remain in use, yet underutilized due to the autoscaler being unable to deallocate these resources.
Andreadis et al.~\cite{DBLP:conf/sc/AndreadisVMI18} demonstrate that scheduler components, including the allocation and provisioning policies, are systematically underspecified. 
Underspecification of such policies and components can lead to significant differences in performance, hampering reproducibility.

Understanding both how resources are used through allocation and how they are provisioned are vital in creating a well-balanced system.
Work such as that of Malawski et al.~\cite{DBLP:journals/fgcs/MalawskiJDN15} investigate scheduling techniques that perform both resource allocation and provisioning.
We conjecture that this interplay is important to investigate in order to improve resource efficiency and understanding systems better.

\subsection{Future Directions}

Emerging areas present plenty of opportunities for resource provisioning research.
With edge datacenters becoming emergent in the edge/fog domain, resource provisioning policies should start taking these types of resources into account.
Typically, for latency sensitive applications, the cost vs. benefit ratio can play an important role.

The rise in popularity of containerized applications through e.g. Docker is also gaining in popularity.
Already products such as Docker Swarm and Kubernetes for container orchestrations are widely adopted by both academia and industry.
Especially in the area of \gls{faas} resource provisioning is an important aspect.
Starting a \gls{vm} or container incurs significant delay in function turnaround time as \glspl{vm} or containers with specific libraries and/or versions have to be booted.
Already work such as that of Aumala et al.~\cite{DBLP:conf/ccgrid/AumalaBOTA19} focus on package aware load balancing to speedup function deployment.

Moghaddam et al.~\cite{DBLP:journals/csur/Kardani-Moghaddam19} provide an extensive survey on resource provisioning and performance management.
Several directions for future work are included in their work.

Another item for future work is reviewing and improving the interplay of the allocation and the provisioning policies. Improving this may lead to improved resource utilization and reduced resource consumption.
\section{Applications and Services}
\label{sct:applications-and-services}

\begin{figure}[htb]
	\adjustbox{max width=\linewidth}{
		\includegraphics{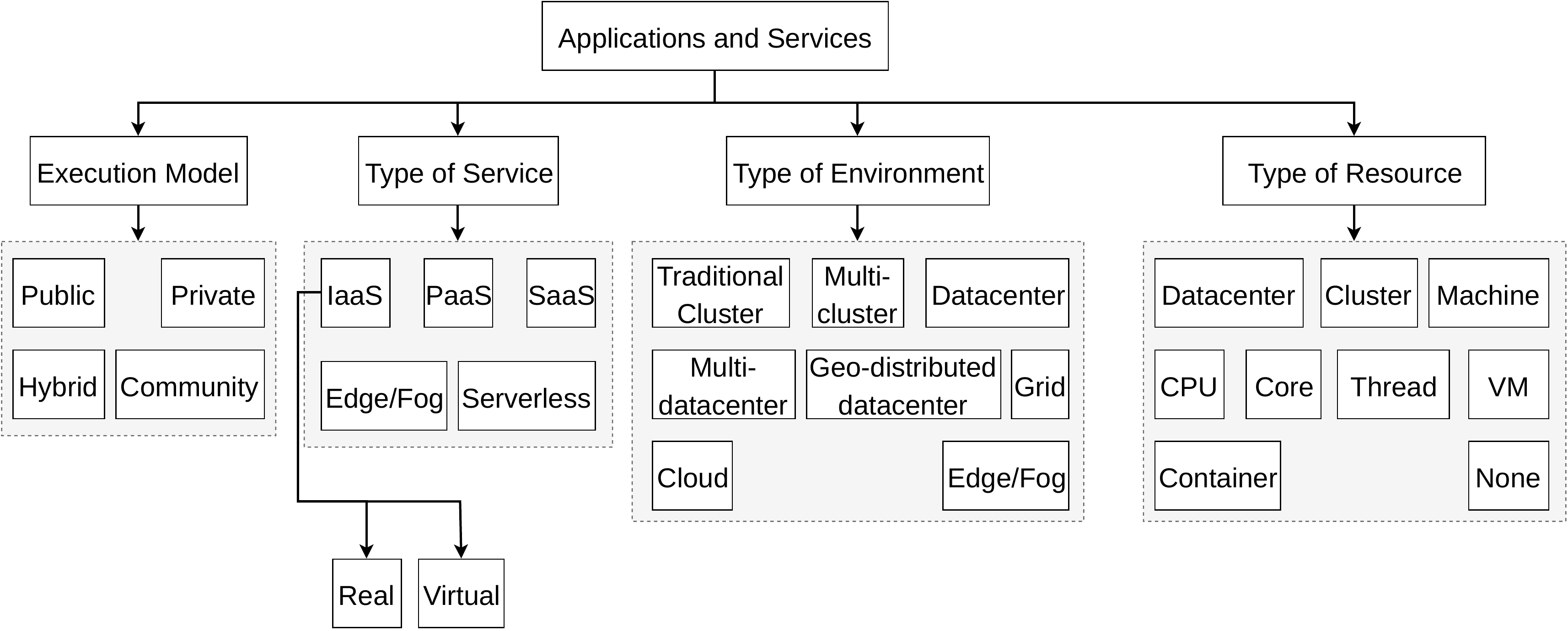}
	}
	\caption{Taxonomy of cloud computing services. Grey boxes depict a group of entities that fall under the same category.} \label{tax:cloud_computing_services}
\end{figure}

\begin{figure}
	\centering
	\begin{subfigure}[t]{0.49\textwidth}
		\includegraphics[width=\textwidth]{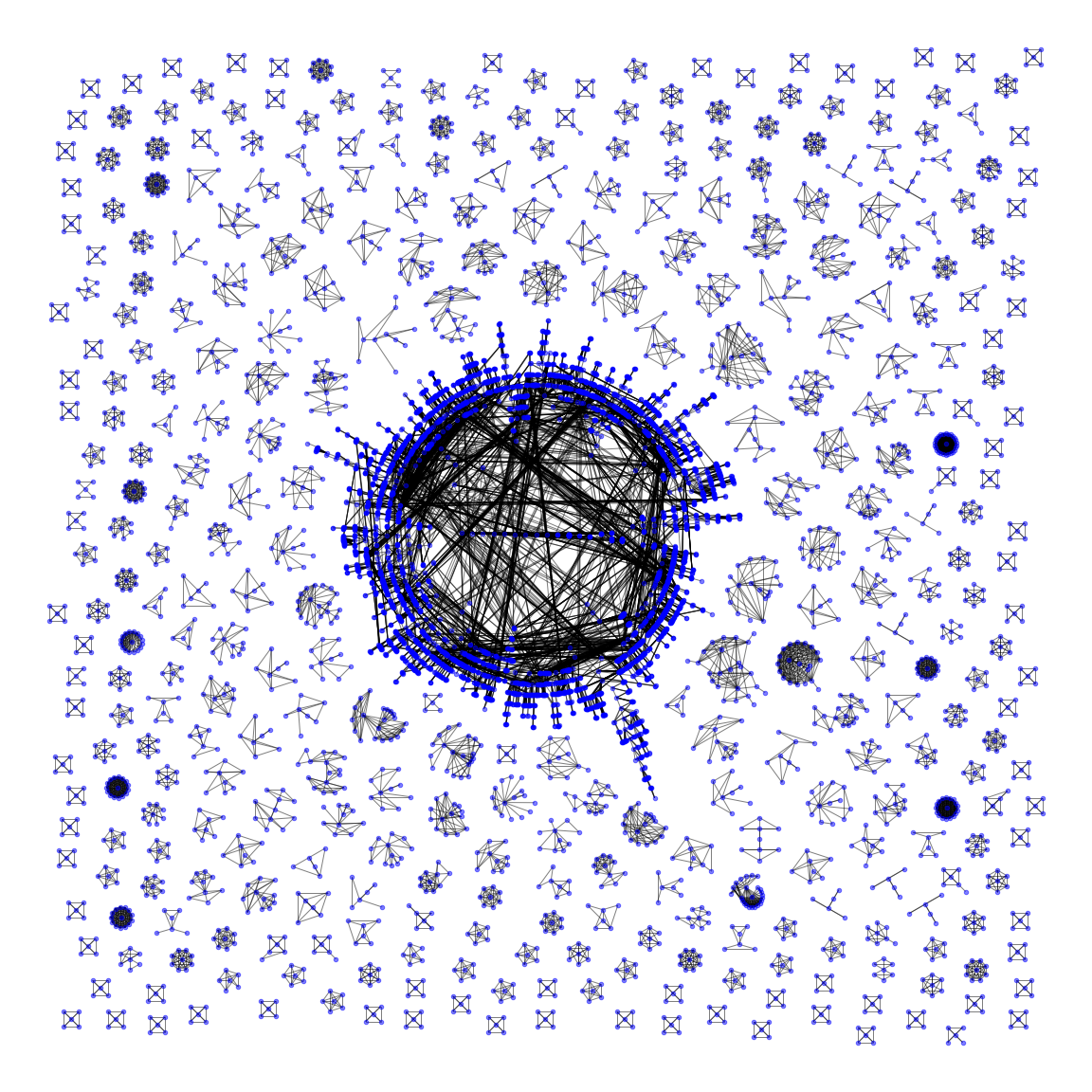}
		\caption{A visual representation of the applications and services community. Only components with cardinality 5 or higher are shown.}
		\label{fig:overview-cloud-computing-services-community}
	\end{subfigure}~
	\begin{subfigure}[t]{0.49\textwidth}
		\includegraphics[width=\textwidth]{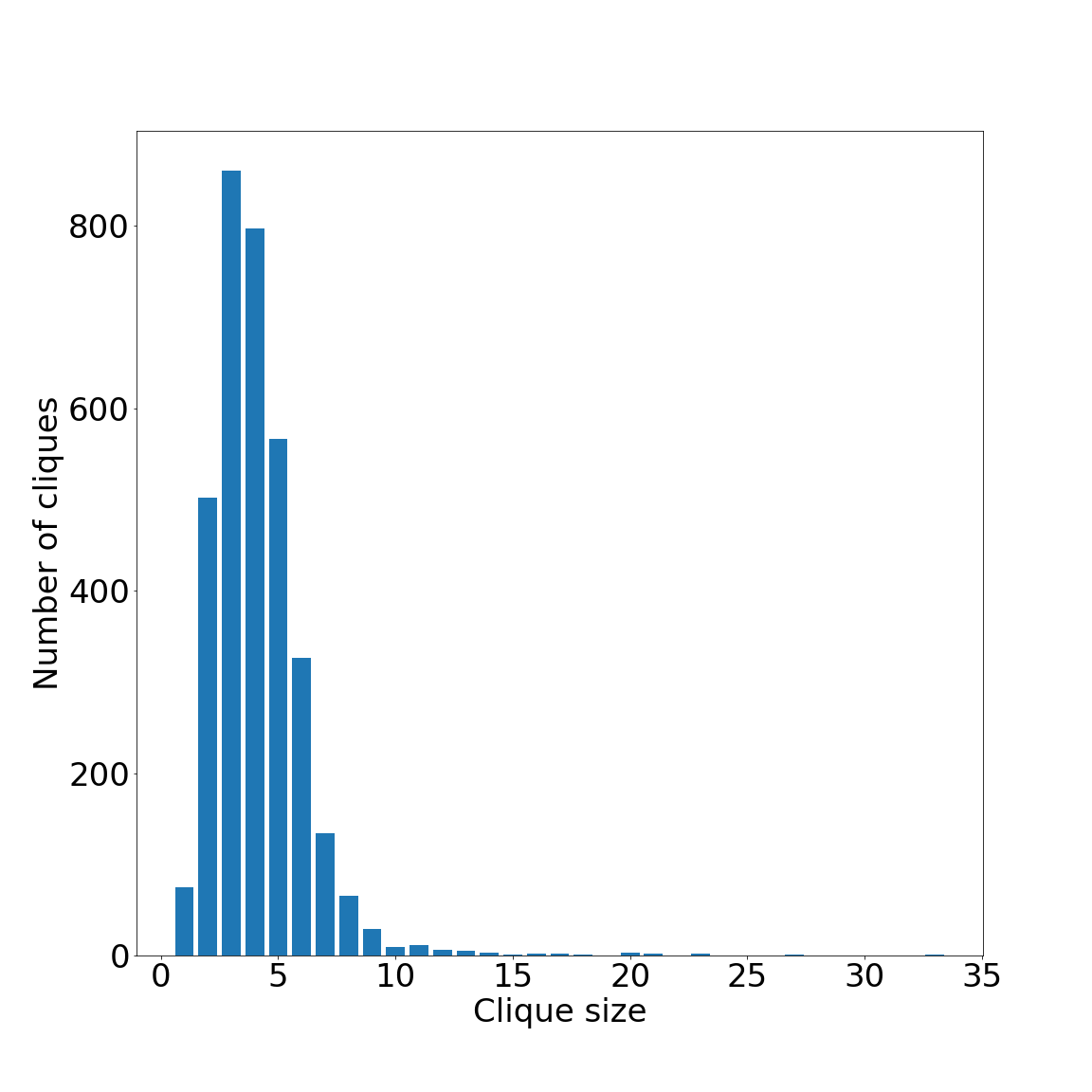}
		\caption{A bar plot depicting the size and number of cliques within the applications and services community.}
		\label{fig:cliques-cloud-computing-services-community}
	\end{subfigure}
	\caption{An overview of the cloud computing services community.}\label{fig:cloud-computing-services-community}
\end{figure}

\begin{figure}
	\centering
	\begin{subfigure}[t]{0.66\textwidth}
		\includegraphics[width=\textwidth,keepaspectratio]{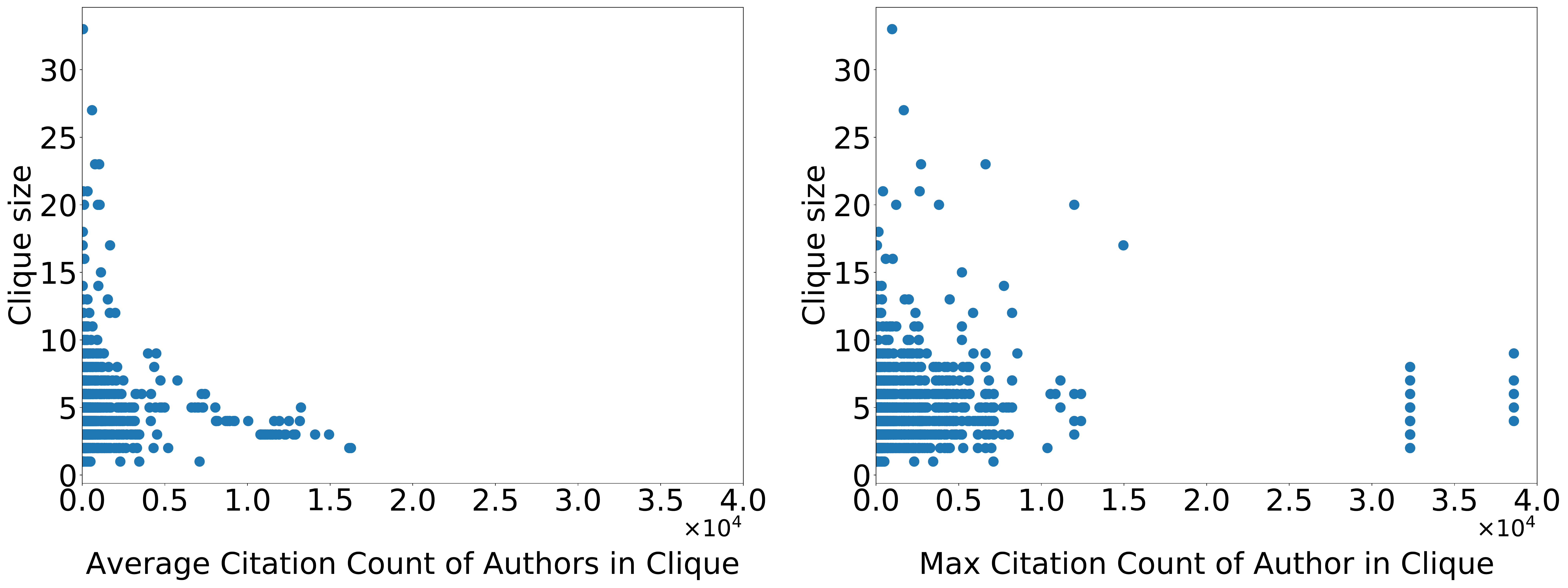}
		\caption{A scatterplot depicting the average (left) and maximum (right) citations of the authors for each clique in the applications and services community.}
		\label{fig:citations-vs-clique-size-applications-and-services-community}
	\end{subfigure}~
	\begin{subfigure}[t]{0.33\textwidth}
		\includegraphics[width=\textwidth,keepaspectratio]{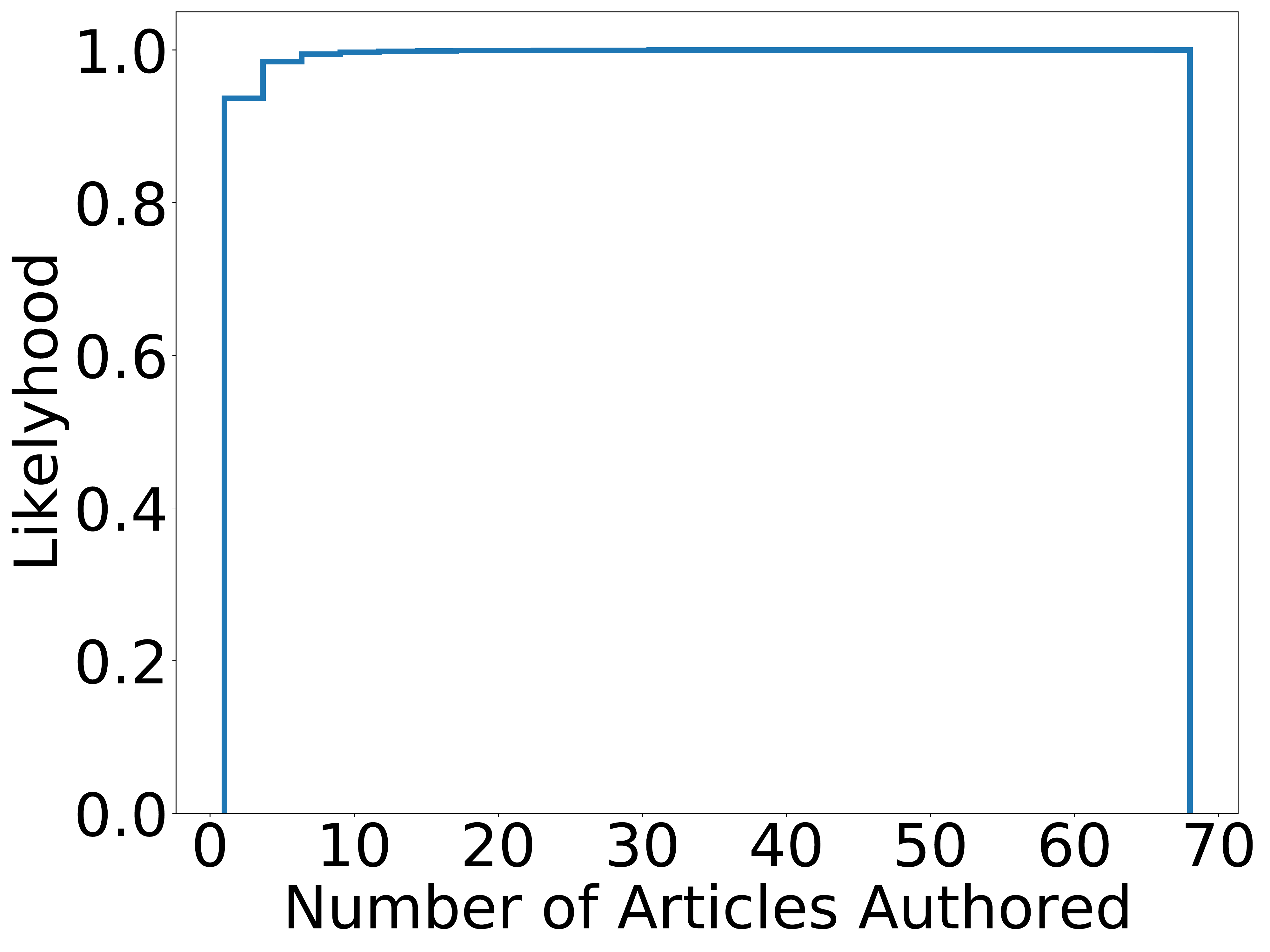}
		\caption{A CDF of number of articles authored per author in the applications and services community.}
		\label{fig:number-articles-co-authored-applications-and-services-community}
	\end{subfigure}
	\caption{Clique size vs. author citation count (average and max) and a CDF of number of articles authored.}
\end{figure}

\begin{figure}[thb]
	\includegraphics[max width=\linewidth]{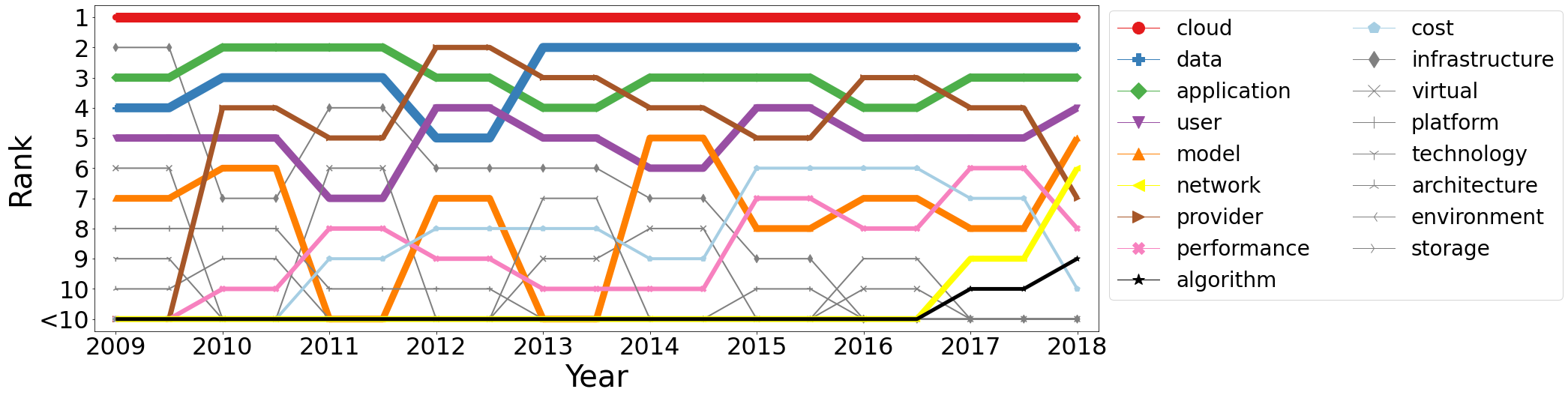}
	\caption{Top-10 keywords in application and services articles in the past decade per year.}
	\label{fig:top-10-application-and-services-per-year}
	\vcutS
\end{figure}

Cloud providers offer several kinds of different services nowadays, most of which still eventually translate into running workflow applications.
To this end, we divide this space using the taxonomy in Figure~\ref{tax:cloud_computing_services}.
Each of these branches are covered in this section.

\executedqueries{query-cloud-computing-services-articles}{SELECT * FROM publications WHERE year BETWEEN 2009 AND 2018 AND (lower(title) LIKE '\%cloud\%' OR lower(abstract) LIKE '\%cloud\%') AND (lower(title) LIKE '\%service\%' OR lower(abstract) LIKE '\%service\%')}

\refquery{query-cloud-computing-services-articles} is used to find articles related to cloud computing services, which in turn are used to verify the completeness of our taxonomies.

\subsection{Community Analysis}
\begin{description}
	\observation{applications-and-services-relations-one-time}{86.6\% of co-authorship relations are one time.}
	\observation{applications-and-services-cost-performance-cloud}{The cost and performance keywords became more important in the last five years in the span 2009-2018. The term \enquote{cloud} has been rising in importance as well.}
	\observation{applications-and-services-larger-cliques-smaller-avg}{Also for the applications and services communities holds that members of large cliques tend to have a lower total citation count on average and in maxima.}
	\observation{applications-and-services-mostly-singly-publication}{More than 90\% of authors authored a single article in the timespan 2009-2018. This is surprising given the scope of this community (and our query).}
\end{description}

\begin{table}[]
\caption{An overview of the applications and services community in numbers.}
\label{tbl:applications-and-services-community-overview}
\begin{tabular}{@{}rrrrrr@{}}
\toprule
Articles & Authors & Co-authorship Relations & Unique relations & Cliques & Largest Clique \\ \midrule
15535 & 9920     & 30901      & 26759              & 3406      & 33             \\ \bottomrule
\end{tabular}
\end{table}

We summarize the main characteristics of the community in Table~\ref{tbl:applications-and-services-community-overview}.
As we can observe, this domain is huge.
More than 15 thousand articles are published in the community, with close to ten thousand authors.
As can be observe from Figures~\ref{fig:overview-cloud-computing-services-community} and~\ref{fig:cliques-cloud-computing-services-community}, the community has a strong collaborating core, with also large cliques running up to 33.
The community features many larger groups, yet 86.6\% of the relationships are one-time.
The size and dynamics of the community show it is a healthy and diverse community.

In Figure~\ref{fig:citations-vs-clique-size-resource-provisioning-community} we plot the average and maximum number of citations per clique in the application and services community.
Given that over three thousand cliques are checked, we can more reliably draw conclusions from this data.
The pattern is very similar to that of the previous inspected communities. 
The small and moderately sized cliques have the highest citation count on average and at maximum, with some outliers in the higher segment too.
This community has the same trend as the previous two investigated where larger cliques tend to have less cited authors both on average and at maximum.

Figure~\ref{fig:number-articles-co-authored-resource-provisioning-community} presents a CDF of the number of articles that get published within the resource provisioning community per author.
From this figure we observe over 90\% of the authors in this community author an article once.
This is surprising given the broad scope of this community and our query.
With over fifteen thousand articles checked, having such a high number of single authorships is surprising.
On the other hand, the tail is heavy too; in the timespan 2009-2018 that was inspected, authors authored up tens and even close to 70 articles.

\subsubsection{Analysis of Trends over Time}
We use the method detailed in Section~\ref{ssct:trends-over-time-workflow-scheduling} combined with the output of \refquery{query-cloud-computing-services-articles}.
The output of the trend analysis is visible in Figure~\ref{fig:top-10-application-and-services-per-year}.
From this figure, we observe the focus of this community is quite condensed.
In total, we find eighteen keywords.
Interestingly, \enquote{data} is a keyword that is consistently in the top-5, indicating the importance of this aspect, possibly in conjunction with keywords such as \enquote{network}, \enquote{infrastructure}, \enquote{performance}, \enquote{cost}, and \enquote{storage}.
The \enquote{platform}, \enquote{technology}, \enquote{architecture}, and \enquote{environment} are also often regarded as important elements, which make sense given the variety of services offered nowadays.

\subsubsection{Emerging Trends}
We use the method detailed in Section~\ref{ssct:emerging-trends} combined with the output of \refquery{query-cloud-computing-services-articles}.
We find the following \textit{new keywords}.
\begin{tcolorbox}[colback=blue!15, colframe=blue!90!black,enhanced,sharp corners,boxsep=-1mm]%
	\enquote{cost}, \enquote{performance}
\end{tcolorbox}

This indicates that the focus on cost and performance has risen for applications and services field.
Especially with the many pricing strategies that can be followed using cloud resources, the trade-off between these two variables becomes interesting.

When looking at \textit{rising keywords}, we obtain the following output.
\begin{tcolorbox}[colback=blue!15, colframe=blue!90!black,enhanced,sharp corners,boxsep=-1mm]%
	\enquote{cloud}
\end{tcolorbox}

As perceived in general, but also in this community, the cloud as environment is deemed important.  

\subsection{Service Types}

\subsubsection{\acrshort{iaas}}
\label{ssct:iaas}
\gls{iaas} is the notion of renting resources from a cloud operator.
These resources can either be virtualized or real.
Typically, \gls{iaas} resources come with an clean OS on which dependencies, libraries, and packages, etc. have to be installed by the client.
Until recently, resources were leased per hour, however most major vendors including Amazon and Microsoft now offer a per-second leasing granularity.

\subsubsection{\acrshort{paas}}
\label{ssct:paas}
\glsreset{paas} 

\gls{paas} is the category of cloud services that allow users to install, configure, deploy, run, and manage their own applications without having to deal with any underlying infrastructure.
It was derived from \gls{saas}~\cite{guardianPaaS}.
The deployment, maintenance, and upgrading of infrastructure are outsourced to the cloud provider.
This service enables e.g. specific versioning or configuration of software when compared to \gls{saas}.

\subsubsection{\acrshort{saas}}
\label{ssct:saas}

\gls{saas} is a more restrictive form of both \gls{iaas} and \gls{paas} that encapsulates a model where applications are offered as a service.
Rather than having to install and set-up their own software, the hosting and installations are provided transparently by the cloud provider, eliminating any hosting intermediary.
It is therefore that the cloud provider, hosting service, developer, and maintainer of the software are usually the same entity.

\subsubsection{Edge/Fog Computing}
\label{ssct:edge-fog-computing}

Edge computing is an emerging paradigm where micro-datacenters and/or devices are put closer to the customer, often referred to as the edge of the network.
This is also referred to as Fog computing~\cite{DBLP:journals/jpdc/MahmudSRB19}.
By introducing such micro-datacenters, latency is reduced compared to sending data to the larger datacenters, further away.
The general consensus is that such micro-datacenters are more expensive to use, as the datacenter operator must perform more management, often in various locations.
In particular, IoT applications and mobile offloading strategies benefit from this new paradigm, enabling real-time processing and streaming of data.

For work done on edge/fog computing, the International Conference on Fog and Edge Computing (ICFEC) provides a good starting point.
Surveys also provide starting points for open challenges and introduction to different concepts and applications, for example see~\cite{yi2015survey} and~\cite{mahmud2018fog}.
Topics within edge/fog computing range from applications such as video streaming to resource consumption methods such energy-efficient scheduling, much alike traditional cloud topics.

\subsubsection{Serverless}
\label{ssct:serverless}
Serverless is an emerging paradigm where clients can choose not to (temporarily) own, or manage resources.
In most cases, resource requirements still have to be specified, yet do not manually have to be provisioned and managed.
This area recently gathered a lot of attention from the (cloud) community.
The perceived benefits lie in the flexibility, cost-effectiveness, and availability properties.
Several articles introduce both problems and opportunities for this new paradigm~\cite{baldini2017serverless, fox2017status, DBLP:conf/wosp/EykIAGE18}.

While emerging, the domain is growing fast with different areas of the domain being explored.
Published articles range from historical~\cite{DBLP:journals/internet/EykTTVUI18} to frameworks.~\cite{DBLP:conf/IEEEcloud/PerezRNCM19}, and from exploration and characterization~\cite{DBLP:conf/usenix/WangLZRS18} to caching~\cite{abad2018package}.

Examples of proposed applications using serverless are graph processing\cite{toadergraphless}, chat bots~\cite{yan2016building}, image processing~\cite{akkus2018sand, oakes2018sock}, and data analytics~\cite{nastic2017serverless}.

\subsection{Type of Environments}

Computing services can be offered on different type of environments.
Traditionally, (local) \textit{clusters} are used for additional computing.
These are generally managed by a single department within a company/institution.
\textit{Multi-cluster} environments such as the Dutch DAS5~\cite{DBLP:journals/computer/BalELNRSSW16} offer resources in often geographically distributed clusters.
These clusters can be managed by a single department or by the different institutions hosting them.
\textit{Datacenters} often comprise multiple clusters within a single location. 
These clusters can belong to a single or multiple entities, but generally, a datacenter consists of clusters belonging to multiple entities, either leased or bought.
\textit{Geo-distributed datacenters} are datacenters that are geographically distributed, often for fault-tolerance or legislation purposes.
The different environments covered so far often have well-defined architectures and the hardware and infrastructure is known.
\textit{Grids}, \textit{Clouds}, and \textit{Edge/Fog} environments are more vague.
Clouds are often composed of geographically distributed datacenters, where you can rent virtual machines in different physical locations using the now popular pay-as-you-go model.
What makes the environments vague is that cloud providers rarely describe (in detail) the underlying hardware, schedulers, policies, and protocols in place.
Grids are a mixture of hardware as they were often composed of a mixture of commodity and state-of-the-art hardware.
This makes it difficult to assess the accessible hardware, nor were there any guarantees that a machine connected to the grid wouldn't suddenly become unavailable.
Finally, the Edge/Fog consists of many different devices \enquote{at the edge} of the network.
These can be micro-datacenters, routers, mobile devices, smart devices, etc.
Additionally, the communication established between these devices may be arbitrary.

\subsection{Execution Model}

As Smanchat et al.~\cite{DBLP:journals/fgcs/SmanchatV15} describe, when using computing services, the execution model can vary.
\textit{Public} resources are available to the public and can be leased from cloud providers.
\textit{Private} resources are only available to a single entity, e.g. company, it is not possible to execute work on these resources if you do not have (private) access.
A \textit{hybrid} model combines both public and private resources. Often when additional compute power is required, an entity can run (part of) the workload on a public cloud.
\textit{Community} resources such as the earlier mentioned DAS5~\cite{DBLP:journals/computer/BalELNRSSW16} enable a community to share maintain resources collectively.

\subsection{Taxonomy of Resource Types}

\begin{figure}[htb]
	\adjustbox{max width=.85\linewidth}{
		\includegraphics{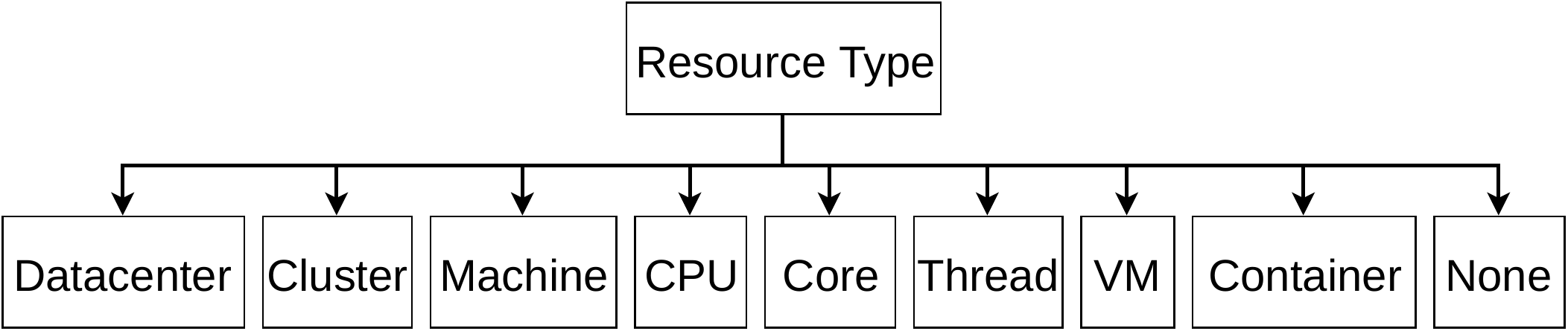}
	}
	\caption{Taxonomy of resource types.} \label{tax:resource_types}
	\vcutM
\end{figure}

When scheduling, different algorithms may consider different resources as the working unit.
Figure~\ref{tax:resource_types} depicts the taxonomy of resource types, which significantly extends the taxonomy presented in~\cite{DBLP:journals/fgcs/SmanchatV15}.
With different granularity possible and the heterogeneity of today's systems, plenty of work differentiates in resource considered.
Typically, literature focuses on cores~\cite{lee2015resource}, \glspl{vm}~\cite{wang2019multi}, and CPU~\cite{DBLP:journals/tpds/TopcuogluHW02}. 
Especially one task per CPU or \gls{vm} is common~\cite{seo2008energy, DBLP:conf/ccgrid/IlyushkinGE15}.
Ma et al.~\cite{DBLP:conf/icac/MaISI17} consider as resource types threads while scheduling tasks of an industrial IoT environment.
Containers started receiving attention as possible alternatives to \glspl{vm}.
As a container does not include an operating system, the overhead can be reduced when using containers.
With the popularity of Docker and Kubernetes, articles are investigating the use cases of containers~\cite{gerlach2014skyport, liu2016flexible}. 
Schedulers considering machines are used in e.g. parallel workflow computing~\cite{rajakumar2004workflow}.
Cluster schedulers were quite ubiquitous in Grid environments~\cite{yu2008workflow}, and can be found in cloud environments as well~\cite{verma2015large, DBLP:conf/sc/PollardJHB18}.
The section \enquote{none} covers situations such as serverless (see Section~\ref{ssct:serverless}) where cloud users need not consider the resources used.
Naturally, resources are still consumed, yet these are entirely managed by the cloud operator and hidden from clients.
Scheduling at the entire scale of a datacenter is common when dealing with super computers.
Several applications that run on (a part) of a super computer can for example be found in the super computing community.
For example, the algorithm for particle simulation that was deployed across a large part of the CORAL supercomputers~\cite{DBLP:conf/sc/BerkowitzCGMNRV18}.

\subsection{Taxonomy of Scheduling Dynamicity}

\begin{figure}[htb]
	\adjustbox{max width=\linewidth, max height=2.3cm}{
		\includegraphics{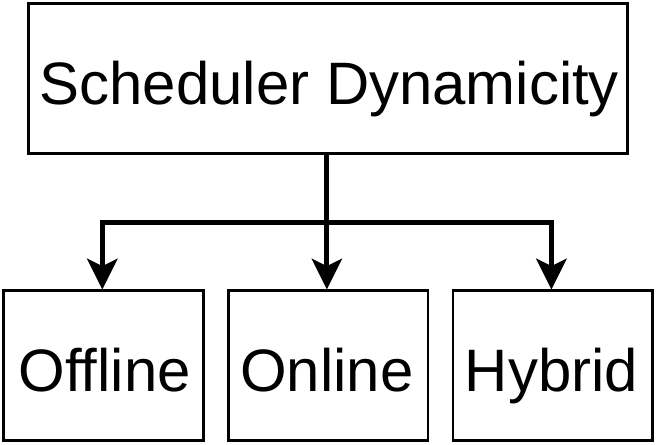}
	}
	\caption{Scheduling dynamicity taxonomy.} \label{tax:scheduling_dynamicity}
	\vcutM
\end{figure}

Scheduling dynamicity describes how allocation policies plan their allocations.
Figure~\ref{tax:scheduling_dynamicity} presents the taxonomy of scheduling dynamicity.
Offline policies construct a total plan for all tasks yet to be scheduled in the system.
The system must then adhere strictly to this plan.
Any issues or runtime variations (e.g. stragglers) are not taking into account at runtime, while some plans do include some \textit{slack} in their schedules.
Online policies construct plans responsively.
Incoming tasks are appended to the plan, or the plan is (partially) reconstructed to optimize for the optimization goal.
Hybrid solutions use a combination of offline and online dynamicity.
Hybrid policies often create an initial execution plan that is then changed proactively.

\section{Mapping Allocation and Provisioning Policies}
\label{sct:workflow-allocation-policies}

Scheduling workflows consists of two main parts: the workflows and its tasks that need to be places on resources, and the resources themselves.
Managing these two parts can be done separately using agnostic policies, or can be done in synergy where the two policies work together.

To demonstrate the different policies for both allocation and provisioning can be mapped to our taxonomies, in this section we for these two categories a number of recent state-of-the-art and well-known policies.
We first describe our method used to obtain these lists of policies followed by an enumeration of their main properties.

\subsection{Method}
To obtain our list of policies, we use a systematic approach.
We first create a query to filter for allocation and provisioning policies, respectively.
Using this query, we extract sixteen policies per category: ten described in the most recent articles according to our database (using citation count as tie-breaker) and the six most popular by citation count.
Articles that are a false positive, i.e., do not describe a policy are skipped.

To highlight recent policies still map to our taxonomies, for the recent policies, we include also articles from more recent years if available in our database, e.g., 2019.

For allocation policies, we note the the optimization goal, strategy, target, and technique used as well as if the policy computes the scheduling plan offline ahead of time, or dynamically at runtime.
For provisioning policies, we note the type of decision making, dynamic provisioning method, and provisioning model used.
Additionally, for each article we also list the number of citations to provide a rough indication of popularity.

\subsection{Allocation Policies}
\begin{tabularx}{\linewidth}{lrllXl}
	\caption{Overview of allocation policies.\newline 
		Header: ST = scheduling technique, OG = optimization goal, OS = optimization strategy.\newline
		ST column: G = Greedy, H = Heuristic, MH = Meta-heuristic, R=Random. }
	\label{tbl:overview-allocation-policies}\vcutM \\
	\toprule
	Name & \#Citations & Year & ST & OG & OS \\
	\midrule
	Titan~\cite{cao2003gridflow} & 522 & 2003 & H & Makespan, Deadline, Resource Utilization & Multi/Min/Global \\ 
	SCS~\cite{DBLP:conf/sc/MaoH11} & 433 & 2011 & G & Cost & Multi/Min/Global \\
	Min-min task~\cite{DBLP:conf/ccgrid/BlytheJDGVMK05} & 383 & 2005 & H & Runtime & Single/Min/Local \\
	Blythe et al.~\cite{DBLP:conf/ccgrid/BlytheJDGVMK05} & 383 & 2005 & G+R & Runtime & Single/Min/Local \\
	Weighted min-min~\cite{DBLP:conf/ccgrid/BlytheJDGVMK05} & 383 & 2005 & H & Runtime & Single/Min/Local \\
	IC-PCP~\cite{DBLP:journals/fgcs/AbrishamiNE13} & 360 & 2012 & H & Deadline, Cost & Multi/Min/Global \\ \midrule
	Qureshi~\cite{qureshi2019profile} & 5 & 2019 & H & Cost, Power, Resource Utilization & Multi/Min/Local \\
	FDHEFT~\cite{zhou2019minimizing} & 5 & 2019 & H & Cost, Makespan & Multi/Min/Global \\
	WSADF~\cite{momenzadeh2019workflow} & 3 & 2019 & H & Bandwidth, Resource Utilization, Throughput & Multi/Min/Global \\
	Stavrinides et al.~\cite{stavrinides2019energy} & 3 & 2019 &  H & Power Consumption, Cost & Multi/Min/Local \\
	BDAS~\cite{DBLP:journals/tpds/ArabnejadBN19} & 2 & 2019 & H & Cost, Deadline & Multi/Min/Local \\
	MOGA~\cite{rehman2019multi} & 2 & 2019 & MH & Makespan, Cost, Deadline, Power Consumption & Multi/Min/Global \\ 
	PSFS~\cite{pietri2019pareto} & 1 & 2019 & H & Deadline, Cost & Multi/Opt/Global \\
	CSFS-H~\cite{pietri2019pareto} & 1 & 2019 & H & Cost & Single/Min/Global \\
	DNCSPO~\cite{DBLP:journals/fgcs/XieZWCXSYY19} & 1 & 2019 & MH & Makespan, Cost & Multi/Opt/Global \\
	Wu et al.~\cite{DBLP:journals/fgcs/WuHCW19} & 1 & 2019 & MH & Makespan, Cost & Multi/Opt/Global \\
	\bottomrule
	\vcutL
\end{tabularx}

In this section we provide an overview of well-cited and state-of-the-art workflow allocation policies.
Table~\ref{tbl:overview-allocation-policies} presents a list of the six most cited and ten recently used/introduced allocation policies, sorted by citations.
We focus on the scheduling technique used, the optimization goal and scheduling strategy.

To obtain articles based on citation count we use the following query:

\executedqueries{allocation-policies-citation-count}{
	SELECT * 
	FROM publications 
	WHERE (lower(title) LIKE '\%workflow\%' OR lower(abstract) LIKE '\%workflow\%') 
	AND ((lower(title) LIKE '\%allocat\%' OR lower(abstract) LIKE '\%allocat\%') 
	OR (lower(title) LIKE '\%schedul\%' OR lower(abstract) LIKE '\%schedul\%') 
	OR (lower(title) LIKE '\%plan\%' OR lower(abstract) LIKE '\%plan\%')) 
	ORDER BY n\_citations DESC
}

To obtain the most recent policies, we use the following query:

\executedqueries{allocation-policies-year}{SELECT * 
	FROM publications 
	WHERE (lower(title) LIKE '\%workflow\%' OR lower(abstract) LIKE '\%workflow\%') 
	AND ((lower(title) LIKE '\%allocat\%' OR lower(abstract) LIKE '\%allocat\%') 
	OR (lower(title) LIKE '\%schedul\%' OR lower(abstract) LIKE '\%schedul\%') 
	OR (lower(title) LIKE '\%plan\%' OR lower(abstract) LIKE '\%plan\%')) 
	ORDER BY year DESC, n\_citations DESC}

\subsection{Provisioning Policies}
\begin{tabularx}{\linewidth}{lrllXl}
	\caption{Overview of provisioning policies.\newline 
		Header: IS = Information Source, ToI = Timeliness of Information, L = Level.}
	\label{tbl:overview-provisioning-policies}\vcutM \\
	\toprule
	Name & \#Citations & Year & IS & ToI & L \\
	\midrule
	DPDS~\cite{DBLP:journals/fgcs/MalawskiJDN15} & 236 & 2015 & Resource & Short-term & Single-tier \\
	WA-DPDS~\cite{DBLP:journals/fgcs/MalawskiJDN15} & 236 & 2015 & Resource, Job & Short-term & Single-tier \\
	SPSS~\cite{DBLP:journals/fgcs/MalawskiJDN15} & 236 & 2015 & Job & Current & Single-tier \\
	D\"ornemann et al.~\cite{DBLP:conf/ccgrid/DornemannJF09} & 173 & 2009 & Resource & Current & Single-tier \\
	PBTS~\cite{DBLP:journals/fgcs/ByunKKM11} & 153 & 2011 & Job & Current & Single-tier \\
	Singh et al.~\cite{DBLP:conf/hpdc/SinghKD07} & 107 & 2007 & Job & Current & Single-tier \\ \midrule
	CLUES~\cite{DBLP:conf/IEEEcloud/PerezRNCM19} & 0 & 2019 & Resource & Current & Single-tier \\
	Prob~\cite{DBLP:conf/icpp/ZhouXHIC19} & 0 & 2019 & Job & Current & Single-tier \\
	EPSM~\cite{DBLP:journals/fgcs/RodriguezB18} & 15 & 2018 & Job & Current & Single-tier \\
	React~\cite{DBLP:journals/tompecs/IlyushkinAHBPEI18} & 10 & 2018 & Resource & Short-term & Single-tier \\
	Adapt~\cite{DBLP:journals/tompecs/IlyushkinAHBPEI18} & 10 & 2018 & Resource & Short-term & Single-tier \\
	Hist~\cite{DBLP:journals/tompecs/IlyushkinAHBPEI18} & 10 & 2018 & Resource & Long-term & Multi-tier \\
	Reg~\cite{DBLP:journals/tompecs/IlyushkinAHBPEI18} & 10 & 2018 & Resource & Long-term & Single-tier \\
	ConPaaS~\cite{DBLP:journals/tompecs/IlyushkinAHBPEI18} & 10 & 2018 & Resource & Long-term & Single-tier \\
	PPDPS~\cite{DBLP:journals/fgcs/SinghGJ18} & 8 & 2018 & Job & Current & Single-tier \\
	BioCloud~\cite{DBLP:journals/fgcs/SenturkBAKMC18} & 6 & 2018 & Job & Current & Single-tier\\
	\bottomrule
	\vcutL
\end{tabularx}

In this section we provide an overview of well-cited and state-of-the-art resource provisioning policies.
Table~\ref{tbl:overview-allocation-policies} presents a list of the six most cited and ten recently used/introduced provisioning policies, sorted by citations.
We focus on the information source used, the timeliness of information on which decisions are based, and the level at which the autoscaler operates.

To obtain articles based on citation count we use the following query:

\executedqueries{provision-policies-citation-count}{
	SELECT * FROM publications WHERE (lower(title) LIKE '\%workflow\%' OR lower(abstract) LIKE '\%workflow\%') AND (lower(title) LIKE '\%provision\%' OR lower(abstract) LIKE '\%provision\%' OR lower(title) LIKE '\%autoscal\%' OR lower(abstract) LIKE '\%autoscal\%') ORDER BY n\_citations DESC
}

To obtain the most recent policies, we use the following query:

\executedqueries{provision-policies-year}{SELECT * FROM publications WHERE (lower(title) LIKE '\%workflow\%' OR lower(abstract) LIKE '\%workflow\%') AND (lower(title) LIKE '\%provision\%' OR lower(abstract) LIKE '\%provision\%' OR lower(title) LIKE '\%autoscal\%' OR lower(abstract) LIKE '\%autoscal\%')
	ORDER BY year DESC, n\_citations DESC}
\section{Related Work}
\label{sct:related-work}

There are several surveys that either overlap this work or that present formalisms which we extend.
As we mention and cite articles of which we use elements or extend directly in the sections themselves, below we will mention shortly the topic of each related article and where our work differs.

Alkhanak et al.~\cite{alkhanak2015cost} focus on cost-aware approaches when scheduling workflows in clouds. 
While their work strongly focuses on the cost aspect of scheduling, we expand more on the other aspects related to workflow scheduling including the optimization targets, allocation techniques, formalisms, etc.
Poola et al.~\cite{poola2017taxonomy} focus on the fault-tolerance aspect of scheduling workflows in clouds.
They provide an in-depth view of fault-tolerance in workflow scheduling and discuss failure models.
Our survey is more general and significantly extends their taxonomies on scheduling targets and techniques.
Khorsand et al.~\cite{khorsand2017taxonomy} cover workflow partitioning of manual (business) workflows and automated execution.
Similar to our work, it features a strong focus on the language and formalism.
While the authors go in-depth in workflow partitioning,in particular in decentralized environments, we dive more in-depth in the allocation and provisioning side of workflow scheduling in both centralized and decentralize environments.
The work of Rodriguez et al.~\cite{rodriguez2016taxonomy} is similar to ours.
They focus on \gls{vm}-mapping, resource provisioning, scheduling targets, and scheduling strategies.
While we do not focus on \gls{vm}-mapping, we focus on the community aspect and workflow formalisms. 
Additionally, we significantly extend their proposed taxonomies on scheduling targets and scheduling strategies which are backed by a systematic literature survey.
Smanchat et al.~\cite{DBLP:journals/fgcs/SmanchatV15} discuss both workflow scheduling and provisioning in their survey.
They present some preliminary taxonomies, most of which we combine with other related work and significantly extend.
Deelman et al.~\cite{DBLP:journals/fgcs/DeelmanGST09} discuss several workflow formalisms used by scientific workflow tools.
Their survey has a different scope as they discuss how scientists use such tools.
Wieczorek et al.~\cite{DBLP:journals/fgcs/WieczorekHP09} discuss extensively workflow scheduling criteria and execution diversity in Grid environments.
While different in scope, we use several elements from their taxonomies and apply it to ours in, e.g., our optimization strategy taxonomy.
Wu et al.~\cite{DBLP:journals/tjs/WuWT15} cover both workflow scheduling and resource provisioning.
Different from our work, they focus on static versus dynamic scheduling and discuss robustness (scheduling) in clouds.
Moghaddam et al.~\cite{DBLP:journals/csur/Kardani-Moghaddam19} focus in their work on performance aware resource provisioning.
Different from our work, their focus lies more in anomaly and performance regression detection and mitigation strategies.
Hilman et al.~\cite{DBLP:journals/corr/abs-1809-05574} focus on multi-tenant systems in which workflows are executed simultaneously.
Different from our article, their focus is on multi-tenant and multi-workflow scheduling.
Our survey on resource provisioning covers more on the topics decision making, elasticity, and the provisioning model.

Additionally, our survey contains some novel elements none of the mentioned related work performs.
The first element is the survey of the community as we present in this article.
Collaborative relationships have been investigated before, but not at the granularity and on the topics that we cover in this survey.
We feel there is room for additional work in this direction, but due to limits in scope and pages, we leave that as future work.
A second element is investigation of important keywords per topic.
Related work does mention emerging trends and recommend future directions, yet rarely is this based on techniques from the information retrieval domain.
Extracting keywords from sparse text such as paper meta-data is a challenging problem due to limited length of the text and 
\section{Threats to Validity}
\label{sct:threats-to-validity}

In this section we cover several challenges to the validity of this work.
In particular, we acknowledge the queries used in this work, the mapping of our policies to validate our taxonomies, and the use of \gls{tfidf} as possible threats to the validity of this article.

\subsection{Queries Used}
The queries used to obtain relevant articles from the databases are instrumental in this work.
The selection of articles returned for each query forms the basis of the community, keyword, and trend analysis per area.
For each query, it's important to include as many relevant keywords as possible, without getting too many false positives.
Our approach to creating queries was to first start with one or two keywords and investigate how many and how many relevant articles were returned.
We then tried out other keywords and investigated how many additional additional (both positive and negative) articles were added.
We believe the current set of queries samples the total set of articles per topic adequately.
Naturally, we might have missed synonyms.
By providing both the database and queries used, readers can rerun our queries on the database and judge the selection of articles themselves.
In fact, we encourage the community to refine our queries and possibly better capture current communities and possibly improve the analysis done in this work.

\subsection{Mapping of Policies}
We mapped several elements of sixteen allocation and sixteen provisioning policies to our taxonomies to validate them being able to map both recent and well-cited policies.
We did not map a significantly higher number of policies to keep this survey from growing too large.
Already, we identified missing elements in taxonomies of relevant work and thus significantly extended existing taxonomies.
Going even more in-depth into a single area, e.g., resource provisioning deserves an article on its own.
We recommend for future work such an approach.

\subsection{The use of \gls{tfidf}}
In several instances, the outcome of \gls{tfidf} produced noise, i.e., false positives.
This is mainly due to titles and abstracts being sparse text.
Extracting meaningful keywords automatically form sparse text is difficult.
We believe we got meaningful results, yet approaches using Latent Dirichlet Allocation or text clustering approaches might yield interesting insights too.
Ideally, obtaining the full article text would lead to improved results.
However, extracting text from PDFs is a tough problem\footnote{See for example \url{https://www.filingdb.com/pdf-text-extraction}.}.

\section{Conclusion}
Clouds and other infrastructures have been widely adopted to run workloads on.
In particular, workflows are a popular workload model nowadays as they support the workload of many domains.
The mixture of \gls{qos} requirements, different scheduling targets from both the user and resource provider makes workflow scheduling a complex problem.
Getting insight into research done in this topic can be a daunting task, with the high-volume of publications in this research area, which is likely to intensify in the future.

Surveys are an excellent way to learn about the current status of a field, emerging trends, and open challenges.
Unfortunately, surveys rarely publish the data on which their survey is based.
Moreover, surveys rarely focus on the community itself; the structure, authors relationships, and citation information can provide insight into their health and operations, which can be interesting to community leaders and organizers.

In this work, we address these issues by performing various types of analyses.

We start by introducing and making open-source our instrument used to gather, filter, and unify article meta-data.
Using this meta-data, we obtain insights into the workflow scheduling community and four related areas.
For each, we analyze in-depth the community, look at important keywords both overall and per year, and identify emerging keywords.
Additionally, using this meta-data we were able to perform a systematic literature survey to construct and validate our taxonomies.

We observe that for all areas, 80+\% of the authors only author a single article in the timespan 2009-2018.
Furthermore, our meta-data suggests that larger cliques tend to have members with a lower average and maximum citation count when compared to small and moderately sized cliques.
This may indicate that the likelihood junior researchers are part of such cliques is higher, and that well-cited authors do not engage with all members of large communities.
However, more research in this direction is required to draw more definitive conclusions.
Our provided open-source instrument provides an excellent start for such future work.

Finally, using our instrument we map the most recent and top-cited allocation and provisioning policies to our taxonomies to demonstrate their completeness.

All software and artifacts that we introduce to obtain paper meta-data and visualizations are made open-source.
We believe these tools are valuable to the community for finding related work (we already experienced this multiple times first-hand), reproduce or perform a survey similar to this study, or redo this study in the future to observe new trends.

Besides the directions for future work that were provided in each section, we believe more directions can be investigated and surveyed.
In particular, our taxonomies can be extended and integrated in other taxonomies; several surveys that we marked as related work expand in different directions with respect to our work.
Deeper analysis into the communities using more data, different angles, and statistical methods may provide additional insights.

\section*{Availability of data and software artifacts}
All data and instruments used in this work are available as open-access, FAIR data.
The database on which this article is based can be found at \url{https://atlarge-research.com/data/2020_csur_aip.db}, AIP and other tools used to generate all floats in this article can be found at \url{https://github.com/atlarge-research/AIP}.

\section*{Acknowledgments}
We thank Erwin van Eyk, Oana Inel, Benno Kruit, and Alexandru Uta for their helpful comments. This work is sponsored by COMMIT and Vidi MagnaData.

\bibliographystyle{ACM-Reference-Format}
\bibliography{references/cites}


\begin{thebibliography}{185}


\ifx \showCODEN    \undefined \def \showCODEN     #1{\unskip}     \fi
\ifx \showDOI      \undefined \def \showDOI       #1{#1}\fi
\ifx \showISBNx    \undefined \def \showISBNx     #1{\unskip}     \fi
\ifx \showISBNxiii \undefined \def \showISBNxiii  #1{\unskip}     \fi
\ifx \showISSN     \undefined \def \showISSN      #1{\unskip}     \fi
\ifx \showLCCN     \undefined \def \showLCCN      #1{\unskip}     \fi
\ifx \shownote     \undefined \def \shownote      #1{#1}          \fi
\ifx \showarticletitle \undefined \def \showarticletitle #1{#1}   \fi
\ifx \showURL      \undefined \def \showURL       {\relax}        \fi
\providecommand\bibfield[2]{#2}
\providecommand\bibinfo[2]{#2}
\providecommand\natexlab[1]{#1}
\providecommand\showeprint[2][]{arXiv:#2}

\bibitem[\protect\citeauthoryear{??}{phy}{[n. d.]}]%
        {physamazon}
 \bibinfo{year}{[n. d.]}\natexlab{}.
\newblock \bibinfo{title}{Amazon plans wind farm to power its datacenters}.
\newblock
  \bibinfo{howpublished}{\url{https://phys.org/news/2015-01-amazon-farm-power-datacenters.html}}.
\newblock
\newblock
\shownote{Accessed: 2019-07-22.}


\bibitem[\protect\citeauthoryear{??}{gua}{[n. d.]}]%
        {guardianPaaS}
 \bibinfo{year}{[n. d.]}\natexlab{}.
\newblock \bibinfo{title}{Google angles for business users with 'platform as a
  service'}.
\newblock
  \bibinfo{howpublished}{\url{https://www.theguardian.com/technology/2008/apr/17/google.software}}.
\newblock
\newblock
\shownote{Accessed: 2018-06-14.}


\bibitem[\protect\citeauthoryear{??}{dat}{[n. d.]}]%
        {datacenterknowledgeenergy}
 \bibinfo{year}{[n. d.]}\natexlab{}.
\newblock \bibinfo{title}{Here's How Much Energy All US Data Centers Consume}.
\newblock
  \bibinfo{howpublished}{\url{https://www.datacenterknowledge.com/archives/2016/06/27/heres-how-much-energy-all-us-data-centers-consume}}.
\newblock
\newblock
\shownote{Accessed: 2019-07-22.}


\bibitem[\protect\citeauthoryear{??}{goo}{[n. d.]}]%
        {google2019renewable}
 \bibinfo{year}{[n. d.]}\natexlab{}.
\newblock \bibinfo{title}{Renewable energy - Data Centers}.
\newblock
  \bibinfo{howpublished}{\url{https://www.google.com/about/datacenters/renewable/}}.
\newblock
\newblock
\shownote{Accessed: 2019-07-22.}


\bibitem[\protect\citeauthoryear{??}{mic}{[n. d.]}]%
        {microsoft2019with}
 \bibinfo{year}{[n. d.]}\natexlab{}.
\newblock \bibinfo{title}{With our latest energy deal, Microsoft's Cheyenne
  datacenter will now be powered entirely by wind energy, keeping us on course
  to build a greener, more responsible cloud}.
\newblock
  \bibinfo{howpublished}{\url{https://blogs.microsoft.com/on-the-issues/2016/11/14/latest-energy-deal-microsofts-cheyenne-datacenter-will-now-powered-entirely-wind-energy-keeping-us-course-build-greener-responsible-cloud}}.
\newblock
\newblock
\shownote{Accessed: 2019-07-22.}


\bibitem[\protect\citeauthoryear{Abad, Boza, and Van~Eyk}{Abad
  et~al\mbox{.}}{2018}]%
        {abad2018package}
\bibfield{author}{\bibinfo{person}{Cristina~L Abad}, \bibinfo{person}{Edwin~F
  Boza}, {and} \bibinfo{person}{Erwin Van~Eyk}.}
  \bibinfo{year}{2018}\natexlab{}.
\newblock \showarticletitle{Package-Aware Scheduling of FaaS Functions}. In
  \bibinfo{booktitle}{\emph{Companion of the 2018 ACM/SPEC International
  Conference on Performance Engineering}}. ACM, \bibinfo{pages}{101--106}.
\newblock

 \begin{quotation}\noindent Introduce a package-aware scheduling algorithm for
  Function-as-a-Service. The algorithm tries to assign functions to workers
  with libraries already loaded to reduce latency. \end{quotation}
\bibitem[\protect\citeauthoryear{Abrishami, Naghibzadeh, and Epema}{Abrishami
  et~al\mbox{.}}{2013a}]%
        {abrishami2013deadline}
\bibfield{author}{\bibinfo{person}{Saeid Abrishami}, \bibinfo{person}{Mahmoud
  Naghibzadeh}, {and} \bibinfo{person}{Dick~HJ Epema}.}
  \bibinfo{year}{2013}\natexlab{a}.
\newblock \showarticletitle{Deadline-constrained workflow scheduling algorithms
  for infrastructure as a service clouds}.
\newblock \bibinfo{journal}{\emph{Future Generation Computer Systems}}
  \bibinfo{volume}{29}, \bibinfo{number}{1} (\bibinfo{year}{2013}),
  \bibinfo{pages}{158--169}.
\newblock

 \begin{quotation}\noindent Presents an adaptation of the partial critical path
  scheduling algorithm for cloud environments to develop two workflow
  scheduling algorithms. \end{quotation}
\bibitem[\protect\citeauthoryear{Abrishami, Naghibzadeh, and Epema}{Abrishami
  et~al\mbox{.}}{2012}]%
        {DBLP:journals/tpds/AbrishamiNE12}
\bibfield{author}{\bibinfo{person}{Saeid Abrishami}, \bibinfo{person}{Mahmoud
  Naghibzadeh}, {and} \bibinfo{person}{Dick H.~J. Epema}.}
  \bibinfo{year}{2012}\natexlab{}.
\newblock \showarticletitle{Cost-Driven Scheduling of Grid Workflows Using
  Partial Critical Paths}.
\newblock \bibinfo{journal}{\emph{{IEEE} Trans. Parallel Distrib. Syst.}}
  \bibinfo{volume}{23}, \bibinfo{number}{8} (\bibinfo{year}{2012}),
  \bibinfo{pages}{1400--1414}.
\newblock
\urldef\tempurl%
\url{https://doi.org/10.1109/TPDS.2011.303}
\showDOI{\tempurl}

 \begin{quotation}\noindent Introduces a new scheduling algorithm based on
  Patrial Critical Paths which tries to minimize the cost of execution while
  meeting a user-defined deadline. The algorithm recursively assigns
  sub-deadlines to tasks and then it assigns the cheapest service to each task
  while still meeting the sub-deadline. The assignment of sub-deadlines is done
  through policies of which three are proposed. Results show that this
  algorithm outperforms Deadline-MDP. \end{quotation}
\bibitem[\protect\citeauthoryear{Abrishami, Naghibzadeh, and Epema}{Abrishami
  et~al\mbox{.}}{2013b}]%
        {DBLP:journals/fgcs/AbrishamiNE13}
\bibfield{author}{\bibinfo{person}{Saeid Abrishami}, \bibinfo{person}{Mahmoud
  Naghibzadeh}, {and} \bibinfo{person}{Dick H.~J. Epema}.}
  \bibinfo{year}{2013}\natexlab{b}.
\newblock \showarticletitle{Deadline-constrained workflow scheduling algorithms
  for Infrastructure as a Service Clouds}.
\newblock \bibinfo{journal}{\emph{Future Generation Comp. Syst.}}
  \bibinfo{volume}{29}, \bibinfo{number}{1} (\bibinfo{year}{2013}),
  \bibinfo{pages}{158--169}.
\newblock
\urldef\tempurl%
\url{https://doi.org/10.1016/j.future.2012.05.004}
\showDOI{\tempurl}

 \begin{quotation}\noindent Extends the work proposed in
  \cite{DBLP:journals/tpds/AbrishamiNE12} for use in cloud environments. Two
  scheduling algorithms are introduced: IC-PCP and IC-PCPD2. IC-PCP has a
  one-phase algorithm that tries to find an (existing or new) instance capable
  of executing the entire path. IC-PCPD2 has a two-phase algorithm that assigns
  sub-deadlines in the first phase and then attempts to find existing machines
  capable of executing the path before attempting to lease new machines in the
  second phase. \end{quotation}
\bibitem[\protect\citeauthoryear{Afzal, Darlington, and McGough}{Afzal
  et~al\mbox{.}}{2006}]%
        {afzal2006qos}
\bibfield{author}{\bibinfo{person}{Ali Afzal}, \bibinfo{person}{John
  Darlington}, {and} \bibinfo{person}{A McGough}.}
  \bibinfo{year}{2006}\natexlab{}.
\newblock \showarticletitle{Qos-constrained stochastic workflow scheduling in
  enterprise and scientific grids}. In \bibinfo{booktitle}{\emph{Proceedings of
  the 7th IEEE/ACM International Conference on Grid Computing}}. IEEE Computer
  Society, \bibinfo{pages}{1--8}.
\newblock

 \begin{quotation}\noindent Introduce a scheduling mechanism for entire grid
  workloads that focuses on minimizing global costs. \end{quotation}
\bibitem[\protect\citeauthoryear{Akkus, Chen, Rimac, Stein, Satzke, Beck,
  Aditya, and Hilt}{Akkus et~al\mbox{.}}{2018}]%
        {akkus2018sand}
\bibfield{author}{\bibinfo{person}{Istemi~Ekin Akkus},
  \bibinfo{person}{Ruichuan Chen}, \bibinfo{person}{Ivica Rimac},
  \bibinfo{person}{Manuel Stein}, \bibinfo{person}{Klaus Satzke},
  \bibinfo{person}{Andre Beck}, \bibinfo{person}{Paarijaat Aditya}, {and}
  \bibinfo{person}{Volker Hilt}.} \bibinfo{year}{2018}\natexlab{}.
\newblock \showarticletitle{$\{$SAND$\}$: Towards High-Performance Serverless
  Computing}. In \bibinfo{booktitle}{\emph{2018 $\{$USENIX$\}$ Annual Technical
  Conference ($\{$USENIX$\}$$\{$ATC$\}$ 18)}}. \bibinfo{pages}{923--935}.
\newblock

 \begin{quotation}\noindent Introduces {SAND}, a serverless computing systems
  that uses application-level sandboxing and a hierarchical message bus to
  improve latency. \end{quotation}
\bibitem[\protect\citeauthoryear{Alkhanak, Lee, and Khan}{Alkhanak
  et~al\mbox{.}}{2015}]%
        {alkhanak2015cost}
\bibfield{author}{\bibinfo{person}{Ehab~Nabiel Alkhanak},
  \bibinfo{person}{Sai~Peck Lee}, {and} \bibinfo{person}{Saif Ur~Rehman Khan}.}
  \bibinfo{year}{2015}\natexlab{}.
\newblock \showarticletitle{Cost-aware challenges for workflow scheduling
  approaches in cloud computing environments: Taxonomy and opportunities}.
\newblock \bibinfo{journal}{\emph{Future Generation Computer Systems}}
  \bibinfo{volume}{50} (\bibinfo{year}{2015}), \bibinfo{pages}{3--21}.
\newblock

 \begin{quotation}\noindent Presents an overview and taxonomies of cost-aware
  workflow scheduling in cloud environments. Both challenges and opportunities
  are covered. \end{quotation}
\bibitem[\protect\citeauthoryear{Alnemr, Cayirci, Dalla~Corte, Garaga, Leenes,
  Mhungu, Pearson, Reed, de~Oliveira, Stefanatou, et~al\mbox{.}}{Alnemr
  et~al\mbox{.}}{2015}]%
        {alnemr2015data}
\bibfield{author}{\bibinfo{person}{Rehab Alnemr}, \bibinfo{person}{Erdal
  Cayirci}, \bibinfo{person}{Lorenzo Dalla~Corte}, \bibinfo{person}{Alexandr
  Garaga}, \bibinfo{person}{Ronald Leenes}, \bibinfo{person}{Rodney Mhungu},
  \bibinfo{person}{Siani Pearson}, \bibinfo{person}{Chris Reed},
  \bibinfo{person}{Anderson~Santana de Oliveira}, \bibinfo{person}{Dimitra
  Stefanatou}, {et~al\mbox{.}}} \bibinfo{year}{2015}\natexlab{}.
\newblock \showarticletitle{A data protection impact assessment methodology for
  cloud}. In \bibinfo{booktitle}{\emph{Annual Privacy Forum}}. Springer,
  \bibinfo{pages}{60--92}.
\newblock

 \begin{quotation}\noindent Introduces a data protection impact assessment
  methodology for cloud environments. In particular it focuses on risk
  assessment of data when facing legislations such as GDPR. \end{quotation}
\bibitem[\protect\citeauthoryear{Ammar, Groeneveld, Bhagavatula, Beltagy,
  Crawford, Downey, Dunkelberger, Elgohary, Feldman, Ha, Kinney, Kohlmeier, Lo,
  Murray, Ooi, Peters, Power, Skjonsberg, Wang, Wilhelm, Yuan, van Zuylen, and
  Etzioni}{Ammar et~al\mbox{.}}{2018}]%
        {ammar:18}
\bibfield{author}{\bibinfo{person}{Waleed Ammar}, \bibinfo{person}{Dirk
  Groeneveld}, \bibinfo{person}{Chandra Bhagavatula}, \bibinfo{person}{Iz
  Beltagy}, \bibinfo{person}{Miles Crawford}, \bibinfo{person}{Doug Downey},
  \bibinfo{person}{Jason Dunkelberger}, \bibinfo{person}{Ahmed Elgohary},
  \bibinfo{person}{Sergey Feldman}, \bibinfo{person}{Vu Ha},
  \bibinfo{person}{Rodney Kinney}, \bibinfo{person}{Sebastian Kohlmeier},
  \bibinfo{person}{Kyle Lo}, \bibinfo{person}{Tyler Murray},
  \bibinfo{person}{Hsu-Han Ooi}, \bibinfo{person}{Matthew Peters},
  \bibinfo{person}{Joanna Power}, \bibinfo{person}{Sam Skjonsberg},
  \bibinfo{person}{Lucy~Lu Wang}, \bibinfo{person}{Chris Wilhelm},
  \bibinfo{person}{Zheng Yuan}, \bibinfo{person}{Madeleine van Zuylen}, {and}
  \bibinfo{person}{Oren Etzioni}.} \bibinfo{year}{2018}\natexlab{}.
\newblock \showarticletitle{Construction of the Literature Graph in Semantic
  Scholar}. In \bibinfo{booktitle}{\emph{NAACL}}.
\newblock
\urldef\tempurl%
\url{https://www.semanticscholar.org/paper/09e3cf5704bcb16e6657f6ceed70e93373a54618}
\showURL{%
\tempurl}


\bibitem[\protect\citeauthoryear{Amstutz, Andeer, Chapman, Chilton, Crusoe,
  Valls~Guimera, Carrasco~Hernandez, Ivkovic, Kartashov, Kern,
  et~al\mbox{.}}{Amstutz et~al\mbox{.}}{2016}]%
        {amstutz2016common}
\bibfield{author}{\bibinfo{person}{Peter Amstutz}, \bibinfo{person}{Robin
  Andeer}, \bibinfo{person}{Brad Chapman}, \bibinfo{person}{John Chilton},
  \bibinfo{person}{Michael~R Crusoe}, \bibinfo{person}{Roman Valls~Guimera},
  \bibinfo{person}{Guillermo Carrasco~Hernandez}, \bibinfo{person}{Sinisa
  Ivkovic}, \bibinfo{person}{Andrey Kartashov}, \bibinfo{person}{John Kern},
  {et~al\mbox{.}}} \bibinfo{year}{2016}\natexlab{}.
\newblock \showarticletitle{Common Workflow Language, Draft 3}.
\newblock  (\bibinfo{year}{2016}).
\newblock


\bibitem[\protect\citeauthoryear{Amvrosiadis, Park, Ganger, Gibson, Baseman,
  and DeBardeleben}{Amvrosiadis et~al\mbox{.}}{2018}]%
        {DBLP:conf/usenix/AmvrosiadisPGGB18}
\bibfield{author}{\bibinfo{person}{George Amvrosiadis},
  \bibinfo{person}{Jun~Woo Park}, \bibinfo{person}{Gregory~R. Ganger},
  \bibinfo{person}{Garth~A. Gibson}, \bibinfo{person}{Elisabeth Baseman}, {and}
  \bibinfo{person}{Nathan DeBardeleben}.} \bibinfo{year}{2018}\natexlab{}.
\newblock \showarticletitle{On the diversity of cluster workloads and its
  impact on research results}. In \bibinfo{booktitle}{\emph{2018 {USENIX}
  Annual Technical Conference, {USENIX} {ATC} 2018, Boston, MA, USA, July
  11-13, 2018.}} \bibinfo{pages}{533--546}.
\newblock
\urldef\tempurl%
\url{https://www.usenix.org/conference/atc18/presentation/amvrosiadis}
\showURL{%
\tempurl}

 \begin{quotation}\noindent Demonstrates that the popular Google cluster
  traces, often used as example of a representative workload, are more of an
  outlier than a typical production trace. Using traces from Los Alamos
  National LAB (LANL) and Two Sigma, a high-frequency trading company, they
  show using various metrics the (significant) differences between these
  traces. \end{quotation}
\bibitem[\protect\citeauthoryear{Andrade, Cirne, Brasileiro, and
  Roisenberg}{Andrade et~al\mbox{.}}{2003}]%
        {DBLP:conf/jsspp/AndradeCBR03}
\bibfield{author}{\bibinfo{person}{Nazareno Andrade}, \bibinfo{person}{Walfredo
  Cirne}, \bibinfo{person}{Francisco~Vilar Brasileiro}, {and}
  \bibinfo{person}{Paulo Roisenberg}.} \bibinfo{year}{2003}\natexlab{}.
\newblock \showarticletitle{OurGrid: An Approach to Easily Assemble Grids with
  Equitable Resource Sharing}. In \bibinfo{booktitle}{\emph{Job Scheduling
  Strategies for Parallel Processing, 9th International Workshop, {JSSPP} 2003,
  Seattle, WA, USA, June 24, 2003, Revised Papers}}. \bibinfo{pages}{61--86}.
\newblock
\urldef\tempurl%
\url{https://doi.org/10.1007/10968987\_4}
\showDOI{\tempurl}

 \begin{quotation}\noindent Introduces OurGrid, a peer-to-peer method to share
  sites with resources to form a grid environment. \end{quotation}
\bibitem[\protect\citeauthoryear{Andreadis, Versluis, Mastenbroek, and
  Iosup}{Andreadis et~al\mbox{.}}{2018}]%
        {DBLP:conf/sc/AndreadisVMI18}
\bibfield{author}{\bibinfo{person}{Georgios Andreadis},
  \bibinfo{person}{Laurens Versluis}, \bibinfo{person}{Fabian Mastenbroek},
  {and} \bibinfo{person}{Alexandru Iosup}.} \bibinfo{year}{2018}\natexlab{}.
\newblock \showarticletitle{A reference architecture for datacenter scheduling:
  design, validation, and experiments}. In
  \bibinfo{booktitle}{\emph{Proceedings of the International Conference for
  High Performance Computing, Networking, Storage, and Analysis, {SC} 2018,
  Dallas, TX, USA, November 11-16, 2018}}. \bibinfo{pages}{37:1--37:15}.
\newblock
\urldef\tempurl%
\url{http://dl.acm.org/citation.cfm?id=3291706}
\showURL{%
\tempurl}

 \begin{quotation}\noindent Introduce a reference architecture for datacenter
  schedulers. The authors demonstrate that schedulers are systematically
  underspecified in literature and demonstrate the impact of not specifying
  components on the reproducibility of this literature. \end{quotation}
\bibitem[\protect\citeauthoryear{Arabnejad, Bubendorfer, and Ng}{Arabnejad
  et~al\mbox{.}}{2019}]%
        {DBLP:journals/tpds/ArabnejadBN19}
\bibfield{author}{\bibinfo{person}{Vahid Arabnejad}, \bibinfo{person}{Kris
  Bubendorfer}, {and} \bibinfo{person}{Bryan Ng}.}
  \bibinfo{year}{2019}\natexlab{}.
\newblock \showarticletitle{Budget and Deadline Aware e-Science Workflow
  Scheduling in Clouds}.
\newblock \bibinfo{journal}{\emph{{IEEE} Trans. Parallel Distrib. Syst.}}
  \bibinfo{volume}{30}, \bibinfo{number}{1} (\bibinfo{year}{2019}),
  \bibinfo{pages}{29--44}.
\newblock
\urldef\tempurl%
\url{https://doi.org/10.1109/TPDS.2018.2849396}
\showDOI{\tempurl}

 \begin{quotation}\noindent Introduces BDAS: a budget and deadline aware
  scheduling technique. The cost-time tradeoff is tunable in this algorithm.
  \end{quotation}
\bibitem[\protect\citeauthoryear{Aumala, Boza, Ortiz{-}Avil{\'{e}}s, Totoy, and
  Abad}{Aumala et~al\mbox{.}}{2019}]%
        {DBLP:conf/ccgrid/AumalaBOTA19}
\bibfield{author}{\bibinfo{person}{Gabriel Aumala}, \bibinfo{person}{Edwin~F.
  Boza}, \bibinfo{person}{Luis Ortiz{-}Avil{\'{e}}s}, \bibinfo{person}{Gustavo
  Totoy}, {and} \bibinfo{person}{Cristina Abad}.}
  \bibinfo{year}{2019}\natexlab{}.
\newblock \showarticletitle{Beyond Load Balancing: Package-Aware Scheduling for
  Serverless Platforms}. In \bibinfo{booktitle}{\emph{19th {IEEE/ACM}
  International Symposium on Cluster, Cloud and Grid Computing, {CCGRID} 2019,
  Larnaca, Cyprus, May 14-17, 2019}}. \bibinfo{pages}{282--291}.
\newblock
\urldef\tempurl%
\url{https://doi.org/10.1109/CCGRID.2019.00042}
\showDOI{\tempurl}

 \begin{quotation}\noindent Introduce a package aware load balancer for FaaS
  execution. The authors implement their policy in the OpenLambda framework and
  observe an average speedup of 1.3x up to 23x at the 80th percentile.
  \end{quotation}
\bibitem[\protect\citeauthoryear{Bal, Epema, de~Laat, van Nieuwpoort, Romein,
  Seinstra, Snoek, and Wijshoff}{Bal et~al\mbox{.}}{2016}]%
        {DBLP:journals/computer/BalELNRSSW16}
\bibfield{author}{\bibinfo{person}{Henri~E. Bal}, \bibinfo{person}{Dick H.~J.
  Epema}, \bibinfo{person}{Cees de Laat}, \bibinfo{person}{Rob van Nieuwpoort},
  \bibinfo{person}{John~W. Romein}, \bibinfo{person}{Frank~J. Seinstra},
  \bibinfo{person}{Cees Snoek}, {and} \bibinfo{person}{Harry A.~G. Wijshoff}.}
  \bibinfo{year}{2016}\natexlab{}.
\newblock \showarticletitle{A Medium-Scale Distributed System for Computer
  Science Research: Infrastructure for the Long Term}.
\newblock \bibinfo{journal}{\emph{{IEEE} Computer}} \bibinfo{volume}{49},
  \bibinfo{number}{5} (\bibinfo{year}{2016}), \bibinfo{pages}{54--63}.
\newblock
\urldef\tempurl%
\url{https://doi.org/10.1109/MC.2016.127}
\showDOI{\tempurl}

 \begin{quotation}\noindent Introduces and describes the DAS5 Dutch national
  multi-cluster supercomputer. \end{quotation}
\bibitem[\protect\citeauthoryear{Bala and Chana}{Bala and Chana}{2012}]%
        {bala2012fault}
\bibfield{author}{\bibinfo{person}{Anju Bala} {and} \bibinfo{person}{Inderveer
  Chana}.} \bibinfo{year}{2012}\natexlab{}.
\newblock \showarticletitle{Fault tolerance-challenges, techniques and
  implementation in cloud computing}.
\newblock \bibinfo{journal}{\emph{International Journal of Computer Science
  Issues (IJCSI)}} \bibinfo{volume}{9}, \bibinfo{number}{1}
  (\bibinfo{year}{2012}), \bibinfo{pages}{288}.
\newblock

 \begin{quotation}\noindent Discusses the existing fault tolerance techniques
  in cloud environments based on the policies and tools used, and research
  challenges. \end{quotation}
\bibitem[\protect\citeauthoryear{Baldini, Castro, Chang, Cheng, Fink, Ishakian,
  Mitchell, Muthusamy, Rabbah, Slominski, et~al\mbox{.}}{Baldini
  et~al\mbox{.}}{2017}]%
        {baldini2017serverless}
\bibfield{author}{\bibinfo{person}{Ioana Baldini}, \bibinfo{person}{Paul
  Castro}, \bibinfo{person}{Kerry Chang}, \bibinfo{person}{Perry Cheng},
  \bibinfo{person}{Stephen Fink}, \bibinfo{person}{Vatche Ishakian},
  \bibinfo{person}{Nick Mitchell}, \bibinfo{person}{Vinod Muthusamy},
  \bibinfo{person}{Rodric Rabbah}, \bibinfo{person}{Aleksander Slominski},
  {et~al\mbox{.}}} \bibinfo{year}{2017}\natexlab{}.
\newblock \showarticletitle{Serverless computing: Current trends and open
  problems}.
\newblock In \bibinfo{booktitle}{\emph{Research Advances in Cloud Computing}}.
  \bibinfo{publisher}{Springer}, \bibinfo{pages}{1--20}.
\newblock

 \begin{quotation}\noindent Surveys existing serverless platforms to identify
  key characteristics and use cases and describe challenges and open problems.
  \end{quotation}
\bibitem[\protect\citeauthoryear{Barr}{Barr}{2018}]%
        {barr_2018}
\bibfield{author}{\bibinfo{person}{Jeff Barr}.}
  \bibinfo{year}{2018}\natexlab{}.
\newblock \bibinfo{title}{New – Per-Second Billing for EC2 Instances and EBS
  Volumes | Amazon Web Services}.
\newblock
\newblock
\urldef\tempurl%
\url{https://aws.amazon.com/blogs/aws/new-per-second-billing-for-ec2-instances-and-ebs-volumes/}
\showURL{%
\tempurl}


\bibitem[\protect\citeauthoryear{Bastos, Braga, and Gomes}{Bastos
  et~al\mbox{.}}{2015}]%
        {bastos2015scientific}
\bibfield{author}{\bibinfo{person}{Bruno~Fernandes Bastos},
  \bibinfo{person}{Regina Maria~Maciel Braga}, {and}
  \bibinfo{person}{Ant{\^o}nio Tadeu~Azevedo Gomes}.}
  \bibinfo{year}{2015}\natexlab{}.
\newblock \showarticletitle{Scientific workflow interchanging through patterns:
  Reversals and lessons learned}. In \bibinfo{booktitle}{\emph{2015 IEEE 11th
  International Conference on e-Science}}. IEEE, \bibinfo{pages}{557--564}.
\newblock

 \begin{quotation}\noindent Discusses the use of workflow patterns combined
  with software architecture concepts to capture semantics in workflows. The
  authors look at different workflow formalisms to inspect interchanging
  specification between workflow management systems. \end{quotation}
\bibitem[\protect\citeauthoryear{Bauer, Herbst, Spinner, Ali-Eldin, and
  Kounev}{Bauer et~al\mbox{.}}{2018}]%
        {bauer2018chameleon}
\bibfield{author}{\bibinfo{person}{Andr{\'e} Bauer}, \bibinfo{person}{Nikolas
  Herbst}, \bibinfo{person}{Simon Spinner}, \bibinfo{person}{Ahmed Ali-Eldin},
  {and} \bibinfo{person}{Samuel Kounev}.} \bibinfo{year}{2018}\natexlab{}.
\newblock \showarticletitle{Chameleon: A Hybrid, Proactive Auto-Scaling
  Mechanism on a Level-Playing Field}.
\newblock \bibinfo{journal}{\emph{IEEE Transactions on Parallel and Distributed
  Systems}} (\bibinfo{year}{2018}).
\newblock

 \begin{quotation}\noindent Introduces an autoscaler based on workload
  forecasting using machine learning. \end{quotation}
\bibitem[\protect\citeauthoryear{Bauer, Lesch, Versluis, Ilyushkin, Herbst, and
  Kounev}{Bauer et~al\mbox{.}}{2019}]%
        {bauer2019chamulteon}
\bibfield{author}{\bibinfo{person}{Andr{\'e} Bauer}, \bibinfo{person}{Veronika
  Lesch}, \bibinfo{person}{Laurens Versluis}, \bibinfo{person}{Alexey
  Ilyushkin}, \bibinfo{person}{Nikolas Herbst}, {and} \bibinfo{person}{Samuel
  Kounev}.} \bibinfo{year}{2019}\natexlab{}.
\newblock \showarticletitle{Chamulteon: Coordinated auto-scaling of
  micro-services}. In \bibinfo{booktitle}{\emph{2019 IEEE 39th International
  Conference on Distributed Computing Systems (ICDCS)}}. IEEE,
  \bibinfo{pages}{2015--2025}.
\newblock

 \begin{quotation}\noindent A redesign of the Chameleon autoscaler aimed at
  multi-tier applications. \end{quotation}
\bibitem[\protect\citeauthoryear{Benoit, {\c{C}}ataly{\"u}rek, Robert, and
  Saule}{Benoit et~al\mbox{.}}{2013}]%
        {benoit2013survey}
\bibfield{author}{\bibinfo{person}{Anne Benoit}, \bibinfo{person}{{\"U}mit~V
  {\c{C}}ataly{\"u}rek}, \bibinfo{person}{Yves Robert}, {and}
  \bibinfo{person}{Erik Saule}.} \bibinfo{year}{2013}\natexlab{}.
\newblock \showarticletitle{A survey of pipelined workflow scheduling: Models
  and algorithms}.
\newblock \bibinfo{journal}{\emph{ACM Computing Surveys (CSUR)}}
  \bibinfo{volume}{45}, \bibinfo{number}{4} (\bibinfo{year}{2013}),
  \bibinfo{pages}{50}.
\newblock

 \begin{quotation}\noindent Presents a survey on workflow scheduling models and
  algorithms primarily focused on pipelined workflow scheduling.
  \end{quotation}
\bibitem[\protect\citeauthoryear{Benoit, Renaud{-}Goud, and Robert}{Benoit
  et~al\mbox{.}}{2010}]%
        {DBLP:conf/ipps/BenoitRR10}
\bibfield{author}{\bibinfo{person}{Anne Benoit}, \bibinfo{person}{Paul
  Renaud{-}Goud}, {and} \bibinfo{person}{Yves Robert}.}
  \bibinfo{year}{2010}\natexlab{}.
\newblock \showarticletitle{Performance and energy optimization of concurrent
  pipelined applications}. In \bibinfo{booktitle}{\emph{24th {IEEE}
  International Symposium on Parallel and Distributed Processing, {IPDPS} 2010,
  Atlanta, Georgia, USA, 19-23 April 2010 - Conference Proceedings}}.
  \bibinfo{pages}{1--12}.
\newblock
\urldef\tempurl%
\url{https://doi.org/10.1109/IPDPS.2010.5470483}
\showDOI{\tempurl}

 \begin{quotation}\noindent Studies the problem of optimizing performance and
  energy consumption on homo- and heterogeneous resources by using multiple
  mapping strategies and energy preserving techniques. \end{quotation}
\bibitem[\protect\citeauthoryear{Berkowitz, Clark, Gambhir, McElvain,
  Nicholson, Rinaldi, Vranas, Walker{-}Loud, Chang, Jo{\'{o}}, Kurth, and
  Orginos}{Berkowitz et~al\mbox{.}}{2018}]%
        {DBLP:conf/sc/BerkowitzCGMNRV18}
\bibfield{author}{\bibinfo{person}{Evan Berkowitz}, \bibinfo{person}{Michael~A.
  Clark}, \bibinfo{person}{Arjun~Singh Gambhir}, \bibinfo{person}{Kenneth
  McElvain}, \bibinfo{person}{Amy Nicholson}, \bibinfo{person}{Enrico Rinaldi},
  \bibinfo{person}{Pavlos Vranas}, \bibinfo{person}{Andr{\'{e}} Walker{-}Loud},
  \bibinfo{person}{Chia{-}Cheng Chang}, \bibinfo{person}{B{\'{a}}lint
  Jo{\'{o}}}, \bibinfo{person}{Thorsten Kurth}, {and} \bibinfo{person}{Kostas
  Orginos}.} \bibinfo{year}{2018}\natexlab{}.
\newblock \showarticletitle{Simulating the \emph{weak} death of the Neutron in
  a femtoscale universe with near-exascale computing}. In
  \bibinfo{booktitle}{\emph{Proceedings of the International Conference for
  High Performance Computing, Networking, Storage, and Analysis, {SC} 2018,
  Dallas, TX, USA, November 11-16, 2018}}. \bibinfo{pages}{55:1--55:9}.
\newblock
\urldef\tempurl%
\url{http://dl.acm.org/citation.cfm?id=3291730}
\showURL{%
\tempurl}

 \begin{quotation}\noindent Presents an algorithm that exponentially decreases
  the time-to-solution for particle simulations. It is deployed across a large
  part of the {CORAL} supercomputer. \end{quotation}
\bibitem[\protect\citeauthoryear{Berl, Gelenbe, Di~Girolamo, Giuliani, De~Meer,
  Dang, and Pentikousis}{Berl et~al\mbox{.}}{2010}]%
        {berl2010energy}
\bibfield{author}{\bibinfo{person}{Andreas Berl}, \bibinfo{person}{Erol
  Gelenbe}, \bibinfo{person}{Marco Di~Girolamo}, \bibinfo{person}{Giovanni
  Giuliani}, \bibinfo{person}{Hermann De~Meer}, \bibinfo{person}{Minh~Quan
  Dang}, {and} \bibinfo{person}{Kostas Pentikousis}.}
  \bibinfo{year}{2010}\natexlab{}.
\newblock \showarticletitle{Energy-efficient cloud computing}.
\newblock \bibinfo{journal}{\emph{The computer journal}} \bibinfo{volume}{53},
  \bibinfo{number}{7} (\bibinfo{year}{2010}), \bibinfo{pages}{1045--1051}.
\newblock

 \begin{quotation}\noindent Surveys several approaches for energy efficient
  cloud computing. Both hardware and software approaches are investigated.
  \end{quotation}
\bibitem[\protect\citeauthoryear{Bilgaiyan, Sagnika, and Das}{Bilgaiyan
  et~al\mbox{.}}{2014}]%
        {bilgaiyan2014workflow}
\bibfield{author}{\bibinfo{person}{Saurabh Bilgaiyan},
  \bibinfo{person}{Santwana Sagnika}, {and} \bibinfo{person}{Madhabananda
  Das}.} \bibinfo{year}{2014}\natexlab{}.
\newblock \showarticletitle{Workflow scheduling in cloud computing environment
  using cat swarm optimization}. In \bibinfo{booktitle}{\emph{Advance Computing
  Conference (IACC), 2014 IEEE International}}. IEEE,
  \bibinfo{pages}{680--685}.
\newblock

 \begin{quotation}\noindent Present a Cat Swarm Optimization algorithm for
  scheduling workflows on resources. \end{quotation}
\bibitem[\protect\citeauthoryear{Bittencourt and Madeira}{Bittencourt and
  Madeira}{2010}]%
        {DBLP:journals/grid/BittencourtM10}
\bibfield{author}{\bibinfo{person}{Luiz~Fernando Bittencourt} {and}
  \bibinfo{person}{Edmundo R.~M. Madeira}.} \bibinfo{year}{2010}\natexlab{}.
\newblock \showarticletitle{Towards the Scheduling of Multiple Workflows on
  Computational Grids}.
\newblock \bibinfo{journal}{\emph{J. Grid Comput.}} \bibinfo{volume}{8},
  \bibinfo{number}{3} (\bibinfo{year}{2010}), \bibinfo{pages}{419--441}.
\newblock
\urldef\tempurl%
\url{https://doi.org/10.1007/s10723-009-9144-1}
\showDOI{\tempurl}

 \begin{quotation}\noindent Describes four initial approaches for scheduling
  multiple workflows on grids. Each of these are reviewed in terms of fairness
  and produces schedule length (makespan). The proposed algorithms use greedy
  and heuristic approaches. \end{quotation}
\bibitem[\protect\citeauthoryear{Bittencourt and Madeira}{Bittencourt and
  Madeira}{2011}]%
        {bittencourt2011hcoc}
\bibfield{author}{\bibinfo{person}{Luiz~Fernando Bittencourt} {and}
  \bibinfo{person}{Edmundo Roberto~Mauro Madeira}.}
  \bibinfo{year}{2011}\natexlab{}.
\newblock \showarticletitle{HCOC: a cost optimization algorithm for workflow
  scheduling in hybrid clouds}.
\newblock \bibinfo{journal}{\emph{Journal of Internet Services and
  Applications}} \bibinfo{volume}{2}, \bibinfo{number}{3}
  (\bibinfo{year}{2011}), \bibinfo{pages}{207--227}.
\newblock

 \begin{quotation}\noindent Introduces HCOC, an workflow scheduling algorithm
  for hybrid clouds that attempts to minimize costs while adhering to
  deadlines. \end{quotation}
\bibitem[\protect\citeauthoryear{Blythe, Jain, Deelman, Gil, Vahi, Mandal, and
  Kennedy}{Blythe et~al\mbox{.}}{2005}]%
        {DBLP:conf/ccgrid/BlytheJDGVMK05}
\bibfield{author}{\bibinfo{person}{James Blythe}, \bibinfo{person}{S. Jain},
  \bibinfo{person}{Ewa Deelman}, \bibinfo{person}{Yolanda Gil},
  \bibinfo{person}{Karan Vahi}, \bibinfo{person}{Anirban Mandal}, {and}
  \bibinfo{person}{Ken Kennedy}.} \bibinfo{year}{2005}\natexlab{}.
\newblock \showarticletitle{Task scheduling strategies for workflow-based
  applications in grids}. In \bibinfo{booktitle}{\emph{5th International
  Symposium on Cluster Computing and the Grid (CCGrid 2005), 9-12 May, 2005,
  Cardiff, {UK}}}. \bibinfo{pages}{759--767}.
\newblock
\urldef\tempurl%
\url{https://doi.org/10.1109/CCGRID.2005.1558639}
\showDOI{\tempurl}

 \begin{quotation}\noindent Compares the behavior of task-based algorithms and
  workflow based algorithms. Workflows are derived from one real application,
  using varying ratios of computation cost to data transfer cost. Experiments
  are conducted using a grid simulator based on NS-2, using 6 sites. Each site
  consists of a single host (varying computational speeds) and storage unit.
  Allows for multiple workflows to be scheduled simultaneously. Performance is
  compared between the task-based and workflow-based approach. \end{quotation}
\bibitem[\protect\citeauthoryear{Bonvin, Papaioannou, and Aberer}{Bonvin
  et~al\mbox{.}}{2011}]%
        {DBLP:conf/ccgrid/BonvinPA11}
\bibfield{author}{\bibinfo{person}{Nicolas Bonvin},
  \bibinfo{person}{Thanasis~G. Papaioannou}, {and} \bibinfo{person}{Karl
  Aberer}.} \bibinfo{year}{2011}\natexlab{}.
\newblock \showarticletitle{Autonomic SLA-Driven Provisioning for Cloud
  Applications}. In \bibinfo{booktitle}{\emph{11th {IEEE/ACM} International
  Symposium on Cluster, Cloud and Grid Computing, CCGrid 2011, Newport Beach,
  CA, USA, May 23-26, 2011}}. \bibinfo{pages}{434--443}.
\newblock
\urldef\tempurl%
\url{https://doi.org/10.1109/CCGrid.2011.24}
\showDOI{\tempurl}

 \begin{quotation}\noindent Introduces a SLA-driven (based on latency)
  provisioning system for Web-service workflow applications. \end{quotation}
\bibitem[\protect\citeauthoryear{B{\"o}rger}{B{\"o}rger}{2012}]%
        {borger2012approaches}
\bibfield{author}{\bibinfo{person}{Egon B{\"o}rger}.}
  \bibinfo{year}{2012}\natexlab{}.
\newblock \showarticletitle{Approaches to modeling business processes: a
  critical analysis of BPMN, workflow patterns and YAWL}.
\newblock \bibinfo{journal}{\emph{Software \& Systems Modeling}}
  \bibinfo{volume}{11}, \bibinfo{number}{3} (\bibinfo{year}{2012}),
  \bibinfo{pages}{305--318}.
\newblock

 \begin{quotation}\noindent Investigate three approaches for describing
  business processes. The authors demonstrate all approaches fail to provide
  means to capture business scenarios and to analyze, communicate and manage
  resulting models. \end{quotation}
\bibitem[\protect\citeauthoryear{Byun, Kee, Kim, and Maeng}{Byun
  et~al\mbox{.}}{2011}]%
        {DBLP:journals/fgcs/ByunKKM11}
\bibfield{author}{\bibinfo{person}{Eun{-}Kyu Byun}, \bibinfo{person}{Yang{-}Suk
  Kee}, \bibinfo{person}{Jin{-}Soo Kim}, {and} \bibinfo{person}{Seungryoul
  Maeng}.} \bibinfo{year}{2011}\natexlab{}.
\newblock \showarticletitle{Cost optimized provisioning of elastic resources
  for application workflows}.
\newblock \bibinfo{journal}{\emph{Future Generation Comp. Syst.}}
  \bibinfo{volume}{27}, \bibinfo{number}{8} (\bibinfo{year}{2011}),
  \bibinfo{pages}{1011--1026}.
\newblock
\urldef\tempurl%
\url{https://doi.org/10.1016/j.future.2011.05.001}
\showDOI{\tempurl}

 \begin{quotation}\noindent Introduces PBTS, an extension of the BTS
  provisioning algorithm in which workflows are partitioned and scheduled on
  cloud resources. \end{quotation}
\bibitem[\protect\citeauthoryear{Cao, Jarvis, Saini, and Nudd}{Cao
  et~al\mbox{.}}{2003}]%
        {cao2003gridflow}
\bibfield{author}{\bibinfo{person}{Junwei Cao}, \bibinfo{person}{Stephen~A
  Jarvis}, \bibinfo{person}{Subhash Saini}, {and} \bibinfo{person}{Graham~R
  Nudd}.} \bibinfo{year}{2003}\natexlab{}.
\newblock \showarticletitle{Gridflow: Workflow management for grid computing}.
  In \bibinfo{booktitle}{\emph{CCGrid 2003. 3rd IEEE/ACM International
  Symposium on Cluster Computing and the Grid, 2003. Proceedings.}} IEEE,
  \bibinfo{pages}{198--205}.
\newblock

 \begin{quotation}\noindent Presents Gridflow, a workflow management system for
  grid computing that offers a user portal and services for both global and
  local workflow scheduling. \end{quotation}
\bibitem[\protect\citeauthoryear{Cardoso, Sheth, and Miller}{Cardoso
  et~al\mbox{.}}{2002}]%
        {cardoso2002workflow}
\bibfield{author}{\bibinfo{person}{Jorge Cardoso}, \bibinfo{person}{Amit
  Sheth}, {and} \bibinfo{person}{John Miller}.}
  \bibinfo{year}{2002}\natexlab{}.
\newblock \showarticletitle{Workflow quality of service}. In
  \bibinfo{booktitle}{\emph{International Conference on Enterprise Integration
  and Modeling Technology}}. Springer, \bibinfo{pages}{303--311}.
\newblock

 \begin{quotation}\noindent Introduces a workflow QoS specification and methods
  to predict, analyze, and monitor QoS. \end{quotation}
\bibitem[\protect\citeauthoryear{Cesario, Lackovic, Talia, and Trunfio}{Cesario
  et~al\mbox{.}}{2013}]%
        {DBLP:journals/concurrency/CesarioLTT13}
\bibfield{author}{\bibinfo{person}{Eugenio Cesario}, \bibinfo{person}{Marco
  Lackovic}, \bibinfo{person}{Domenico Talia}, {and} \bibinfo{person}{Paolo
  Trunfio}.} \bibinfo{year}{2013}\natexlab{}.
\newblock \showarticletitle{Programming knowledge discovery workflows in
  service-oriented distributed systems}.
\newblock \bibinfo{journal}{\emph{Concurrency and Computation: Practice and
  Experience}} \bibinfo{volume}{25}, \bibinfo{number}{10}
  (\bibinfo{year}{2013}), \bibinfo{pages}{1482--1504}.
\newblock
\urldef\tempurl%
\url{https://doi.org/10.1002/cpe.2936}
\showDOI{\tempurl}

 \begin{quotation}\noindent Introduces a workflow formalism and a workflow
  processing framework named DIS3GNO. DIS3GNO is aimed at designing and running
  distributed knowledge discovery processes in the Knowledge Gridsystem.
  \end{quotation}
\bibitem[\protect\citeauthoryear{Chen and Zhang}{Chen and Zhang}{2009}]%
        {DBLP:journals/tsmc/ChenZ09}
\bibfield{author}{\bibinfo{person}{Wei{-}neng Chen} {and} \bibinfo{person}{Jun
  Zhang}.} \bibinfo{year}{2009}\natexlab{}.
\newblock \showarticletitle{An Ant Colony Optimization Approach to a Grid
  Workflow Scheduling Problem With Various QoS Requirements}.
\newblock \bibinfo{journal}{\emph{{IEEE} Trans. Systems, Man, and Cybernetics,
  Part {C}}} \bibinfo{volume}{39}, \bibinfo{number}{1} (\bibinfo{year}{2009}),
  \bibinfo{pages}{29--43}.
\newblock
\urldef\tempurl%
\url{https://doi.org/10.1109/TSMCC.2008.2001722}
\showDOI{\tempurl}

 \begin{quotation}\noindent Proses an Ant Colony Optimization (ACO) algorithm
  to schedule large workflows with multiple QoS parameters in grids. The
  algorithm allows users to specify QoS preferences as well as a minimum QoS
  threshold. The algorithm attempts to find a solution that satisfies all QoS
  constraints and optimizes the user-preferred QoS parameters. To this end,
  seven new heuristics are proposed for the ACO approach. \end{quotation}
\bibitem[\protect\citeauthoryear{Choi, Adufu, and Kim}{Choi
  et~al\mbox{.}}{2017}]%
        {choi2017data}
\bibfield{author}{\bibinfo{person}{Jieun Choi}, \bibinfo{person}{Theodora
  Adufu}, {and} \bibinfo{person}{Yoonhee Kim}.}
  \bibinfo{year}{2017}\natexlab{}.
\newblock \showarticletitle{Data-locality aware scientific workflow scheduling
  methods in HPC cloud environments}.
\newblock \bibinfo{journal}{\emph{International Journal of Parallel
  Programming}} \bibinfo{volume}{45}, \bibinfo{number}{5}
  (\bibinfo{year}{2017}), \bibinfo{pages}{1128--1141}.
\newblock

 \begin{quotation}\noindent Introduces D-LAWS: a data-locality aware workflow
  scheduling technique and a locality-aware resource management method for
  data-intensive scientific workflows. Using VM consolidation, exploiting task
  parallelism, and considering bandwidth and transfer times, performance
  improvements are reported. \end{quotation}
\bibitem[\protect\citeauthoryear{dblp team: dblp computer~science
  bibliography.}{dblp team: dblp computer~science bibliography.}{2019}]%
        {dblpSnapshotNovember2019}
\bibfield{author}{\bibinfo{person}{The dblp team: dblp computer~science
  bibliography.}} \bibinfo{year}{2019}\natexlab{}.
\newblock \bibinfo{title}{Monthly snapshot release of November 2019.}
\newblock
\newblock
\urldef\tempurl%
\url{https://dblp.org/xml/release/dblp-2019-11-01.xml.gz}
\showURL{%
\tempurl}


\bibitem[\protect\citeauthoryear{Deelman, Gannon, Shields, and Taylor}{Deelman
  et~al\mbox{.}}{2009}]%
        {DBLP:journals/fgcs/DeelmanGST09}
\bibfield{author}{\bibinfo{person}{Ewa Deelman}, \bibinfo{person}{Dennis
  Gannon}, \bibinfo{person}{Matthew~S. Shields}, {and} \bibinfo{person}{Ian~J.
  Taylor}.} \bibinfo{year}{2009}\natexlab{}.
\newblock \showarticletitle{Workflows and e-Science: An overview of workflow
  system features and capabilities}.
\newblock \bibinfo{journal}{\emph{Future Generation Comp. Syst.}}
  \bibinfo{volume}{25}, \bibinfo{number}{5} (\bibinfo{year}{2009}),
  \bibinfo{pages}{528--540}.
\newblock
\urldef\tempurl%
\url{https://doi.org/10.1016/j.future.2008.06.012}
\showDOI{\tempurl}

 \begin{quotation}\noindent Discusses and presents a taxonomy of how
  researchers use scientific workflow tools. \end{quotation}
\bibitem[\protect\citeauthoryear{Deelman, Livny, Mehta, Pavlo, Singh, Su, Vahi,
  and Wenger}{Deelman et~al\mbox{.}}{2006}]%
        {deelman2006pegasus}
\bibfield{author}{\bibinfo{person}{Ewa Deelman}, \bibinfo{person}{Miron Livny},
  \bibinfo{person}{Gaurang Mehta}, \bibinfo{person}{Andrew Pavlo},
  \bibinfo{person}{Gurmeet Singh}, \bibinfo{person}{Mei-Hui Su},
  \bibinfo{person}{Karan Vahi}, {and} \bibinfo{person}{R~Kent Wenger}.}
  \bibinfo{year}{2006}\natexlab{}.
\newblock \showarticletitle{Pegasus and DAGMan From Concept to Execution:
  Mapping Scientific Workflows onto Today's Cyberinfrastructure.}. In
  \bibinfo{booktitle}{\emph{High Performance Computing Workshop}}.
  \bibinfo{pages}{56--74}.
\newblock

 \begin{quotation}\noindent Discusses the synergy between Pegasus and DAGMan
  when executing scientific workflows. \end{quotation}
\bibitem[\protect\citeauthoryear{Deng, Liu, and Sun}{Deng
  et~al\mbox{.}}{1997}]%
        {deng1997scheme}
\bibfield{author}{\bibinfo{person}{Zhong Deng}, \bibinfo{person}{JW-S Liu},
  {and} \bibinfo{person}{J Sun}.} \bibinfo{year}{1997}\natexlab{}.
\newblock \showarticletitle{A scheme for scheduling hard real-time applications
  in open system environment}. In \bibinfo{booktitle}{\emph{Real-Time Systems,
  1997. Proceedings., Ninth Euromicro Workshop on}}. IEEE,
  \bibinfo{pages}{191--199}.
\newblock

 \begin{quotation}\noindent Described a two-level hierarchical scheme for
  scheduling applications. The scheme supports priority-driven applications.
  \end{quotation}
\bibitem[\protect\citeauthoryear{Deng and Liu}{Deng and Liu}{1997}]%
        {DBLP:conf/rtss/DengL97}
\bibfield{author}{\bibinfo{person}{Zhong Deng} {and}
  \bibinfo{person}{Jane~W.{-}S. Liu}.} \bibinfo{year}{1997}\natexlab{}.
\newblock \showarticletitle{Scheduling real-time applications in an open
  environment}. In \bibinfo{booktitle}{\emph{Proceedings of the 18th {IEEE}
  Real-Time Systems Symposium {(RTSS} '97), December 3-5, 1997, San Francisco,
  CA, {USA}}}. \bibinfo{pages}{308--319}.
\newblock
\urldef\tempurl%
\url{https://doi.org/10.1109/REAL.1997.641292}
\showDOI{\tempurl}

 \begin{quotation}\noindent Presents a hierarchical scheduling method for
  scheduling both real-time and non-real-time applications in open
  environments. \end{quotation}
\bibitem[\protect\citeauthoryear{D{\"{o}}rnemann, Juhnke, and
  Freisleben}{D{\"{o}}rnemann et~al\mbox{.}}{2009}]%
        {DBLP:conf/ccgrid/DornemannJF09}
\bibfield{author}{\bibinfo{person}{Tim D{\"{o}}rnemann}, \bibinfo{person}{Ernst
  Juhnke}, {and} \bibinfo{person}{Bernd Freisleben}.}
  \bibinfo{year}{2009}\natexlab{}.
\newblock \showarticletitle{On-Demand Resource Provisioning for {BPEL}
  Workflows Using Amazon's Elastic Compute Cloud}. In
  \bibinfo{booktitle}{\emph{9th {IEEE/ACM} International Symposium on Cluster
  Computing and the Grid, CCGrid 2009, Shanghai, China, 18-21 May 2009}}.
  \bibinfo{pages}{140--147}.
\newblock
\urldef\tempurl%
\url{https://doi.org/10.1109/CCGRID.2009.30}
\showDOI{\tempurl}

 \begin{quotation}\noindent Introduces an autoscaler for BPEL workflow
  execution on cloud resources by extending the BPEL engine. \end{quotation}
\bibitem[\protect\citeauthoryear{Dougherty, White, and Schmidt}{Dougherty
  et~al\mbox{.}}{2012}]%
        {DBLP:journals/fgcs/DoughertyWS12}
\bibfield{author}{\bibinfo{person}{Brian Dougherty}, \bibinfo{person}{Jules
  White}, {and} \bibinfo{person}{Douglas~C. Schmidt}.}
  \bibinfo{year}{2012}\natexlab{}.
\newblock \showarticletitle{Model-driven auto-scaling of green cloud computing
  infrastructure}.
\newblock \bibinfo{journal}{\emph{Future Generation Comp. Syst.}}
  \bibinfo{volume}{28}, \bibinfo{number}{2} (\bibinfo{year}{2012}),
  \bibinfo{pages}{371--378}.
\newblock
\urldef\tempurl%
\url{https://doi.org/10.1016/j.future.2011.05.009}
\showDOI{\tempurl}

 \begin{quotation}\noindent Introduces a model-driven approach for selecting a
  resource configuration to reduce energy consumption and costs.
  \end{quotation}
\bibitem[\protect\citeauthoryear{Duan, Prodan, and Li}{Duan
  et~al\mbox{.}}{2014}]%
        {duan2014multi}
\bibfield{author}{\bibinfo{person}{Rubing Duan}, \bibinfo{person}{Radu Prodan},
  {and} \bibinfo{person}{Xiaorong Li}.} \bibinfo{year}{2014}\natexlab{}.
\newblock \showarticletitle{Multi-objective game theoretic scheduling of
  bag-of-tasks workflows on hybrid clouds}.
\newblock \bibinfo{journal}{\emph{IEEE Transactions on Cloud Computing}}
  \bibinfo{volume}{2}, \bibinfo{number}{1} (\bibinfo{year}{2014}),
  \bibinfo{pages}{29--42}.
\newblock

 \begin{quotation}\noindent Presents a scheduling algorithm for homogeneous and
  parallel bags-of-tasks that minimizes execution time and costs while adhering
  to bandwidth and storage requirements. \end{quotation}
\bibitem[\protect\citeauthoryear{Durillo, Nae, and Prodan}{Durillo
  et~al\mbox{.}}{2013}]%
        {durillo2013multi}
\bibfield{author}{\bibinfo{person}{Juan~J Durillo}, \bibinfo{person}{Vlad Nae},
  {and} \bibinfo{person}{Radu Prodan}.} \bibinfo{year}{2013}\natexlab{}.
\newblock \showarticletitle{Multi-objective workflow scheduling: An analysis of
  the energy efficiency and makespan tradeoff}. In
  \bibinfo{booktitle}{\emph{Cluster, Cloud and Grid Computing (CCGrid), 2013
  13th IEEE/ACM International Symposium on}}. IEEE, \bibinfo{pages}{203--210}.
\newblock

 \begin{quotation}\noindent Investigate the tradeoff solutions computed by a
  bi-objective optimization algorithm called MOHEFT in terms of
  energy-efficiency and makespan. The authors report up to 85\% less energy
  consumption while incurring a 3.3\% makespan increase. \end{quotation}
\bibitem[\protect\citeauthoryear{Durillo, Nae, and Prodan}{Durillo
  et~al\mbox{.}}{2014}]%
        {durillo2014multi}
\bibfield{author}{\bibinfo{person}{Juan~J Durillo}, \bibinfo{person}{Vlad Nae},
  {and} \bibinfo{person}{Radu Prodan}.} \bibinfo{year}{2014}\natexlab{}.
\newblock \showarticletitle{Multi-objective energy-efficient workflow
  scheduling using list-based heuristics}.
\newblock \bibinfo{journal}{\emph{Future Generation Computer Systems}}
  \bibinfo{volume}{36} (\bibinfo{year}{2014}), \bibinfo{pages}{221--236}.
\newblock

 \begin{quotation}\noindent Introduces a workflow scheduling algorithm that
  computes trade-off optimal solutions in terms of makespan and energy
  consumption. The approach is based on empirical models captured from
  real-world systems. \end{quotation}
\bibitem[\protect\citeauthoryear{Duro, Garc{\'\i}a~Blas, Isaila,
  Carretero~P{\'e}rez, Wozniak, and Ross}{Duro et~al\mbox{.}}{2014}]%
        {duro2014exploiting}
\bibfield{author}{\bibinfo{person}{Rodrigo Duro}, \bibinfo{person}{Javier
  Garc{\'\i}a~Blas}, \bibinfo{person}{Flor{\'\i}n~Daniel Isaila},
  \bibinfo{person}{Jes{\'u}s Carretero~P{\'e}rez}, \bibinfo{person}{Justin~M
  Wozniak}, {and} \bibinfo{person}{Rob Ross}.} \bibinfo{year}{2014}\natexlab{}.
\newblock \showarticletitle{Exploiting data locality in Swift/T workflows using
  Hercules}.
\newblock  (\bibinfo{year}{2014}).
\newblock

 \begin{quotation}\noindent Explores the tradeoffs of data-locality and load
  balancing by using swift/T in combination with the Hercules in-memory store
  to execute workflows in distributed systems. \end{quotation}
\bibitem[\protect\citeauthoryear{El-Zarif and Awad}{El-Zarif and Awad}{2012}]%
        {el2012ordinal}
\bibfield{author}{\bibinfo{person}{Nader El-Zarif} {and}
  \bibinfo{person}{Mariette Awad}.} \bibinfo{year}{2012}\natexlab{}.
\newblock \showarticletitle{An ordinal optimization like GA for improved OFDMA
  system carrier allocations}. In \bibinfo{booktitle}{\emph{2012 6th IEEE
  International Conference Intelligent Systems}}. IEEE,
  \bibinfo{pages}{412--418}.
\newblock

 \begin{quotation}\noindent Proposes to enhance the traditional genetic
  algorithm with ordinal optimization to improve paramter selection.
  \end{quotation}
\bibitem[\protect\citeauthoryear{Eyk, Iosup, Abad, Grohmann, and Eismann}{Eyk
  et~al\mbox{.}}{2018}]%
        {DBLP:conf/wosp/EykIAGE18}
\bibfield{author}{\bibinfo{person}{Erwin~Van Eyk}, \bibinfo{person}{Alexandru
  Iosup}, \bibinfo{person}{Cristina~L. Abad}, \bibinfo{person}{Johannes
  Grohmann}, {and} \bibinfo{person}{Simon Eismann}.}
  \bibinfo{year}{2018}\natexlab{}.
\newblock \showarticletitle{A {SPEC} {RG} Cloud Group's Vision on the
  Performance Challenges of FaaS Cloud Architectures}. In
  \bibinfo{booktitle}{\emph{Companion of the 2018 {ACM/SPEC} International
  Conference on Performance Engineering, {ICPE} 2018, Berlin, Germany, April
  09-13, 2018}}. \bibinfo{pages}{21--24}.
\newblock
\urldef\tempurl%
\url{https://doi.org/10.1145/3185768.3186308}
\showDOI{\tempurl}

 \begin{quotation}\noindent Identifies six performance-related issues in
  state-of-the-art FaaS platforms and present their roadmap for tackling these.
  \end{quotation}
\bibitem[\protect\citeauthoryear{Fahringer, Qin, and Hainzer}{Fahringer
  et~al\mbox{.}}{2005}]%
        {fahringer2005specification}
\bibfield{author}{\bibinfo{person}{Thomas Fahringer}, \bibinfo{person}{Jun
  Qin}, {and} \bibinfo{person}{Stefan Hainzer}.}
  \bibinfo{year}{2005}\natexlab{}.
\newblock \showarticletitle{Specification of grid workflow applications with
  AGWL: an Abstract Grid Workflow Language}. In
  \bibinfo{booktitle}{\emph{Cluster Computing and the Grid, 2005. CCGrid 2005.
  IEEE International Symposium on}}, Vol.~\bibinfo{volume}{2}. IEEE,
  \bibinfo{pages}{676--685}.
\newblock

 \begin{quotation}\noindent Presents AGWL: a language for describing grid
  workflow applications. AGWL is the main interface for the ASKALON grid
  environment. \end{quotation}
\bibitem[\protect\citeauthoryear{Fox, Ishakian, Muthusamy, and Slominski}{Fox
  et~al\mbox{.}}{2017}]%
        {fox2017status}
\bibfield{author}{\bibinfo{person}{Geoffrey~C Fox}, \bibinfo{person}{Vatche
  Ishakian}, \bibinfo{person}{Vinod Muthusamy}, {and}
  \bibinfo{person}{Aleksander Slominski}.} \bibinfo{year}{2017}\natexlab{}.
\newblock \showarticletitle{Status of serverless computing and
  function-as-a-service (faas) in industry and research}.
\newblock \bibinfo{journal}{\emph{arXiv preprint arXiv:1708.08028}}
  (\bibinfo{year}{2017}).
\newblock

 \begin{quotation}\noindent Summarizes the issues raised during the first
  edition of the WoSC serverless workshop. \end{quotation}
\bibitem[\protect\citeauthoryear{Genez, Bittencourt, and Madeira}{Genez
  et~al\mbox{.}}{2013}]%
        {genez2013using}
\bibfield{author}{\bibinfo{person}{Thiago~AL Genez}, \bibinfo{person}{Luiz~F
  Bittencourt}, {and} \bibinfo{person}{Edmundo~RM Madeira}.}
  \bibinfo{year}{2013}\natexlab{}.
\newblock \showarticletitle{Using time discretization to schedule scientific
  workflows in multiple cloud providers}. In \bibinfo{booktitle}{\emph{2013
  IEEE Sixth International Conference on Cloud Computing}}. IEEE,
  \bibinfo{pages}{123--130}.
\newblock

 \begin{quotation}\noindent Introduces a scheduling policy based on integer
  linear programming to minimize cost when scheduling workflow application
  across multiple cloud providers. \end{quotation}
\bibitem[\protect\citeauthoryear{Genez, Bittencourt, Sakellariou, and
  Madeira}{Genez et~al\mbox{.}}{2016}]%
        {genez2016flexible}
\bibfield{author}{\bibinfo{person}{Thiago~AL Genez}, \bibinfo{person}{Luiz~F
  Bittencourt}, \bibinfo{person}{Rizos Sakellariou}, {and}
  \bibinfo{person}{Edmundo~RM Madeira}.} \bibinfo{year}{2016}\natexlab{}.
\newblock \showarticletitle{A flexible scheduler for workflow ensembles}. In
  \bibinfo{booktitle}{\emph{Proceedings of the 9th International Conference on
  Utility and Cloud Computing}}. ACM, \bibinfo{pages}{55--62}.
\newblock

 \begin{quotation}\noindent Introduce a scheduling algorithm based on the
  particle swarm optimization technique that schedules according to a user
  defined objective function, e.g., makespan, fairness, or cost.
  \end{quotation}
\bibitem[\protect\citeauthoryear{Gerlach, Tang, Keegan, Harrison, Wilke,
  Bischof, D'Souza, Devoid, Murphy-Olson, Desai, et~al\mbox{.}}{Gerlach
  et~al\mbox{.}}{2014}]%
        {gerlach2014skyport}
\bibfield{author}{\bibinfo{person}{Wolfgang Gerlach}, \bibinfo{person}{Wei
  Tang}, \bibinfo{person}{Kevin Keegan}, \bibinfo{person}{Travis Harrison},
  \bibinfo{person}{Andreas Wilke}, \bibinfo{person}{Jared Bischof},
  \bibinfo{person}{Mark D'Souza}, \bibinfo{person}{Scott Devoid},
  \bibinfo{person}{Daniel Murphy-Olson}, \bibinfo{person}{Narayan Desai},
  {et~al\mbox{.}}} \bibinfo{year}{2014}\natexlab{}.
\newblock \showarticletitle{Skyport: container-based execution environment
  management for multi-cloud scientific workflows}. In
  \bibinfo{booktitle}{\emph{Proceedings of the 5th International Workshop on
  Data-Intensive Computing in the Clouds}}. IEEE Press,
  \bibinfo{pages}{25--32}.
\newblock

 \begin{quotation}\noindent Introduces Skyport, a container-based execution
  environment that reduces the complexity of setting up an environment for
  executing complex workflows. \end{quotation}
\bibitem[\protect\citeauthoryear{Ghodsi, Zaharia, Hindman, Konwinski, Shenker,
  and Stoica}{Ghodsi et~al\mbox{.}}{2011}]%
        {ghodsi2011dominant}
\bibfield{author}{\bibinfo{person}{Ali Ghodsi}, \bibinfo{person}{Matei
  Zaharia}, \bibinfo{person}{Benjamin Hindman}, \bibinfo{person}{Andy
  Konwinski}, \bibinfo{person}{Scott Shenker}, {and} \bibinfo{person}{Ion
  Stoica}.} \bibinfo{year}{2011}\natexlab{}.
\newblock \showarticletitle{Dominant Resource Fairness: Fair Allocation of
  Multiple Resource Types}. In \bibinfo{booktitle}{\emph{Nsdi}},
  Vol.~\bibinfo{volume}{11}. \bibinfo{pages}{24--24}.
\newblock

 \begin{quotation}\noindent Introduces a policy that improves fairness and
  throughput in systems where users share resources. The policy has been
  implemented in Mesos to demonstrate its effectiveness. \end{quotation}
\bibitem[\protect\citeauthoryear{Goecks, Nekrutenko, and Taylor}{Goecks
  et~al\mbox{.}}{2010}]%
        {goecks2010galaxy}
\bibfield{author}{\bibinfo{person}{Jeremy Goecks}, \bibinfo{person}{Anton
  Nekrutenko}, {and} \bibinfo{person}{James Taylor}.}
  \bibinfo{year}{2010}\natexlab{}.
\newblock \showarticletitle{Galaxy: a comprehensive approach for supporting
  accessible, reproducible, and transparent computational research in the life
  sciences}.
\newblock \bibinfo{journal}{\emph{Genome biology}} \bibinfo{volume}{11},
  \bibinfo{number}{8} (\bibinfo{year}{2010}), \bibinfo{pages}{R86}.
\newblock

 \begin{quotation}\noindent Introduces Galaxy, a platform for genomic research
  that tracks and manages data provenance. \end{quotation}
\bibitem[\protect\citeauthoryear{Gog, Schwarzkopf, Gleave, Watson, and
  Hand}{Gog et~al\mbox{.}}{2016}]%
        {DBLP:conf/osdi/GogSGWH16}
\bibfield{author}{\bibinfo{person}{Ionel Gog}, \bibinfo{person}{Malte
  Schwarzkopf}, \bibinfo{person}{Adam Gleave}, \bibinfo{person}{Robert N.~M.
  Watson}, {and} \bibinfo{person}{Steven Hand}.}
  \bibinfo{year}{2016}\natexlab{}.
\newblock \showarticletitle{Firmament: Fast, Centralized Cluster Scheduling at
  Scale}. In \bibinfo{booktitle}{\emph{12th {USENIX} Symposium on Operating
  Systems Design and Implementation, {OSDI} 2016, Savannah, GA, USA, November
  2-4, 2016.}} \bibinfo{pages}{99--115}.
\newblock
\urldef\tempurl%
\url{https://www.usenix.org/conference/osdi16/technical-sessions/presentation/gog}
\showURL{%
\tempurl}

 \begin{quotation}\noindent Presents Firmament, a centralized cluster scheduler
  with low task placement latency. Unlike heuristics, it finds the
  policy-optimal assignment. The scheduler matches the latency of distributed
  solutions for workloads of short tasks and improves batch task response times
  by six times, beating four widely-used centralized and distributed schedulers
  on a real-word cluster. \end{quotation}
\bibitem[\protect\citeauthoryear{Goiri, Beauchea, Le, Nguyen, Haque, Guitart,
  Torres, and Bianchini}{Goiri et~al\mbox{.}}{2011}]%
        {goiri2011greenslot}
\bibfield{author}{\bibinfo{person}{{\'I}{\~n}igo Goiri}, \bibinfo{person}{Ryan
  Beauchea}, \bibinfo{person}{Kien Le}, \bibinfo{person}{Thu~D Nguyen},
  \bibinfo{person}{Md~E Haque}, \bibinfo{person}{Jordi Guitart},
  \bibinfo{person}{Jordi Torres}, {and} \bibinfo{person}{Ricardo Bianchini}.}
  \bibinfo{year}{2011}\natexlab{}.
\newblock \showarticletitle{Greenslot: scheduling energy consumption in green
  datacenters}. In \bibinfo{booktitle}{\emph{SC'11: Proceedings of 2011
  International Conference for High Performance Computing, Networking, Storage
  and Analysis}}. IEEE, \bibinfo{pages}{1--11}.
\newblock

 \begin{quotation}\noindent Introduces GreenSlot, a parallel batch job
  scheduler for datacenters equipped with solar panels. The scheduler predicts
  energy available in the near future and attempts to maximize the consumption
  of green energy over using the traditional grid. \end{quotation}
\bibitem[\protect\citeauthoryear{Guo, Hong, Chafi, Iosup, and Epema}{Guo
  et~al\mbox{.}}{2017}]%
        {guo2017modeling}
\bibfield{author}{\bibinfo{person}{Yong Guo}, \bibinfo{person}{Sungpack Hong},
  \bibinfo{person}{Hassan Chafi}, \bibinfo{person}{Alexandru Iosup}, {and}
  \bibinfo{person}{Dick Epema}.} \bibinfo{year}{2017}\natexlab{}.
\newblock \showarticletitle{Modeling, analysis, and experimental comparison of
  streaming graph-partitioning policies}.
\newblock \bibinfo{journal}{\emph{J. Parallel and Distrib. Comput.}}
  \bibinfo{volume}{108} (\bibinfo{year}{2017}), \bibinfo{pages}{106--121}.
\newblock

 \begin{quotation}\noindent Studies the impact of load imbalance and
  communication overhead in graph processing systems. The authors create a
  model of the execution time of stream graph processing and propose new
  partitioning policies to avoid performance degradation. \end{quotation}
\bibitem[\protect\citeauthoryear{Guo, Fox, and Zhou}{Guo et~al\mbox{.}}{2012}]%
        {guo2012investigation}
\bibfield{author}{\bibinfo{person}{Zhenhua Guo}, \bibinfo{person}{Geoffrey
  Fox}, {and} \bibinfo{person}{Mo Zhou}.} \bibinfo{year}{2012}\natexlab{}.
\newblock \showarticletitle{Investigation of data locality in mapreduce}. In
  \bibinfo{booktitle}{\emph{Proceedings of the 2012 12th IEEE/ACM International
  Symposium on Cluster, Cloud and Grid Computing (ccgrid 2012)}}. IEEE Computer
  Society, \bibinfo{pages}{419--426}.
\newblock

 \begin{quotation}\noindent Introduce a mathematical model of data locality for
  parallel systems running applications such as MapReduce/GFS. Using this
  model, the authors reason about the impact of data locality and introduce a
  scheduling technique to schedule multiple tasks simultaneously on a node to
  achieve optimal data locality. \end{quotation}
\bibitem[\protect\citeauthoryear{Gupta, Lehal, et~al\mbox{.}}{Gupta
  et~al\mbox{.}}{2009}]%
        {gupta2009survey}
\bibfield{author}{\bibinfo{person}{Vishal Gupta}, \bibinfo{person}{Gurpreet~S
  Lehal}, {et~al\mbox{.}}} \bibinfo{year}{2009}\natexlab{}.
\newblock \showarticletitle{A survey of text mining techniques and
  applications}.
\newblock \bibinfo{journal}{\emph{Journal of emerging technologies in web
  intelligence}} \bibinfo{volume}{1}, \bibinfo{number}{1}
  (\bibinfo{year}{2009}), \bibinfo{pages}{60--76}.
\newblock


\bibitem[\protect\citeauthoryear{Heinze, Pappalardo, Jerzak, and Fetzer}{Heinze
  et~al\mbox{.}}{2014}]%
        {DBLP:conf/debs/HeinzePJF14}
\bibfield{author}{\bibinfo{person}{Thomas Heinze}, \bibinfo{person}{Valerio
  Pappalardo}, \bibinfo{person}{Zbigniew Jerzak}, {and}
  \bibinfo{person}{Christof Fetzer}.} \bibinfo{year}{2014}\natexlab{}.
\newblock \showarticletitle{Auto-scaling techniques for elastic data stream
  processing}. In \bibinfo{booktitle}{\emph{The 8th {ACM} International
  Conference on Distributed Event-Based Systems, {DEBS} '14, Mumbai, India, May
  26-29, 2014}}. \bibinfo{pages}{318--321}.
\newblock
\urldef\tempurl%
\url{https://doi.org/10.1145/2611286.2611314}
\showDOI{\tempurl}

 \begin{quotation}\noindent Investigates autoscaling strategies employed in
  their work and in existing elastic data stream processing systems. Four
  requirements are specified for using auto-scaling techniques. \end{quotation}
\bibitem[\protect\citeauthoryear{Hilman, Rodriguez, and Buyya}{Hilman
  et~al\mbox{.}}{2018}]%
        {DBLP:journals/corr/abs-1809-05574}
\bibfield{author}{\bibinfo{person}{Muhammad~Hafizhuddin Hilman},
  \bibinfo{person}{Maria~Alejandra Rodriguez}, {and} \bibinfo{person}{Rajkumar
  Buyya}.} \bibinfo{year}{2018}\natexlab{}.
\newblock \showarticletitle{Multiple Workflows Scheduling in Multi-tenant
  Distributed Systems: {A} Taxonomy and Future Directions}.
\newblock \bibinfo{journal}{\emph{CoRR}}  \bibinfo{volume}{abs/1809.05574}
  (\bibinfo{year}{2018}).
\newblock
\showeprint[arxiv]{1809.05574}
\urldef\tempurl%
\url{http://arxiv.org/abs/1809.05574}
\showURL{%
\tempurl}

 \begin{quotation}\noindent Presents a survey on multi-tenant systems in which
  multiple workflows are executed simultaneously. \end{quotation}
\bibitem[\protect\citeauthoryear{Hoheisel and Der}{Hoheisel and Der}{2003}]%
        {hoheisel2003dynamic}
\bibfield{author}{\bibinfo{person}{Andreas Hoheisel} {and} \bibinfo{person}{Uwe
  Der}.} \bibinfo{year}{2003}\natexlab{}.
\newblock \showarticletitle{Dynamic workflows for grid applications}. In
  \bibinfo{booktitle}{\emph{Proceedings of the Cracow Grid Workshop}},
  Vol.~\bibinfo{volume}{3}.
\newblock

 \begin{quotation}\noindent Introduces GJobDL, a Petri net based language for
  defining grid workflows. The language converts abstract workflows into
  concrete workflows at runtime, based on required elements at runtime.
  \end{quotation}
\bibitem[\protect\citeauthoryear{Ilyushkin, Ali{-}Eldin, Herbst, Bauer,
  Papadopoulos, Epema, and Iosup}{Ilyushkin et~al\mbox{.}}{2018}]%
        {DBLP:journals/tompecs/IlyushkinAHBPEI18}
\bibfield{author}{\bibinfo{person}{Alexey Ilyushkin}, \bibinfo{person}{Ahmed
  Ali{-}Eldin}, \bibinfo{person}{Nikolas Herbst}, \bibinfo{person}{Andr{\'{e}}
  Bauer}, \bibinfo{person}{Alessandro~Vittorio Papadopoulos},
  \bibinfo{person}{Dick H.~J. Epema}, {and} \bibinfo{person}{Alexandru Iosup}.}
  \bibinfo{year}{2018}\natexlab{}.
\newblock \showarticletitle{An Experimental Performance Evaluation of
  Autoscalers for Complex Workflows}.
\newblock \bibinfo{journal}{\emph{{TOMPECS}}} \bibinfo{volume}{3},
  \bibinfo{number}{2} (\bibinfo{year}{2018}), \bibinfo{pages}{8:1--8:32}.
\newblock
\urldef\tempurl%
\url{https://doi.org/10.1145/3164537}
\showDOI{\tempurl}

 \begin{quotation}\noindent Evaluates seven state-of-the-art autoscalers using
  two scientific workloads of workflows and various novel elasticity, workflow,
  and system metrics. \end{quotation}
\bibitem[\protect\citeauthoryear{Ilyushkin, Ghit, and Epema}{Ilyushkin
  et~al\mbox{.}}{2015}]%
        {DBLP:conf/ccgrid/IlyushkinGE15}
\bibfield{author}{\bibinfo{person}{Alexey Ilyushkin}, \bibinfo{person}{Bogdan
  Ghit}, {and} \bibinfo{person}{Dick H.~J. Epema}.}
  \bibinfo{year}{2015}\natexlab{}.
\newblock \showarticletitle{Scheduling Workloads of Workflows with Unknown Task
  Runtimes}. In \bibinfo{booktitle}{\emph{15th {IEEE/ACM} International
  Symposium on Cluster, Cloud and Grid Computing, CCGrid 2015, Shenzhen, China,
  May 4-7, 2015}}. \bibinfo{pages}{606--616}.
\newblock
\urldef\tempurl%
\url{https://doi.org/10.1109/CCGrid.2015.27}
\showDOI{\tempurl}

 \begin{quotation}\noindent Presents four scheduling policies for scheduling
  workloads of workflows with unknown task runtimes in an online setting. As
  workloads can fluctuate in their resource requirements during runtime, the
  policies reserve different amount of processors to deal with these
  fluctuations. \end{quotation}
\bibitem[\protect\citeauthoryear{Iosup}{Iosup}{2009}]%
        {iosup2009framework}
\bibfield{author}{\bibinfo{person}{Alexandru Iosup}.}
  \bibinfo{year}{2009}\natexlab{}.
\newblock \showarticletitle{A framework for the study of grid inter-operation
  mechanisms}.
\newblock  (\bibinfo{year}{2009}).
\newblock

 \begin{quotation}\noindent The doctoral thesis of Alexandru Iosup on grid
  mechanisms. \end{quotation}
\bibitem[\protect\citeauthoryear{Iosup, Tannenbaum, Farrellee, Epema, and
  Livny}{Iosup et~al\mbox{.}}{2008}]%
        {DBLP:journals/sp/IosupTFEL08}
\bibfield{author}{\bibinfo{person}{Alexandru Iosup}, \bibinfo{person}{Todd
  Tannenbaum}, \bibinfo{person}{Matthew Farrellee}, \bibinfo{person}{Dick H.~J.
  Epema}, {and} \bibinfo{person}{Miron Livny}.}
  \bibinfo{year}{2008}\natexlab{}.
\newblock \showarticletitle{Inter-operating grids through Delegated
  MatchMaking}.
\newblock \bibinfo{journal}{\emph{Scientific Programming}}
  \bibinfo{volume}{16}, \bibinfo{number}{2-3} (\bibinfo{year}{2008}),
  \bibinfo{pages}{233--253}.
\newblock
\urldef\tempurl%
\url{https://doi.org/10.3233/SPR-2008-0246}
\showDOI{\tempurl}

 \begin{quotation}\noindent Proposes DMM, a solution to the problem of
  inter-Grid resource selection by combining hierarchical and decentralized
  approaches to interconnect Grids. Peer-to-peer connections are created
  between sites under the same administrative tool to realize delegated
  matchmaking. \end{quotation}
\bibitem[\protect\citeauthoryear{Jaeger, Prakash, and Ishikawa}{Jaeger
  et~al\mbox{.}}{1994}]%
        {jaeger1994framework}
\bibfield{author}{\bibinfo{person}{Trent Jaeger}, \bibinfo{person}{Atul
  Prakash}, {and} \bibinfo{person}{Masayuki Ishikawa}.}
  \bibinfo{year}{1994}\natexlab{}.
\newblock \showarticletitle{A framework for automatic improvement of workflows
  to meet performance goals}. In \bibinfo{booktitle}{\emph{Proceedings Sixth
  International Conference on Tools with Artificial Intelligence. TAI 94}}.
  IEEE, \bibinfo{pages}{640--646}.
\newblock

 \begin{quotation}\noindent Uses the notion of satisficing to create a
  framework to automatically modify a model of a business information system
  that executes business workflows to meet performance goals. The modifications
  are generated using AI techniques. \end{quotation}
\bibitem[\protect\citeauthoryear{Jayadivya, Nirmala, and Bhanu}{Jayadivya
  et~al\mbox{.}}{2012}]%
        {jayadivya2012fault}
\bibfield{author}{\bibinfo{person}{SK Jayadivya}, \bibinfo{person}{Jaya~S
  Nirmala}, {and} \bibinfo{person}{Mary Saira~S Bhanu}.}
  \bibinfo{year}{2012}\natexlab{}.
\newblock \showarticletitle{Fault tolerant workflow scheduling based on
  replication and resubmission of tasks in Cloud Computing}.
\newblock \bibinfo{journal}{\emph{International Journal on Computer Science and
  Engineering}} \bibinfo{volume}{4}, \bibinfo{number}{6}
  (\bibinfo{year}{2012}), \bibinfo{pages}{996}.
\newblock

 \begin{quotation}\noindent Introduces FTWS, a workflow scheduling algorithm
  that is fault tolerant to failures through replication and resubmission of
  tasks. \end{quotation}
\bibitem[\protect\citeauthoryear{Jones and Nitzberg}{Jones and
  Nitzberg}{1999}]%
        {DBLP:conf/ipps/JonesN99}
\bibfield{author}{\bibinfo{person}{James~Patton Jones} {and}
  \bibinfo{person}{Bill Nitzberg}.} \bibinfo{year}{1999}\natexlab{}.
\newblock \showarticletitle{Scheduling for Parallel Supercomputing: {A}
  Historical Perspective of Achievable Utilization}. In
  \bibinfo{booktitle}{\emph{Job Scheduling Strategies for Parallel Processing,
  IPPS/SPDP'99 Workshop, JSSPP'99, San Juan, Puerto Rico, April 16, 1999,
  Proceedings}}. \bibinfo{pages}{1--16}.
\newblock
\urldef\tempurl%
\url{https://doi.org/10.1007/3-540-47954-6\_1}
\showDOI{\tempurl}

 \begin{quotation}\noindent Discusses possible utilization levels for super
  computers based on experience gained from running jobs on NASA super
  computers. \end{quotation}
\bibitem[\protect\citeauthoryear{Jr. and Graham}{Jr. and Graham}{1972}]%
        {DBLP:journals/acta/CoffmanG72}
\bibfield{author}{\bibinfo{person}{Edward G.~Coffman Jr.} {and}
  \bibinfo{person}{Ronald~L. Graham}.} \bibinfo{year}{1972}\natexlab{}.
\newblock \showarticletitle{Optimal Scheduling for Two-Processor Systems}.
\newblock \bibinfo{journal}{\emph{Acta Inf.}}  \bibinfo{volume}{1}
  (\bibinfo{year}{1972}), \bibinfo{pages}{200--213}.
\newblock
\urldef\tempurl%
\url{https://doi.org/10.1007/BF00288685}
\showDOI{\tempurl}

 \begin{quotation}\noindent Introduces a simple model for workflows used to
  schedule on two-processor systems. \end{quotation}
\bibitem[\protect\citeauthoryear{Kardani{-}Moghaddam, Buyya, and
  Ramamohanarao}{Kardani{-}Moghaddam et~al\mbox{.}}{2019}]%
        {DBLP:journals/csur/Kardani-Moghaddam19}
\bibfield{author}{\bibinfo{person}{Sara Kardani{-}Moghaddam},
  \bibinfo{person}{Rajkumar Buyya}, {and} \bibinfo{person}{Kotagiri
  Ramamohanarao}.} \bibinfo{year}{2019}\natexlab{}.
\newblock \showarticletitle{Performance-Aware Management of Cloud Resources:
  {A} Taxonomy and Future Directions}.
\newblock \bibinfo{journal}{\emph{{ACM} Comput. Surv.}} \bibinfo{volume}{52},
  \bibinfo{number}{4} (\bibinfo{year}{2019}), \bibinfo{pages}{84:1--84:37}.
\newblock
\urldef\tempurl%
\url{https://doi.org/10.1145/3337956}
\showDOI{\tempurl}

 \begin{quotation}\noindent Presents a survey on cloud resource management. In
  particular, the authors focus on performance management. \end{quotation}
\bibitem[\protect\citeauthoryear{Kaur and Mehta}{Kaur and Mehta}{2017}]%
        {kaur2017resource}
\bibfield{author}{\bibinfo{person}{Parmeet Kaur} {and} \bibinfo{person}{Shikha
  Mehta}.} \bibinfo{year}{2017}\natexlab{}.
\newblock \showarticletitle{Resource provisioning and work flow scheduling in
  clouds using augmented Shuffled Frog Leaping Algorithm}.
\newblock \bibinfo{journal}{\emph{J. Parallel and Distrib. Comput.}}
  \bibinfo{volume}{101} (\bibinfo{year}{2017}), \bibinfo{pages}{41--50}.
\newblock

 \begin{quotation}\noindent Introduce a resource algorithm based on the
  shuffled frog leaping algorithm that aim to minimze costs while meeting
  specified deadlines. \end{quotation}
\bibitem[\protect\citeauthoryear{Khorsand, Ghobaei-Arani, and
  Ramezanpour}{Khorsand et~al\mbox{.}}{2018}]%
        {khorsand2018fahp}
\bibfield{author}{\bibinfo{person}{Reihaneh Khorsand}, \bibinfo{person}{Mostafa
  Ghobaei-Arani}, {and} \bibinfo{person}{Mohammadreza Ramezanpour}.}
  \bibinfo{year}{2018}\natexlab{}.
\newblock \showarticletitle{FAHP approach for autonomic resource provisioning
  of multitier applications in cloud computing environments}.
\newblock \bibinfo{journal}{\emph{Software: Practice and Experience}}
  \bibinfo{volume}{48}, \bibinfo{number}{12} (\bibinfo{year}{2018}),
  \bibinfo{pages}{2147--2173}.
\newblock

 \begin{quotation}\noindent Introduces FAHP: an autoscaling approach for
  autoscaling multi-tier applications in cloud environments. \end{quotation}
\bibitem[\protect\citeauthoryear{Khorsand, Safi-Esfahani, Nematbakhsh, and
  Mohsenzade}{Khorsand et~al\mbox{.}}{2017}]%
        {khorsand2017taxonomy}
\bibfield{author}{\bibinfo{person}{Reihaneh Khorsand},
  \bibinfo{person}{Faramarz Safi-Esfahani}, \bibinfo{person}{Naser
  Nematbakhsh}, {and} \bibinfo{person}{Mehran Mohsenzade}.}
  \bibinfo{year}{2017}\natexlab{}.
\newblock \showarticletitle{Taxonomy of workflow partitioning problems and
  methods in distributed environments}.
\newblock \bibinfo{journal}{\emph{Journal of Systems and Software}}
  \bibinfo{volume}{132} (\bibinfo{year}{2017}), \bibinfo{pages}{253--271}.
\newblock

 \begin{quotation}\noindent Presents a survey and taxonomies on workflow
  partitioning in distributed environments. The article covers both business
  (manual) and compute (automated) workflows and the language aspect of
  defining such workflows. \end{quotation}
\bibitem[\protect\citeauthoryear{Law}{Law}{[n. d.]}]%
        {lawamendments}
\bibfield{author}{\bibinfo{person}{RF~Federal Law}.} \bibinfo{year}{[n.
  d.]}\natexlab{}.
\newblock \bibinfo{title}{On Amendments to certain legislative acts of the
  Russian Federation in connection with the improvement of the legal status of
  the State (municipal) institutions" of May 08, 2010 № 83-FZ}.
\newblock
\newblock


\bibitem[\protect\citeauthoryear{Lee, Han, Zomaya, and Yousif}{Lee
  et~al\mbox{.}}{2015}]%
        {lee2015resource}
\bibfield{author}{\bibinfo{person}{Young~Choon Lee}, \bibinfo{person}{Hyuck
  Han}, \bibinfo{person}{Albert~Y Zomaya}, {and} \bibinfo{person}{Mazin
  Yousif}.} \bibinfo{year}{2015}\natexlab{}.
\newblock \showarticletitle{Resource-efficient workflow scheduling in clouds}.
\newblock \bibinfo{journal}{\emph{Knowledge-Based Systems}}
  \bibinfo{volume}{80} (\bibinfo{year}{2015}), \bibinfo{pages}{153--162}.
\newblock

 \begin{quotation}\noindent Presents the Maximum Effective Reduction (MER)
  algorithm, which trades some of a workflows' makespan for resource efficiency
  by consolidating tasks. The authors demonstrate that, on average, the
  makespan is increased by less than 10\% while resource usage decreased by
  54\%. \end{quotation}
\bibitem[\protect\citeauthoryear{{LeVeque}, {Mitchell}, and
  {Stodden}}{{LeVeque} et~al\mbox{.}}{2012}]%
        {6171147}
\bibfield{author}{\bibinfo{person}{R.~J. {LeVeque}}, \bibinfo{person}{I.~M.
  {Mitchell}}, {and} \bibinfo{person}{V. {Stodden}}.}
  \bibinfo{year}{2012}\natexlab{}.
\newblock \showarticletitle{Reproducible research for scientific computing:
  Tools and strategies for changing the culture}.
\newblock \bibinfo{journal}{\emph{Computing in Science Engineering}}
  \bibinfo{volume}{14}, \bibinfo{number}{4} (\bibinfo{date}{July}
  \bibinfo{year}{2012}), \bibinfo{pages}{13--17}.
\newblock
\showISSN{1558-366X}
\urldef\tempurl%
\url{https://doi.org/10.1109/MCSE.2012.38}
\showDOI{\tempurl}

 \begin{quotation}\noindent Discusses the emerging need for reproducibility in
  science and mentions tools and strategies to improve sharing. \end{quotation}
\bibitem[\protect\citeauthoryear{Ley}{Ley}{2009}]%
        {ley2009dblp}
\bibfield{author}{\bibinfo{person}{Michael Ley}.}
  \bibinfo{year}{2009}\natexlab{}.
\newblock \showarticletitle{DBLP: some lessons learned}.
\newblock \bibinfo{journal}{\emph{Proceedings of the VLDB Endowment}}
  \bibinfo{volume}{2}, \bibinfo{number}{2} (\bibinfo{year}{2009}),
  \bibinfo{pages}{1493--1500}.
\newblock

 \begin{quotation}\noindent Reviews the evolution of DBLP and discusses
  problems and challenges that arose over the years. \end{quotation}
\bibitem[\protect\citeauthoryear{Li, Peng, Lei, and Zhang}{Li
  et~al\mbox{.}}{2011}]%
        {li2011energy}
\bibfield{author}{\bibinfo{person}{Jiandun Li}, \bibinfo{person}{Junjie Peng},
  \bibinfo{person}{Zhou Lei}, {and} \bibinfo{person}{Wu Zhang}.}
  \bibinfo{year}{2011}\natexlab{}.
\newblock \showarticletitle{An energy-efficient scheduling approach based on
  private clouds}.
\newblock \bibinfo{journal}{\emph{Journal of Information \&computational
  Science}} \bibinfo{volume}{8}, \bibinfo{number}{4} (\bibinfo{year}{2011}),
  \bibinfo{pages}{716--724}.
\newblock

 \begin{quotation}\noindent Explores characteristics related to VM workflow
  scheduling in private clouds. An energy-efficient scheduling approach is
  proposed that aims at saving user time, lowering energy consumption, and
  improves load-balancing across the compute nodes. \end{quotation}
\bibitem[\protect\citeauthoryear{Li, Zhang, Wu, Li, and Zhao}{Li
  et~al\mbox{.}}{2012}]%
        {li2012trust}
\bibfield{author}{\bibinfo{person}{Wenjuan Li}, \bibinfo{person}{Qifei Zhang},
  \bibinfo{person}{Jiyi Wu}, \bibinfo{person}{Jing Li}, {and}
  \bibinfo{person}{Haili Zhao}.} \bibinfo{year}{2012}\natexlab{}.
\newblock \showarticletitle{Trust-based and qos demand clustering analysis
  customizable cloud workflow scheduling strategies}. In
  \bibinfo{booktitle}{\emph{Cluster Computing Workshops (CLUSTER WORKSHOPS),
  2012 IEEE International Conference on}}. IEEE, \bibinfo{pages}{111--119}.
\newblock

 \begin{quotation}\noindent Introduces a workflow management system that
  supports non-DAGs, i.e., iteration in workflows using a new heuristic
  algorithm. \end{quotation}
\bibitem[\protect\citeauthoryear{Li, Ge, Yang, Huang, Hu, Hu, and Luo}{Li
  et~al\mbox{.}}{2016}]%
        {li2016security}
\bibfield{author}{\bibinfo{person}{Zhongjin Li}, \bibinfo{person}{Jidong Ge},
  \bibinfo{person}{Hongji Yang}, \bibinfo{person}{Liguo Huang},
  \bibinfo{person}{Haiyang Hu}, \bibinfo{person}{Hao Hu}, {and}
  \bibinfo{person}{Bin Luo}.} \bibinfo{year}{2016}\natexlab{}.
\newblock \showarticletitle{A security and cost aware scheduling algorithm for
  heterogeneous tasks of scientific workflow in clouds}.
\newblock \bibinfo{journal}{\emph{Future Generation Computer Systems}}
  \bibinfo{volume}{65} (\bibinfo{year}{2016}), \bibinfo{pages}{140--152}.
\newblock

 \begin{quotation}\noindent Introduces SCAS: a security and cost aware
  scheduler based on particle swarm optimization. \end{quotation}
\bibitem[\protect\citeauthoryear{Liu, Aida, Yokoyama, and Masatani}{Liu
  et~al\mbox{.}}{2016}]%
        {liu2016flexible}
\bibfield{author}{\bibinfo{person}{Kai Liu}, \bibinfo{person}{Kento Aida},
  \bibinfo{person}{Shigetoshi Yokoyama}, {and} \bibinfo{person}{Yoshinobu
  Masatani}.} \bibinfo{year}{2016}\natexlab{}.
\newblock \showarticletitle{Flexible container-based computing platform on
  cloud for scientific workflows}. In \bibinfo{booktitle}{\emph{2016
  International Conference on Cloud Computing Research and Innovations
  (ICCCRI)}}. IEEE, \bibinfo{pages}{56--63}.
\newblock

 \begin{quotation}\noindent Introduces a flexible computing platform for
  executing scientific workflows in cloud environments. By integrating Galaxy,
  the authors demonstrate their platform has negligible overhead.
  \end{quotation}
\bibitem[\protect\citeauthoryear{Liu, Chen, Yang, and Jin}{Liu
  et~al\mbox{.}}{2008}]%
        {liu2008throughput}
\bibfield{author}{\bibinfo{person}{Ke Liu}, \bibinfo{person}{Jinjun Chen},
  \bibinfo{person}{Yun Yang}, {and} \bibinfo{person}{Hai Jin}.}
  \bibinfo{year}{2008}\natexlab{}.
\newblock \showarticletitle{A throughput maximization strategy for scheduling
  transaction-intensive workflows on SwinDeW-G}.
\newblock \bibinfo{journal}{\emph{Concurrency and Computation: Practice and
  Experience}} \bibinfo{volume}{20}, \bibinfo{number}{15}
  (\bibinfo{year}{2008}), \bibinfo{pages}{1807--1820}.
\newblock

 \begin{quotation}\noindent Introduces two scheduling algorithms for
  transaction-intensive workflows at the instance and task level, respectively.
  The algorithms focus on maximizing throughput. \end{quotation}
\bibitem[\protect\citeauthoryear{Liu, Bollen, Nelson, and Van~de Sompel}{Liu
  et~al\mbox{.}}{2005}]%
        {liu2005co}
\bibfield{author}{\bibinfo{person}{Xiaoming Liu}, \bibinfo{person}{Johan
  Bollen}, \bibinfo{person}{Michael~L Nelson}, {and} \bibinfo{person}{Herbert
  Van~de Sompel}.} \bibinfo{year}{2005}\natexlab{}.
\newblock \showarticletitle{Co-authorship networks in the digital library
  research community}.
\newblock \bibinfo{journal}{\emph{Information processing \& management}}
  \bibinfo{volume}{41}, \bibinfo{number}{6} (\bibinfo{year}{2005}),
  \bibinfo{pages}{1462--1480}.
\newblock

 \begin{quotation}\noindent Introduces AuthorRank, a new metric that can be
  used to rank authors given a co-authorship network. \end{quotation}
\bibitem[\protect\citeauthoryear{Ma, Ilyushkin, Stegehuis, and Iosup}{Ma
  et~al\mbox{.}}{2017}]%
        {DBLP:conf/icac/MaISI17}
\bibfield{author}{\bibinfo{person}{Shenjun Ma}, \bibinfo{person}{Alexey
  Ilyushkin}, \bibinfo{person}{Alexander Stegehuis}, {and}
  \bibinfo{person}{Alexandru Iosup}.} \bibinfo{year}{2017}\natexlab{}.
\newblock \showarticletitle{Ananke: {A} Q-Learning-Based Portfolio Scheduler
  for Complex Industrial Workflows}. In \bibinfo{booktitle}{\emph{2017 {IEEE}
  International Conference on Autonomic Computing, {ICAC} 2017, Columbus, OH,
  USA, July 17-21, 2017}}. \bibinfo{pages}{227--232}.
\newblock
\urldef\tempurl%
\url{https://doi.org/10.1109/ICAC.2017.21}
\showDOI{\tempurl}

 \begin{quotation}\noindent Presents a workflow scheduler that employs a
  Q-learning based mechanism to reduce deadline violations in an IoT production
  environment at Shell. \end{quotation}
\bibitem[\protect\citeauthoryear{Mahmud, Kotagiri, and Buyya}{Mahmud
  et~al\mbox{.}}{2018}]%
        {mahmud2018fog}
\bibfield{author}{\bibinfo{person}{Redowan Mahmud},
  \bibinfo{person}{Ramamohanarao Kotagiri}, {and} \bibinfo{person}{Rajkumar
  Buyya}.} \bibinfo{year}{2018}\natexlab{}.
\newblock \showarticletitle{Fog computing: A taxonomy, survey and future
  directions}.
\newblock In \bibinfo{booktitle}{\emph{Internet of everything}}.
  \bibinfo{publisher}{Springer}, \bibinfo{pages}{103--130}.
\newblock

 \begin{quotation}\noindent This survey presents taxonomies, challenges, and
  future directions for the Fog computing domain. \end{quotation}
\bibitem[\protect\citeauthoryear{Mahmud, Srirama, Ramamohanarao, and
  Buyya}{Mahmud et~al\mbox{.}}{2019}]%
        {DBLP:journals/jpdc/MahmudSRB19}
\bibfield{author}{\bibinfo{person}{Redowan Mahmud},
  \bibinfo{person}{Satish~Narayana Srirama}, \bibinfo{person}{Kotagiri
  Ramamohanarao}, {and} \bibinfo{person}{Rajkumar Buyya}.}
  \bibinfo{year}{2019}\natexlab{}.
\newblock \showarticletitle{Quality of Experience (QoE)-aware placement of
  applications in Fog computing environments}.
\newblock \bibinfo{journal}{\emph{J. Parallel Distrib. Comput.}}
  \bibinfo{volume}{132} (\bibinfo{year}{2019}), \bibinfo{pages}{190--203}.
\newblock
\urldef\tempurl%
\url{https://doi.org/10.1016/j.jpdc.2018.03.004}
\showDOI{\tempurl}


\bibitem[\protect\citeauthoryear{Malawski, Juve, Deelman, and
  Nabrzyski}{Malawski et~al\mbox{.}}{2015}]%
        {DBLP:journals/fgcs/MalawskiJDN15}
\bibfield{author}{\bibinfo{person}{Maciej Malawski}, \bibinfo{person}{Gideon
  Juve}, \bibinfo{person}{Ewa Deelman}, {and} \bibinfo{person}{Jarek
  Nabrzyski}.} \bibinfo{year}{2015}\natexlab{}.
\newblock \showarticletitle{Algorithms for cost- and deadline-constrained
  provisioning for scientific workflow ensembles in IaaS clouds}.
\newblock \bibinfo{journal}{\emph{Future Generation Comp. Syst.}}
  \bibinfo{volume}{48} (\bibinfo{year}{2015}), \bibinfo{pages}{1--18}.
\newblock
\urldef\tempurl%
\url{https://doi.org/10.1016/j.future.2015.01.004}
\showDOI{\tempurl}

 \begin{quotation}\noindent Presents and analyzes three scheduling policies
  (DPDS, WA-DPDS, and SPSS) that both allocate tasks and provision resources in
  simulation. \end{quotation}
\bibitem[\protect\citeauthoryear{Malik, Tullsen, and Homayoun}{Malik
  et~al\mbox{.}}{2017}]%
        {malik2017co}
\bibfield{author}{\bibinfo{person}{Maria Malik}, \bibinfo{person}{Dean~M
  Tullsen}, {and} \bibinfo{person}{Houman Homayoun}.}
  \bibinfo{year}{2017}\natexlab{}.
\newblock \showarticletitle{Co-locating and concurrent fine-tuning MapReduce
  applications on microservers for energy efficiency}. In
  \bibinfo{booktitle}{\emph{Workload Characterization (IISWC), 2017 IEEE
  International Symposium on}}. IEEE, \bibinfo{pages}{22--31}.
\newblock

 \begin{quotation}\noindent Investigates the interplay between MapReduce
  configurations together with the application and architecture level
  parameters. New opportunities to co-locate MapReduce applications at the node
  level are discovered. \end{quotation}
\bibitem[\protect\citeauthoryear{Mao and Humphrey}{Mao and Humphrey}{2011}]%
        {DBLP:conf/sc/MaoH11}
\bibfield{author}{\bibinfo{person}{Ming Mao} {and} \bibinfo{person}{Marty
  Humphrey}.} \bibinfo{year}{2011}\natexlab{}.
\newblock \showarticletitle{Auto-scaling to minimize cost and meet application
  deadlines in cloud workflows}. In \bibinfo{booktitle}{\emph{Conference on
  High Performance Computing Networking, Storage and Analysis, {SC} 2011,
  Seattle, WA, USA, November 12-18, 2011}}. \bibinfo{pages}{49:1--49:12}.
\newblock
\urldef\tempurl%
\url{https://doi.org/10.1145/2063384.2063449}
\showDOI{\tempurl}

 \begin{quotation}\noindent Presents an auto-scaling mechanism that focusses on
  both user performance requirements and budget concerns. Using virtual
  machines as the basic computing elements, jobs specified as workflows, and
  user performance requirements assigned as (soft) deadlines, jobs are
  scheduled in such as way to meet all deadlines while minimizing costs. To
  minimize costs, virtual machines are (de)allocated dynamically.
  \end{quotation}
\bibitem[\protect\citeauthoryear{Marozzo, Talia, and Trunfio}{Marozzo
  et~al\mbox{.}}{2015}]%
        {DBLP:journals/concurrency/MarozzoTT15}
\bibfield{author}{\bibinfo{person}{Fabrizio Marozzo}, \bibinfo{person}{Domenico
  Talia}, {and} \bibinfo{person}{Paolo Trunfio}.}
  \bibinfo{year}{2015}\natexlab{}.
\newblock \showarticletitle{JS4Cloud: script-based workflow programming for
  scalable data analysis on cloud platforms}.
\newblock \bibinfo{journal}{\emph{Concurrency and Computation: Practice and
  Experience}} \bibinfo{volume}{27}, \bibinfo{number}{17}
  (\bibinfo{year}{2015}), \bibinfo{pages}{5214--5237}.
\newblock
\urldef\tempurl%
\url{https://doi.org/10.1002/cpe.3563}
\showDOI{\tempurl}

 \begin{quotation}\noindent Presents JS4Cloud, a workflow formalism for
  defining and executing data analysis workflows. The formalism is implemented
  in DMCF and performance results are shown. \end{quotation}
\bibitem[\protect\citeauthoryear{McGough, Young, Afzal, Newhouse, and
  Darlington}{McGough et~al\mbox{.}}{2004}]%
        {mcgough2004workflow}
\bibfield{author}{\bibinfo{person}{Stephen McGough}, \bibinfo{person}{Laurie
  Young}, \bibinfo{person}{Ali Afzal}, \bibinfo{person}{Steven Newhouse}, {and}
  \bibinfo{person}{John Darlington}.} \bibinfo{year}{2004}\natexlab{}.
\newblock \showarticletitle{Workflow enactment in ICENI}. In
  \bibinfo{booktitle}{\emph{UK e-Science All Hands Meeting}},
  Vol.~\bibinfo{volume}{9}. \bibinfo{pages}{894--900}.
\newblock

 \begin{quotation}\noindent Discusses how abstract workflows are made concrete
  in the ICENI workflow management system. \end{quotation}
\bibitem[\protect\citeauthoryear{Momenzadeh and Safi-Esfahani}{Momenzadeh and
  Safi-Esfahani}{2019}]%
        {momenzadeh2019workflow}
\bibfield{author}{\bibinfo{person}{Zahra Momenzadeh} {and}
  \bibinfo{person}{Faramarz Safi-Esfahani}.} \bibinfo{year}{2019}\natexlab{}.
\newblock \showarticletitle{Workflow scheduling applying adaptable and dynamic
  fragmentation (WSADF) based on runtime conditions in cloud computing}.
\newblock \bibinfo{journal}{\emph{Future Generation Computer Systems}}
  \bibinfo{volume}{90} (\bibinfo{year}{2019}), \bibinfo{pages}{327--346}.
\newblock

 \begin{quotation}\noindent Introduce WSADF: a workflow fragmentation and
  scheduling combination to reduce bandwidth and improve throughput, makespan,
  and costs, etc. \end{quotation}
\bibitem[\protect\citeauthoryear{Nastic, Rausch, Scekic, Dustdar, Gusev,
  Koteska, Kostoska, Jakimovski, Ristov, and Prodan}{Nastic
  et~al\mbox{.}}{2017}]%
        {nastic2017serverless}
\bibfield{author}{\bibinfo{person}{Stefan Nastic}, \bibinfo{person}{Thomas
  Rausch}, \bibinfo{person}{Ognjen Scekic}, \bibinfo{person}{Schahram Dustdar},
  \bibinfo{person}{Marjan Gusev}, \bibinfo{person}{Bojana Koteska},
  \bibinfo{person}{Magdalena Kostoska}, \bibinfo{person}{Boro Jakimovski},
  \bibinfo{person}{Sasko Ristov}, {and} \bibinfo{person}{Radu Prodan}.}
  \bibinfo{year}{2017}\natexlab{}.
\newblock \showarticletitle{A serverless real-time data analytics platform for
  edge computing}.
\newblock \bibinfo{journal}{\emph{IEEE Internet Computing}}
  \bibinfo{volume}{21}, \bibinfo{number}{4} (\bibinfo{year}{2017}),
  \bibinfo{pages}{64--71}.
\newblock

 \begin{quotation}\noindent Introduce an analytics platform and model for
  serverless edge computing applications. Main design challenges and
  requirements based on real-life scenarios are discussed. \end{quotation}
\bibitem[\protect\citeauthoryear{Oakes, Yang, Zhou, Houck, Harter,
  Arpaci-Dusseau, and Arpaci-Dusseau}{Oakes et~al\mbox{.}}{2018}]%
        {oakes2018sock}
\bibfield{author}{\bibinfo{person}{Edward Oakes}, \bibinfo{person}{Leon Yang},
  \bibinfo{person}{Dennis Zhou}, \bibinfo{person}{Kevin Houck},
  \bibinfo{person}{Tyler Harter}, \bibinfo{person}{Andrea Arpaci-Dusseau},
  {and} \bibinfo{person}{Remzi Arpaci-Dusseau}.}
  \bibinfo{year}{2018}\natexlab{}.
\newblock \showarticletitle{$\{$SOCK$\}$: Rapid Task Provisioning with
  Serverless-Optimized Containers}. In \bibinfo{booktitle}{\emph{2018
  $\{$USENIX$\}$ Annual Technical Conference ($\{$USENIX$\}$$\{$ATC$\}$ 18)}}.
  \bibinfo{pages}{57--70}.
\newblock

 \begin{quotation}\noindent Introduces SOCK, a container system for serverless
  workloads. Scalability bottlenecks are carefully avoided leading to a
  significant speedup over e.g. Docker. \end{quotation}
\bibitem[\protect\citeauthoryear{Onan, Koruko{\u{g}}lu, and Bulut}{Onan
  et~al\mbox{.}}{2016}]%
        {onan2016ensemble}
\bibfield{author}{\bibinfo{person}{Aytu{\u{g}} Onan}, \bibinfo{person}{Serdar
  Koruko{\u{g}}lu}, {and} \bibinfo{person}{Hasan Bulut}.}
  \bibinfo{year}{2016}\natexlab{}.
\newblock \showarticletitle{Ensemble of keyword extraction methods and
  classifiers in text classification}.
\newblock \bibinfo{journal}{\emph{Expert Systems with Applications}}
  \bibinfo{volume}{57} (\bibinfo{year}{2016}), \bibinfo{pages}{232--247}.
\newblock

 \begin{quotation}\noindent Enumerates and performs an empirical analysis of
  five different keyword extraction methods and ensembles. \end{quotation}
\bibitem[\protect\citeauthoryear{P{\'{e}}rez, Risco, Naranjo, Caballer, and
  Molt{\'{o}}}{P{\'{e}}rez et~al\mbox{.}}{2019}]%
        {DBLP:conf/IEEEcloud/PerezRNCM19}
\bibfield{author}{\bibinfo{person}{Alfonso P{\'{e}}rez},
  \bibinfo{person}{Sebasti{\'{a}}n Risco}, \bibinfo{person}{Diana~M. Naranjo},
  \bibinfo{person}{Miguel Caballer}, {and} \bibinfo{person}{Germ{\'{a}}n
  Molt{\'{o}}}.} \bibinfo{year}{2019}\natexlab{}.
\newblock \showarticletitle{On-Premises Serverless Computing for Event-Driven
  Data Processing Applications}. In \bibinfo{booktitle}{\emph{12th {IEEE}
  International Conference on Cloud Computing, {CLOUD} 2019, Milan, Italy, July
  8-13, 2019}}. \bibinfo{pages}{414--421}.
\newblock
\urldef\tempurl%
\url{https://doi.org/10.1109/CLOUD.2019.00073}
\showDOI{\tempurl}

 \begin{quotation}\noindent Present a framework for FaaS computing on
  on-premise platforms. Virtual machines from different cloud providers can be
  used to horizontally scale Kubernetes nodes. \end{quotation}
\bibitem[\protect\citeauthoryear{Peterson}{Peterson}{2017}]%
        {peterson_2017}
\bibfield{author}{\bibinfo{person}{Becky Peterson}.}
  \bibinfo{year}{2017}\natexlab{}.
\newblock \bibinfo{title}{Google takes a page from Amazon Web Services and adds
  per-second billing for its cloud}.
\newblock
\newblock
\urldef\tempurl%
\url{http://uk.businessinsider.com/google-cloud-matches-amazon-web-services-with-per-second-billing-2017-9}
\showURL{%
\tempurl}


\bibitem[\protect\citeauthoryear{Pietri and Sakellariou}{Pietri and
  Sakellariou}{2014}]%
        {pietri2014energy}
\bibfield{author}{\bibinfo{person}{Ilia Pietri} {and} \bibinfo{person}{Rizos
  Sakellariou}.} \bibinfo{year}{2014}\natexlab{}.
\newblock \showarticletitle{Energy-aware workflow scheduling using frequency
  scaling}. In \bibinfo{booktitle}{\emph{2014 43rd International Conference on
  Parallel Processing Workshops}}. IEEE, \bibinfo{pages}{104--113}.
\newblock

 \begin{quotation}\noindent Introduces a workflow scheduling algorithm that
  uses dynamic frequency and voltage scaling to reduce energy consumption while
  meeting deadlines. \end{quotation}
\bibitem[\protect\citeauthoryear{Pietri and Sakellariou}{Pietri and
  Sakellariou}{2019}]%
        {pietri2019pareto}
\bibfield{author}{\bibinfo{person}{Ilia Pietri} {and} \bibinfo{person}{Rizos
  Sakellariou}.} \bibinfo{year}{2019}\natexlab{}.
\newblock \showarticletitle{A Pareto-based approach for CPU provisioning of
  scientific workflows on clouds}.
\newblock \bibinfo{journal}{\emph{Future Generation Computer Systems}}
  \bibinfo{volume}{94} (\bibinfo{year}{2019}), \bibinfo{pages}{479--487}.
\newblock

 \begin{quotation}\noindent Introduces PSFS and CSFS-H. PSFS identifies and
  selects Pareto-efficient resources with respect to cost and makespan. The
  policy makes a trade-off of CPUs and CPU-frequency versus workflow makespan.
  CSFS-H is a variation of the CSFS-Max algorithm to minimize costs by lowering
  CPU frequency. \end{quotation}
\bibitem[\protect\citeauthoryear{Pollard, Jain, Herbein, and Bhatele}{Pollard
  et~al\mbox{.}}{2018}]%
        {DBLP:conf/sc/PollardJHB18}
\bibfield{author}{\bibinfo{person}{Samuel~D. Pollard}, \bibinfo{person}{Nikhil
  Jain}, \bibinfo{person}{Stephen Herbein}, {and} \bibinfo{person}{Abhinav
  Bhatele}.} \bibinfo{year}{2018}\natexlab{}.
\newblock \showarticletitle{Evaluation of an interference-free node allocation
  policy on fat-tree clusters}. In \bibinfo{booktitle}{\emph{Proceedings of the
  International Conference for High Performance Computing, Networking, Storage,
  and Analysis, {SC} 2018, Dallas, TX, USA, November 11-16, 2018}}.
  \bibinfo{pages}{26:1--26:13}.
\newblock
\urldef\tempurl%
\url{http://dl.acm.org/citation.cfm?id=3291691}
\showURL{%
\tempurl}

 \begin{quotation}\noindent Introduce an allocation policy for fat-tree
  clusters that isolate jobs to eliminate inter-job interference. The authors
  test their solution in a production setting where they demonstrate QoS
  requirements are not negatively impacted. \end{quotation}
\bibitem[\protect\citeauthoryear{Poola, Ramamohanarao, and Buyya}{Poola
  et~al\mbox{.}}{2014}]%
        {poola2014fault}
\bibfield{author}{\bibinfo{person}{Deepak Poola}, \bibinfo{person}{Kotagiri
  Ramamohanarao}, {and} \bibinfo{person}{Rajkumar Buyya}.}
  \bibinfo{year}{2014}\natexlab{}.
\newblock \showarticletitle{Fault-tolerant workflow scheduling using spot
  instances on clouds}.
\newblock \bibinfo{journal}{\emph{Procedia Computer Science}}
  \bibinfo{volume}{29} (\bibinfo{year}{2014}), \bibinfo{pages}{523--533}.
\newblock

 \begin{quotation}\noindent Proposes a scheduling algorithm that schedules
  workloads on on-demand and spot cloud instances to reduce costs while meeting
  deadlines. The scheduling algorithm is robust against performance variations
  and premature instance termination. \end{quotation}
\bibitem[\protect\citeauthoryear{Poola, Salehi, Ramamohanarao, and Buyya}{Poola
  et~al\mbox{.}}{2017}]%
        {poola2017taxonomy}
\bibfield{author}{\bibinfo{person}{Deepak Poola}, \bibinfo{person}{Mohsen~Amini
  Salehi}, \bibinfo{person}{Kotagiri Ramamohanarao}, {and}
  \bibinfo{person}{Rajkumar Buyya}.} \bibinfo{year}{2017}\natexlab{}.
\newblock \showarticletitle{A taxonomy and survey of fault-tolerant workflow
  management systems in cloud and distributed computing environments}.
\newblock In \bibinfo{booktitle}{\emph{Software Architecture for Big Data and
  the Cloud}}. \bibinfo{publisher}{Elsevier}, \bibinfo{pages}{285--320}.
\newblock

 \begin{quotation}\noindent Presents a survey and taxonomies on fault-tolerance
  on workflow management in clouds and other distirbuted environments.
  \end{quotation}
\bibitem[\protect\citeauthoryear{Quang, Kim, Rho, Kim, Kim, Hwang, Medernach,
  and Breton}{Quang et~al\mbox{.}}{2015}]%
        {DBLP:conf/ccgrid/QuangKRKKHMB15}
\bibfield{author}{\bibinfo{person}{Bui~The Quang}, \bibinfo{person}{Jik{-}Soo
  Kim}, \bibinfo{person}{Seungwoo Rho}, \bibinfo{person}{Seoyoung Kim},
  \bibinfo{person}{Sangwan Kim}, \bibinfo{person}{Soonwook Hwang},
  \bibinfo{person}{Emmanuel Medernach}, {and} \bibinfo{person}{Vincent
  Breton}.} \bibinfo{year}{2015}\natexlab{}.
\newblock \showarticletitle{A Comparative Analysis of Scheduling Mechanisms for
  Virtual Screening Workflow in a Shared Resource Environment}. In
  \bibinfo{booktitle}{\emph{15th {IEEE/ACM} International Symposium on Cluster,
  Cloud and Grid Computing, CCGrid 2015, Shenzhen, China, May 4-7, 2015}}.
  \bibinfo{pages}{853--862}.
\newblock
\urldef\tempurl%
\url{https://doi.org/10.1109/CCGrid.2015.123}
\showDOI{\tempurl}

 \begin{quotation}\noindent Presents a comparative study of two policies for
  the fair scheduling of virtual screening workflows using the same
  infrastructure. They focus on fairness, overall system throughput, and
  response time. \end{quotation}
\bibitem[\protect\citeauthoryear{Qureshi}{Qureshi}{2019}]%
        {qureshi2019profile}
\bibfield{author}{\bibinfo{person}{Basit Qureshi}.}
  \bibinfo{year}{2019}\natexlab{}.
\newblock \showarticletitle{Profile-based power-aware workflow scheduling
  framework for energy-efficient data centers}.
\newblock \bibinfo{journal}{\emph{Future Generation Computer Systems}}
  \bibinfo{volume}{94} (\bibinfo{year}{2019}), \bibinfo{pages}{453--467}.
\newblock

 \begin{quotation}\noindent Presents a framework and workflow scheduling
  mechanism to minimize cost and power consumption through improving resource
  utilization. \end{quotation}
\bibitem[\protect\citeauthoryear{Rajakumar, Arunachalam, and
  Selladurai}{Rajakumar et~al\mbox{.}}{2004}]%
        {rajakumar2004workflow}
\bibfield{author}{\bibinfo{person}{S Rajakumar}, \bibinfo{person}{VP
  Arunachalam}, {and} \bibinfo{person}{V Selladurai}.}
  \bibinfo{year}{2004}\natexlab{}.
\newblock \showarticletitle{Workflow balancing strategies in parallel machine
  scheduling}.
\newblock \bibinfo{journal}{\emph{The International Journal of Advanced
  Manufacturing Technology}} \bibinfo{volume}{23}, \bibinfo{number}{5-6}
  (\bibinfo{year}{2004}), \bibinfo{pages}{366--374}.
\newblock

 \begin{quotation}\noindent Investigates tree strategies (random, shortest
  processing time, and longest processing time) for assigning parallel jobs to
  machines. \end{quotation}
\bibitem[\protect\citeauthoryear{Ramakrishnan, Koelbel, Kee, Wolski, Nurmi,
  Gannon, Obertelli, YarKhan, Mandal, Huang, et~al\mbox{.}}{Ramakrishnan
  et~al\mbox{.}}{2009}]%
        {ramakrishnan2009vgrads}
\bibfield{author}{\bibinfo{person}{Lavanya Ramakrishnan},
  \bibinfo{person}{Charles Koelbel}, \bibinfo{person}{Yang-Suk Kee},
  \bibinfo{person}{Rich Wolski}, \bibinfo{person}{Daniel Nurmi},
  \bibinfo{person}{Dennis Gannon}, \bibinfo{person}{Graziano Obertelli},
  \bibinfo{person}{Asim YarKhan}, \bibinfo{person}{Anirban Mandal},
  \bibinfo{person}{T~Mark Huang}, {et~al\mbox{.}}}
  \bibinfo{year}{2009}\natexlab{}.
\newblock \showarticletitle{VGrADS: enabling e-Science workflows on grids and
  clouds with fault tolerance}. In \bibinfo{booktitle}{\emph{Proceedings of the
  Conference on High Performance Computing Networking, Storage and Analysis}}.
  ACM, \bibinfo{pages}{47}.
\newblock

 \begin{quotation}\noindent Introduces VGrADS, a virtual grid execution system
  that offers resources abstracted over grids and clouds. The article reports
  on experiences, usage, and fault-tolerance. \end{quotation}
\bibitem[\protect\citeauthoryear{Rehman, Hussain, ur~Rehman, Zia, and
  Shamshirband}{Rehman et~al\mbox{.}}{2019}]%
        {rehman2019multi}
\bibfield{author}{\bibinfo{person}{Attiqa Rehman}, \bibinfo{person}{Syed~S
  Hussain}, \bibinfo{person}{Zia ur Rehman}, \bibinfo{person}{Seemal Zia},
  {and} \bibinfo{person}{Shahaboddin Shamshirband}.}
  \bibinfo{year}{2019}\natexlab{}.
\newblock \showarticletitle{Multi-objective approach of energy efficient
  workflow scheduling in cloud environments}.
\newblock \bibinfo{journal}{\emph{Concurrency and Computation: Practice and
  Experience}} \bibinfo{volume}{31}, \bibinfo{number}{8}
  (\bibinfo{year}{2019}), \bibinfo{pages}{e4949}.
\newblock

 \begin{quotation}\noindent Presents MOGA, a multi-objective genetic algorithm
  approach for workflow scheduling. The solution focuses on makespan, budget,
  deadline, and power consumption optimization. \end{quotation}
\bibitem[\protect\citeauthoryear{Rodriguez and Buyya}{Rodriguez and
  Buyya}{2016}]%
        {rodriguez2016taxonomy}
\bibfield{author}{\bibinfo{person}{Maria~Alejandra Rodriguez} {and}
  \bibinfo{person}{Rajkumar Buyya}.} \bibinfo{year}{2016}\natexlab{}.
\newblock \showarticletitle{A taxonomy and survey on scheduling algorithms for
  scientific workflows in IaaS cloud computing environments}.
\newblock \bibinfo{journal}{\emph{Concurrency and Computation: Practice and
  Experience}} (\bibinfo{year}{2016}).
\newblock

 \begin{quotation}\noindent Studies the challenges that arise from scheduling
  workflows in cloud computing environments. Existing algorithms are studied
  from the perspective of the scheduling model and a taxonomy that focuses on
  features particular to clouds is presented. \end{quotation}
\bibitem[\protect\citeauthoryear{Rodriguez and Buyya}{Rodriguez and
  Buyya}{2018}]%
        {DBLP:journals/fgcs/RodriguezB18}
\bibfield{author}{\bibinfo{person}{Maria~Alejandra Rodriguez} {and}
  \bibinfo{person}{Rajkumar Buyya}.} \bibinfo{year}{2018}\natexlab{}.
\newblock \showarticletitle{Scheduling dynamic workloads in multi-tenant
  scientific workflow as a service platforms}.
\newblock \bibinfo{journal}{\emph{Future Generation Comp. Syst.}}
  \bibinfo{volume}{79} (\bibinfo{year}{2018}), \bibinfo{pages}{739--750}.
\newblock
\urldef\tempurl%
\url{https://doi.org/10.1016/j.future.2017.05.009}
\showDOI{\tempurl}

 \begin{quotation}\noindent Introduces a resource provisioning strategy for
  Workflow-as-a-Service platforms that attempts to minimize cost while meeting
  deadlines. \end{quotation}
\bibitem[\protect\citeauthoryear{Ron Gill~CMA}{Ron Gill~CMA}{2011}]%
        {ron2011cloud}
\bibfield{author}{\bibinfo{person}{CFM Ron Gill~CMA}.}
  \bibinfo{year}{2011}\natexlab{}.
\newblock \showarticletitle{Why cloud computing matters to finance}.
\newblock \bibinfo{journal}{\emph{Strategic Finance}} \bibinfo{volume}{92},
  \bibinfo{number}{7} (\bibinfo{year}{2011}), \bibinfo{pages}{43}.
\newblock

 \begin{quotation}\noindent Based on a survey of roughly 800 IMA members, the
  meaning of cloud services to finance is explained. \end{quotation}
\bibitem[\protect\citeauthoryear{Schwarzkopf, Konwinski, Abd-El-Malek, and
  Wilkes}{Schwarzkopf et~al\mbox{.}}{2013}]%
        {schwarzkopf2013omega}
\bibfield{author}{\bibinfo{person}{Malte Schwarzkopf}, \bibinfo{person}{Andy
  Konwinski}, \bibinfo{person}{Michael Abd-El-Malek}, {and}
  \bibinfo{person}{John Wilkes}.} \bibinfo{year}{2013}\natexlab{}.
\newblock \showarticletitle{Omega: flexible, scalable schedulers for large
  compute clusters}.
\newblock  (\bibinfo{year}{2013}).
\newblock

 \begin{quotation}\noindent Introduces Omega: a large-scale compute cluster
  scheduler that uses a lock-free, optimistic concurrency control approach to
  schedule a mixture of jobs types. The goals of this scheduler is to achieve
  flexibility in terms of implementation extensibility and performance
  scalability. \end{quotation}
\bibitem[\protect\citeauthoryear{Senturk, Balakrishnan, Abu{-}Doleh, Kaya,
  Malluhi, and {\c{C}}ataly{\"{u}}rek}{Senturk et~al\mbox{.}}{2018}]%
        {DBLP:journals/fgcs/SenturkBAKMC18}
\bibfield{author}{\bibinfo{person}{Izzet~F. Senturk},
  \bibinfo{person}{Ponnuraman Balakrishnan}, \bibinfo{person}{Anas
  Abu{-}Doleh}, \bibinfo{person}{Kamer Kaya}, \bibinfo{person}{Qutaibah~M.
  Malluhi}, {and} \bibinfo{person}{{\"{U}}mit~V. {\c{C}}ataly{\"{u}}rek}.}
  \bibinfo{year}{2018}\natexlab{}.
\newblock \showarticletitle{A resource provisioning framework for
  bioinformatics applications in multi-cloud environments}.
\newblock \bibinfo{journal}{\emph{Future Gener. Comput. Syst.}}
  \bibinfo{volume}{78} (\bibinfo{year}{2018}), \bibinfo{pages}{379--391}.
\newblock
\urldef\tempurl%
\url{https://doi.org/10.1016/j.future.2016.06.008}
\showDOI{\tempurl}

 \begin{quotation}\noindent Introduce BioCloud: a resource provisioning
  framework for running biology applications in multi-cloud environments.
  \end{quotation}
\bibitem[\protect\citeauthoryear{Seo, Jeong, Park, and Lee}{Seo
  et~al\mbox{.}}{2008}]%
        {seo2008energy}
\bibfield{author}{\bibinfo{person}{Euiseong Seo}, \bibinfo{person}{Jinkyu
  Jeong}, \bibinfo{person}{Seonyeong Park}, {and} \bibinfo{person}{Joonwon
  Lee}.} \bibinfo{year}{2008}\natexlab{}.
\newblock \showarticletitle{Energy efficient scheduling of real-time tasks on
  multicore processors}.
\newblock \bibinfo{journal}{\emph{IEEE transactions on parallel and distributed
  systems}} \bibinfo{volume}{19}, \bibinfo{number}{11} (\bibinfo{year}{2008}),
  \bibinfo{pages}{1540--1552}.
\newblock

 \begin{quotation}\noindent Introduces a strategy that takes into account that
  dynamic voltage scaling requires the same voltage for all cores on the chip.
  Additionally, cores are turned off during execution to improve power
  consumption. \end{quotation}
\bibitem[\protect\citeauthoryear{Shi, Cao, Zhang, Li, and Xu}{Shi
  et~al\mbox{.}}{2016}]%
        {shi2016edge}
\bibfield{author}{\bibinfo{person}{Weisong Shi}, \bibinfo{person}{Jie Cao},
  \bibinfo{person}{Quan Zhang}, \bibinfo{person}{Youhuizi Li}, {and}
  \bibinfo{person}{Lanyu Xu}.} \bibinfo{year}{2016}\natexlab{}.
\newblock \showarticletitle{Edge computing: Vision and challenges}.
\newblock \bibinfo{journal}{\emph{IEEE Internet of Things Journal}}
  \bibinfo{volume}{3}, \bibinfo{number}{5} (\bibinfo{year}{2016}),
  \bibinfo{pages}{637--646}.
\newblock

 \begin{quotation}\noindent Presents a vision, challenges, and opportunities
  for the edge computing paradigm. \end{quotation}
\bibitem[\protect\citeauthoryear{Shishido, Estrella, Toledo, and
  Reiff-Marganiec}{Shishido et~al\mbox{.}}{2018}]%
        {shishido2018cloudsim}
\bibfield{author}{\bibinfo{person}{Henrique~Yoshikazu Shishido},
  \bibinfo{person}{J{\'u}lio~Cezar Estrella}, \bibinfo{person}{Claudio
  Fabiano~Motta Toledo}, {and} \bibinfo{person}{Stephan Reiff-Marganiec}.}
  \bibinfo{year}{2018}\natexlab{}.
\newblock \showarticletitle{A CloudSim Extension for Evaluating Security
  Overhead in Workflow Execution in Clouds}. In \bibinfo{booktitle}{\emph{2018
  Sixth International Symposium on Computing and Networking (CANDAR)}}. IEEE,
  \bibinfo{pages}{174--180}.
\newblock

 \begin{quotation}\noindent Proposes an extension to CloudSim where the
  overhead of security methods are measured. In particular, the authors measure
  the impact of authentication, integrity, and encryption. \end{quotation}
\bibitem[\protect\citeauthoryear{Shoaib and Das}{Shoaib and Das}{2014}]%
        {DBLP:journals/corr/ShoaibD14a}
\bibfield{author}{\bibinfo{person}{Yasir Shoaib} {and} \bibinfo{person}{Olivia
  Das}.} \bibinfo{year}{2014}\natexlab{}.
\newblock \showarticletitle{Performance-oriented Cloud Provisioning: Taxonomy
  and Survey}.
\newblock \bibinfo{journal}{\emph{CoRR}}  \bibinfo{volume}{abs/1411.5077}
  (\bibinfo{year}{2014}).
\newblock
\showeprint[arxiv]{1411.5077}
\urldef\tempurl%
\url{http://arxiv.org/abs/1411.5077}
\showURL{%
\tempurl}

 \begin{quotation}\noindent Provides a survey and taxonomy of resource
  provisioning in the cloud. \end{quotation}
\bibitem[\protect\citeauthoryear{Simon}{Simon}{1956}]%
        {simon1956rational}
\bibfield{author}{\bibinfo{person}{Herbert~A Simon}.}
  \bibinfo{year}{1956}\natexlab{}.
\newblock \showarticletitle{Rational choice and the structure of the
  environment.}
\newblock \bibinfo{journal}{\emph{Psychological review}} \bibinfo{volume}{63},
  \bibinfo{number}{2} (\bibinfo{year}{1956}), \bibinfo{pages}{129}.
\newblock

 \begin{quotation}\noindent Introduces the term "satisficing" from an organic
  point of view. This term is also adopted by the system's community to
  generate sub-optimal, but good enough solutions to trade-off with speed or
  costs. \end{quotation}
\bibitem[\protect\citeauthoryear{Singh, Kesselman, and Deelman}{Singh
  et~al\mbox{.}}{2007}]%
        {DBLP:conf/hpdc/SinghKD07}
\bibfield{author}{\bibinfo{person}{Gurmeet Singh}, \bibinfo{person}{Carl
  Kesselman}, {and} \bibinfo{person}{Ewa Deelman}.}
  \bibinfo{year}{2007}\natexlab{}.
\newblock \showarticletitle{A provisioning model and its comparison with
  best-effort for performance-cost optimization in grids}. In
  \bibinfo{booktitle}{\emph{Proceedings of the 16th International Symposium on
  High-Performance Distributed Computing {(HPDC-16} 2007), 25-29 June 2007,
  Monterey, California, {USA}}}. \bibinfo{pages}{117--126}.
\newblock
\urldef\tempurl%
\url{https://doi.org/10.1145/1272366.1272382}
\showDOI{\tempurl}

 \begin{quotation}\noindent Investigates the performance of a multi-objective
  genetic algorithm to trade-off resources provisioning versus workflow
  makespan. Using simulation, the authors investigate the impact of such
  trade-offs using several different workflows. \end{quotation}
\bibitem[\protect\citeauthoryear{Singh, Gupta, and Jana}{Singh
  et~al\mbox{.}}{2018}]%
        {DBLP:journals/fgcs/SinghGJ18}
\bibfield{author}{\bibinfo{person}{Vishakha Singh}, \bibinfo{person}{Indrajeet
  Gupta}, {and} \bibinfo{person}{Prasanta~K. Jana}.}
  \bibinfo{year}{2018}\natexlab{}.
\newblock \showarticletitle{A novel cost-efficient approach for
  deadline-constrained workflow scheduling by dynamic provisioning of
  resources}.
\newblock \bibinfo{journal}{\emph{Future Gener. Comput. Syst.}}
  \bibinfo{volume}{79} (\bibinfo{year}{2018}), \bibinfo{pages}{95--110}.
\newblock
\urldef\tempurl%
\url{https://doi.org/10.1016/j.future.2017.09.054}
\showDOI{\tempurl}

 \begin{quotation}\noindent Introduces PPDPS, a cost-efficient and
  deadline-aware scheduling and provisioning algorithm. \end{quotation}
\bibitem[\protect\citeauthoryear{Smanchat and Viriyapant}{Smanchat and
  Viriyapant}{2015}]%
        {DBLP:journals/fgcs/SmanchatV15}
\bibfield{author}{\bibinfo{person}{Sucha Smanchat} {and}
  \bibinfo{person}{Kanchana Viriyapant}.} \bibinfo{year}{2015}\natexlab{}.
\newblock \showarticletitle{Taxonomies of workflow scheduling problem and
  techniques in the cloud}.
\newblock \bibinfo{journal}{\emph{Future Generation Comp. Syst.}}
  \bibinfo{volume}{52} (\bibinfo{year}{2015}), \bibinfo{pages}{1--12}.
\newblock
\urldef\tempurl%
\url{https://doi.org/10.1016/j.future.2015.04.019}
\showDOI{\tempurl}

 \begin{quotation}\noindent Proposes several taxonomies of workflow scheduling
  problems and techniques used in cloud environments. \end{quotation}
\bibitem[\protect\citeauthoryear{Song, Kwok, and Hwang}{Song
  et~al\mbox{.}}{2005}]%
        {DBLP:conf/ipps/SongKH05}
\bibfield{author}{\bibinfo{person}{Shanshan Song}, \bibinfo{person}{Yu{-}Kwong
  Kwok}, {and} \bibinfo{person}{Kai Hwang}.} \bibinfo{year}{2005}\natexlab{}.
\newblock \showarticletitle{Security-Driven Heuristics and {A} Fast Genetic
  Algorithm for Trusted Grid Job Scheduling}. In \bibinfo{booktitle}{\emph{19th
  International Parallel and Distributed Processing Symposium {(IPDPS} 2005),
  {CD-ROM} / Abstracts Proceedings, 4-8 April 2005, Denver, CO, {USA}}}.
\newblock
\urldef\tempurl%
\url{https://doi.org/10.1109/IPDPS.2005.397}
\showDOI{\tempurl}

 \begin{quotation}\noindent Proposes an enhancement of the Min-Min and
  Sufferage heuristics under three risk modes driven by security
  considerations. Additionally, this work introduces a new Space-Time Genetic
  Algorithm for trusted job scheduling which outperforms the enhanced Min-min
  and Sufferage heuristics. \end{quotation}
\bibitem[\protect\citeauthoryear{Stavrinides and Karatza}{Stavrinides and
  Karatza}{2019}]%
        {stavrinides2019energy}
\bibfield{author}{\bibinfo{person}{Georgios~L Stavrinides} {and}
  \bibinfo{person}{Helen~D Karatza}.} \bibinfo{year}{2019}\natexlab{}.
\newblock \showarticletitle{An energy-efficient, QoS-aware and cost-effective
  scheduling approach for real-time workflow applications in cloud computing
  systems utilizing DVFS and approximate computations}.
\newblock \bibinfo{journal}{\emph{Future Generation Computer Systems}}
  \bibinfo{volume}{96} (\bibinfo{year}{2019}), \bibinfo{pages}{216--226}.
\newblock

 \begin{quotation}\noindent Proposes a scheduling technique using DVFS to fill
  up schedule gaps in an energy-efficient way. By trading off fidelity, yet
  keeping track of input errors, less energy is consumed while the output is
  within reasonable set bounds. \end{quotation}
\bibitem[\protect\citeauthoryear{Stein}{Stein}{2010}]%
        {stein2010case}
\bibfield{author}{\bibinfo{person}{Lincoln~D Stein}.}
  \bibinfo{year}{2010}\natexlab{}.
\newblock \showarticletitle{The case for cloud computing in genome
  informatics}.
\newblock \bibinfo{journal}{\emph{Genome biology}} \bibinfo{volume}{11},
  \bibinfo{number}{5} (\bibinfo{year}{2010}), \bibinfo{pages}{207}.
\newblock

 \begin{quotation}\noindent Advocates that genome informatics should move to
  cloud computing. \end{quotation}
\bibitem[\protect\citeauthoryear{Tanaka and Tatebe}{Tanaka and Tatebe}{2012}]%
        {tanaka2012workflow}
\bibfield{author}{\bibinfo{person}{Masahiro Tanaka} {and}
  \bibinfo{person}{Osamu Tatebe}.} \bibinfo{year}{2012}\natexlab{}.
\newblock \showarticletitle{Workflow scheduling to minimize data movement using
  multi-constraint graph partitioning}. In
  \bibinfo{booktitle}{\emph{Proceedings of the 2012 12th IEEE/ACM International
  Symposium on Cluster, Cloud and Grid Computing (ccgrid 2012)}}. IEEE Computer
  Society, \bibinfo{pages}{65--72}.
\newblock

 \begin{quotation}\noindent Attempts to improve data locality (or data movement
  as the authors call it) by using a Multi-Constraint Graph Partitioning
  approach. \end{quotation}
\bibitem[\protect\citeauthoryear{Tang, Zhang, Yao, Li, Zhang, and Su}{Tang
  et~al\mbox{.}}{2008}]%
        {DBLP:conf/kdd/TangZYLZS08}
\bibfield{author}{\bibinfo{person}{Jie Tang}, \bibinfo{person}{Jing Zhang},
  \bibinfo{person}{Limin Yao}, \bibinfo{person}{Juanzi Li}, \bibinfo{person}{Li
  Zhang}, {and} \bibinfo{person}{Zhong Su}.} \bibinfo{year}{2008}\natexlab{}.
\newblock \showarticletitle{ArnetMiner: extraction and mining of academic
  social networks}. In \bibinfo{booktitle}{\emph{Proceedings of the 14th {ACM}
  {SIGKDD} International Conference on Knowledge Discovery and Data Mining, Las
  Vegas, Nevada, USA, August 24-27, 2008}}. \bibinfo{pages}{990--998}.
\newblock
\urldef\tempurl%
\url{https://doi.org/10.1145/1401890.1402008}
\showDOI{\tempurl}

 \begin{quotation}\noindent Discusses Aminer, a platform where academic social
  networks and paper meta-data are parsed, analyzed, and offered as datasets.
  \end{quotation}
\bibitem[\protect\citeauthoryear{Toader, Uta, Musaafir, and Iosup}{Toader
  et~al\mbox{.}}{2019}]%
        {toadergraphless}
\bibfield{author}{\bibinfo{person}{Lucian Toader}, \bibinfo{person}{Alexandru
  Uta}, \bibinfo{person}{Ahmed Musaafir}, {and} \bibinfo{person}{Alexandru
  Iosup}.} \bibinfo{year}{2019}\natexlab{}.
\newblock \showarticletitle{Graphless: Toward Serverless Graph Processing}.
\newblock  (\bibinfo{year}{2019}), \bibinfo{pages}{66--73}.
\newblock


\bibitem[\protect\citeauthoryear{Topcuoglu, Hariri, and Wu}{Topcuoglu
  et~al\mbox{.}}{2002}]%
        {DBLP:journals/tpds/TopcuogluHW02}
\bibfield{author}{\bibinfo{person}{Haluk Topcuoglu}, \bibinfo{person}{Salim
  Hariri}, {and} \bibinfo{person}{Min{-}You Wu}.}
  \bibinfo{year}{2002}\natexlab{}.
\newblock \showarticletitle{Performance-Effective and Low-Complexity Task
  Scheduling for Heterogeneous Computing}.
\newblock \bibinfo{journal}{\emph{{IEEE} Trans. Parallel Distrib. Syst.}}
  \bibinfo{volume}{13}, \bibinfo{number}{3} (\bibinfo{year}{2002}),
  \bibinfo{pages}{260--274}.
\newblock
\urldef\tempurl%
\url{https://doi.org/10.1109/71.993206}
\showDOI{\tempurl}

 \begin{quotation}\noindent Presents two scheduling algorithms for a bounded
  number of heterogeneous processors: HEFT and CPOP. Both algorithms focus on
  high performance and fast scheduling time. \end{quotation}
\bibitem[\protect\citeauthoryear{van Beek, Oikonomou, and Iosup}{van Beek
  et~al\mbox{.}}{2019}]%
        {vancpu}
\bibfield{author}{\bibinfo{person}{Vincent van Beek}, \bibinfo{person}{Giorgos
  Oikonomou}, {and} \bibinfo{person}{Alexandru Iosup}.}
  \bibinfo{year}{2019}\natexlab{}.
\newblock \showarticletitle{A CPU Contention Predictor for Business-Critical
  Workloads in Cloud Datacenters}.
\newblock  (\bibinfo{year}{2019}), \bibinfo{pages}{56--61}.
\newblock


\bibitem[\protect\citeauthoryear{Van Der~Aalst and Ter~Hofstede}{Van Der~Aalst
  and Ter~Hofstede}{2005}]%
        {van2005yawl}
\bibfield{author}{\bibinfo{person}{Wil~MP Van Der~Aalst} {and}
  \bibinfo{person}{Arthur~HM Ter~Hofstede}.} \bibinfo{year}{2005}\natexlab{}.
\newblock \showarticletitle{YAWL: yet another workflow language}.
\newblock \bibinfo{journal}{\emph{Information systems}} \bibinfo{volume}{30},
  \bibinfo{number}{4} (\bibinfo{year}{2005}), \bibinfo{pages}{245--275}.
\newblock

 \begin{quotation}\noindent Introduces YAWL: a language for constructing
  workflows. YAWL was designed with Petri nets as starting point, and extended
  to overcome limitations encountered with Petri nets. \end{quotation}
\bibitem[\protect\citeauthoryear{van~der Aalst}{van~der Aalst}{1998}]%
        {DBLP:journals/jcsc/Aalst98}
\bibfield{author}{\bibinfo{person}{Wil M.~P. van~der Aalst}.}
  \bibinfo{year}{1998}\natexlab{}.
\newblock \showarticletitle{The Application of Petri Nets to Workflow
  Management}.
\newblock \bibinfo{journal}{\emph{Journal of Circuits, Systems, and Computers}}
  \bibinfo{volume}{8}, \bibinfo{number}{1} (\bibinfo{year}{1998}),
  \bibinfo{pages}{21--66}.
\newblock
\urldef\tempurl%
\url{https://doi.org/10.1142/S0218126698000043}
\showDOI{\tempurl}

 \begin{quotation}\noindent Discusses the use of Petri nets in the context of
  business workflow management. Work on the verification of workflows and
  workflow tools is presented. \end{quotation}
\bibitem[\protect\citeauthoryear{van~der Aalst, Van~Hee, and Houben}{van~der
  Aalst et~al\mbox{.}}{1994}]%
        {van1994modelling}
\bibfield{author}{\bibinfo{person}{Willibrordus Martinus~Pancratius van~der
  Aalst}, \bibinfo{person}{KM Van~Hee}, {and} \bibinfo{person}{GJ Houben}.}
  \bibinfo{year}{1994}\natexlab{}.
\newblock \showarticletitle{Modelling and analysing workflow using a Petri-net
  based approach}. In \bibinfo{booktitle}{\emph{Proceedings of the second
  Workshop on Computer-Supported Cooperative Work, Petri nets and related
  formalisms}}. of, \bibinfo{pages}{31--50}.
\newblock

 \begin{quotation}\noindent Formalizes the main concepts when using Petri nets
  to model and execute workflows in offices and management. \end{quotation}
\bibitem[\protect\citeauthoryear{van Eyk, Toader, Talluri, Versluis, Uta, and
  Iosup}{van Eyk et~al\mbox{.}}{2018a}]%
        {van2018serverless}
\bibfield{author}{\bibinfo{person}{Erwin van Eyk}, \bibinfo{person}{Lucian
  Toader}, \bibinfo{person}{Sacheendra Talluri}, \bibinfo{person}{Laurens
  Versluis}, \bibinfo{person}{Alexandru Uta}, {and} \bibinfo{person}{Alexandru
  Iosup}.} \bibinfo{year}{2018}\natexlab{a}.
\newblock \showarticletitle{Serverless is More: From PaaS to Present Cloud
  Computing}.
\newblock \bibinfo{journal}{\emph{IEEE Internet Computing}}
  \bibinfo{volume}{22} (\bibinfo{year}{2018}).
\newblock

 \begin{quotation}\noindent The article provides an understanding of the early
  days of serverless computing and it's path to where it is now. Obstacles and
  opportunities in this area are provided. \end{quotation}
\bibitem[\protect\citeauthoryear{van Eyk, Toader, Talluri, Versluis, Uta, and
  Iosup}{van Eyk et~al\mbox{.}}{2018b}]%
        {DBLP:journals/internet/EykTTVUI18}
\bibfield{author}{\bibinfo{person}{Erwin van Eyk}, \bibinfo{person}{Lucian
  Toader}, \bibinfo{person}{Sacheendra Talluri}, \bibinfo{person}{Laurens
  Versluis}, \bibinfo{person}{Alexandru Uta}, {and} \bibinfo{person}{Alexandru
  Iosup}.} \bibinfo{year}{2018}\natexlab{b}.
\newblock \showarticletitle{Serverless is More: From PaaS to Present Cloud
  Computing}.
\newblock \bibinfo{journal}{\emph{{IEEE} Internet Computing}}
  \bibinfo{volume}{22}, \bibinfo{number}{5} (\bibinfo{year}{2018}).
\newblock
\newblock
\shownote{Sep/Oct issue.}

 \begin{quotation}\noindent Introduces the serverless paradigm; its origin, the
  current state, and the obstacles and opportunities it presents.
  \end{quotation}
\bibitem[\protect\citeauthoryear{Vasan, Sivasubramaniam, Shimpi, Sivabalan, and
  Subbiah}{Vasan et~al\mbox{.}}{2010}]%
        {DBLP:conf/hpca/VasanSSSS10}
\bibfield{author}{\bibinfo{person}{Arunchandar Vasan}, \bibinfo{person}{Anand
  Sivasubramaniam}, \bibinfo{person}{Vikrant Shimpi}, \bibinfo{person}{T.
  Sivabalan}, {and} \bibinfo{person}{Rajesh Subbiah}.}
  \bibinfo{year}{2010}\natexlab{}.
\newblock \showarticletitle{Worth their watts? - an empirical study of
  datacenter servers}. In \bibinfo{booktitle}{\emph{16th International
  Conference on High-Performance Computer Architecture {(HPCA-16} 2010), 9-14
  January 2010, Bangalore, India}}. \bibinfo{pages}{1--10}.
\newblock
\urldef\tempurl%
\url{https://doi.org/10.1109/HPCA.2010.5463056}
\showDOI{\tempurl}

 \begin{quotation}\noindent Identifies several areas where improvements can be
  made regarding power consumption in datacenters. \end{quotation}
\bibitem[\protect\citeauthoryear{Verma, Pedrosa, Korupolu, Oppenheimer, Tune,
  and Wilkes}{Verma et~al\mbox{.}}{2015}]%
        {verma2015large}
\bibfield{author}{\bibinfo{person}{Abhishek Verma}, \bibinfo{person}{Luis
  Pedrosa}, \bibinfo{person}{Madhukar Korupolu}, \bibinfo{person}{David
  Oppenheimer}, \bibinfo{person}{Eric Tune}, {and} \bibinfo{person}{John
  Wilkes}.} \bibinfo{year}{2015}\natexlab{}.
\newblock \showarticletitle{Large-scale cluster management at Google with
  Borg}. In \bibinfo{booktitle}{\emph{Proceedings of the Tenth European
  Conference on Computer Systems}}. ACM, \bibinfo{pages}{18}.
\newblock

 \begin{quotation}\noindent Presents a summary on Borg, the cluster management
  system run at Google. Borg uses admission control, task-packing,
  over-commitment, and machine sharing to achieve high utilization levels.
  Borg's design, architecture, and features are described, a quantitative
  analysis is presented, and an examination of lessons learned. \end{quotation}
\bibitem[\protect\citeauthoryear{Versluis, Eyk, and Iosup}{Versluis
  et~al\mbox{.}}{2018a}]%
        {DBLP:conf/wosp/VersluisEI18}
\bibfield{author}{\bibinfo{person}{Laurens Versluis},
  \bibinfo{person}{Erwin~Van Eyk}, {and} \bibinfo{person}{Alexandru Iosup}.}
  \bibinfo{year}{2018}\natexlab{a}.
\newblock \showarticletitle{An Analysis of Workflow Formalisms for Workflows
  with Complex Non-Functional Requirements}. In
  \bibinfo{booktitle}{\emph{Companion of the 2018 {ACM/SPEC} International
  Conference on Performance Engineering, {ICPE} 2018, Berlin, Germany, April
  09-13, 2018}}. \bibinfo{pages}{107--112}.
\newblock
\urldef\tempurl%
\url{https://doi.org/10.1145/3185768.3186297}
\showDOI{\tempurl}

 \begin{quotation}\noindent Surveys the three most popular formalisms used in
  literature to construct workflows with. Compares these formalisms using a
  library of workflows and a set criteria to evaluate their fit for task based
  non-function requirements. \end{quotation}
\bibitem[\protect\citeauthoryear{{Versluis}, {Math{\'a}}, {Talluri}, {Hegeman},
  {Prodan}, {Deelman}, and {Iosup}}{{Versluis} et~al\mbox{.}}{2019}]%
        {2019arXiv190607471V}
\bibfield{author}{\bibinfo{person}{Laurens {Versluis}}, \bibinfo{person}{Roland
  {Math{\'a}}}, \bibinfo{person}{Sacheendra {Talluri}}, \bibinfo{person}{Tim
  {Hegeman}}, \bibinfo{person}{Radu {Prodan}}, \bibinfo{person}{Ewa {Deelman}},
  {and} \bibinfo{person}{Alexand~ru {Iosup}}.} \bibinfo{year}{2019}\natexlab{}.
\newblock \showarticletitle{{The Workflow Trace Archive: Open-Access Data from
  Public and Private Computing Infrastructures -- Technical Report}}.
\newblock \bibinfo{journal}{\emph{arXiv e-prints}}, Article
  \bibinfo{articleno}{arXiv:1906.07471} (\bibinfo{date}{Jun}
  \bibinfo{year}{2019}), \bibinfo{numpages}{arXiv:1906.07471}~pages.
\newblock
\showeprint[arxiv]{cs.DC/1906.07471}

 \begin{quotation}\noindent Presents the Workflow Trace Archive, an archive
  containing realistic open-access workflow traces from various domains,
  including industry. The authors show through a survey that realistic workflow
  traces are used infrequent and realistic and open-access traces are even
  rarer. The contents of the archive are systematically characterized.
  \end{quotation}
\bibitem[\protect\citeauthoryear{Versluis, Neacsu, and Iosup}{Versluis
  et~al\mbox{.}}{2018b}]%
        {DBLP:conf/ccgrid/VersluisNI18}
\bibfield{author}{\bibinfo{person}{Laurens Versluis}, \bibinfo{person}{Mihai
  Neacsu}, {and} \bibinfo{person}{Alexandru Iosup}.}
  \bibinfo{year}{2018}\natexlab{b}.
\newblock \showarticletitle{A Trace-Based Performance Study of Autoscaling
  Workloads of Workflows in Datacenters}. In \bibinfo{booktitle}{\emph{18th
  {IEEE/ACM} International Symposium on Cluster, Cloud and Grid Computing,
  {CCGRID} 2018, Washington, DC, USA, May 1-4, 2018}}.
  \bibinfo{pages}{223--232}.
\newblock
\urldef\tempurl%
\url{https://doi.org/10.1109/CCGRID.2018.00037}
\showDOI{\tempurl}

 \begin{quotation}\noindent Compares seven state-of-the-art autoscalers using
  both state-of-the-art and well-established metrics via four distinct
  experiments. \end{quotation}
\bibitem[\protect\citeauthoryear{Vukmirovi{\'c}, Erdeljan, Imre, and
  {\v{C}}apko}{Vukmirovi{\'c} et~al\mbox{.}}{2012}]%
        {vukmirovic2012optimal}
\bibfield{author}{\bibinfo{person}{S Vukmirovi{\'c}}, \bibinfo{person}{A
  Erdeljan}, \bibinfo{person}{L Imre}, {and} \bibinfo{person}{D {\v{C}}apko}.}
  \bibinfo{year}{2012}\natexlab{}.
\newblock \showarticletitle{Optimal workflow scheduling in critical
  infrastructure systems with neural networks}.
\newblock \bibinfo{journal}{\emph{Journal of applied research and technology}}
  \bibinfo{volume}{10}, \bibinfo{number}{2} (\bibinfo{year}{2012}),
  \bibinfo{pages}{114--121}.
\newblock

 \begin{quotation}\noindent Introduces an architecture for workflow executions
  in a critical infrastructure. \end{quotation}
\bibitem[\protect\citeauthoryear{Waldspurger and Weihl}{Waldspurger and
  Weihl}{1994}]%
        {waldspurger1994lottery}
\bibfield{author}{\bibinfo{person}{Carl~A Waldspurger} {and}
  \bibinfo{person}{William~E Weihl}.} \bibinfo{year}{1994}\natexlab{}.
\newblock \showarticletitle{Lottery scheduling: Flexible proportional-share
  resource management}. In \bibinfo{booktitle}{\emph{Proceedings of the 1st
  USENIX conference on Operating Systems Design and Implementation}}. USENIX
  Association, \bibinfo{pages}{1}.
\newblock

 \begin{quotation}\noindent Introduces the lottery scheduling algorithm for
  assigning resources to tasks. \end{quotation}
\bibitem[\protect\citeauthoryear{Wang, Li, Zhang, Ristenpart, and Swift}{Wang
  et~al\mbox{.}}{2018}]%
        {DBLP:conf/usenix/WangLZRS18}
\bibfield{author}{\bibinfo{person}{Liang Wang}, \bibinfo{person}{Mengyuan Li},
  \bibinfo{person}{Yinqian Zhang}, \bibinfo{person}{Thomas Ristenpart}, {and}
  \bibinfo{person}{Michael~M. Swift}.} \bibinfo{year}{2018}\natexlab{}.
\newblock \showarticletitle{Peeking Behind the Curtains of Serverless
  Platforms}. In \bibinfo{booktitle}{\emph{2018 {USENIX} Annual Technical
  Conference, {USENIX} {ATC} 2018, Boston, MA, USA, July 11-13, 2018.}}
  \bibinfo{pages}{133--146}.
\newblock
\urldef\tempurl%
\url{https://www.usenix.org/conference/atc18/presentation/wang-liang}
\showURL{%
\tempurl}

 \begin{quotation}\noindent Conducts a large measurement study by executing
  50,00 function instances across three services. Each platform's performance
  is characterized and discussed. \end{quotation}
\bibitem[\protect\citeauthoryear{Wang, Li, Liang, and Li}{Wang
  et~al\mbox{.}}{2016a}]%
        {DBLP:conf/sc/WangLLL16}
\bibfield{author}{\bibinfo{person}{Wei Wang}, \bibinfo{person}{Baochun Li},
  \bibinfo{person}{Ben Liang}, {and} \bibinfo{person}{Jun Li}.}
  \bibinfo{year}{2016}\natexlab{a}.
\newblock \showarticletitle{Multi-resource fair sharing for datacenter jobs
  with placement constraints}. In \bibinfo{booktitle}{\emph{Proceedings of the
  International Conference for High Performance Computing, Networking, Storage
  and Analysis, {SC} 2016, Salt Lake City, UT, USA, November 13-18, 2016}}.
  \bibinfo{pages}{1003--1014}.
\newblock
\urldef\tempurl%
\url{https://doi.org/10.1109/SC.2016.85}
\showDOI{\tempurl}

 \begin{quotation}\noindent Proposes the Task Share Fairness (TSF) policy. This
  policy enables multi-resource fair sharing. The authors show 60
  sped up over existing fair schedulers. \end{quotation}
\bibitem[\protect\citeauthoryear{Wang, Zhu, Ying, Tan, and Zhang}{Wang
  et~al\mbox{.}}{2016b}]%
        {wang2016maptask}
\bibfield{author}{\bibinfo{person}{Weina Wang}, \bibinfo{person}{Kai Zhu},
  \bibinfo{person}{Lei Ying}, \bibinfo{person}{Jian Tan}, {and}
  \bibinfo{person}{Li Zhang}.} \bibinfo{year}{2016}\natexlab{b}.
\newblock \showarticletitle{Maptask scheduling in mapreduce with data locality:
  Throughput and heavy-traffic optimality}.
\newblock \bibinfo{journal}{\emph{IEEE/ACM Transactions on Networking (TON)}}
  \bibinfo{volume}{24}, \bibinfo{number}{1} (\bibinfo{year}{2016}),
  \bibinfo{pages}{190--203}.
\newblock

 \begin{quotation}\noindent Discusses the limitations of data-locality aware
  scheduling for MapReduce and focuses on balancing data-locality with load
  balancing to improve throughput and minimize delay. The authors prove their
  proposed algorithm is heavy-traffic optimal, so that it minimizes the number
  of backlogged tasks. \end{quotation}
\bibitem[\protect\citeauthoryear{Wang, Liu, Zheng, Xia, Li, Chen, Guo, and
  Xie}{Wang et~al\mbox{.}}{2019}]%
        {wang2019multi}
\bibfield{author}{\bibinfo{person}{Yuandou Wang}, \bibinfo{person}{Hang Liu},
  \bibinfo{person}{Wanbo Zheng}, \bibinfo{person}{Yunni Xia},
  \bibinfo{person}{Yawen Li}, \bibinfo{person}{Peng Chen},
  \bibinfo{person}{Kunyin Guo}, {and} \bibinfo{person}{Hong Xie}.}
  \bibinfo{year}{2019}\natexlab{}.
\newblock \showarticletitle{Multi-objective workflow scheduling with
  Deep-Q-network-based Multi-agent Reinforcement Learning}.
\newblock \bibinfo{journal}{\emph{IEEE Access}}  \bibinfo{volume}{7}
  (\bibinfo{year}{2019}), \bibinfo{pages}{39974--39982}.
\newblock

 \begin{quotation}\noindent Uses a Deep-Q-Network in a multi-agent
  reinforcement setting to schedule workflows. The approach focuses on
  optimizing towards both makespan and cost. \end{quotation}
\bibitem[\protect\citeauthoryear{Whitney and Delforge}{Whitney and
  Delforge}{2014}]%
        {whitney2014data}
\bibfield{author}{\bibinfo{person}{Josh Whitney} {and} \bibinfo{person}{Pierre
  Delforge}.} \bibinfo{year}{2014}\natexlab{}.
\newblock \showarticletitle{Data center efficiency assessment}.
\newblock \bibinfo{journal}{\emph{Issue paper on NRDC (The Natural Resource
  Defense Council)}} (\bibinfo{year}{2014}).
\newblock


\bibitem[\protect\citeauthoryear{Wieczorek, Hoheisel, and Prodan}{Wieczorek
  et~al\mbox{.}}{2009}]%
        {DBLP:journals/fgcs/WieczorekHP09}
\bibfield{author}{\bibinfo{person}{Marek Wieczorek}, \bibinfo{person}{Andreas
  Hoheisel}, {and} \bibinfo{person}{Radu Prodan}.}
  \bibinfo{year}{2009}\natexlab{}.
\newblock \showarticletitle{Towards a general model of the multi-criteria
  workflow scheduling on the grid}.
\newblock \bibinfo{journal}{\emph{Future Generation Comp. Syst.}}
  \bibinfo{volume}{25}, \bibinfo{number}{3} (\bibinfo{year}{2009}),
  \bibinfo{pages}{237--256}.
\newblock
\urldef\tempurl%
\url{https://doi.org/10.1016/j.future.2008.09.002}
\showDOI{\tempurl}

 \begin{quotation}\noindent Proposes several taxonomies considering five
  topics: workflow model, scheduling criteria, scheduling process, resource
  model and task model. Relevant work is surveyed and classified according the
  proposed taxonomies. \end{quotation}
\bibitem[\protect\citeauthoryear{Wilkes}{Wilkes}{2011}]%
        {clusterdata:Wilkes2011}
\bibfield{author}{\bibinfo{person}{John Wilkes}.}
  \bibinfo{year}{2011}\natexlab{}.
\newblock \bibinfo{title}{More {Google} cluster data}.
\newblock \bibinfo{howpublished}{Google research blog}.
\newblock
\newblock
\shownote{Posted at
  \url{http://googleresearch.blogspot.com/2011/11/more-google-cluster-data.html}.}


\bibitem[\protect\citeauthoryear{Wu, Hao, Cai, and Wang}{Wu
  et~al\mbox{.}}{2019}]%
        {DBLP:journals/fgcs/WuHCW19}
\bibfield{author}{\bibinfo{person}{Binghong Wu}, \bibinfo{person}{Kuangrong
  Hao}, \bibinfo{person}{Xin Cai}, {and} \bibinfo{person}{Tong Wang}.}
  \bibinfo{year}{2019}\natexlab{}.
\newblock \showarticletitle{An integrated algorithm for multi-agent
  fault-tolerant scheduling based on {MOEA}}.
\newblock \bibinfo{journal}{\emph{Future Gener. Comput. Syst.}}
  \bibinfo{volume}{94} (\bibinfo{year}{2019}), \bibinfo{pages}{51--61}.
\newblock
\urldef\tempurl%
\url{https://doi.org/10.1016/j.future.2018.11.001}
\showDOI{\tempurl}

 \begin{quotation}\noindent Introduces a workflow scheduling algorithm that
  takes failures into account. The algorithm is based on the IPIREM
  rescheduling model and an improved NSGA-III algorithm. \end{quotation}
\bibitem[\protect\citeauthoryear{Wu, Wu, and Tan}{Wu et~al\mbox{.}}{2015}]%
        {DBLP:journals/tjs/WuWT15}
\bibfield{author}{\bibinfo{person}{Fuhui Wu}, \bibinfo{person}{Qingbo Wu},
  {and} \bibinfo{person}{Yusong Tan}.} \bibinfo{year}{2015}\natexlab{}.
\newblock \showarticletitle{Workflow scheduling in cloud: a survey}.
\newblock \bibinfo{journal}{\emph{The Journal of Supercomputing}}
  \bibinfo{volume}{71}, \bibinfo{number}{9} (\bibinfo{year}{2015}),
  \bibinfo{pages}{3373--3418}.
\newblock
\urldef\tempurl%
\url{https://doi.org/10.1007/s11227-015-1438-4}
\showDOI{\tempurl}

 \begin{quotation}\noindent Presents a survey and taxonomies on static and
  dynamic workflow plan computation, resource provisioning, and robust and
  fault-tolerance scheduling. \end{quotation}
\bibitem[\protect\citeauthoryear{Wu, Tang, and Li}{Wu et~al\mbox{.}}{2012}]%
        {wu2012priority}
\bibfield{author}{\bibinfo{person}{Hu Wu}, \bibinfo{person}{Zhuo Tang}, {and}
  \bibinfo{person}{Renfa Li}.} \bibinfo{year}{2012}\natexlab{}.
\newblock \showarticletitle{A priority constrained scheduling strategy of
  multiple workflows for cloud computing}. In
  \bibinfo{booktitle}{\emph{Advanced Communication Technology (ICACT), 2012
  14th International Conference on}}. IEEE, \bibinfo{pages}{1086--1089}.
\newblock

 \begin{quotation}\noindent Introduces PISA, an algorithm that focusses both on
  fairness and consumer's priority. It is capable of scheduling multiple
  workflows where tasks have different priority weights. \end{quotation}
\bibitem[\protect\citeauthoryear{Wu, Ni, Gu, and Liu}{Wu et~al\mbox{.}}{2010}]%
        {wu2010revised}
\bibfield{author}{\bibinfo{person}{Zhangjun Wu}, \bibinfo{person}{Zhiwei Ni},
  \bibinfo{person}{Lichuan Gu}, {and} \bibinfo{person}{Xiao Liu}.}
  \bibinfo{year}{2010}\natexlab{}.
\newblock \showarticletitle{A revised discrete particle swarm optimization for
  cloud workflow scheduling}. In \bibinfo{booktitle}{\emph{2010 International
  Conference on Computational Intelligence and Security}}. IEEE,
  \bibinfo{pages}{184--188}.
\newblock

 \begin{quotation}\noindent Introduces a Revised Discrete Particle Swarm
  Optimization for cloud workflow execution that takes both data transfer costs
  and execution costs into account. \end{quotation}
\bibitem[\protect\citeauthoryear{W{\"u}st, B{\"u}tikofer, Spielberger, and
  Sigrist}{W{\"u}st et~al\mbox{.}}{2015}]%
        {wust2015generation}
\bibfield{author}{\bibinfo{person}{Raimond~Matthias W{\"u}st},
  \bibinfo{person}{Stephan B{\"u}tikofer}, \bibinfo{person}{J{\"u}rgen
  Spielberger}, {and} \bibinfo{person}{J{\"o}rg Sigrist}.}
  \bibinfo{year}{2015}\natexlab{}.
\newblock \showarticletitle{Generation of interactive questionnaires using
  YAWL-based workflow models}.
\newblock \bibinfo{journal}{\emph{Management Studies}} \bibinfo{volume}{3},
  \bibinfo{number}{11/12} (\bibinfo{year}{2015}), \bibinfo{pages}{273--280}.
\newblock

 \begin{quotation}\noindent Introduce an algorithm to dynamically generate
  interactive questionnaires for the configuration of workflow systems and the
  consistency of the workflow model used. \end{quotation}
\bibitem[\protect\citeauthoryear{Xiang, Zhang, and Zhang}{Xiang
  et~al\mbox{.}}{2017}]%
        {xiang2017greedy}
\bibfield{author}{\bibinfo{person}{Bin Xiang}, \bibinfo{person}{Bibo Zhang},
  {and} \bibinfo{person}{Lin Zhang}.} \bibinfo{year}{2017}\natexlab{}.
\newblock \showarticletitle{Greedy-ant: ant colony system-inspired workflow
  scheduling for heterogeneous computing}.
\newblock \bibinfo{journal}{\emph{IEEE Access}}  \bibinfo{volume}{5}
  (\bibinfo{year}{2017}), \bibinfo{pages}{11404--11412}.
\newblock

 \begin{quotation}\noindent Introduces an algorithm based on the ant colony
  heuristic that attempts to minimize the total workflow execution time in
  heterogeneous environments. \end{quotation}
\bibitem[\protect\citeauthoryear{Xie, Yin, Ruan, Ding, Tian, Majors,
  Manzanares, and Qin}{Xie et~al\mbox{.}}{2010}]%
        {xie2010improving}
\bibfield{author}{\bibinfo{person}{Jiong Xie}, \bibinfo{person}{Shu Yin},
  \bibinfo{person}{Xiaojun Ruan}, \bibinfo{person}{Zhiyang Ding},
  \bibinfo{person}{Yun Tian}, \bibinfo{person}{James Majors},
  \bibinfo{person}{Adam Manzanares}, {and} \bibinfo{person}{Xiao Qin}.}
  \bibinfo{year}{2010}\natexlab{}.
\newblock \showarticletitle{Improving mapreduce performance through data
  placement in heterogeneous hadoop clusters}. In
  \bibinfo{booktitle}{\emph{2010 IEEE International Symposium on Parallel \&
  Distributed Processing, Workshops and Phd Forum (IPDPSW)}}. IEEE,
  \bibinfo{pages}{1--9}.
\newblock

 \begin{quotation}\noindent Introduce a data placement scheme to improve the
  performance of IO-heavy MapReduce applications through data-locality on
  heterogeneous nodes. Through they scheme, data is rebalanced across nodes
  \end{quotation}
\bibitem[\protect\citeauthoryear{Xie, Zhu, Wang, Cheng, Xu, Sani, Yuan, and
  Yang}{Xie et~al\mbox{.}}{2019}]%
        {DBLP:journals/fgcs/XieZWCXSYY19}
\bibfield{author}{\bibinfo{person}{Ying Xie}, \bibinfo{person}{Yuanwei Zhu},
  \bibinfo{person}{Yeguo Wang}, \bibinfo{person}{Yongliang Cheng},
  \bibinfo{person}{Rongbin Xu}, \bibinfo{person}{Abubakar~Sadiq Sani},
  \bibinfo{person}{Dong Yuan}, {and} \bibinfo{person}{Yun Yang}.}
  \bibinfo{year}{2019}\natexlab{}.
\newblock \showarticletitle{A novel directional and non-local-convergent
  particle swarm optimization based workflow scheduling in cloud-edge
  environment}.
\newblock \bibinfo{journal}{\emph{Future Gener. Comput. Syst.}}
  \bibinfo{volume}{97} (\bibinfo{year}{2019}), \bibinfo{pages}{361--378}.
\newblock
\urldef\tempurl%
\url{https://doi.org/10.1016/j.future.2019.03.005}
\showDOI{\tempurl}

 \begin{quotation}\noindent Introduces DNCPSO, an extension of the default
  particle Swarm Optimization policy focused on reducing makespan and cost.
  \end{quotation}
\bibitem[\protect\citeauthoryear{Yaghoobi, Fanian, Khajemohammadi, and
  Gulliver}{Yaghoobi et~al\mbox{.}}{2013}]%
        {yaghoobi2013non}
\bibfield{author}{\bibinfo{person}{Mansoure Yaghoobi}, \bibinfo{person}{Ali
  Fanian}, \bibinfo{person}{Hassan Khajemohammadi}, {and}
  \bibinfo{person}{T~Aaron Gulliver}.} \bibinfo{year}{2013}\natexlab{}.
\newblock \showarticletitle{A non-cooperative game theory approach to optimize
  workflow scheduling in grid computing}. In \bibinfo{booktitle}{\emph{2013
  IEEE Pacific Rim Conference on Communications, Computers and Signal
  Processing (PACRIM)}}. IEEE, \bibinfo{pages}{108--113}.
\newblock

 \begin{quotation}\noindent Introduces a non-cooperative game theory scheduling
  approach to minimize cost and makespan. \end{quotation}
\bibitem[\protect\citeauthoryear{Yan, Castro, Cheng, and Ishakian}{Yan
  et~al\mbox{.}}{2016}]%
        {yan2016building}
\bibfield{author}{\bibinfo{person}{Mengting Yan}, \bibinfo{person}{Paul
  Castro}, \bibinfo{person}{Perry Cheng}, {and} \bibinfo{person}{Vatche
  Ishakian}.} \bibinfo{year}{2016}\natexlab{}.
\newblock \showarticletitle{Building a chatbot with serverless computing}. In
  \bibinfo{booktitle}{\emph{Proceedings of the 1st International Workshop on
  Mashups of Things and APIs}}. ACM, \bibinfo{pages}{5}.
\newblock

 \begin{quotation}\noindent Introduces a chatbot built using serverless
  computing. \end{quotation}
\bibitem[\protect\citeauthoryear{Yassa, Chelouah, Kadima, and Granado}{Yassa
  et~al\mbox{.}}{2013}]%
        {yassa2013multi}
\bibfield{author}{\bibinfo{person}{Sonia Yassa}, \bibinfo{person}{Rachid
  Chelouah}, \bibinfo{person}{Hubert Kadima}, {and} \bibinfo{person}{Bertrand
  Granado}.} \bibinfo{year}{2013}\natexlab{}.
\newblock \showarticletitle{Multi-objective approach for energy-aware workflow
  scheduling in cloud computing environments}.
\newblock \bibinfo{journal}{\emph{The Scientific World Journal}}
  \bibinfo{volume}{2013} (\bibinfo{year}{2013}).
\newblock

 \begin{quotation}\noindent Introduces a dynamic voltage and frequency scaling
  adaptation to the multi-objective discrete particle swarm optimization
  heuristic to lower makespan, energy consumption, and costs. \end{quotation}
\bibitem[\protect\citeauthoryear{Yi, Li, and Li}{Yi et~al\mbox{.}}{2015}]%
        {yi2015survey}
\bibfield{author}{\bibinfo{person}{Shanhe Yi}, \bibinfo{person}{Cheng Li},
  {and} \bibinfo{person}{Qun Li}.} \bibinfo{year}{2015}\natexlab{}.
\newblock \showarticletitle{A survey of fog computing: concepts, applications
  and issues}. In \bibinfo{booktitle}{\emph{Proceedings of the 2015 workshop on
  mobile big data}}. ACM, \bibinfo{pages}{37--42}.
\newblock

 \begin{quotation}\noindent This survey discussed concepts, applications, and
  challenges in the Fog computing domain. \end{quotation}
\bibitem[\protect\citeauthoryear{Young, McGough, Newhouse, and
  Darlington}{Young et~al\mbox{.}}{2003}]%
        {Scholar:conf/ahm/YoungL2003}
\bibfield{author}{\bibinfo{person}{Laurie Young}, \bibinfo{person}{Stephen
  McGough}, \bibinfo{person}{Steven Newhouse}, {and} \bibinfo{person}{John
  Darlington}.} \bibinfo{year}{2003}\natexlab{}.
\newblock \showarticletitle{Scheduling architecture and algorithms within the
  ICENI grid middleware}. In \bibinfo{booktitle}{\emph{UK e-science all hands
  meeting}}. Citeseer, \bibinfo{pages}{5--12}.
\newblock

 \begin{quotation}\noindent Compares four heuristic scheduling algorithms on
  time and cost optimisation: game theory, simulated annealing, random and best
  of \emph{n} random. Out of these four, the simulated annealing scheduler
  performs best. Likewise, this paper demonstrates a uncooperative game theory
  approach is worse than random. \end{quotation}
\bibitem[\protect\citeauthoryear{Yu and Buyya}{Yu and Buyya}{2005}]%
        {DBLP:journals/sigmod/YuB05}
\bibfield{author}{\bibinfo{person}{Jia Yu} {and} \bibinfo{person}{Rajkumar
  Buyya}.} \bibinfo{year}{2005}\natexlab{}.
\newblock \showarticletitle{A taxonomy of scientific workflow systems for grid
  computing}.
\newblock \bibinfo{journal}{\emph{{SIGMOD} Record}} \bibinfo{volume}{34},
  \bibinfo{number}{3} (\bibinfo{year}{2005}), \bibinfo{pages}{44--49}.
\newblock
\urldef\tempurl%
\url{https://doi.org/10.1145/1084805.1084814}
\showDOI{\tempurl}

 \begin{quotation}\noindent Presents a taxonomy of approaches for building and
  executing workflows in Grid environments. Additionaly, an overview of Grid
  workflow systems is provided. \end{quotation}
\bibitem[\protect\citeauthoryear{Yu and Buyya}{Yu and Buyya}{2006}]%
        {yu2006scheduling}
\bibfield{author}{\bibinfo{person}{Jia Yu} {and} \bibinfo{person}{Rajkumar
  Buyya}.} \bibinfo{year}{2006}\natexlab{}.
\newblock \showarticletitle{Scheduling scientific workflow applications with
  deadline and budget constraints using genetic algorithms}.
\newblock \bibinfo{journal}{\emph{Scientific Programming}}
  \bibinfo{volume}{14}, \bibinfo{number}{3-4} (\bibinfo{year}{2006}),
  \bibinfo{pages}{217--230}.
\newblock

 \begin{quotation}\noindent Introduces a genetic algorithm for workflow
  scheduling under deadline and budget constraints. \end{quotation}
\bibitem[\protect\citeauthoryear{Yu, Buyya, and Ramamohanarao}{Yu
  et~al\mbox{.}}{2008}]%
        {yu2008workflow}
\bibfield{author}{\bibinfo{person}{Jia Yu}, \bibinfo{person}{Rajkumar Buyya},
  {and} \bibinfo{person}{Kotagiri Ramamohanarao}.}
  \bibinfo{year}{2008}\natexlab{}.
\newblock \showarticletitle{Workflow scheduling algorithms for grid computing}.
\newblock In \bibinfo{booktitle}{\emph{Metaheuristics for scheduling in
  distributed computing environments}}. \bibinfo{publisher}{Springer},
  \bibinfo{pages}{173--214}.
\newblock

 \begin{quotation}\noindent Surveys and categorizes exisiting workflow
  scheduling algorithms for Grid environments. The main two categories selected
  for categorization are best-effort and QoS constrained. \end{quotation}
\bibitem[\protect\citeauthoryear{Yu, Ramamohanarao, and Buyya}{Yu
  et~al\mbox{.}}{2009}]%
        {yu2009deadline}
\bibfield{author}{\bibinfo{person}{Jia Yu}, \bibinfo{person}{Kotagiri
  Ramamohanarao}, {and} \bibinfo{person}{Rajkumar Buyya}.}
  \bibinfo{year}{2009}\natexlab{}.
\newblock \showarticletitle{Deadline/budget-based scheduling of workflows on
  utility grids}.
\newblock \bibinfo{journal}{\emph{Market-Oriented Grid and Utility Computing}}
  \bibinfo{volume}{200}, \bibinfo{number}{9} (\bibinfo{year}{2009}),
  \bibinfo{pages}{427--450}.
\newblock

 \begin{quotation}\noindent Introduces two scheduling heuristics for workflow
  scheduling under deadline and budget constraints. \end{quotation}
\bibitem[\protect\citeauthoryear{Yu and Shi}{Yu and Shi}{2008}]%
        {DBLP:conf/icppw/YuS08}
\bibfield{author}{\bibinfo{person}{Zhifeng Yu} {and} \bibinfo{person}{Weisong
  Shi}.} \bibinfo{year}{2008}\natexlab{}.
\newblock \showarticletitle{A Planner-Guided Scheduling Strategy for Multiple
  Workflow Applications}. In \bibinfo{booktitle}{\emph{37th International
  Conference on Parallel Processing - Workshops, 8-12 September 2008, Portland,
  Oregon, {USA}}}. \bibinfo{pages}{1--8}.
\newblock
\urldef\tempurl%
\url{https://doi.org/10.1109/ICPP-W.2008.10}
\showDOI{\tempurl}

 \begin{quotation}\noindent Presents a planner-guided scheduling strategy for
  the dynamic scheduling of multiple workflows in a cluster environment,
  without the requirement of merging a priori. This is done by leveraging job
  dependence information and execution time estimation. Jobs are scheduled
  independently, preventing the need to merge workflows. \end{quotation}
\bibitem[\protect\citeauthoryear{Zhan and Huo}{Zhan and Huo}{2012}]%
        {zhan2012improved}
\bibfield{author}{\bibinfo{person}{Shaobin Zhan} {and}
  \bibinfo{person}{Hongying Huo}.} \bibinfo{year}{2012}\natexlab{}.
\newblock \showarticletitle{Improved PSO-based task scheduling algorithm in
  cloud computing}.
\newblock \bibinfo{journal}{\emph{Journal of Information \& Computational
  Science}} \bibinfo{volume}{9}, \bibinfo{number}{13} (\bibinfo{year}{2012}),
  \bibinfo{pages}{3821--3829}.
\newblock

 \begin{quotation}\noindent Proposes an improved particle swarm optimization
  algorithm for resource scheduling in cloud computing environments.
  \end{quotation}
\bibitem[\protect\citeauthoryear{Zhang, Cao, Hwang, and Wu}{Zhang
  et~al\mbox{.}}{2011}]%
        {DBLP:conf/cloudcom/ZhangCHW11}
\bibfield{author}{\bibinfo{person}{Fan Zhang}, \bibinfo{person}{Junwei Cao},
  \bibinfo{person}{Kai Hwang}, {and} \bibinfo{person}{Cheng Wu}.}
  \bibinfo{year}{2011}\natexlab{}.
\newblock \showarticletitle{Ordinal Optimized Scheduling of Scientific
  Workflows in Elastic Compute Clouds}. In \bibinfo{booktitle}{\emph{{IEEE} 3rd
  International Conference on Cloud Computing Technology and Science, CloudCom
  2011, Athens, Greece, November 29 - December 1, 2011}}.
  \bibinfo{pages}{9--17}.
\newblock
\urldef\tempurl%
\url{https://doi.org/10.1109/CloudCom.2011.12}
\showDOI{\tempurl}

 \begin{quotation}\noindent Propose to extend ordinal optimization with
  iteration to significantly reduce the search space and lower overhead for
  scheduling scientific workflow in elastic cloud environments. The authors
  compare with a Monte Carlo simulation and blind pick approach.
  \end{quotation}
\bibitem[\protect\citeauthoryear{Zhang, Yoshida, and Tang}{Zhang
  et~al\mbox{.}}{2008}]%
        {zhang2008tfidf}
\bibfield{author}{\bibinfo{person}{Wen Zhang}, \bibinfo{person}{Taketoshi
  Yoshida}, {and} \bibinfo{person}{Xijin Tang}.}
  \bibinfo{year}{2008}\natexlab{}.
\newblock \showarticletitle{TFIDF, LSI and multi-word in information retrieval
  and text categorization}. In \bibinfo{booktitle}{\emph{2008 IEEE
  International Conference on Systems, Man and Cybernetics}}. IEEE,
  \bibinfo{pages}{108--113}.
\newblock

 \begin{quotation}\noindent Compares the performance of TF-IDF, LSI, and
  multi-word approaches for obtaining keywords for text categorization.
  \end{quotation}
\bibitem[\protect\citeauthoryear{Zhao and Sakellariou}{Zhao and
  Sakellariou}{2003}]%
        {zhao2003experimental}
\bibfield{author}{\bibinfo{person}{Henan Zhao} {and} \bibinfo{person}{Rizos
  Sakellariou}.} \bibinfo{year}{2003}\natexlab{}.
\newblock \showarticletitle{An experimental investigation into the rank
  function of the heterogeneous earliest finish time scheduling algorithm}. In
  \bibinfo{booktitle}{\emph{European Conference on Parallel Processing}}.
  Springer, \bibinfo{pages}{189--194}.
\newblock

 \begin{quotation}\noindent Investigates which weights should be put on
  vertices and edges of a directed acyclic graph when using the HEFT scheduling
  algorithm. \end{quotation}
\bibitem[\protect\citeauthoryear{Zhao and Sakellariou}{Zhao and
  Sakellariou}{2006}]%
        {DBLP:conf/ipps/ZhaoS06}
\bibfield{author}{\bibinfo{person}{Henan Zhao} {and} \bibinfo{person}{Rizos
  Sakellariou}.} \bibinfo{year}{2006}\natexlab{}.
\newblock \showarticletitle{Scheduling multiple DAGs onto heterogeneous
  systems}. In \bibinfo{booktitle}{\emph{20th International Parallel and
  Distributed Processing Symposium {(IPDPS} 2006), Proceedings, 25-29 April
  2006, Rhodes Island, Greece}}.
\newblock
\urldef\tempurl%
\url{https://doi.org/10.1109/IPDPS.2006.1639387}
\showDOI{\tempurl}

 \begin{quotation}\noindent Presents six policies for scheduling multiple DAGs
  on heterogeneous machines. In particular two scheduling policies focus on
  fairness (in terms of slowdown) while minimizing makespan. \end{quotation}
\bibitem[\protect\citeauthoryear{Zhao}{Zhao}{2004}]%
        {Zhao2004ResultVA}
\bibfield{author}{\bibinfo{person}{Shanyu Zhao}.}
  \bibinfo{year}{2004}\natexlab{}.
\newblock \showarticletitle{Result Verification and Trust-based Scheduling in
  Open Peer-to-Peer Cycle Sharing Systems}.
\newblock

 \begin{quotation}\noindent Introdces a Quiz system to verify if a resource
  does not return malicious results. By using a trust reputation system,
  trusted scheduling can be obtained to a certain level of accuracy, without
  requiring much overhead for verification. \end{quotation}
\bibitem[\protect\citeauthoryear{Zhou, Xiao, He, Ibrahim, and Cheng}{Zhou
  et~al\mbox{.}}{2019a}]%
        {DBLP:conf/icpp/ZhouXHIC19}
\bibfield{author}{\bibinfo{person}{Amelie~Chi Zhou}, \bibinfo{person}{Yao
  Xiao}, \bibinfo{person}{Bingsheng He}, \bibinfo{person}{Shadi Ibrahim}, {and}
  \bibinfo{person}{Reynold Cheng}.} \bibinfo{year}{2019}\natexlab{a}.
\newblock \showarticletitle{Incorporating Probabilistic Optimizations for
  Resource Provisioning of Data Processing Workflows}. In
  \bibinfo{booktitle}{\emph{Proceedings of the 48th International Conference on
  Parallel Processing, {ICPP} 2019, Kyoto, Japan, August 05-08, 2019}}.
  \bibinfo{pages}{6:1--6:10}.
\newblock
\urldef\tempurl%
\url{https://doi.org/10.1145/3337821.3337847}
\showDOI{\tempurl}

 \begin{quotation}\noindent Introduces a resource provisioning strategy where
  system variances in, e.g., IO and network speed are incorporated as
  time-dependent random variables. \end{quotation}
\bibitem[\protect\citeauthoryear{Zhou, Zhang, Sun, Zhou, Wei, and Hu}{Zhou
  et~al\mbox{.}}{2019b}]%
        {zhou2019minimizing}
\bibfield{author}{\bibinfo{person}{Xiumin Zhou}, \bibinfo{person}{Gongxuan
  Zhang}, \bibinfo{person}{Jin Sun}, \bibinfo{person}{Junlong Zhou},
  \bibinfo{person}{Tongquan Wei}, {and} \bibinfo{person}{Shiyan Hu}.}
  \bibinfo{year}{2019}\natexlab{b}.
\newblock \showarticletitle{Minimizing cost and makespan for workflow
  scheduling in cloud using fuzzy dominance sort based HEFT}.
\newblock \bibinfo{journal}{\emph{Future Generation Computer Systems}}
  \bibinfo{volume}{93} (\bibinfo{year}{2019}), \bibinfo{pages}{278--289}.
\newblock

 \begin{quotation}\noindent Introduces FDHEFT, a fuzzy dominance addition to
  the popular HEFT scheduling algorithm to minimize cost and makespan of
  workflows running in IaaS environments. \end{quotation}
\bibitem[\protect\citeauthoryear{Zhu, Wang, Guo, Zhu, Yang, and Liu}{Zhu
  et~al\mbox{.}}{2016}]%
        {zhu2016fault}
\bibfield{author}{\bibinfo{person}{Xiaomin Zhu}, \bibinfo{person}{Ji Wang},
  \bibinfo{person}{Hui Guo}, \bibinfo{person}{Dakai Zhu},
  \bibinfo{person}{Laurence~T Yang}, {and} \bibinfo{person}{Ling Liu}.}
  \bibinfo{year}{2016}\natexlab{}.
\newblock \showarticletitle{Fault-tolerant scheduling for real-time scientific
  workflows with elastic resource provisioning in virtualized clouds}.
\newblock \bibinfo{journal}{\emph{IEEE Transactions on Parallel and Distributed
  Systems}} \bibinfo{volume}{27}, \bibinfo{number}{12} (\bibinfo{year}{2016}),
  \bibinfo{pages}{3501--3517}.
\newblock

 \begin{quotation}\noindent Introduces FASTER, a dynamic fault-tolerant
  workflow scheduling algorithm for clouds that handles failures in both
  message passing and task allocation. \end{quotation}
\bibitem[\protect\citeauthoryear{Zhu and Jiang}{Zhu and Jiang}{2016}]%
        {DBLP:journals/tpds/ZhuJ16}
\bibfield{author}{\bibinfo{person}{Zhongma Zhu} {and} \bibinfo{person}{Rui
  Jiang}.} \bibinfo{year}{2016}\natexlab{}.
\newblock \showarticletitle{A Secure Anti-Collusion Data Sharing Scheme for
  Dynamic Groups in the Cloud}.
\newblock \bibinfo{journal}{\emph{{IEEE} Trans. Parallel Distrib. Syst.}}
  \bibinfo{volume}{27}, \bibinfo{number}{1} (\bibinfo{year}{2016}),
  \bibinfo{pages}{40--50}.
\newblock
\urldef\tempurl%
\url{https://doi.org/10.1109/TPDS.2015.2388446}
\showDOI{\tempurl}

 \begin{quotation}\noindent Introduce a scheme for secure key sharing within
  cloud settings, where fine-grained access can be managed. Participants whose
  key is revoked cannot access the original data file even if they perform a
  collusion attack. \end{quotation}
\end{thebibliography}

\appendix

\end{document}